\documentclass[useAMS,usenatbib]{mn2e}
\usepackage{graphics,epsfig,amsmath,amssymb,amstext,txfonts,array,color}
\usepackage{graphicx}
\usepackage[T1]{fontenc}

\DeclareMathAlphabet\mbi{OML}{cmm}{b}{it}
\DeclareGraphicsExtensions{.jpg, .png , .pdf}

\title[UHECR acceleration at GRB internal shocks]{UHECR acceleration at GRB internal shocks}

\author[N. Globus, D. Allard, R. Mochkovitch, E. Parizot]
  {N.~Globus,$^1$\thanks{E-mail: noemie@wise.tau.ac.il}
  D.~Allard,$^2$ R.~ Mochkovitch,$^3$ E.~ Parizot$^2$ \\
  $^1$School of Physics \& Astronomy, Tel Aviv University, Tel Aviv 69978, Israel\\
  $^2$Laboratoire Astroparticule et Cosmologie, Universit\'e Paris Diderot/CNRS, 10 rue A. Domon et L. Duquet, 75205 Paris Cedex 13, France\\
  $^3$UPMC-CNRS, UMR7095, Institut d'Astrophysique de Paris, 98 bis boulevard Arago, 75014 Paris, France}
\date{Released \today}

\pagerange{\pageref{firstpage}--\pageref{lastpage}} \pubyear{2002}

\def\LaTeX{L\kern-.36em\raise.3ex\hbox{a}\kern-.15em
    T\kern-.1667em\lower.7ex\hbox{E}\kern-.125emX}

\begin{document}
\label{firstpage}
\maketitle

\begin{abstract}
We study the acceleration of cosmic-ray protons and nuclei at GRB internal shocks. Physical quantities (magnetic fields, baryon and photon densities, shock velocity) and their time evolution, relevant to cosmic-ray acceleration and energy losses, are estimated using the internal shock modeling implemented by Daigne \& Mochkovitch (1998). Within this framework, we consider different hypotheses about the way the energy dissipated at internal shocks is shared between accelerated cosmic-rays, electrons and the magnetic field. We model cosmic-ray acceleration at mildly relativistic shocks, using numerical tools inspired by the work of Niemiec \& Ostrowski (2004), including all the significant energy loss processes that might limit cosmic-ray acceleration at GRB internal shocks.\\
We calculate cosmic-ray and neutrino release from single GRBs, for various prompt emission luminosities, assuming that nuclei heavier than protons are present in the relativistic wind at the beginning of the internal shock phase. We find that protons can only reach maximum energies of the order $10^{19.5}$ eV in the most favorable cases, while intermediate and heavy nuclei are able to reach higher values of the order of $10^{20}$ eV and above. The spectra of particles escaping from the acceleration site are found to be very hard for the different nuclear species. In addition a significant and much softer neutron component is present in the cases of intermediate and high luminosity GRBs due to the photodisintegration of accelerated nuclei during the early stages of the shock propagation. As a result, the combined spectrum of protons and neutrons from single GRBs is found to be much softer than those of the other nuclear species.\\
We calculate the diffuse UHECR flux expected on Earth by convoluting the cosmic-ray output from single GRBs of various luminosities by the GRB luminosity function derived by Wanderman \& Piran (2010). We show that only the models assuming that ({\it i}) the prompt emission represent only a very small fraction of the energy dissipated at internal shocks (especially for low and intermediate luminosity bursts), and that ({\it ii}) most of this dissipated energy is communicated to accelerated cosmic-rays, are able to reproduce the magnitude of the UHECR flux observed on Earth. For these models, the observed shape of the UHECR spectrum can be well reproduced above the ankle and the evolution of the composition is compatible with the trend suggested by Auger data. We discuss the implications of the softer proton component (consequence of the neutron emission in the sources) for the phenomenology of the transition from Galactic to extragalactic cosmic-ray, in the light of the recent composition analyses from the 
KASCADE-Grande experiment.\\
Finally, we find that the associated secondary particle diffuse fluxes do not upset any current observational limit or measurement. Diffuse neutrino flux from GRB sources of the order of those we calculated should however be detected with the lifetime of neutrino observatories such as IceCube or KM3Net.
  
\end{abstract}

\section{Introduction}

After decades of observational and theoretical efforts, the question of the origin of ultrahigh-energy cosmic-rays (UHECRs) still remains unanswered. In particular, the nature of their sources and the acceleration mechanism responsible for their colossal energies are unknown. 

In the recent years, the Pierre Auger Observatory (Auger) as well as the High Resolution Fly's eye (HiRes) and the Telescope Array (TA) experiments have provided measurements with unprecedented statistics and resolution in the highest energy range of the cosmic-ray spectrum, above $10^{18}$ eV. Measurements of the all sky UHECR spectrum have shown the presence of a feature above $\sim 10^{19.5} $eV compatible with the so-called Greisen, Zatsepin and Kuz'min (GZK) cut-off, signature of the interaction of protons or heavier nuclei with extragalactic photon backgrounds. Moreover, a first hint of an anisotropy signal at energies larger that $\sim 50$ EeV has been reported by Auger. Although currently weaker than what has been anticipated after the first communication, this signal might give a hint of the overall correlation between UHECRs and nearby extragalactic matter. From the point of view of the composition of UHECRs, Auger data suggest a rather radical transition from a relatively light composition 
around the ankle of the cosmic-ray spectrum (about $3-4\times10^{18}$ eV) to a heavier composition above a few tens of EeV\footnote{Let us note that neither the anisotropy signal nor the composition trend suggested by Auger data has been confirmed by HiRes or TA experiments. These two experiments have however accumulated a lower statistics above $10^{19}$ eV. They are furthermore observing a different portion of the sky.}. Although the spectrum and anisotropy measurements favor extragalactic scenari for the origin of UHECRs, these observations are not constraining enough to reveal the nature of their sources. On the other hand, the composition trend suggested by Auger can, in principle, be understood if complex nuclei are accelerated at energies larger than protons. In particular, in a scenario where the maximum energy reached by protons would be lower than the the GZK energy scale, nuclei accelerated up to the same maximum rigidity would reach energies $Z$ times higher (where $Z$ is the charge of a given nucleus) which would result in a transition toward a gradually heavier UHECR composition. This composition trend, if further supported by higher statistics measurements, would be quite constraining for the acceleration mechanism and the source environment.  The above-mentioned scenario could only hold if ({\it i}) complex nuclei are abundant in the composition at the source and ({\it ii}) the energy loss rate within the acceleration site is sufficiently low for nuclei to be accelerated at energies larger than protons and escape from the source.  

Gamma-ray bursts (GRBs) are among the best candidate sources for UHECRs. Their very luminous prompt and afterglow emissions are thought to be related to the ejection of an ultrarelativistic outflow consecutive to the core collapse of a very massive star (long burst) or the merging of two compact objects (short burst). The fluctuations of the central engine lead to the formation of internal shocks in which much of the jet power is dissipated. Above the photosphere, those shocks become collisionless and provide sites for particle acceleration. Another dissipation and possible acceleration sites are the so-called external shock occuring when the jet encounters the dense interstellar medium (Vietri 1995; see however Gallant and Achterberg 1999). Among the popular scenarios invoked to explain the GRBs prompt emission spectral and temporal properties, the internal shock model (Rees \& M\'esz\'aros 1993) has been the most extensively discussed. The internal shock model relies on the presence of short time scale variability of the Lorentz factor within the relativistic plasma outflow emitted by the central engine. Mildly relativistic shocks are expected to form once the fast layers of plasma catch up with the slower parts, and to accelerate electrons whose subsequent cooling triggers the prompt emission. Shortly after this scenario was proposed to account for GRBs prompt emission, internal shocks also appeared as credible potential candidates for the acceleration of UHECRs. This possibility was first proposed by Waxman (1995) whose line of reasoning was based on the observation that ({\it i}) internal shocks physical parameters were likely to fulfill the "Hillas criterion" (Hillas 1984) for cosmic-rays acceleration above $10^{20}$ eV and ({\it ii}) that the GRB emissivity in gamma-rays during the prompt phase was of the same order as the emissivity required for UHECR sources above $10^{18}$ eV. Although the latter argument has become quite controversial in the past few years (see for instance Berezinsky et al., 2006), the study of protons acceleration at GRB internal shocks has been the purpose of many studies since Waxman's pioneering work (Milgrom \& Usov 1995; B\"{o}ttcher \& Dermer 1998; Waxman \& Bahcall 2000;  Dermer \& Humi 2001;  Gialis \& Pelletier 2003a, 2003b, 2005). Among them, some were dedicated to the possible contribution of galactic GRBs to cosmic-rays at knee and above energies (Atoyan \& Dermer 2006) or more recently to UHECRs (Calvez et al.  2010). The specific case of ultra-high energy (above $10^{19}$ eV) nuclei acceleration was considered in fewer studies (Wang et al.  2008; Murase et al.  2008; Metzger et al.  2011) the latter also considering the possibility of heavy nuclei nucleosynthesis within GRBs relativistic outflows. The important question of nuclei survival during the early phase of GRBs (and its dependence on the physical processes at play during this phase) was recently discussed by Horiuchi et al. (2012). The production of very high energy secondary particles, natural aftermath and possible signature of cosmic-ray acceleration to the highest energies has also been extensively studied. Predictions for very high and ultrahigh energy neutrino fluxes from GRBs can be found for instance in Waxman \& Bahcall (1997),  Guetta et al. (2004), Murase \& Nagataki (2006), Murase et al. (2008), H\"{u}mmer et al. (2009), Ahlers et al. (2011), H\"{u}mmer et al. (2012), He et al. (2012), Baerwald et al. (2014) while the possible contribution of accelerated cosmic-ray induced gamma-rays to the prompt emission was estimated in Asano \& Inoue (2007), Asano et al. (2009), Razzaque et al. (2010), Murase et al. (2012).  
 
In this paper, we reconsider the question of UHECR acceleration at GRBs internal shocks. We base our study on a Monte-Carlo calculation of UHECR protons and nuclei acceleration at mildly relativistic shocks including all the relevant energy loss processes and the associated secondary neutrinos and photons emission. In the next section, we present our modeling of GRBs internal shocks based on previous works by Daigne \& Mochkovitch (1998) and calculate the prompt emission SEDs for different hypotheses on the relativistic outflow physical parameters. These SED will later serve as a photon background during UHECR acceleration. In Sect.~3, we introduce a Monte-Carlo calculation of cosmic-ray acceleration at relativistic shocks following the numerical method introduced by Niemiec and Ostrowski (2004, 2006a, 2006b). We analyze the dependence of the expected accelerated cosmic-ray spectrum and the acceleration time on physical parameters, such as the turbulence spectrum or the shock Lorentz factor, and the cosmic-ray escape from the wind. We then discuss the relevant energy loss processes at play during cosmic-ray acceleration at internal shocks and calculate the corresponding energy loss times, in Sect.~4. The comparison between the expected acceleration time and loss time allows us to give a first estimate of the maximum energy reachable by cosmic-rays protons or nuclei. In Sect.~5, we describe our Monte-Carlo calculation in which the energy loss mechanisms and internal shocks physical conditions derived from Sect.~2 are included to the numerical modeling of particle acceleration. We calculate cosmic-ray spectra at the source, for different GRB luminosities, as well as secondary particles production. We then use the recent GRB luminosity function and cosmological evolution derived by Wanderman \& Piran (2010) to estimate the diffuse UHECR and neutrino fluxes expected for our models in Sect.~6. We finally discuss the physical parameter space (especially the different cases of energy redistribution between electrons, cosmic-rays and the magnetic field) that could allow GRBs internal shocks to be the main source of UHECRs, summarize our results and conclude in Sect.~7.

\section{Modeling of GRB internal shocks}

\subsection{The prompt emission of GRBs}
The origin of the prompt emission of GRBs is still debated. Their very large luminosity and the short time scale variability of the observed light curves constrain the emission to be produced within a relativistic jet  of typical 
Lorentz factor 100 - 1000 to avoid photon annihilation and the production of electron-positron pairs that would generate a large Thomson optical depth (Piran 1999; Lithwick \& Sari 2001;   Hasco\"{e}t et al. 2012). The acceleration of the relativistic outflow from the central engine (an accreting stellar mass black hole or a magnetar) can have either a thermal or magnetic origin. At some radius, a photosphere forms in the flow, its location, luminosity and temperature depending on the respective thermal and magnetic power injected at the base of the jet and on the amount of entrained baryonic matter (Daigne \& Mochkovitch 2002;  Hasco\"{e}t, Daigne \& Mochkovitch 2013). Pure photospheric emission cannot explain GRB spectra, which are non-thermal, having the form of broken power-laws with respective spectral indices $\alpha$ and $\beta$ at low and high energy. The average values of $\alpha$ and $\beta$ are $-1$ and $-2.25$ with a typical break energy of a few hundreds keV (Kaneko et al., 2006).

There are three leading models for creating the prompt emission of GRBs:
({\it i}) sub-photospheric dissipation where various processes can modify the emerging initially thermal spectrum to produce the 
observed broken power-law (Rees \& M\'esz\'aros 2005; Pe'er, M\'esz\'aros \& Rees 2005; Ryde et al. 2011; Beloborodov 2010;  Levinson 2012; Beloborodov 2013; Keren \& Levinson 2014); ({\it ii}) magnetic reconnection in a magnetized ejecta (Giannios 2012; Mc Kinney \& Uzdensky 2012; Yuan \& Zhang 2012; Zhang \& Zhang 2014) and ({\it iii}) internal shocks, which occur if the distribution of Lorentz factor is variable in the outflow so that rapid parts can catch up and collide with slower ones (Rees \& M\'esz\'aros 1994; Kobayshi, Piran \& Sari 1997; Daigne \& Mochkovitch 1998). 
Detailed models of internal shocks have been constructed (Bosnjak \& Daigne 2014) and many of their predictions compare favourably to observations (e.g. hardness duration, hardness intensity and hardness fluence correlations, width of pulses as a function of energy, etc). 
However there are also problems such as a low efficiency (a few percents typically since only the fluctuations and not the bulk of 
the kinetic power is dissipated) and a spectral shape that does not fit the data at low energy (Preece et al 1998; Ghisellini et al. 2000). The predicted spectral slope $\alpha=-1.5$, 
corresponding to synchrotron radiation in the so-called fast cooling regime, is too soft. Some possibilities to produce a harder spectrum (Derishev 2007; Daigne et al. 2011, Uhm \& Zhang 2014) might account for $\alpha$ values up to $-1$ but reaching even larger values from $-1$ to 0 appears quite challenging. 
Also internal shock models require that the flow should be initially magnetically dominated because otherwise the thermal emission of the photosphere might be brighter than that of internal shocks (Daigne \& Mochkovitch 2002) and simultaneously that the ratio $\sigma$ of the magnetic to kinetic energy in the flow should have decreased below 0.1 at the location of internal shocks to allow them to preserve the required efficiency (Mimica \& Aloy 2010; Narayan et al. 2011).
   
Photospheric and reconnection models have been proposed as possible solutions to these problems. They are potentially more efficient than collisionless internal shocks. In photospheric models, the basic radiation mechanism is not synchrotron (but synchrotron emission is nevertheless involved in some cases, e.g. Vurm et al. 2011). Internal shocks, if they occur below the photosphere, are mediated by radiation. They  can reproduce the observed GRB spectra with $\alpha$ values from $-1$ to 0 (Levinson 2012; Keren \& Levinson 2014). However, they are unlikely to accelerate cosmic-rays by Fermi processes because of their large transition layer. Reconnection models appear natural if the flow remains magnetized at large distances from the source. For the moment, these alternatives (especially reconnection) make much fewer predictions than internal shocks that can be compared to observations. As a result, they cannot be tested as thoroughly and it is therefore difficult to draw too definite conclusions. In spite of this somewhat uncertain situation we believe it remains important to explore the ability of the (collisionless) internal shock scenario to account for the production of UHECRs.   

\subsection{A simple model for internal shocks} 

\begin{figure}
{\rotatebox{0}{\includegraphics[scale=0.4]{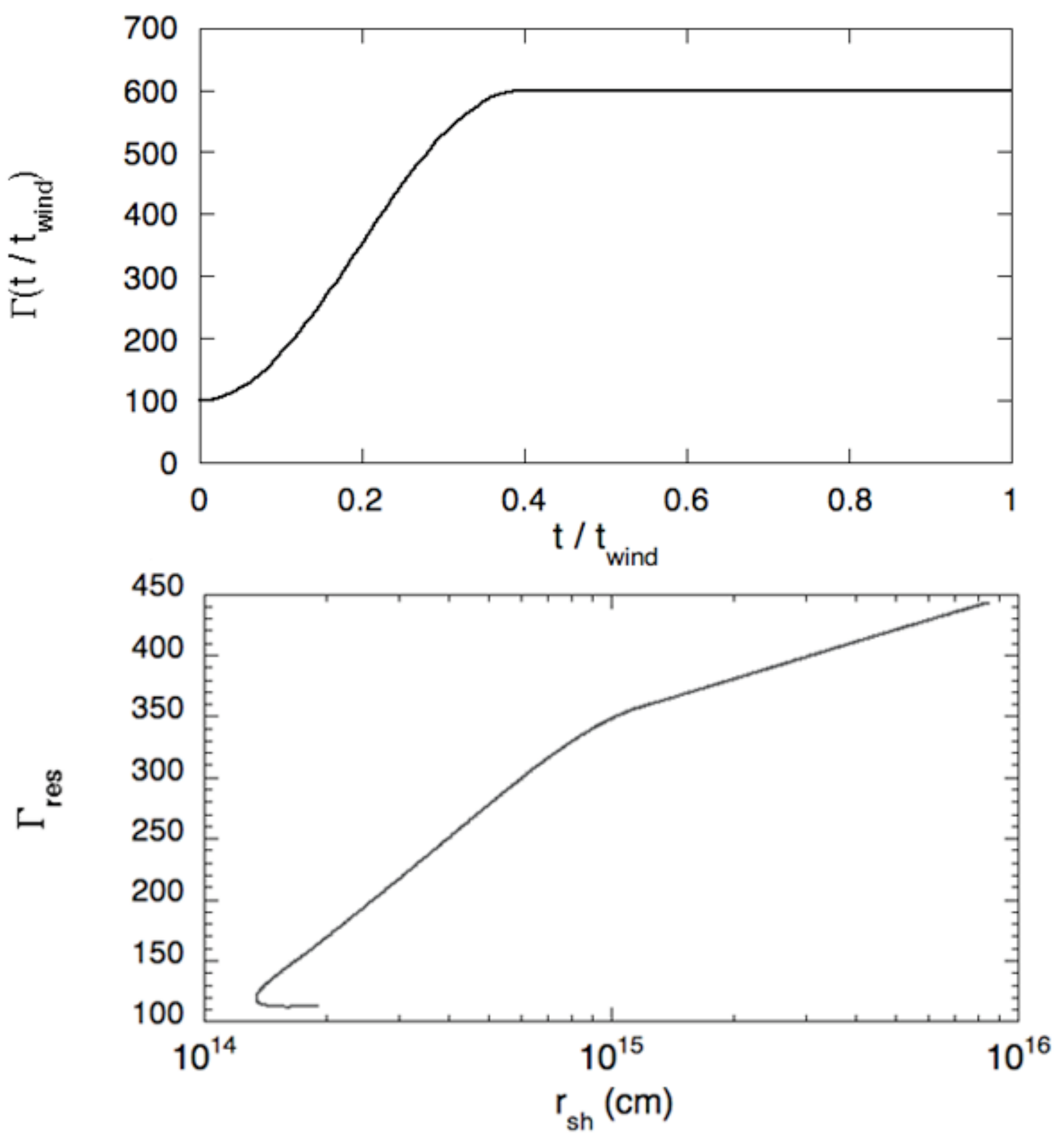}}}
\caption{Upper panel: initial Lorentz factor distribution of the relativistic wind as a function of time. Lower panel:  $r_{\rm sh}$ evolution of the Lorentz factor of the shocked fluid in the central source frame $\Gamma_{\rm res}$ as obtained from  Eq.~\ref{eq:gamma_res}.}
\label{fig:profil}
\end{figure}

We use a simple model where the relativistic outflow emitted by the central engine is represented by a large number of shells that interact by direct collisions only (Daigne \& Mochkovitch 1998). Pressure waves are neglected but this is a reasonable approximation (which has been confirmed by hydrodynamic calculations, see e.g. Daigne \& Mochkovitch 2000) as kinetic energy largely dominates over internal energy in the flow. We prepare an initial set of 1000 shells separated by intervals $\delta t=2$ ms in the source frame, corresponding to a total injection time $t_{\rm wind}=2$ s. The distribution of the Lorentz factor in the flow is given by     
\begin{equation}
\Gamma(t) = \left\lbrace\begin{array}{cl}
& 350-250\ {\cos}\left[\pi\,\frac{t}{0.4\,t_{\rm wind}}\right]\ \ \ {\rm for}\ \ t\leq0.4\,t_{\rm wind}\\
& 600\ \ \ {\rm for}\ \ t>0.4\,t_{\rm wind}\\
\end{array}\right.
\label{eq:profil}
\end{equation}
where $t$ is the emission time of a shell. With such a distribution, the rapid part of the flow ($\Gamma=600$) will be decelerated by the relatively slower part ($\Gamma\lesssim 100$) placed ahead of it. Two shocks are formed starting from the region where the initial gradient of Lorentz factors is the largest, propagating respectively toward the front and the back of the flow. In our simple model, the propagation of these two shocks is discretized by a large number of elementary collisions between individual shells. With our adopted distribution of Lorentz factors, the  shock propagating toward the front of the flow is short lived (we will not consider it in the following calculations) while the other one becomes rapidly dominant and is responsible for the bulk of the dissipation and therefore for the prompt emission (see Daigne \& Mochkovitch 1998 and Daigne \& Mochkovitch 2000, for more details). 
As a result of energy dissipation throughout the propagation of the shock over large distances, one pulse lasting a few second will be obtained in the lightcurve, which is typical in long GRBs. Variability on a much shorter timescale (down to a few milliseconds) is indeed observed but most of the burst energy is generally carried 
by much longer pulses.    

The mass of each individual shell in the flow is obtained assuming a constant injected power $L_{\rm wind}$ so that
\begin{equation}
m_i={{L_{\rm wind}}\delta t\over \Gamma_i c^2}\,.
\end{equation}
We then follow the motion of the shells until a collision between two shells $i$ and $i+1$ occurs. We record the time 
$t_{\rm sh}$ and location $r_{\rm sh}$ (both measured in the central source frame) of the collision and compute the dissipated energy
\begin{equation}
{E_{\rm diss}}=\left(m_i\Gamma_i+m_{i+1}\Gamma_{i+1}-(m_i+m_{i+1})\,\Gamma_{\rm res}\right)c^2\,.
\end{equation}
We estimate $\Gamma_{\rm res}$ (the Lorentz factor of the shocked fluid resulting from the collision, as seen in the central source frame) by considering that most of the energy available in the collision has been released when the less massive of the two shells has swept up a mass comparable to its own mass in the other layer. One then obtains
\begin{equation} 
\Gamma_{\rm res}\simeq\sqrt{\Gamma_i\Gamma_{i+1}}\,.
\label{eq:gamma_res}
\end{equation}
It should be noticed that the dissipated energy represents only a small fraction of the total kinetic energy of the
merging shells, around 15\% for our assumed distribution of Lorentz factors, as the collisions are only mildly relativistic with a relative velocity and Lorentz factor given by
\begin{equation} 
\beta_{\rm rel}={\kappa^2-1\over \kappa^2+1}\ \ \ ,\ \ \ \Gamma_{\rm rel}={\kappa^2+1\over 2\kappa}\,,
\end{equation} 
where $\kappa$ is the contrast of Lorentz factor between two shells that collide at a radius $r_{\rm sh}$ (leading for example to $\beta_{\rm rel}=0.8$ and
$\Gamma_{\rm rel}=1,67$ for $\kappa=3$). In our simple model, $\rm \beta_{\rm rel}$ and $\rm \Gamma_{\rm rel}$ also represent the velocity and Lorentz factor of the shock in the undisturbed fluid rest frame, at a given time during its propagation, we call them hereafter respectively $\beta_{\rm sh}$ and $\Gamma_{\rm sh}$.\\
The dissipated energy will be received by the observer starting at a time 
\begin{equation} 
t_{\rm obs}=t_{\rm sh}-{r_{\rm sh}\over c}
\end{equation} 
and will extend over a duration 
\begin{equation} 
\Delta t_{\rm obs}={r_{\rm sh}\over 2 c\,\Gamma_{\rm res}^2}
\end{equation} 
due to the curvature of the emitting shell. 
After each collision, we continue to follow the evolution of the system of shells until the next collision occurs. 
The calculation ends when the remaining shells are all ordered with the Lorentz factor increasing outwards or, when only one
shell remains. Adding all the elementary contributions from these collisions finally gives the power received by the observer as a function of time as well as the dynamical evolution of all the important physical quantities during the shock propagation.

\subsection{Microphysics}
Our simple shell model mimics the hydrodynamics of the flow and provides, at any given time during the shock propagation, estimates of the post-shock density $\rm \rho$ and the energy dissipated per unit mass $ e_{\rm diss}$ in the comoving frame of the shocked material. The two quantities are estimated using the following relations:
\begin{equation}
\rho \sim {L_{\rm wind}\over 4\pi\,r_{\rm sh}^2\Gamma_{\rm res}^2\,c^3}\,,
\label{eq:rho}
\end{equation}
\begin{equation}
e_{\rm diss}={{E_{\rm diss}}\over (m_i+m_{i+1})\,\Gamma_{\rm res}}\,.
\label{eq:ediss}
\end{equation} 
Once the redistribution parameters of the dissipated energy $\rm \epsilon_{cr}$, $\rm \epsilon_{e}$ and $ \epsilon_{B}$ (corresponding respectively to the fraction of the dissipated energy injected into accelerated cosmic-rays, accelerated electrons and the magnetic field) have been fixed, it is then possible to obtain an estimate of the magnetic field in the shocked medium,
\begin{equation} 
B=(8\pi\,\epsilon_B\,\rho\,e_{\rm diss})^{1/2}\propto L_{\rm wind}^{1/2}\,.
\label{eq:bfield}
\end{equation} 
Moreover, assuming that the distribution of the shock-accelerated electrons is a power-law of index $p$ for $\rm \Gamma_{\rm e,\,min} < \Gamma_e  < \Gamma_{\rm e,\,max}$ one
obtains 
\begin{equation}
\frac{\Gamma_{\rm e,\,min}^{-p+2}-\Gamma_{\rm e,\,max}^{-p+2}}{\Gamma_{\rm e,\,min}^{-p+1}-\Gamma_{\rm e,\,max}^{-p+1}}=\left({\epsilon_{\rm e}\over \zeta}\right)\left({p-2\over p-1}\right)\left({m_{\rm p}\over m_{\rm e}}\right)\,\left({e_{\rm diss}\over c^2}\right)
\label{eq:gametrue}
\end{equation} 
where $m_{\rm p}$ and $m_{\rm e}$ are the proton and electron masses and $\zeta$ is the fraction of electrons which are accelerated. This equation can be solved numerically to estimate  $\rm \Gamma_{e,\,min}$ (once $\rm \Gamma_{e,\,max}$ is known, see below). Let us note that Eq.~\ref{eq:gametrue} reduces to the often used relation:
\begin{equation}
\Gamma_{\rm e,\,min}\simeq\left({ \epsilon_{\rm e}\over \zeta}\right)\left({p-2\over p-1}\right)\left({m_{\rm p}\over m_{\rm e}}\right)\,\left({e_{\rm diss}\over c^2}\right)
\label{eq:game}
\end{equation} 
in the limit $p>2$ and $\rm \Gamma_{\rm e,\,max}\ggg \rm \Gamma_{\rm e,\,min}$. \\
Once the magnetic field and the minimum Lorentz factor of the accelerated electrons are known the synchrotron energy, which
is also the peak energy of the photon spectrum in the fast cooling regime, is readily obtained
\begin{equation}
{E_{\rm peak}}\simeq1.75\, 10^{-3} \left(\frac{B}{\rm G}\right)\,\left(\frac{\Gamma_{\rm e,min}}{10^3}\right)^2 \,\left(\frac{\Gamma_{\rm res}}{100}\right) \ \ {\rm keV}\,.
\label{eq:epeak}
\end{equation}
In practice, to reach typical $E_{\rm peak}$ values from a few keV to a few MeV, small values of $\rm \zeta$ ($\lesssim 0.01$) are required (see discussion in Daigne \& Mochkovitch 1998, and references therein). It is beyond the scope of this paper to discuss the physical relevance of such small values of $\rm \zeta$, as the physics of mildly relativistic shocks (say, $\Gamma_{\rm sh}=1-3$) has not been extensively explored by simulations (see Sect.~3). 

\begin{figure*}
\begin{tabular}{cc}
{\rotatebox{0}{\includegraphics[scale=0.24]{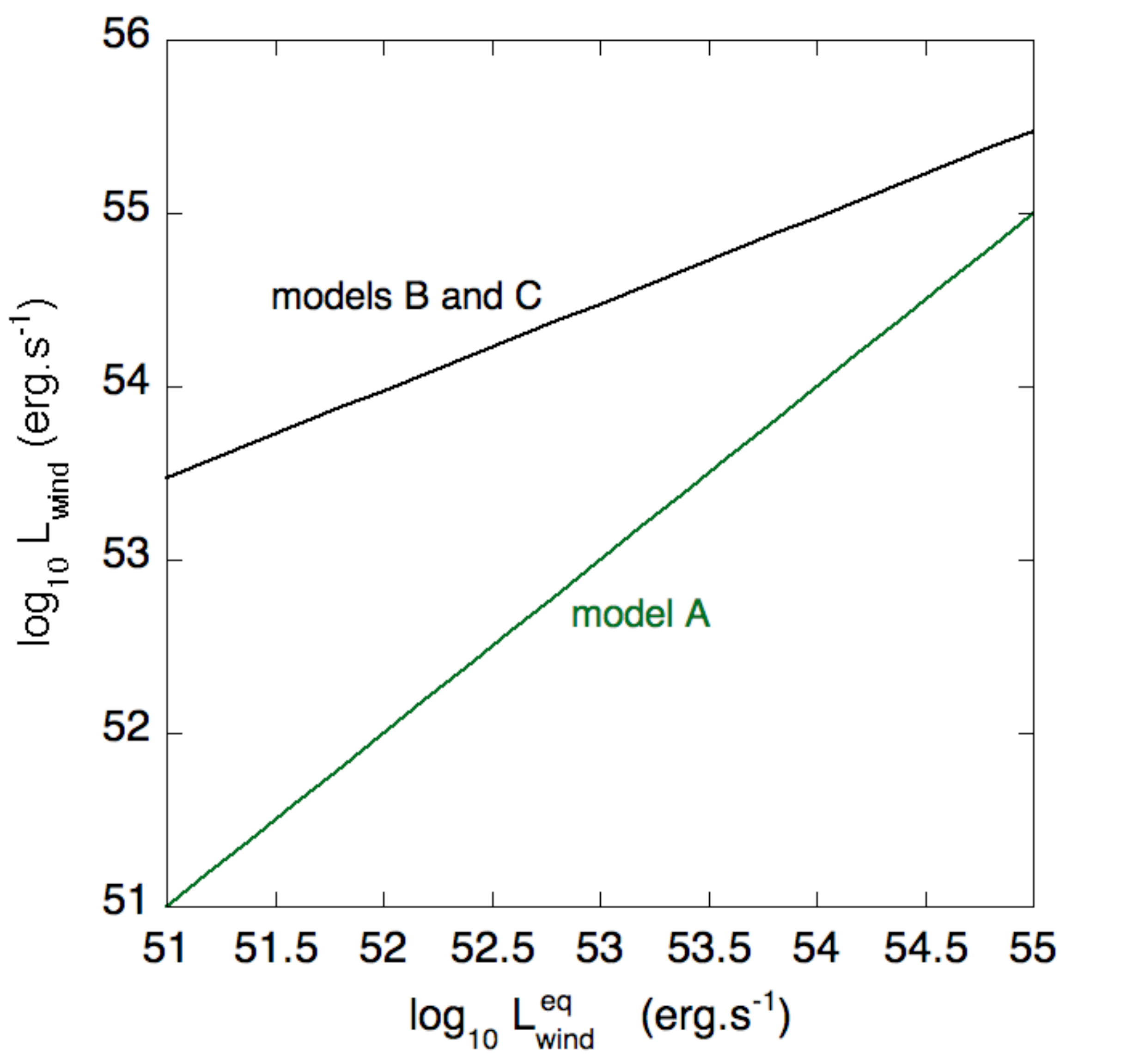}}}
{\rotatebox{0}{\includegraphics[scale=0.24]{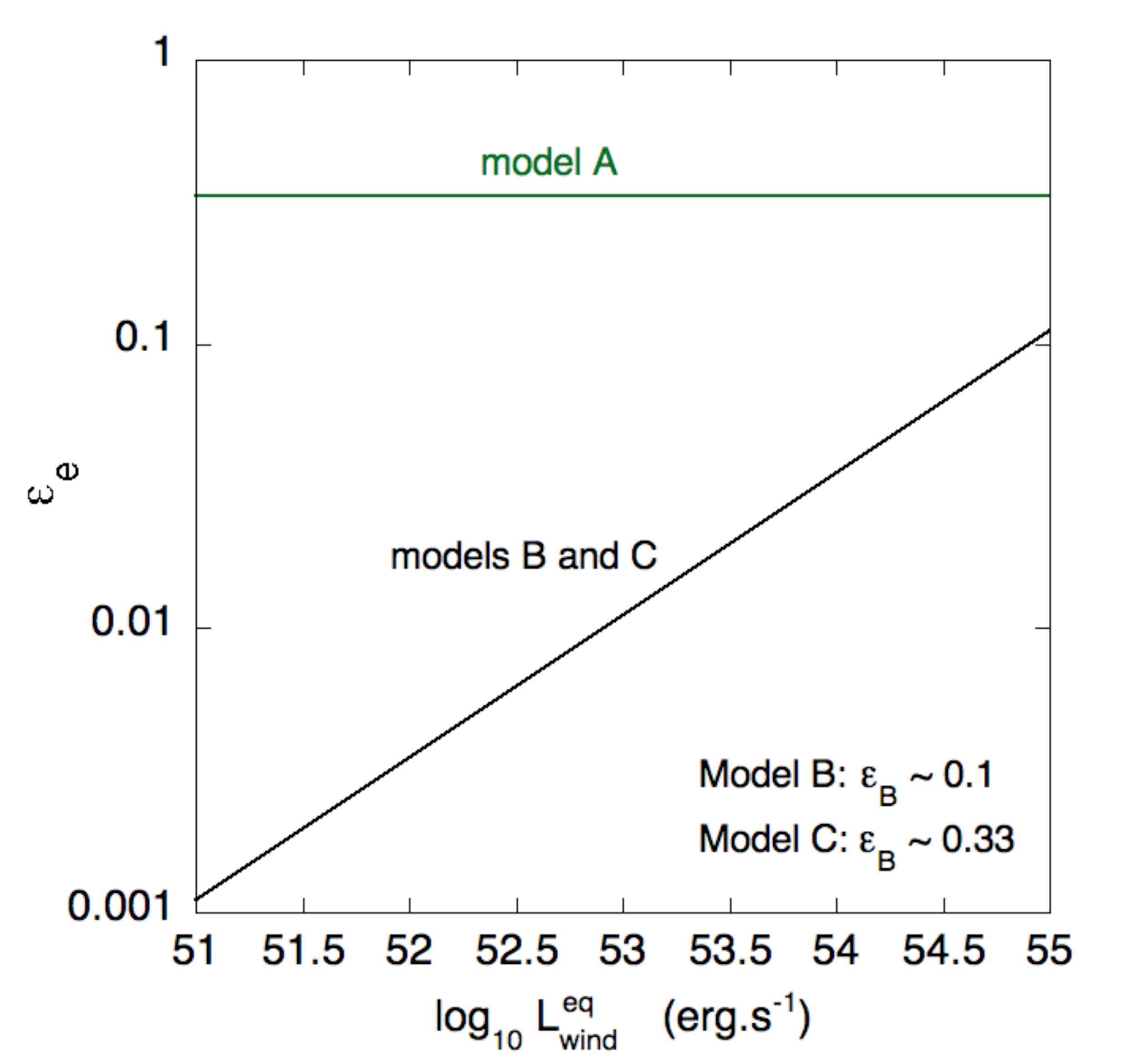}}}
{\rotatebox{0}{\includegraphics[scale=0.24]{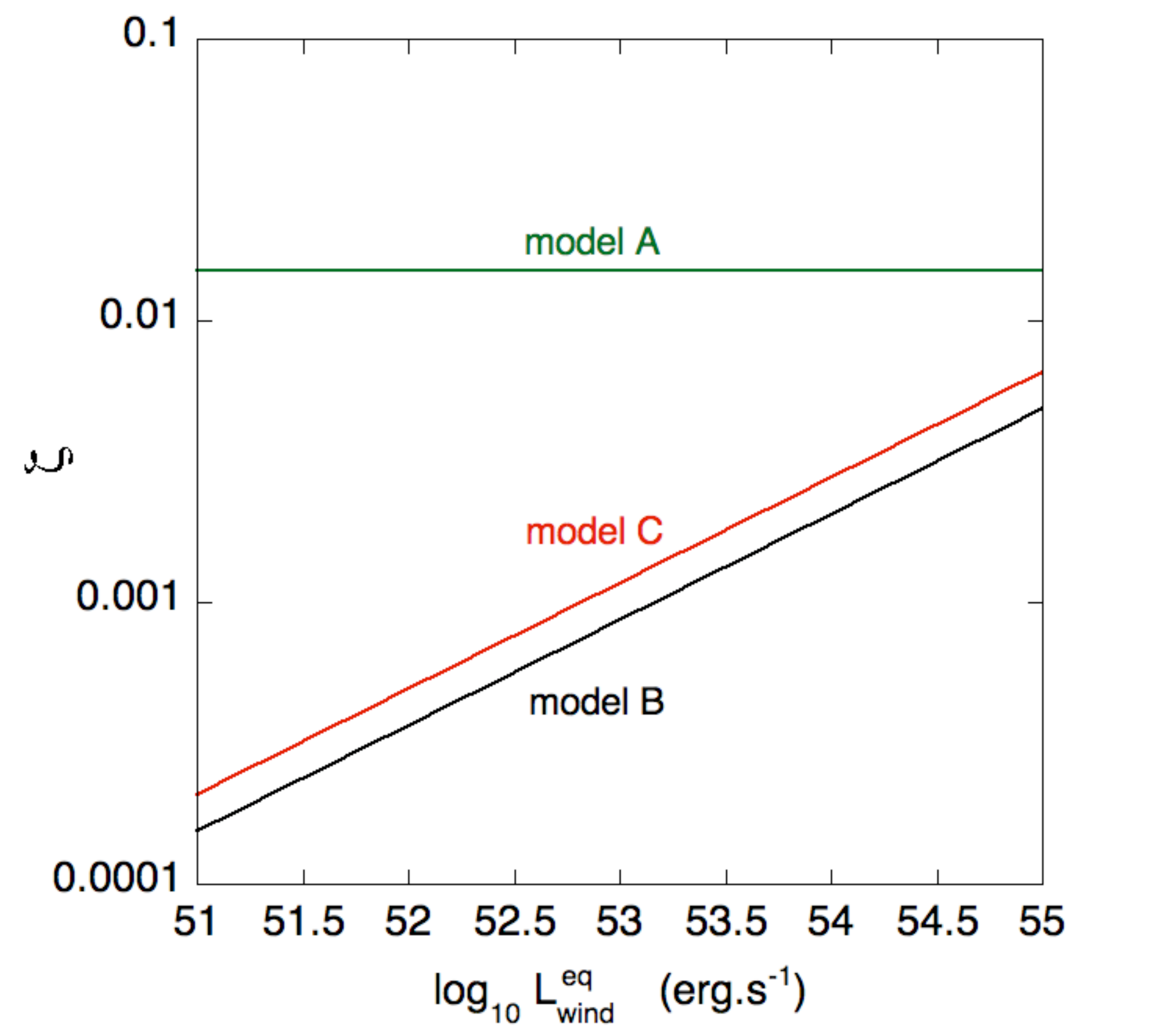}}}
\end{tabular}
\caption{Evolution of the physical parameters of models A, B and C as a function of $L_{\rm wind}^{\rm eq}$ (see text and Eqs.~\ref{eq:lwind}-\ref{eq:zeta}). The evolutions of $L_{\rm wind}$, $\rm \epsilon_e$ and $\rm \zeta$ for the three models are shown respectively in the left, central and right panels.}
\label{fig:lwind}
\end{figure*}

\subsection{Dissipated energy redistribution models}
In the following, we will consider three models corresponding to three different combinations of the dissipated energy redistribution parameters. For all these models, we assume that the prompt emission is dominated by the photon emission consecutive to the cooling of accelerated electrons. For the first model, hereafter model A, we assume equipartition of the dissipated energy between the accelerated cosmic-rays, the accelerated electrons and the magnetic field, i.e $ \epsilon_{\rm cr}=\epsilon_{\rm e}=\epsilon_B=1/3$. As mentioned above, $\sim 15\%$ of the wind energy is dissipated during internal collisions, for the Lorentz factor distribution given in Eq.~\ref{eq:profil}. This means that $\sim5\%$ of the wind energy is communicated to accelerated electrons. The fast cooling of these electrons by synchrotron or inverse Compton losses redistributes almost entirely this energy to the prompt emission photons which, then, ultimately represent $\sim 5\%$ of the energy initially injected in the relativistic 
wind, i.e the prompt emission luminosity $L_\gamma\simeq0.05\times L_{\rm wind}$. As a result, considering wind luminosities between $10^{51}$ and $\rm10^{55}\,\rm erg\,s^{-1}$ one gets prompt emission luminosities $L_\gamma$ between $\sim5\,10^{49}$ and $\rm \sim5\,10^{53}\,\rm erg\,s^{-1}$. By setting the fraction of accelerated electrons $\rm \zeta=1.5\,10^{-2}$, the peak energy $E_{\rm peak}$ obtained lies between a few keV at low luminosities to $\sim 1\, \rm MeV$ for the largest luminosities.

For the two other models, hereafter models B and C, we make a very different assumption on the dissipated energy redistribution parameter for the accelerated electrons $\rm \epsilon_e$ and, as a consequence, on the efficiency of the prompt emission.  We assume that the prompt emission efficiency is lower than in the case of model A and goes from approximately 0.01\% for low values of $L_\gamma$ to 1\% for the highest prompt emission luminosities. It implies larger assumed values of the wind luminosity $L_{\rm wind}$ for a given prompt emission luminosity $L_\gamma$. Practically, to reproduce prompt emission luminosities between $5\,10^{49}$ and $\rm 5\,10^{53}\,\rm erg\,s^{-1}$, as in the case of model A, we assume wind luminosities between $3\,10^{53}$ and $\rm3\,10^{55}\,\rm erg\,s^{-1}$. We use then the following relation between the wind luminosities of the three different models:
\begin{equation}
L_{\rm wind}^{\rm B/C}=3\times\left(\frac{L_{\rm wind}^{\rm eq}}{\rm10^{55}\,\rm erg\,s^{-1}}\right)^{-1/2}\times L_{\rm wind}^{\rm eq}\,,
\label{eq:lwind}
\end{equation}
where $L_{\rm wind}^{\rm B/C}$ refers to the wind luminosity for models B and C and $L_{\rm wind}^{\rm eq}$ to the corresponding wind luminosity for model A. Models B and C differ only by the value of the dissipated energy redistribution parameters $\epsilon_B$ and $\rm \epsilon_{cr}$. For model B, we assume that most of the dissipated energy is communicated into cosmic-rays, $\rm \epsilon_{cr}\simeq0.9$ and $\epsilon_B\simeq0.1$, while model C assumes a larger fraction given to the magnetic fields,  $\rm \epsilon_{cr}\simeq0.66$ and $\epsilon_B\simeq0.33$. The parameters $\rm \epsilon_e$ and $\rm \zeta$ are adjusted for each wind luminosity in order to reproduce the same prompt emission (meaning the same luminosity $L_\gamma$ and the same peak energy $E_{\rm peak}$) as the corresponding case for model A (i.e, the value of $L_{\rm wind}^{\rm eq}$ corresponding to the value of $L_{\rm wind}$ for models B and C as defined in Eq.~\ref{eq:lwind}). The values of $\rm \epsilon_e$ and $\rm \zeta$ for models B and C 
are related to those of model A by the following relations:
\begin{equation}
\epsilon_e=\left(\frac{L_{\rm wind}^{\rm eq}}{L_{\rm wind}}\right)\times \epsilon_e^{\rm eq}
\label{eq:epse}
\end{equation}
and
\begin{equation}
\zeta=\left(\frac{L_{\rm wind}^{\rm eq}}{{  L_{\rm wind}}}\right)^{3/4} \times \left(\frac{\epsilon_B}{\epsilon_B^{\rm eq}}\right)^{1/4}\times \zeta^{\rm eq}
\label{eq:zeta}
\end{equation}
where the index "eq" refers to the values used for model A and $L_{\rm wind}^{\rm eq}$ is related to $L_{\rm wind}$ through Eq.~\ref{eq:lwind}.
Fig.~\ref{fig:lwind} shows values of $L_{\rm wind}$, $\rm \epsilon_e$, $\rm \zeta$ for model A, B and C as a function of $L_{\rm wind}^{\rm eq}$ implied by Eqs.~(\ref{eq:lwind}-\ref{eq:zeta}). As already mentioned, the wind luminosities assumed for models B and C are larger than for model A, especially for low prompt emission luminosities (for $L_\gamma=5\,10^{49}\,\rm erg\,s^{-1}$, i.e $L_{\rm wind}^{\rm eq}=10^{51}\,\rm erg\,s^{-1}$, Eq.~\ref{eq:lwind} implies $L_{\rm wind}=3\,10^{53}\,\rm erg\,s^{-1}$ for models B and C). As a consequence the values of $\rm \epsilon_e$ and $\rm \zeta$ (central and right panels of Fig.~\ref{fig:lwind}) for models B and C are much lower than for model A in order to get the same prompt emission: $\rm \epsilon_e$ goes from $\sim 10^{-3}$ to $\sim 10^{-1}$ while $\rm \zeta$ goes from a few $10^{-4}$ to a few $10^{-3}$ for the range of values of $L_{\rm wind}^{\rm eq}$ we consider. Let us note that we obtain different values of $\rm \zeta$ for models B and C, due to the 
different values of $\epsilon_B$. Model B, which implies lower values of the magnetic field for a given value of  $L_{\rm wind}^{\rm eq}$, requires a lower value of $\rm \zeta$ to reach the same value of $E_{\rm peak}$ (see Eq.~\ref{eq:game} and \ref{eq:epeak}).

\begin{figure*}
\begin{tabular}{cc}
{\includegraphics[width=0.8\textwidth]{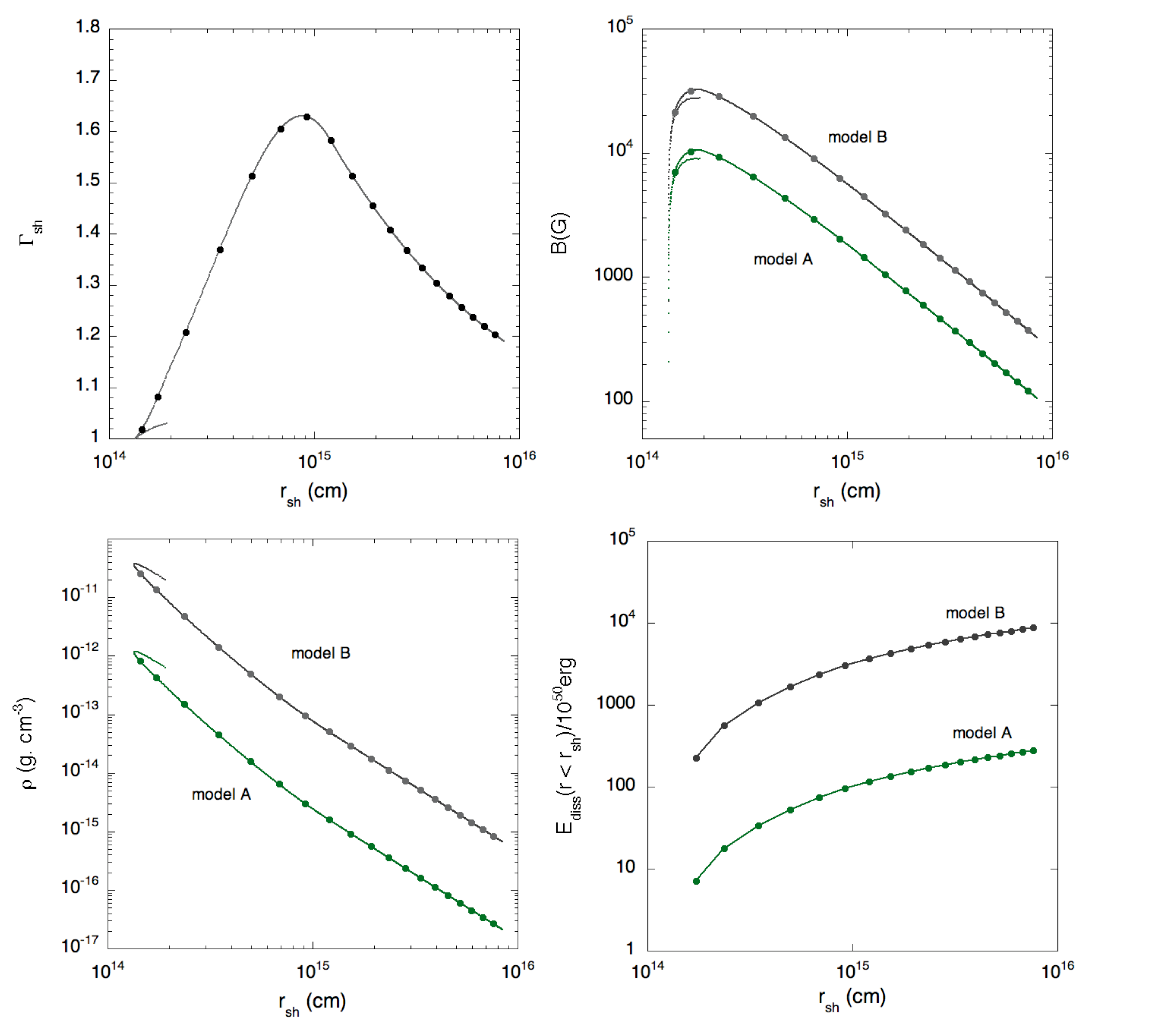} }
\end{tabular}
\caption{Evolution of some physical quantities as a function of $r_{\rm sh}$, the distance of the shock from the central source. The numerical values are calculated with our simple modeling of internal shocks, calculated for models A and B assuming $L_{\rm wind}^{\rm eq}=10^{53}\,\rm erg\,s^{-1}$. Top left : evolution of the shock Lorentz factor in the undisturbed fluid frame. Top right : evolution of the magnetic in the shocked medium. Bottom left : evolution of the density of the shocked medium. Bottom right : evolution of the integrated dissipated energy. }
\label{fig:param}
\end{figure*}

In the following, when comparing predictions of models A, B and C, we will often refer to the value of $L_{\rm wind}^{\rm eq}$, rather than the actual value of $L_{\rm wind}$ (except for model A since by definition $L_{\rm wind}\equiv L_{\rm wind}^{\rm eq}$ for that model). It is important to keep in mind that $L_{\rm wind}$ for models B and C is larger than the corresponding value of $L_{\rm wind}^{\rm eq}$ and that the prompt emission luminosity $L_\gamma$ is related to $L_{\rm wind}^{\rm eq}$ (for our choice of initial Lorentz factors distribution) by the relation $L_\gamma \simeq 0.05 \times L_{\rm wind}^{\rm eq}$ whatever the energy redistribution model. On the other hand the relation between $L_{\rm wind}$ and  $L_\gamma$ (i.e the prompt emission efficiency) depends on the model assumed and also depends on $L_{\rm wind}$ for models B and C.

Fig.~\ref{fig:param} gives some example of the physical quantities which can be estimated with our modeling. It displays the evolution of the shock Lorentz factor (top left panel), the magnetic field in the shocked medium (top right panel), the density of the shocked medium (bottom left panel) and the integrated dissipated energy (bottom right panel) throughout the propagation of the shock (i.e, as a function of the distance of the shock from the central source $\rm r_{\rm sh})$, for models A and B assuming $L_{\rm wind}^{\rm eq}=10^{53}\,\rm erg\,s^{-1}$ (that corresponds to $L_{\rm wind}=3\,10^{54}\,\rm erg\,s^{-1}$ for models B and C). Two branches are visible on the first three graphs, one being largely dominant. They correspond to the two shocks propagating respectively to the front and to the back of the flow from their starting point where the initial Lorentz factor gradient was the largest. As mentioned earlier, it can be seen that one of these shocks (actually the one propagating towards the front 
of the flow) is very short lived and disappears after propagating on a short distance range. The second shock propagates over a much larger distance. For our hypothesis on the initial distribution of Lorentz factors, it forms at $r_{\rm sh}\simeq 10^{14}$ cm from the central source and propagates up to $r_{\rm sh}\simeq 10^{16}$ cm, where only one shell remains. The top left panel of Fig.~\ref{fig:param} shows the evolution of the shock Lorentz factor $\Gamma_{\rm sh}$ (in the undisturbed fluid rest frame) as a function of $r_{\rm sh}$.\\
The evolution of $\Gamma_{\rm sh}$ depends only on the Lorentz factor initial distribution and is then the same for model A, B and C. One sees that the shock is mildly relativistic with a maximum Lorentz factor during the shock propagation around $\Gamma_{\rm sh}\simeq1.65$. Even with a significantly larger contrast in the Lorentz factor distribution, the shock would remain midly relativistic during its propagation. In the case of a Lorentz factor distribution going from 100 to 1000, for instance, the shock Lorentz factor reaches a maximum around $\Gamma_{\rm sh}\simeq2$. Conversely, for a  Lorentz factor distribution going from 100 to 400, the shock would reach at most $\Gamma_{\rm sh}\simeq1.4$.\\
The density and the magnetic field of the shocked fluid are obviously no longer identical for models A, B and C. Since $\rm \rho$ is proportional to the wind luminosity assumed, the density implied for models B and C is then, at any distance form the central source, 30 times larger than for model A. The magnetic field is proportional to $(\epsilon_{ B}\times L_{\rm wind})^{1/2}$ and then is also larger for models B and C, by a factor $\sim \sqrt{10}$ for model B and $\rm \sqrt{30}$ for model C (not displayed) for this particular value of $L_{\rm wind}^{\rm eq}$. The density of the shocked medium decreases slightly slower than $r_{\rm sh}^{-2}$ during the shock propagation (see Eq.~\ref{eq:rho}), due to the increase of $\Gamma_{\rm res}$ with $r_{\rm sh}$ (see the lower panel of Fig.~\ref{fig:profil}) and as a consequence, the magnetic field decreases slightly slower than $r_{\rm sh}^{-1}$. At early stages of the shock propagation, the magnetic field can reach very large values, above $\rm10^4\,G$ for $L_{\rm wind}^{\rm eq}=10^{53}\,\rm erg\,s^{-1}$ and even above $\rm10^5\,G$ for the largest wind luminosity.\\
Finally, the bottom right panel shows the evolution of the integrated dissipated energy (${E_{\rm diss}}(r\leq r_{\rm sh})$) as a function 
of $r_{\rm sh}$ (only the contribution of the long lived shock is shown on this graph). One sees only a relatively small fraction of the energy is dissipated during the early stage of the shock propagation ($\sim 2/3$ of the energy is dissipated at radii larger than $\rm 10^{15}\,cm$). The total amount of energy dissipated at the end of the shock propagation in the flow is $E_{\rm diss}\simeq 2.8\times10^{52}\,\rm erg$ in the case of model A, which represent 14\% of the energy injected in the 2s duration relativistic wind. As the dissipated energy is proportional to the wind luminosity, for a given Lorentz factor distribution,  $E_{\rm diss}$ is 30 times larger for models B and C for this assumed value of $L_{\rm wind}^{\rm eq}$.\\
It is important to note that the simple approach modeling the propagation of the shock as a series of elementary collisions between elementary shells has been validated by the comparison with a more sophisticated hydrodynamical calculation by Daigne \& Mochkovitch (2000). Although the simple model largely underestimate the density of the shocked fluid at early stages of the shock propagation (on a range where only a very small fraction of the energy is dissipated) a good agreement is found between the two approaches, the density estimated with the simple model being only slightly lower (within a factor $\sim$3 to 5) than the one obtained with the hydrodynamical code. Differences of this magnitude do not have any strong impact on the results we obtain in the next sections.\\
Regarding the large wind luminosities required to account for the prompt emission luminosity, in particular in the cases of model B and C, it is important to check that the wind is transparent to Thomson scattering, at least during a significant portion of the shock propagation. The Thomson opacity, at a given distance $r_{\rm sh}$ from the central source, is in good approximation given by (see Daigne \& Mochkovitch 2002) :
\begin{equation}
\tau=\frac{\sigma_T L_{\rm wind}}{8\pi m_{\rm p}\, r_{\rm sh}(c\Gamma_{\rm res})^3}\,.
\label{eq:tau}
\end{equation}
In the following, we only take into account in our calculations the phase of the shock propagation for which $\tau < 0.5$. Even for the largest wind luminosities we assume, the wind remains transparent during most of the shock propagation, and thus most of the energy dissipated at the shock can be used to accelerate electrons and cosmic-rays and produce the prompt emission.\\  
Let us finally note that the different curves shown in Fig.~\ref{fig:param} are made of $\sim$1000 individual shell collisions in order to follow the evolution of the physical quantities during the shock propagation. It is of course not possible to calculate photon emission or cosmic-ray acceleration at 1000 points of the shock propagation. In the following, we will then further discretize the evolution of the shock and make calculations at 18 different points representing 18 stages regularly distributed during the shock propagation. These 18 stages, that we hereafter call the 18 \emph{snapshots}, are represented by full circles in the different graphs of Fig.~\ref{fig:param}. 

\begin{figure*}
\begin{tabular}{cc}
{\rotatebox{0}{\includegraphics[scale=0.23]{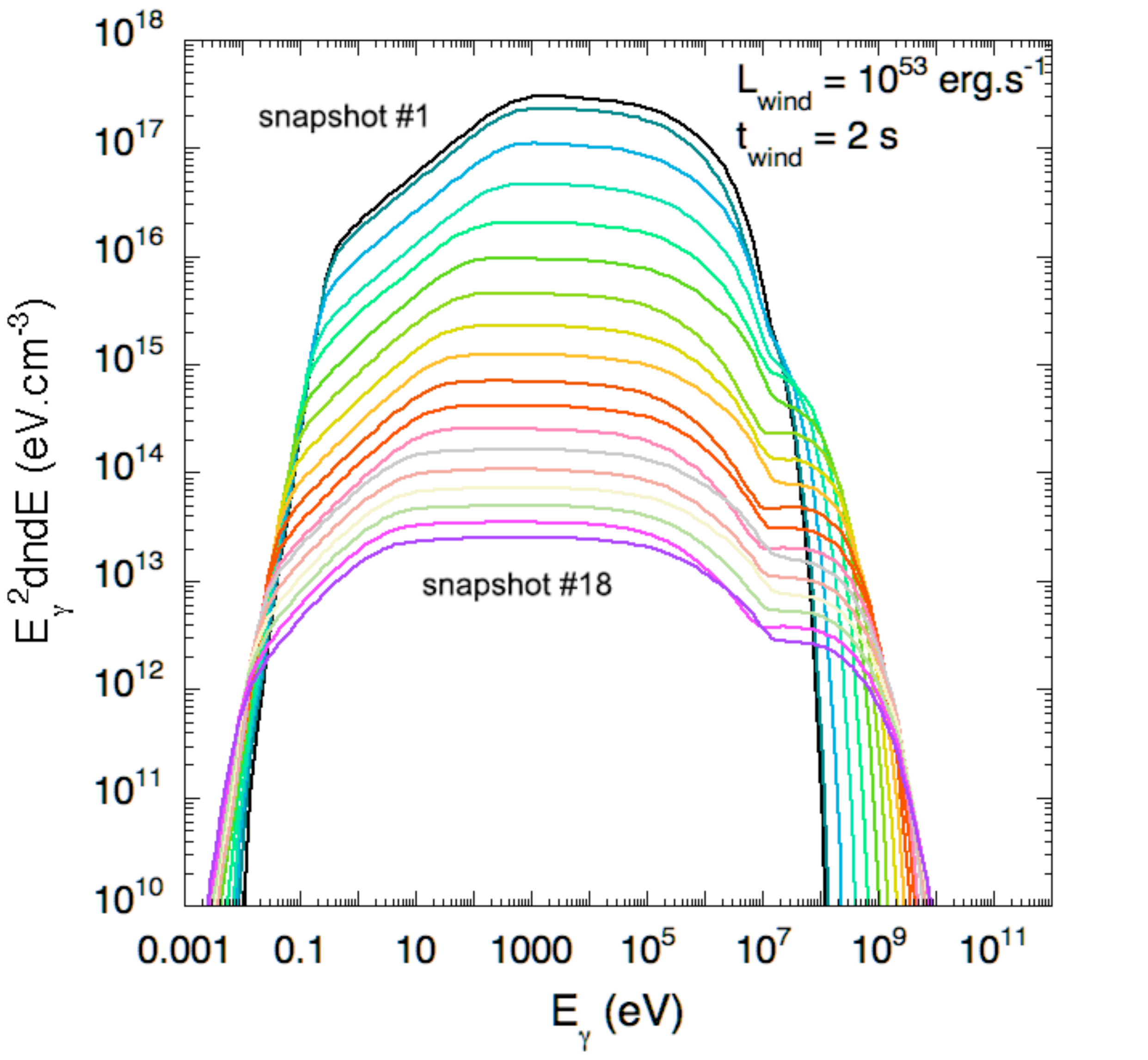}}}
{\rotatebox{0}{\includegraphics[scale=0.23]{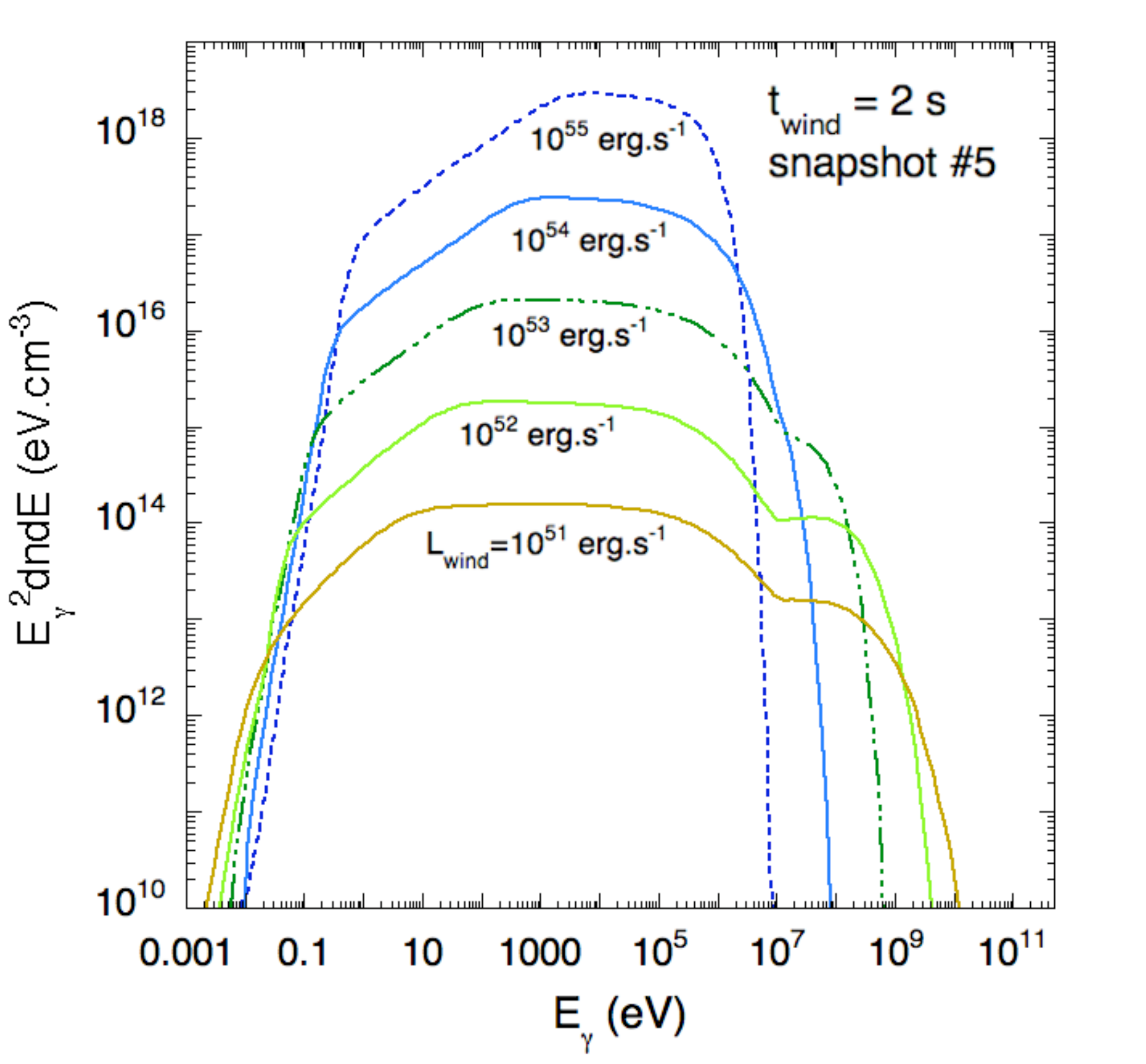}}}
{\rotatebox{0}{\includegraphics[scale=0.23]{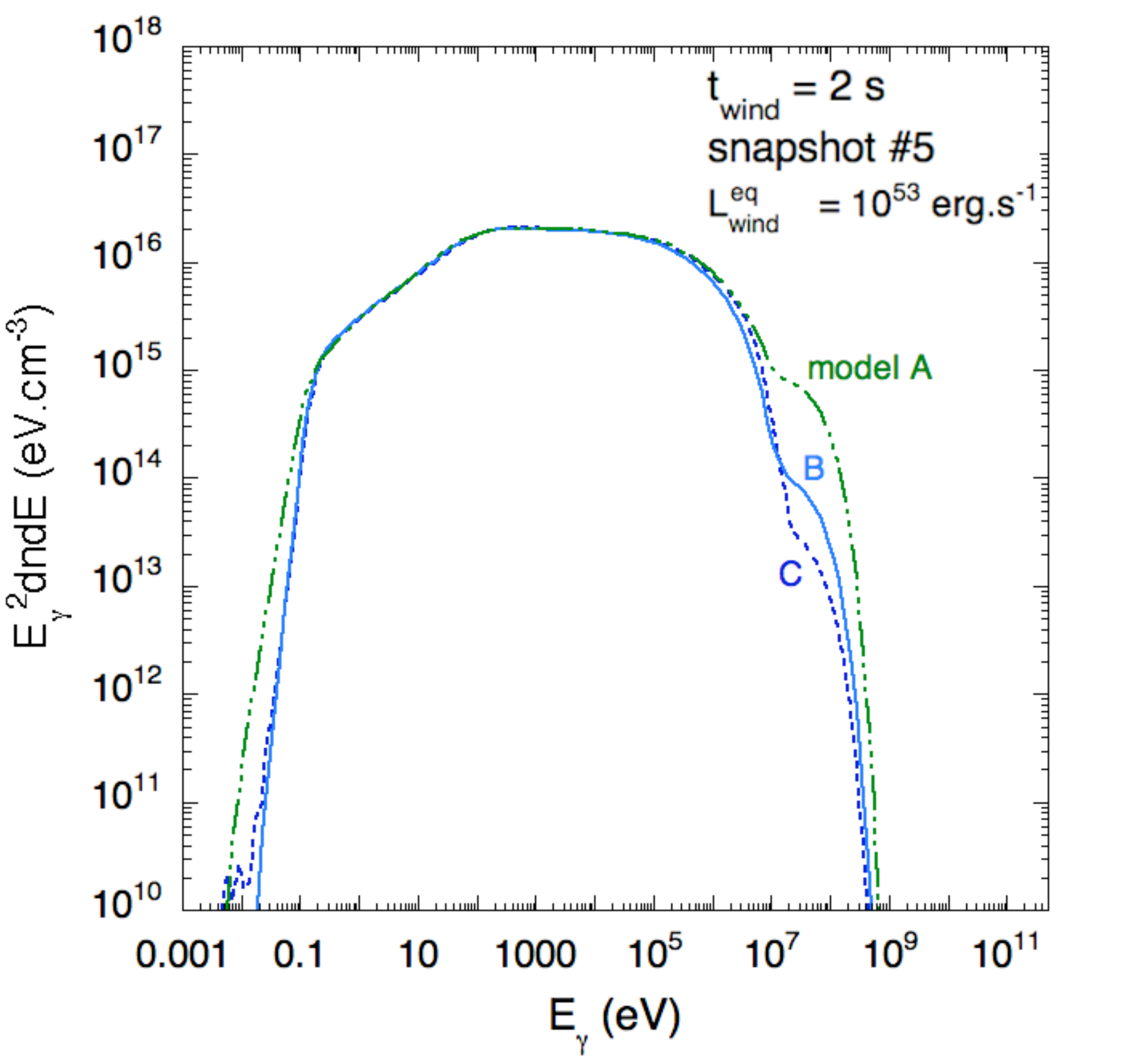}}}
\end{tabular}
\caption{Examples of photon spectra (multiplied by $E^2$) calculated using our modeling of GRB internal shocks and our radiative code. Left : For the 18 snapshots discretising the shock propagation in the case of model A assuming $L_{\rm wind}^{\rm eq}=10^{53}\,\rm erg\,s^{-1}$. Center : in the case of model A assuming $L_{\rm wind}^{\rm eq}=10^{51},\,\,10^{52},\,\,10^{53},\,\,10^{54},\,\,10^{55}\,\,\rm erg\,s^{-1}$, at the position of snapshot \#5 ($r_{\rm sh}\simeq 6.9\,10^{14}\,\rm cm$). Right : For models A, B and C, assuming $L_{\rm wind}^{\rm eq}=10^{53}\,\rm erg\,s^{-1}$ at the position of snapshot \#5.}
\label{fig:SED}
\end{figure*}

\subsection{Calculations of photon spectra from the cooling of accelerated electrons}

All the physical quantities estimated in the previous paragraph are relevant to the calculation of cosmic-ray acceleration at GRB internal shocks. However, at this point, a key ingredient, namely the spectrum of background photons in the acceleration region, is still needed to estimate cosmic-ray energy photo-interactions during their acceleration. In the most popular versions of the internal shock model these background photons are mostly produced by the cooling of accelerated electrons and are, of course, responsible for the bulk of GRB prompt emission.To calculate this prompt emission, we follow step by step the numerical approach developed and fully described in Bosnjak et al. (2009).\\ 
We assume that the accelerated electron spectrum follows a power law of index $p$ (we use $p=2.05$ in our calculations\footnote{This assumed value, well in the range suggested by observations, is not critical for the calculations presented throughout the next sections.}) in the Lorentz factor range between $\rm \Gamma_{e,\,min}$ and $\rm \Gamma_{e,\,max}$. Assuming that the acceleration time of the electrons is proportional to their Larmor time, i.e $t_{\rm acc}{({\rm e^-}, E)}=\kappa_0\times t_{\rm L}{({\rm e^-}, E)}$ (where $\rm \kappa_0$ is the proportionality coefficient we set to 100 in our calculations\footnote{This value is consistent with our results in Sect.~3.},  
$t_{\rm L}{({\rm e^-}, E)}=r_{\rm L}{({\rm e^-}, E)}/c\simeq {E}/ec^2B$ is the Larmor time of the electron of energy $E$  and $r_{\rm L}$ the corresponding Larmor radius), the maximum Lorentz factor reached by accelerated electrons $\rm \Gamma_{e,\,max}$ can be estimated by calculating the energy at which the acceleration time is equal to the energy loss time. \\ In the context of GRB internal shocks, the maximum energy of electrons can either be limited by synchrotron, inverse Compton or adiabatic losses. In practice we only consider the synchrotron and adiabatic timescales to estimate $\rm \Gamma_{e,\,max}$, since the inverse Compton timescale cannot be calculated without prior knowledge of the photon background. Let us note however that the inverse Compton process is never a dominant source of energy losses for high energy electrons for the cases we consider in the following.\\
The timescale for synchrotron losses of an electron with energy $E$, $t_{\rm syn}{({\rm e^-}, E)}$, in the presence of a magnetic field $B$, is given by (in the limit $\rm \beta \rightarrow 1$) 
\begin{equation}
t_{\rm syn}{({\rm e^-}, E)}=\frac{6\pi m_{\rm e}^2 c^3}{\sigma_T B^2 {E}}\simeq { 3.95\times 10^{14}\left(\frac{E}{{\rm eV}}\right)^{-1}}\left(\frac{B}{\rm G}\right)^{-2}\,\rm s\,,
\label{eq:tsyne}
\end{equation}
where $\sigma_T$ is the Thomson cross section.\\
The adiabatic losses time scale $t_{\rm exp}$, assuming spherical expansion, is approximately 
\begin{equation}
t_{\rm exp}\simeq \frac{r_{\rm sh}}{\Gamma_{\rm res}c}
\label{eq:tad}
\end{equation}
where $r_{\rm sh}$ and $\rm \Gamma_{\rm res}$, introduced in the previous paragraph, are respectively the distance of the shock from the source and the Lorentz factor of the shocked fluid (both estimated in the source frame)
One then obtains:
\begin{equation}
\Gamma_{\rm e,\,max}={\rm min} \left (\sqrt{\frac{6\pi e c}{\kappa_0\sigma_TB}}\,;\,\frac{t_{\rm exp}eB}{\kappa_0 m_{\rm e} }\right)
\label{eq:gamemax}
\end{equation}
Once $\rm \Gamma_{e,\,max}$ has been estimated, $\rm \Gamma_{e,\,min}$ is computed solving Eq.~\ref{eq:gametrue}.
In all the cases we consider in the following, synchrotron radiation is the limiting process for electrons acceleration during the whole shock propagation and moreover the electrons will be in the fast cooling regime in the whole energy range.\\
For a given GRB, i.e a given value of $L_{\rm wind}^{\rm eq}$ between $10^{51}$ and $\rm 10^{55}\,\rm erg\,s^{-1}$ and a given energy redistribution model, we calculate $\rm \Gamma_{e,\,min}$ and $\rm \Gamma_{e,\,max}$ for each of the 18 snapshots discretizing the shock propagation. For a given snapshot, the electron power law spectrum between $\rm \Gamma_{e,\,min}$ and $\rm \Gamma_{e,\,max}$ serves as the initial distribution of the accelerated electrons in the shocked fluid frame. We then use a radiative code that solves numerically simultaneously the time evolution of the energy distribution of electrons and photons (see Bosnjak et al. 2009 for more details) taking into account the most relevant processes, namely the adiabatic cooling, the synchrotron cooling and photon emission (Rybicki \& Lightman 1979; Longair 2011), the inverse Compton scattering (Jones 1968), the synchrotron self-absorption (Rybicki \& Lightman 1979) and the $\gamma\gamma$ pair production (Gould \& Shreder 1967). For each snapshot, 
the evolution of the energy distributions is followed during a time corresponding to $t_{\rm exp}(r_{\rm sh})$. In practice, since we are in the fast cooling regime,  most of the evolution, and in particular the formation of the photon energy distribution, occurs on timescales much shorter than $ t_{\rm exp}$.

Fig.~\ref{fig:SED} shows some examples of photon spectra, in the frame of the shocked fluid, calculated using the above described modeling of the internal shocks and radiative code. The left panel of Fig.~\ref{fig:SED} shows the photon spectra obtained for the 18 snapshots of the shock propagation in the case of model A and $L_{\rm wind}^{\rm eq}(=L_{\rm wind})=10^{53}\,\rm erg\,s^{-1}$. This graph shows the evolution of the photon spectrum during the shock propagation. The photon density scales, as expected, approximately as $r_{\rm sh}^{-2}$ and is then $\sim 4$ orders of magnitude larger at the beginning of the shock propagation than at the end. One also sees a transition from hard to soft of the peak energy $E_{\rm peak}$ due to the evolution of the magnetic field and the different quantities impacting  $\rm \Gamma_{e,\,min}$. At high energy, the $\rm \gamma \gamma$ opacity decreases during the shock propagation following the density. The high energy $\gamma \gamma$ cut-off is then shifted to higher 
energies  and the contribution of photons boosted by inverse Compton (or more precisely synchrotron self-Compton) becomes more visible as the shock propagates away from the central source and the magnetic field decreases.\\
The central panel shows the photon spectra calculated at the position of the snapshot \#5 ($r_{\rm sh}\simeq6.9\,10^{14}\,\rm cm$, quite early during the shock propagation) for model A and different values of $L_{\rm wind}^{\rm eq}$: $10^{51}$, $10^{52}$, $10^{53}$, $10^{54}$ and $\rm10^{55}\,\rm erg\,s^{-1}$. As expected for model A, the photon density is approximately proportional to $L_{\rm wind}$ and the peak energy to $L_{\rm wind}^{1/2}$ (that means approximately proportional to the magnetic field). As a result of the photon density evolution with $L_{\rm wind}$, the high energy $\rm \gamma \gamma$ cut-off is shifted to higher energies as the wind luminosity decreases together with a larger contribution of inverse Compton photons (again due to the lower magnetic field).\\
Finally, the right panel shows photon spectra at the position of snapshot \#5, assuming $L_{\rm wind}^{\rm eq}=10^{53}\,\rm erg\,s^{-1}$ for models A, B and C (corresponding to $L_{\rm wind}=3\,10^{54}\,\rm erg\,s^{-1}$ for models B and C). One can see that the photon spectra are quasi identical for the three models which was expected since the values of the parameter $\rm \epsilon_e$ and $\rm \zeta$ for models B and C were tuned for that purpose. A small difference (irrelevant for our calculations in the following) can be seen at high energy just before the spectral cut-off due to the different contributions of inverse Compton photons which depend on the value of the magnetic field (model A, for which the magnetic field is the lowest, has the strongest contribution from inverse Compton photons).

We calculated the evolution of the photon spectrum during the shock propagation at various wind luminosities, corresponding to values of $L_{\rm wind}^{\rm eq}$ between $10^{51}$ and $\rm 10^{55}\,\rm erg\,s^{-1}$ for models A, B and C. In the following, these photons will serve as targets for photo-interactions of accelerated protons and nuclei. It is important to note that we calculated these photon spectra (in particular $\rm \gamma \gamma$ interactions assuming the produced photons are distributed isotropically in the shocked fluid. This is, in principle, not true and Hasco\"{e}t et al. (2012) have shown that a more careful treatment of the photon emission geometry (using the same modeling of internal shocks as we do here) resulted in lower $\gamma \gamma$ opacity at high energies (and then to a higher energy $\gamma \gamma$ cut-off for a given wind Lorentz factor) due to the suppression of head-on $\gamma \gamma$ interaction. The assumption of isotropy of the photon background in the frame of the shocked fluid should also have an influence on the photo-interaction rate of accelerated protons and nuclei. The overestimate of the interaction rate should be however lower than for $\gamma \gamma$ interactions since accelerated protons and nuclei  will be scattered by magnetic fields in the acceleration region allowing for head-on interactions as in the assumption of isotropy. It is however important to remind that it is likely that the photo-interaction rates we calculate in the following are probably slightly overestimated. 

\section{Cosmic-rays acceleration at midly relativistic shocks}

In the previous section we found that GRBs internal shocks are mildly relativistic, with typical Lorentz factors lasting between $\sim 1$ and $\sim 2$, at least for the moderate contrasts in the wind Lorentz factor distribution we considered in Sect.~2. To discuss cosmic-ray acceleration at GRBs internal shocks, it is necessary to model first order Fermi acceleration at (mildly) relativistic shocks.
\subsection{Introduction}
Fermi acceleration at relativistic shocks has been studied for three decades, early studies pointed out (see e.g., the early works by Peacock 1981 or Kirk \& Schneider 1987) the significant anisotropies in the particles distribution function expected in the downstream (shocked fluid) and the upstream (undisturbed fluid) media due to the shock velocity. Various shock configurations in terms of the magnetic field obliquity, the level of turbulence or its power spectrum (Kirk \& Schneider 1987b; Heavens \& Drury 1988; Begelman \& Kirk 1990; Ostrowski 1991, 1993; Bednarz \& Ostrowski 1998; Gallant \& Achterberg 1999; Kirk et al. 2000; Achterberg et al. 2001; Lemoine \& Pelletier 2003) were later studied, using in most cases the "test particle" approach.  Some of these studies suggested in particular that, in the case of a highly turbulent magnetic field and ultrarelativistic shocks, the spectral index of accelerated particles converges to a  value of $\alpha\simeq2.2-2.3$. A more elaborated Monte-Carlo treatment (still using the test particle approach), considering in particular realistic jump conditions and continuous magnetic field lines across the shock front was proposed by Niemiec \& Ostrowski (2004). Their study showed, for mildly relativistic shocks, that the accelerated particles spectral index can highly vary depending on the turbulence type and level or the magnetic field obliquity. They found that, in most cases, particles spectra cannot be well fitted by a single power law but  usually exhibit a quite noticeable curvature. Furthermore, they found steep spectra with early high energy cut-offs for superluminal shocks configurations. This trend was later confirmed by the same authors (Niemiec \& Ostrowski 2006a, 2006b) who pointed out the inefficiency of Fermi acceleration at ultra-relativistic shocks, unless a strong small scale turbulence is present in the downstream medium. The same conclusion was drawn analytically by Lemoine, Pelletier \&  Revenu (2006). As a consequence, special efforts were devoted recently to model the microphysics of ultra-relativistic shocks and, in particular, the growth of strong small scale turbulence (see the review by Lemoine \& Pelletier 2011). Various types of plasma instabilities have been studied analytically (Medvedev \& Loeb 1999; Lyubarsky \& Eichler 2005; Achterberg et al. 2007; Bret et al. 2010; Lemoine \& Pelletier 2010, 2011). Moreover, important progresses in the understanding of the microphysics of ultra-relativistic shocks have been obtained with Particles in Cells (PIC) simulations (Spitkovsky 2008; Hashino 2008; Sironi \& Spitkovsky 2009, 2011). The latter have confirmed that Fermi acceleration should be inefficient at strongly magnetized ultra-relativistic shocks while the growth of small scale turbulence is possible (and then particle acceleration efficient) for much lower magnetizations. In the meantime, the microphysics of mildly relativistic shocks has been much less investigated; therefore we will limit ourselves to relatively simple magnetic field configurations in what follows.

\subsection{Numerical method}
For our simulation of cosmic-ray acceleration at mildly relativistic shocks, we follow the numerical test particle method developed by Niemiec \& Ostrowski (2004, 2006a, 2006b). The shock front is modeled as an infinite planar discontinuity (infinitely thin). In the upstream medium, the magnetic field is assumed to be an isotropic and homogenous turbulence to which can be added a regular component with an inclination $\rm \theta$ with respect to the shock normal. The downstream magnetic field is obtained applying hydrodynamical jump conditions (Synge 1953), which give the shock compression ratio $\rm r=\beta_{up}/\beta_{down}$ ($\rm\beta_{up}$ and $\rm\beta_{down}$ being respectively the velocities of the upstream and downstream fluid in the shock rest frame), to the upstream field (see e.g., Gallant \& Achterberg 1999; Lemoine \& Revenu 2006). The component of the field parallel to the shock normal, $B_\parallel$, is left unmodified by the shock crossing while the perpendicular component, $ B_\perp$, is 
multiplied by a factor $\rm R=\frac{\Gamma_{up}}{\Gamma_{down}}\times r $ ($\rm\Gamma_{up}$ and $\rm\Gamma_{down}$ being respectively the Lorentz factors of the upstream and downstream fluid in the shock rest frame). We then have the following relations :
\begin{equation}
\frac{B_{\parallel,\rm up}}{B_{\parallel,\rm down}}=1 
\label{eq:jump1}
\end{equation}
and
\begin{equation}
\frac{B_{\rm\perp,down}}{B_{\rm\perp,up}}=\frac{\rm\Gamma_{up}}{\rm\Gamma_{down}}\times r=\frac{\rm\beta_{up}\rm\Gamma_{up}}{\rm\beta_{down}\rm\Gamma_{down}}
\label{eq:jump2}
\end{equation}
Likewise, we compress the turbulence maximum length scale $\rm \lambda_{max}$ by a factor R in the direction of the shock propagation following Niemiec \& Ostrowski (2004). We then apply the above-mentioned jump conditions to the upstream field. This method ensures that the turbulence length scale is properly compressed and that the magnetic field is continuous across the shock front.

The turbulent component of the magnetic field $\delta B$ in the upstream medium is described by the summation of sinusoidal static waves. Concretely, the purely turbulent field is represented by a sum of $N_{\mathrm{m}}$ modes as in Giacalone and Jokipii (1999):
\begin{equation}
\delta \vec{B} = \sum_{n=1}^{N_{\mathrm{m}}}A_{k_{n}}\vec\xi_{n}\exp(ik_{n}z^\prime_{n} + i\beta_{n}),
\label{eq:BGC}
\end{equation}
where $\vec\xi_{n} = \cos\alpha_{n}{\vec{x}}^\prime_{n} + i \sin\alpha_{n}\vec{{y}}^\prime_{n}$, $\alpha_{n}$ and $\beta_{n}$ are random phases (chosen once) and $\rm[x^\prime_{n},y^\prime_{n},z^\prime_{n}] = [\mathcal{R}(\theta_{n},\phi_{n})]\times[x,y,z]$ are coordinates obtained by a rotation of the reference frame bringing the z axis in the direction of the $n^\mathrm{th}$ contributing wave (i.e. $\vec{k}_{n}$ is in direction $[\theta_{n},\phi_{n}]$, also chosen randomly). The amplitude $A(k_{n})$ is determined as a function of $\|\vec{k}_{n}\|$ according to a specific assumption on the type of turbulence (here, we will assume in most cases a Kolmogorov spectrum with wavelengths between $\lambda_{\mathrm{min}}$ and $\lambda_{\mathrm{max}}$) and the chosen field variance (see Giacalone and Jokipii 1999, for more details). In the following we assume  $N_{\mathrm{m}}=600$ modes between wavelengths $\rm \lambda_{min}=10^{-6}\times \lambda_{max}$ and $\rm \lambda_{max}$ (the modes are 
separated by a constant logarithmic step of 0.01, that is, 100 modes per wavelength decade). Once the turbulent field is modeled, the spatial transport of charged particles is simply followed by integrating the Lorentz equation, $\gamma\frac{d{\vec{v}}}{dt}=\frac{q}{m}{\vec{v}}\times{\vec{B}}$, which we performed with the fifth order Runge-Kutta method (Press et al. 1993 and references therein). 

In the following paragraph, we discuss cosmic-ray acceleration at mildly relativistic shocks for different combination of physical parameters that are expected to influence the cosmic-ray output after a Fermi-type acceleration: the shock Lorentz factor, the type of turbulence (i.e, the index of the turbulence power spectrum), the obliquity of the regular field component and the turbulence level. For a given set of parameters, we calculate the spectrum of accelerated particles by integrating particles trajectories alternatively in the downstream and in the upstream rest frame during their whole journey in the vicinity of the shock. The characteristics of a particle (the energy, the velocity vector, the position and elapsed time in the acceleration region) are Lorentz transformed accordingly at each shock crossing. Particles are followed until they are either advected far away downstream or manage to escape far upstream.
\begin{figure}
{\rotatebox{0}{\includegraphics[scale=0.45]{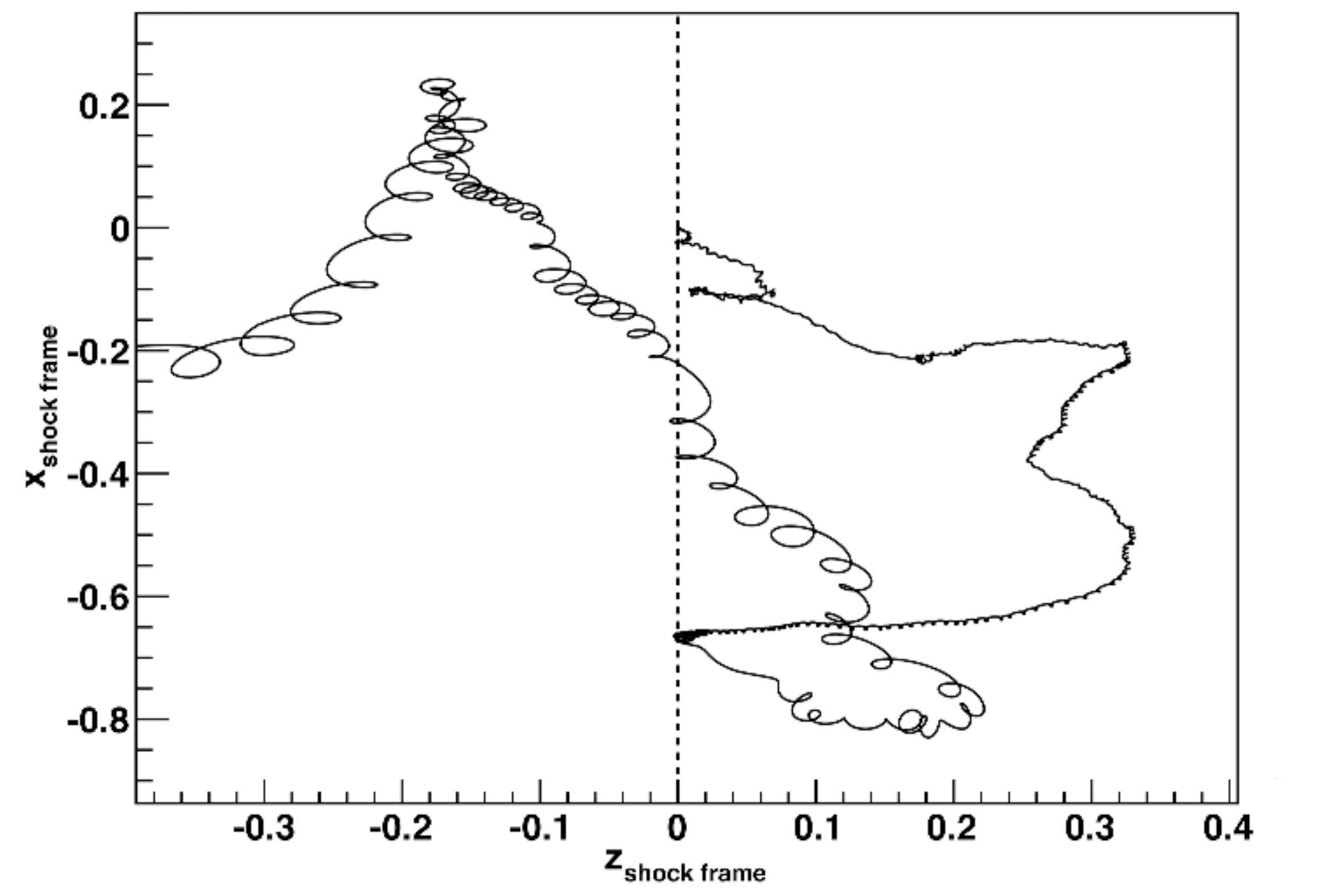}}}
\caption{Exemple of a trajectory of an accelerated particle as seen in the shock rest frame, projected on the (x, z) plane, the shock Lorentz factor is $\Gamma_{\rm sh}=1.1$ (along the z axis), the magnetic magnetic field is purely turbulent with a Kolmogorow power spectrum. The particle with initial energy $E_0=10^{-6}\times E_{\rm max}$ achieves 9 cycles before being advected downstream with a final energy $E_f\simeq70\times E_0$. }
\label{fig:traj}
\end{figure}
An example of a particle trajectory, as seen in the shock frame and projected in the (x, z) plane, is shown in Fig.\ref{fig:traj}.  In the following, we consider particles which are ultra-relativistic already at "injection" and on the whole energy range (the momentum $p \simeq E$ and the rigidity $\varrho \simeq E/Z$ for ultra-relativistic particles). Practically, our simulations start with $N$ particles (for most simulations with take $N$ between $30000$ and $60000$) injected upstream of the shock  with an energy $ E_0=10^{-6}\times E_{\rm max}$,  $E_{\rm max}$ being the energy at which the Larmor (or gyration) radius equals the largest turbulence length scale, i.e.
\begin{equation}
E_{\rm max}:\, r_{\rm L}(E_{\rm max})\equiv\lambda_{\rm max}.
\label{eq:Emax}
\end{equation}
The $N$ particles are initially distributed isotropically upstream and eventually enter the downstream medium after crossing the shock front. Each trajectory is followed in the compressed turbulence in the downstream medium until the particle either returns to the shock or is advected far away, in which case we consider the particle will never come to the shock front again. In practice, we consider a particle is advected far away downstream (and then is lost for the acceleration process) if it has not come back to the shock front after a duration of 2000 Larmor times $t_{\rm L}({E})$ (where $E$ is the current energy of the particle in the downstream rest frame). We checked that changing this duration criterion to larger values did not affect any of the results shown in the following. After the trajectories of all the particles have been computed, the particles that have been advected far away downstream are replaced by duplicating those that managed to return to the shock front following the trajectory 
splitting algorithm of Niemiec \& Ostrowski (2004). The strategy is to divide the weight of the duplicated particle  according to the number of duplications it suffered. The position of the daughter particle is then slightly  modified by a small fraction of the minimum turbulence length scale of the magnetic field. The characteristics of the new set of $N$ particles are then Lorentz transformed to follow each trajectory in the upstream frame until the particle either crosses the shock again or escape far away upstream. For the latter case, we set a free boundary escape at a distance $r=10\,\lambda_{\rm max}$ away from the shock front (the boundary is moving with the shock). Once all the particles trajectories have been computed, the splitting procedure is applied again. Practically, the upstream boundary escape is only relevant for energies approaching $E_{\rm max}$. At lower energies, all the particles end up being outrun by the shock. The whole procedure is applied for the subsequent cycles 
until either all the particles escape at a given step, or the energy of the particles exceed $100\,E_{\rm max}$. Let us note that, in the rest of this section, the energy scale is defined relative to the maximum energy $E_{\rm max}$ defined in Eq. (\ref{eq:Emax}), and similarly the acceleration times are given in units of the Larmor time at this specific energy $t_{\rm L}(E_{\rm max}$). The results we obtain can be rescaled to energy units by specifying explicitly the value of the magnetic field, the maximum turbulence scale and the charge of the particle. In the absence of energy losses (which will be the case in the next paragraph), there is a complete symmetry in the behavior of different nuclear species (different nuclei with different charges $Z$) at a given rigidity $\varrho\simeq E/Z$ with respect to the acceleration process.  

\subsection{Results} 

\subsubsection{Accelerated particles spectra}
\begin{figure}
\centering{\includegraphics[scale=0.5]{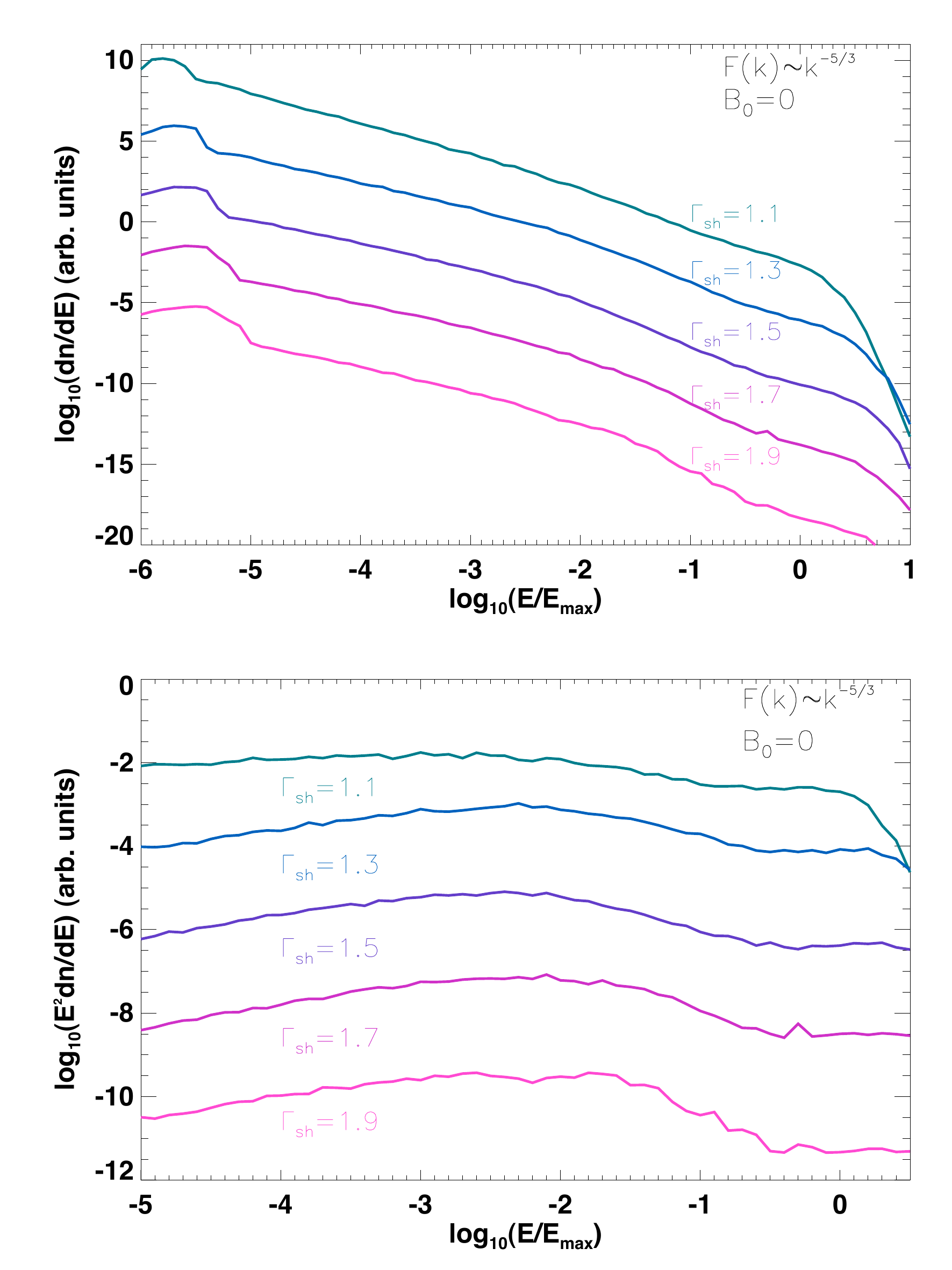}}
\caption{Top : Energy spectra of particles advected downstream for different shock Lorentz factors $\Gamma_{\rm sh}$ : 1.1 ($\beta_{\rm sh}\simeq0.42)$, 1.3 ($\beta_{\rm sh}\simeq0.64)$, 1.5 ($\beta_{\rm sh}\simeq0.75)$, 1.7 ($\beta_{\rm sh}\simeq0.81)$, 1.9 ($\beta_{\rm sh}\simeq0.85)$. In all cases, the magnetic field is assumed to be purely turbulent with a Kolmogorov power spectrum. Bottom : Same as in the upper panel but the spectra are now multiplied by $E^2$.}.
\label{fig:all_gamma}
\end{figure}

We apply the above described simulation procedure to different sets of physical parameters of the shock.  In each case, we calculate the spectra of particles advected downstream and escaping upstream as well as the evolution of the acceleration time with the energy. The quantities displayed in this section (energies and times) are calculated in the shock rest frame. Spectra of particles advected downstream for shock Lorentz factors $\Gamma_{\rm sh}$ between 1.1 and 1.9 (covering most of the range of the shock Lorentz factors during the evolution of the GRB single pulse we consider) are displayed in Fig.~\ref{fig:all_gamma}. We assume a purely turbulent magnetic field,  isotropic in the upstream medium and following a Kolmogorov scaling for the the turbulence power spectrum ($ F(k)\propto k^n$ with $n=-5/3$). We do not consider, however, the dynamics of the magnetic field in a consistent way (i.e, we do not treat wave growth, damping and cosmic-ray back reaction). As mentioned earlier, energies are expressed 
in units of $E_{\rm max}$. All the calculated spectra exhibit an initial bump corresponding to the larger mean energy gain experienced during the first Fermi cycle (see for instance Gallant \& Achterberg 1999) due to the assumed initial isotropy of the particles in the upstream rest frame. This appears as a distinct bump in the spectra because all the particles are artificially injected at the same energy in the calculations. Above these low energy bumps, the spectra deviate from a single power law, as can be seen in the lower panel of Fig.~\ref{fig:all_gamma}. The different curves exhibit a similar trend, a hard spectrum at low energy that get gradually softer as the energy increases before flattening again at  $E\simeq E_{\rm max}$. A similar behavior was already found and discussed in Niemiec \& Ostrowski (2004). At low energy the probability for the particles to be reflected after crossing the shock from upstream to downstream is large due to the compression of the magnetic field behind the shock and 
the low amplitude of the turbulence in the resonant modes for low energy particles. Diffusion across field lines becomes gradually more efficient as the energy increases, leading to a decrease of the reflection probability which results in a softening of the spectra. At higher energy (close to $E_{\rm max}$) the spectra show another hardening of the slope as particles enter the weak scattering regime ($r_{\rm L}(E)\gtrsim \lambda_{\rm max}$). 

The competition between reflexion and transmission has been investigated in great detail in Niemiec \& Ostrowski (2004). We also note that the curvature (deviation from a single power law) of the spectra, is absent when the compression of the magnetic field behind the shock is disabled. In these cases, we found that the spectra of accelerated particles for the same Lorentz factor range can be well fitted by single power laws (with indices between $\sim1.8$ and $\sim2.2$). The above-mentioned effects are less pronounced for shocks with low Lorentz factors (1.1 and below), as the situation gets closer to the non-relativistic regime ($\Gamma_{\rm sh}\rightarrow 1$). At large Lorentz factors ($\sim2.5$ and above) the spectra obtained are much steeper (softer) and display early high-energy cut-offs, confirming that Fermi acceleration becomes progressively inefficient for increasingly relativistic shock velocities (see Niemiec \& Ostrowski 2006a, 2006b; Lemoine, Pelletier \&  Revenu 2006).

\begin{figure}
\centering{\includegraphics[scale=0.47]{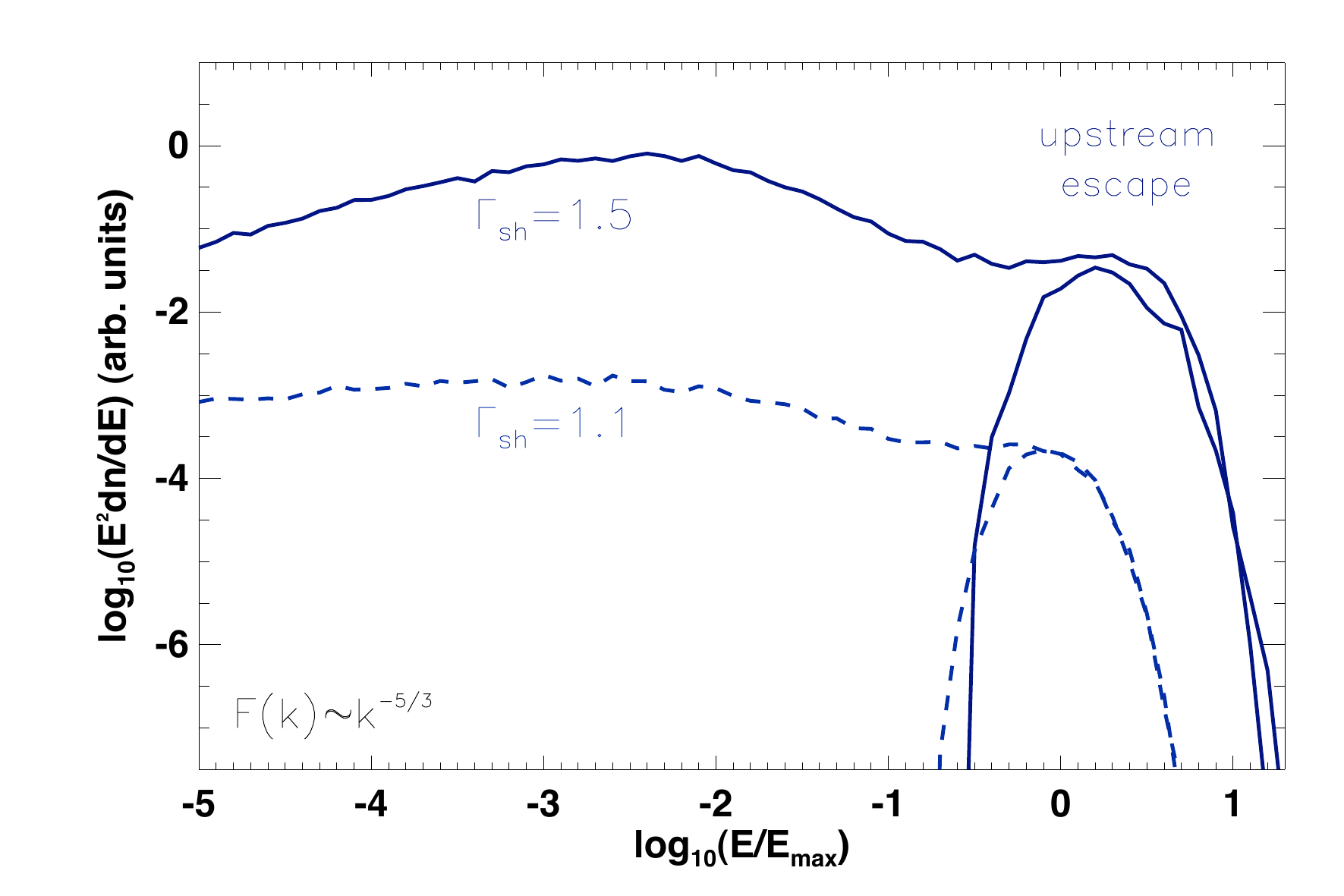}}
\caption{Spectra (multiplied by $E^2$) of accelerated particles advected downstream of the shock (same as in Fig.~\ref{fig:all_gamma}) and of accelerated particles escaping upstream (bell-like shapes) as seen in the shock rest frame. The cases of shock Lorentz factors 1.1 and 1.5 are displayed (the other cases displayed in Fig.~\ref{fig:all_gamma} are omitted for clarity).}
\label{fig:up_down}
\end{figure}

At energies above $E_{\rm max}$, a cut-off is finally observed in all cases due to the efficient escape through the free boundary we placed in the upstream medium. As can be seen in Fig.~\ref{fig:up_down}, the spectra of particles escaping upstream are extremely peaked as the escape upstream of the shock is only efficient for particles with energies close to $E_{\rm max}$ which are only weakly deflected by the magnetic turbulence and can outrun the shock long enough to reach the boundary. The energy range on which the escape upstream is operative mainly depends on two physical parameters, namely the distance of the free boundary and the shock Lorentz factor. The spectrum of particles escaping upstream is expected to be shifted to higher (resp. lower) energies as the free boundary is located further (resp. closer) from the shock (however, we observed that placing the free boundary at a distance $\lambda_{\rm max}$ from the shock instead of $\rm 10\times \lambda_{max}$ only produces a slight shift to lower 
energies). The shock Lorentz factor dependence is also intuitive (and the effect can be observed in Fig.~\ref{fig:up_down}): the faster the shock the higher the energies of escaping particles, since particles of a given energy will be more easily caught up by the shock before reaching the escape boundary if the shock velocity is higher. Let us add that for purely turbulent fields, the escape upstream does not strongly depend on the assumed turbulence power spectrum as particles must be, in all cases, in the weak scattering regime in order to escape through the free boundary. In the presence of a regular magnetic field (hereafter noted $B_0$), the escape upstream can however strongly depend on its obliquity, characterized by the parameter $\theta$ (the angle between $B_0$ and the normal to the shock),  and on the turbulence level (i.e. the ratio  $\delta B/B_0$).

\begin{figure}
{\rotatebox{0}{\includegraphics[scale=0.5]{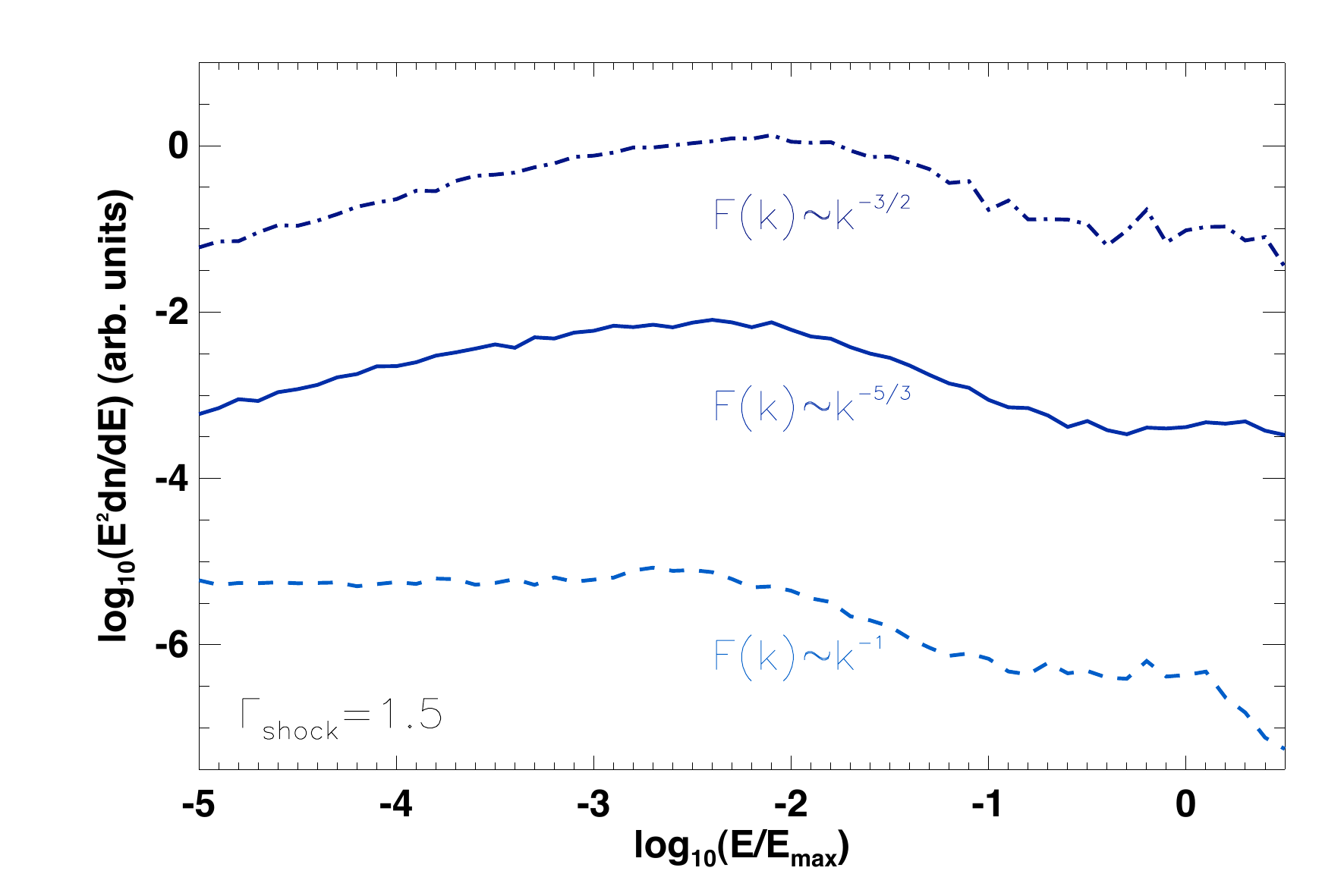}}}
\caption{Spectra (multiplied by $E^2$) of accelerated particles advected downstream of the shock ($\rm \Gamma_{sh}=1.5$) for three different turbulence power spectra : a Kolmogorov turbulence $\rm F(k)\propto k^{-5/3}$, a Kraichnan turbulence ($\rm F(k)\propto k^{-3/2}$) and a hard power spectrum ($\rm F(k)\propto k^{-1}$). } 
\label{fig:turbulence}
\end{figure}

Although we will exclusively use purely turbulent magnetic fields with a Kolmogorov power spectrum in our final calculations (with energy losses), we found it useful to discuss how predictions on the accelerated particle spectrum and acceleration time are affected when assuming different  background magnetic field structures. In the following, we consider the effect of  the turbulence power spectrum, the turbulence level and the presence of a regular magnetic field component with different obliquities. The spectra of accelerated particles advected downstream of the shock, for different type of turbulence, are shown in Fig.~\ref{fig:turbulence}. Three different turbulence power spectra are displayed (in each case, the assumed shock Lorentz factor is $\rm \Gamma_{sh}=1.5$): a Kolmogorov turbulence ($\rm F(k)\propto k^{-5/3}$), a Kraichnan turbulence ($\rm F(k)\propto k^{-3/2}$) and a hard power spectrum turbulence ($\rm F(k)\propto k^{-1}$). In all cases, the spectrum shows the same kind of curvature, as 
discussed above; the slope of the accelerated particles spectrum in the different regimes, and consequently the total available energy communicated to the highest energy particles, appears however to depend somewhat on the turbulence power spectrum. 

\begin{figure}
{\rotatebox{0}{\includegraphics[scale=0.5]{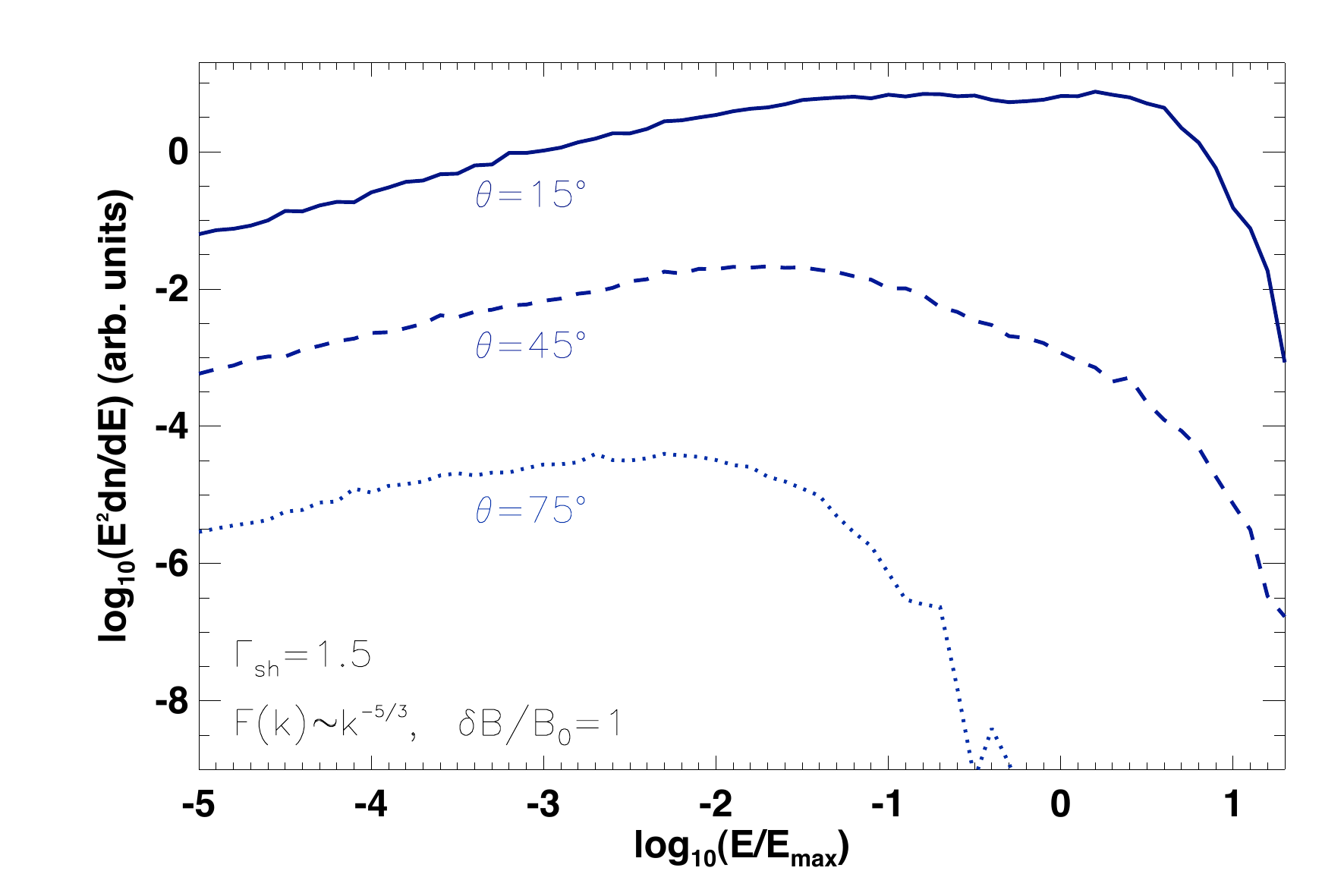}}}
\caption{Spectra (multiplied by $E^2$) of accelerated particles advected downstream, for the cases of shocks with Lorentz factor 1.5, Kolmogorov turbulence power spectrum, turbulence level $\rm \delta B/B_0=1$ and three different obliquities of the magnetic field $\rm \theta=15^\circ$, $\rm \theta=45^\circ$ and $\rm \theta=75^\circ$.}
\label{fig:spectrum_angle}
\end{figure}

We now turn to the case of the presence of a non negligible regular magnetic field $B_0$. We first consider a turbulence level $\delta B/B_0=1$, and three different obliquities $\rm \theta=15^\circ$ (quasi-parallel shock), $\rm \theta=45^\circ$, and $\rm \theta=75^\circ$ (quasi perpendicular shock). The results are displayed in Fig.~\ref{fig:spectrum_angle}. In the quasi-parallel shock configuration, the regime of efficient reflexion lasts longer than in the purely turbulent case due to the presence of the regular field. As a result, the spectrum of particles advected downstream is globally harder and the softening regime less pronounced. The two other cases ($\rm \theta=45^\circ$ and $\rm \theta=75^\circ$) show more pronounced and earlier softenings of the spectrum as one moves to increasingly superluminal (i.e, $\rm \beta_{up}/\cos\theta > 1$) shock configurations ($\rm \beta_{up}/\cos\theta \simeq 3$ (resp. 1.06) for $\rm \theta=75^\circ$ (resp. $\rm \theta=45^\circ$). Let us note in addition that the 
escape upstream (not displayed in Fig.~\ref{fig:spectrum_angle} is favored for the quasi-parallel shock case while highly suppressed (or even non existent) for $\rm \theta=45^\circ$ and $\rm \theta=75^\circ$. 

\begin{figure}
\centering{\includegraphics[scale=0.5]{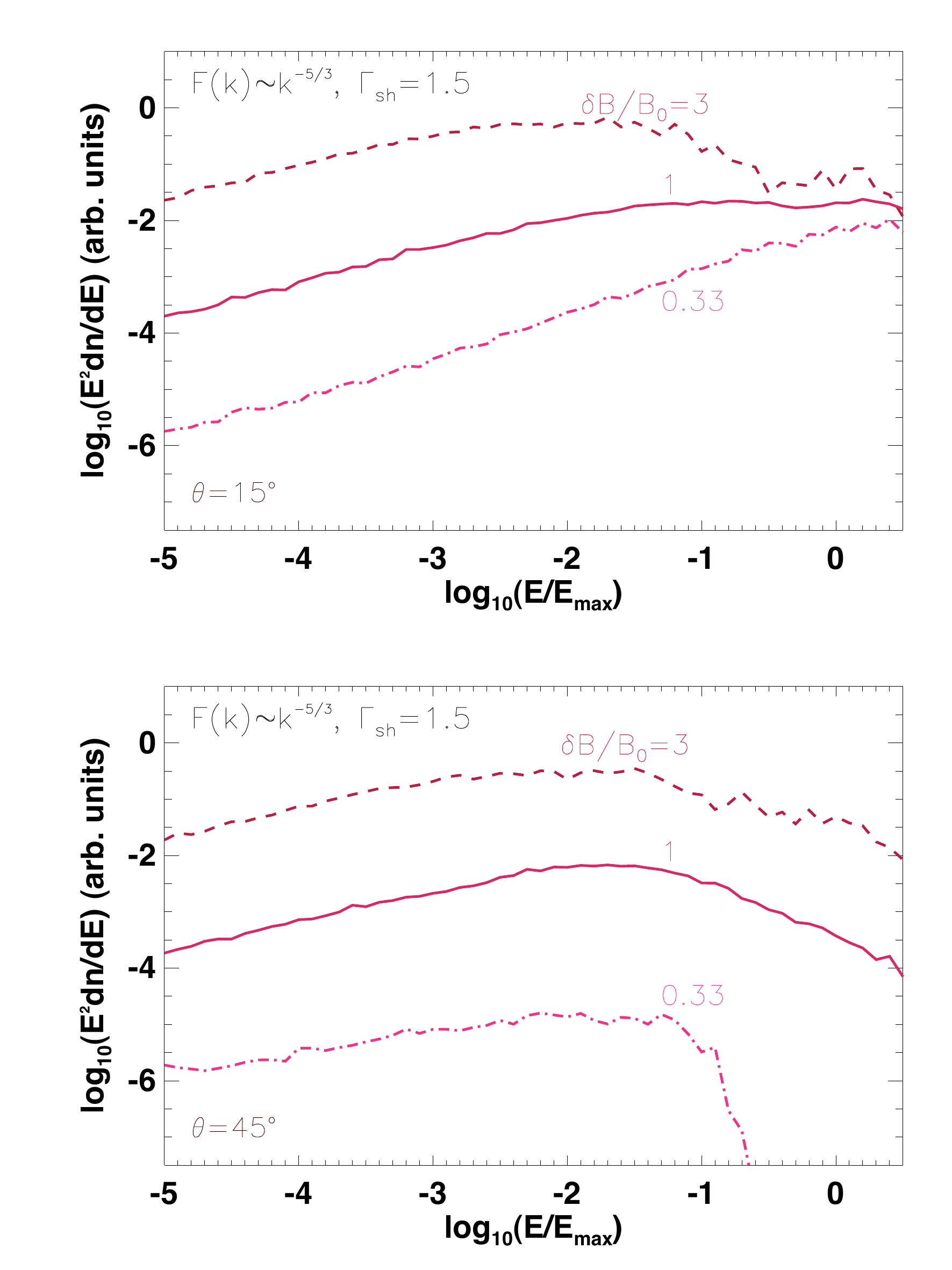}}
\caption{Same as Fig.~\ref{fig:spectrum_angle} Cases of regular fields with given obliquity are now studied for three different turbulence levels : $\delta B/B_0=0.33$, $\delta B/B_0=1$ and $\delta B/B_0=3$. Top : for a magnetic field obliquity of $\rm \theta=15^\circ$. Bottom : for a magnetic field obliquity of $\rm \theta=45^\circ$.}
\label{fig:Breg}
\end{figure}

In addition to the case $\delta B/B_0=1$, we considered two other turbulence levels, $\delta B/B_0=0.33$ and $\delta B/B_0=3$. The results are displayed in Fig.~\ref{fig:Breg}. For the weakest turbulence assumed, the trends observed for the $\delta B/B_0=1$ case are enhanced due to the suppression of the diffusion across field lines (which is favored when the turbulence is strong). For the quasi-parallel shock, the spectrum of particles advected downstream is hard and does not show any visible softening below $E_{\rm max}$, where the spectrum is finally broken due to the very efficient escape upstream. For $\rm \theta=45^\circ$, the spectrum shows an early softening and then a cut-off more pronounced than in the case $\delta B/B_0=1$.  For the highest level of turbulence ($\delta B/B_0=3$),  spectra obtained for obliquities $\rm \theta=15^\circ$, and $\rm \theta=45^\circ$ start to converge toward the purely turbulent case and the two cases look quite similar (this is also true to a lesser extent for the case 
of the quasi-perpendicular shock, $\theta=75^\circ$). The escape upstream is however still favored for the quasi-parallel shock configuration and somewhat suppressed for higher obliquities at this level of turbulence.

Overall, the trends we observe are in good agreement with the more complete study of cosmic-ray acceleration at mildly relativistic shock of Niemiec \& Ostrowski (2004). The fraction of the available energy that can be communicated to very-high or ultra-high energy particles appears to depend on the shock physical parameters;  the optimum case is reached when assuming a Kolmogorov or Kraichnan turbulence power spectrum. The effect of a regular component of the magnetic field is significant at low turbulence levels: the acceleration is more efficient at low obliquities while superluminal shock configurations result in much softer spectra with early cut-offs.

\subsubsection{Acceleration times} 

\begin{figure}
\centering{\includegraphics[scale=0.32]{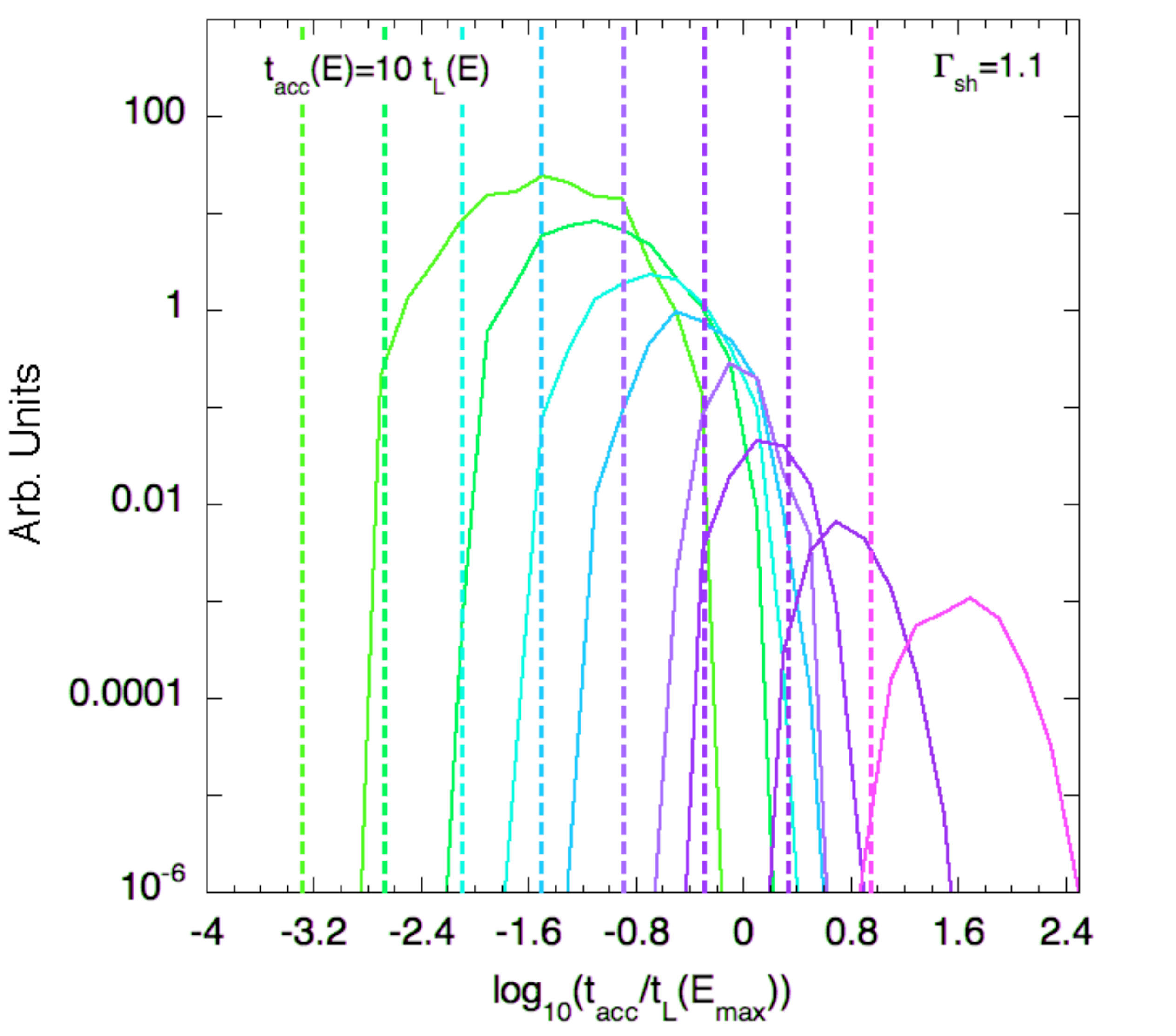}}
\centering{\includegraphics[scale=0.32]{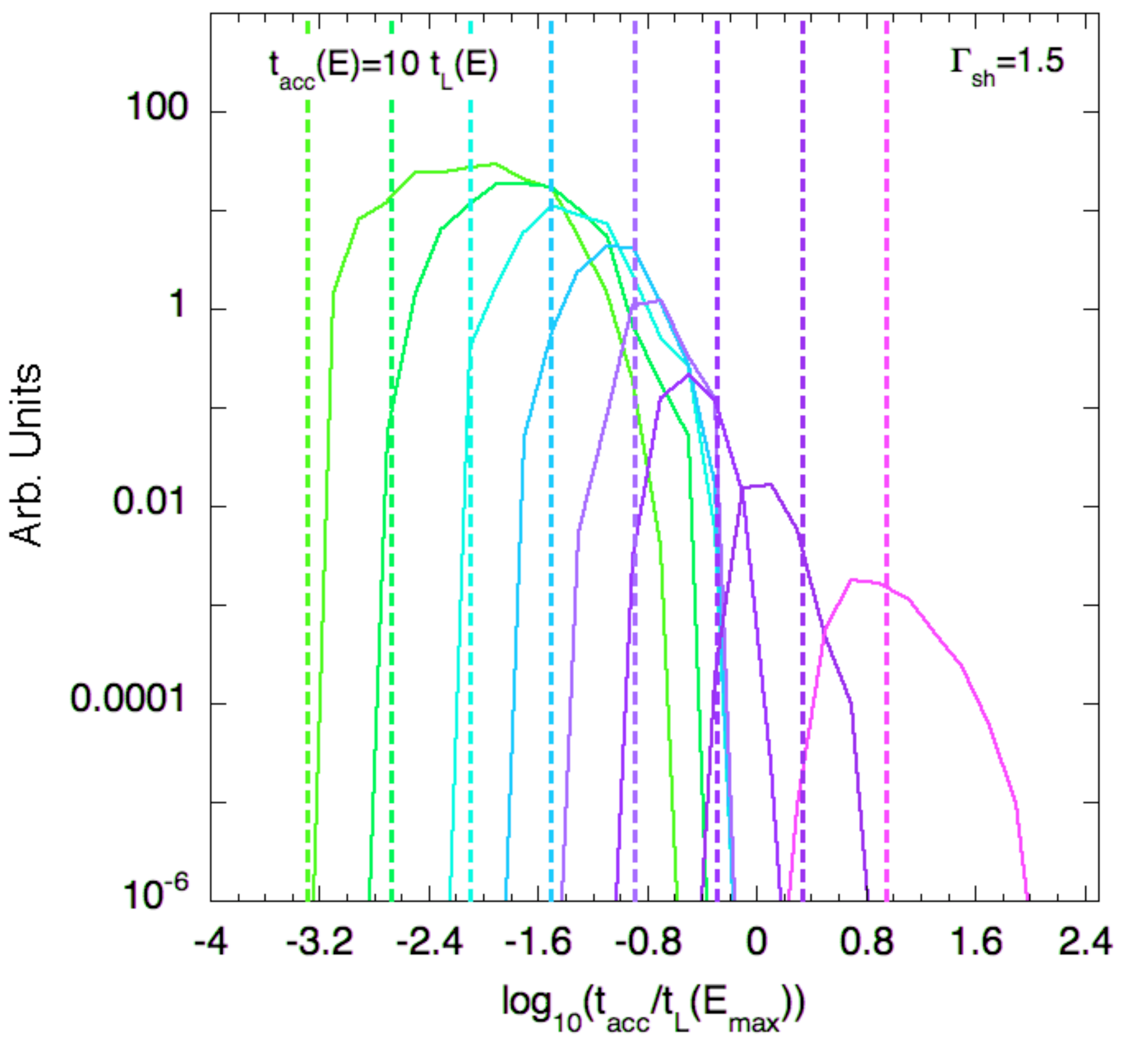}}
\caption{Distribution of acceleration times for different energy bins (in unit $E/E_{\rm max}$), from left to right : $10^{-4.3}$, $10^{-3.7}$, $10^{-3.1}$, $10^{-2.5}$, $10^{-1.9}$, $10^{-1.3}$, $10^{-0.7}$, $10^{-0.1}$. Each distribution is compared to the assumption $t_{\rm acc}({E})=10\times t_{\rm L}(E)$ (vertical lines). The two panels show the cases of  shocks with purely turbulent magnetic fields (with a Kolnogorov power spectrum) and Lorentz factor 1.1 (top) and 1.5 (bottom).}
\label{fig:tacc_distrib}
\end{figure}

The acceleration time is a crucial quantity for cosmic-ray acceleration, as in astrophysical sources, an acceleration process can only be efficient if the acceleration time is smaller or comparable to the energy loss or dynamical time scales. In this paragraph, we briefly discuss the acceleration times corresponding to the spectra we previously obtained. These acceleration times will be used in the next section to make preliminary estimates of the expected maximum energy reachable at GRBs internal shocks and its dependence on various physical parameters. 

\begin{figure}
\centering{\includegraphics[scale=0.37]{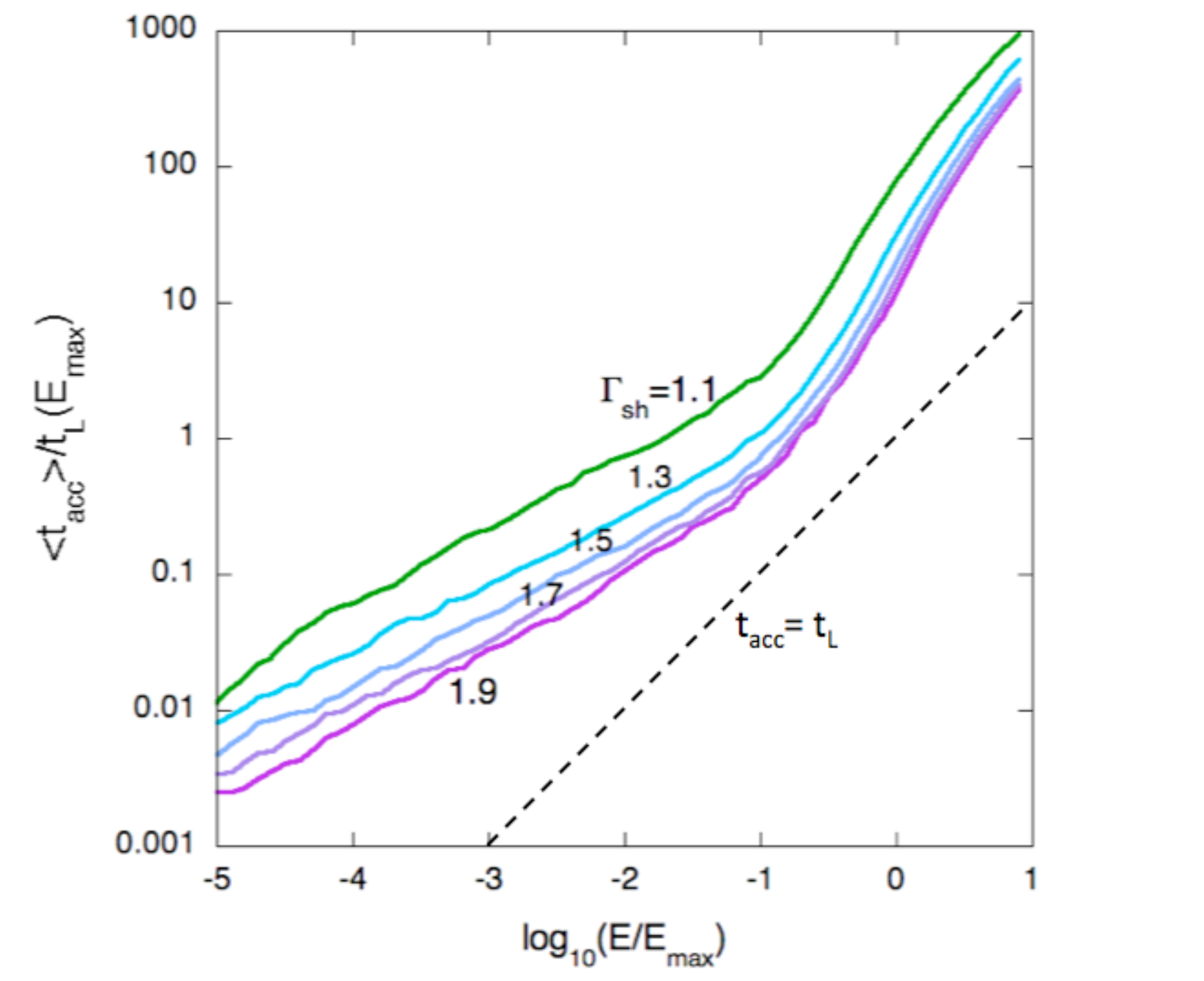}}
\caption{Energy evolution of the average acceleration time for shocks with Lorentz factors between 1.1 and 1.9 and purely turbulent magnetic fields (with a Kolmogorov power spectrum).}
\label{fig:acc_time_gamma}
\end{figure}

We calculate the acceleration time by computing the quantity $E\Delta T/(E_f-E_i)$ (where $E_i$ and $E_f$ are the energies at the beginning and at the end of a cycle respectively, $E=(E_i+E_f)/2$ and $\Delta T$ is the duration of a cycle), for each individual complete cycle downstream$\rightarrow$upstream$\rightarrow$downstream of each simulated particles. We then calculate the mean value in each energy bin to obtain $<t_{acc}(E)>$.\\
We first consider the case of shocks with various Lorentz factors between 1.1 and 1.9 and purely turbulent magnetic fields with a Kolmogorov power spectrum. The energies are given in $E_{\rm max}$ units and the acceleration times in $t_{\rm L}({E_{\rm max}})$ units.  The distribution of acceleration times for different energy bins is shown in Fig.~\ref{fig:tacc_distrib} for two different shock Lorentz factors ($\rm\Gamma_{sh}=$1.1 and 1.5) and compared with the acceleration time expected under the assumption (often used in cosmic-ray acceleration calculations) of proportionality between the acceleration time and the Larmor time of the particle: $t_{\rm acc}({E})=\kappa_0 t_{\rm L}(E)$, with $\rm \kappa_0=10$ in Fig.~\ref{fig:tacc_distrib}. 
Moreover, the energy evolution of the mean value of the acceleration time, $<t_{acc}>$, shown in Fig.~\ref{fig:acc_time_gamma} (for shock Lorentz factors between 1.1 and 1.9 as in Fig.~\ref{fig:all_gamma}) is well described by a broken power law, $<t_{acc}>$ increasing faster as the energy approaches $E_{\rm max}$. One can see, however, a slight softening at the highest energies due to the fact that we considered only in these plots particles that end up being advected downstream. Above $E_{\rm max}$ an increasing fraction of the particles manage to escape through the boundary layer upstream and are not caught back by the shock. Only those which did not escape upstream contribute to the average acceleration time at these energies, explaining the softening\footnote{We checked, in one case, that this softening was removed by placing the free boundary much further away. This procedure is however costly in computation time as high energy particles can spend a very long time upstream before being caught by the 
shock front.}. In addition, one sees that the acceleration time gets shorter for larger Lorentz factors. 
\begin{figure}
\centering{\includegraphics[scale=0.32]{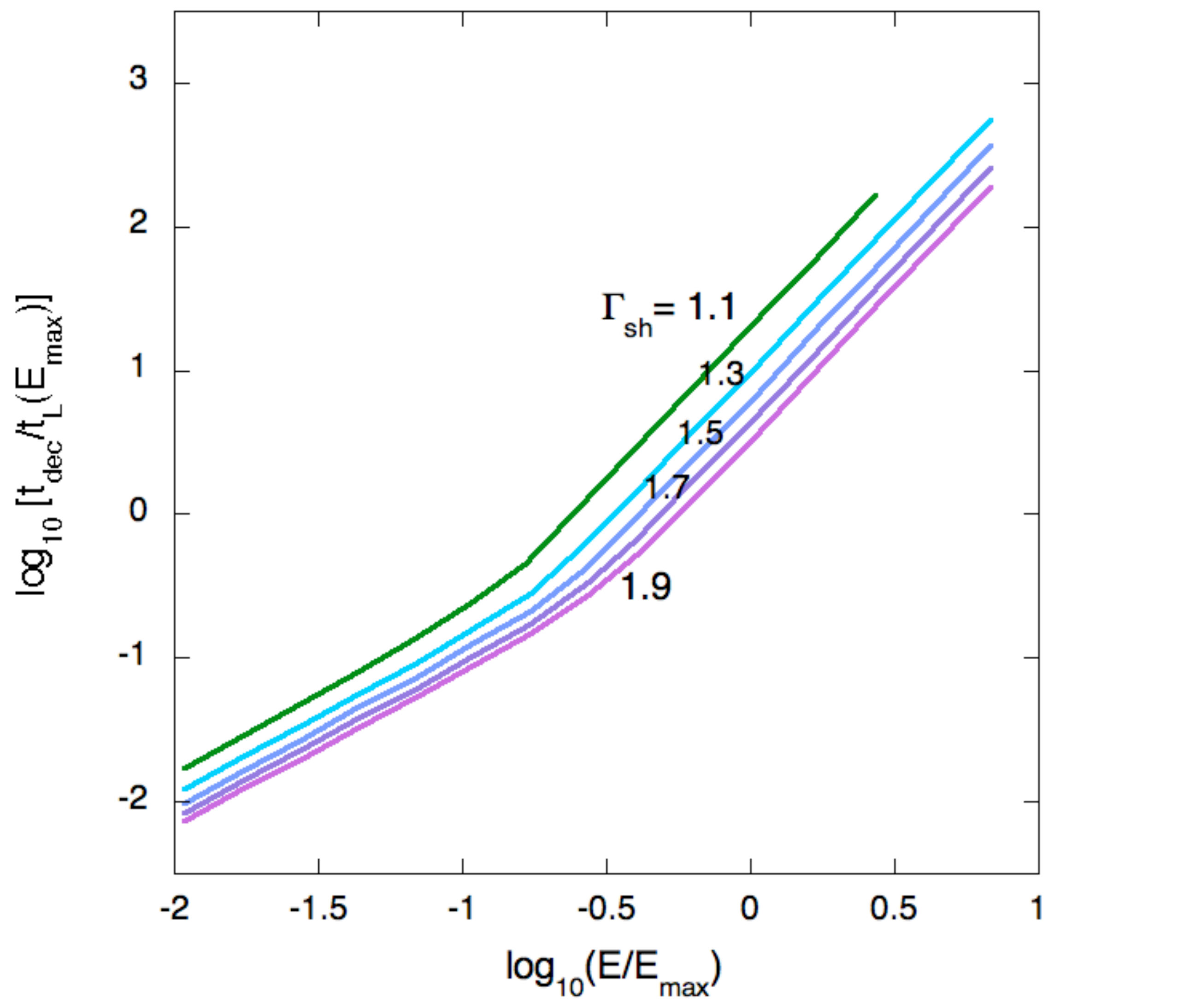}}
\caption{Energy evolution (in units of $E_{\rm max}$) of the average time $t_{\rm dec}$ (in units $t_{\rm L}(E_{\rm max})$) for a particle to suffer deflection of an angle $\beta_{\rm sh}$ (corresponding to the shock velocity upstream) from its initial direction. Different shock Lorentz factors are considered : $\Gamma_{\rm sh}=$1.1, 1.3, 1.5, 1.7 and 1.9.}
\label{fig:decorel}
\end{figure}
A simple qualitative understanding of these trends can be obtained by noticing that (at least  for the cases we studied, with a compressed magnetic field in the downstream medium) particles spend on average more time in the upstream medium (although the average time spent downstream is of the same order of magnitude). The shorter acceleration time at a given energy for larger shock Lorentz factors comes, to some extent\footnote{The larger average energy gain at each cycle also contributes to the shortening of the acceleration time for larger shock Lorentz factors.}, from the fact that, in the upstream frame, a smaller deflexion with respect to the initial direction is needed for a particle to be 
caught back by the shock. One can relate the behavior observed in Fig.~\ref{fig:acc_time_gamma} to the energy evolution of the average time for a particle to be deflected by an angle $\rm \beta_{sh}$ (corresponding to $\Gamma_{\rm sh}$ between 1.1 and 1.9) from its initial direction shown in Fig.~\ref{fig:decorel}. Similar broken power laws are observed, the inflection is located approximately in the same energy range and marks the transition toward the weak scattering regime ($r_{\rm L}(E)\gtrsim \lambda_c$, with $\lambda_c$ the coherence length of the magnetic field\footnote{The coherence length $\lambda_c$ is in general a fraction (typically a few tenth) of the maximum turbulence scale $\rm \lambda_{max}$. The numerical relation between $\lambda_c$ and  $\lambda_{\rm max}$ depends on the turbulence power spectrum ($\lambda_c\simeq 0.2\times \lambda_{\rm max}$ for a Kolmogorov power spectrum) }). 

\begin{figure}
\centering{\rotatebox{0}{\includegraphics[scale=0.32]{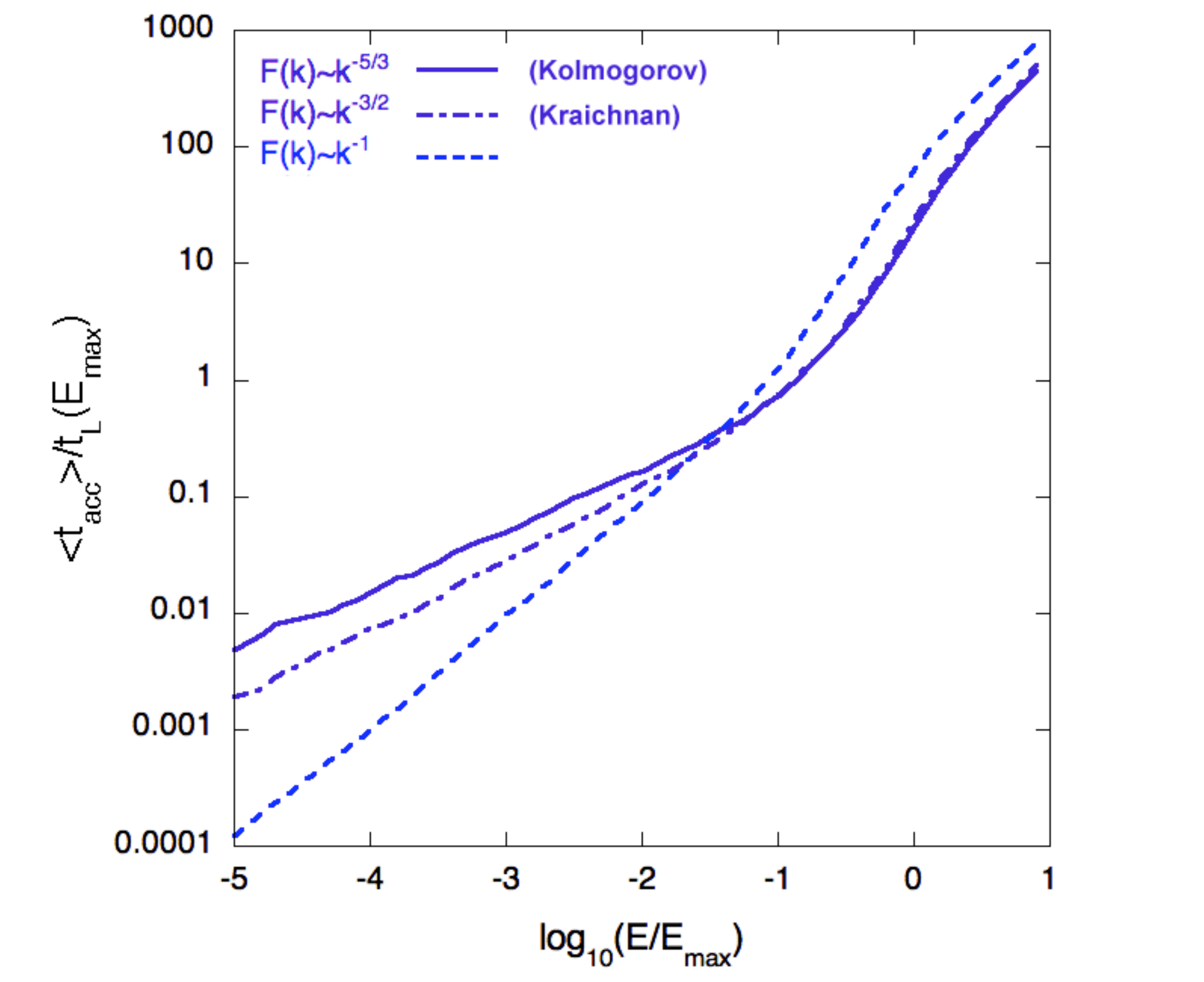}}}
\caption{Energy evolution of the average acceleration time for shocks with Lorentz factor 1.5 and a purely turbulent magnetic field. Three different cases of turbulence power spectrum are displayed : Kolmogorov  ($\rm F(k)\propto k^{-5/3}$),  Kraichnan ($\rm F(k)\propto k^{-3/2}$) and a hard power spectrum ($\rm F(k)\propto k^{-1}$).}
\label{fig:acc_time_turbu}
\end{figure}

The energy evolution of the acceleration time we observe is in any case far from showing a linear relation with the Larmor time, $t_{\rm acc}({E})=\kappa_0 t_{\rm L}(E)$, as already emphasized in Gialis \& Pelletier (2003). At the energy $E=E_{\rm max}$, $\rm\kappa_0\simeq13$ for $\rm \Gamma_{sh}=1.9$,  $\simeq20$ for $\rm \Gamma_{sh}=1.5$ and $\simeq80$ for $\rm \Gamma_{sh}=1.1$. Let us note that the energy evolution of the acceleration time does not depend much on the turbulence power spectrum in the weak scattering regime, as illustrated in Fig.~\ref{fig:acc_time_turbu}, that displays the energy evolution of the mean value of the acceleration time $<t_{\rm acc}>$ for Kolmogorov, Kraichnan and hard turbulence power spectra (for $\rm \Gamma_{sh}=1.5$ as in Fig.~\ref{fig:turbulence}). This is, of course, not the case at lower energy where particles have a resonant interaction with the magnetic turbulence.
\begin{figure}
\centering{\rotatebox{0}{\includegraphics[scale=0.32]{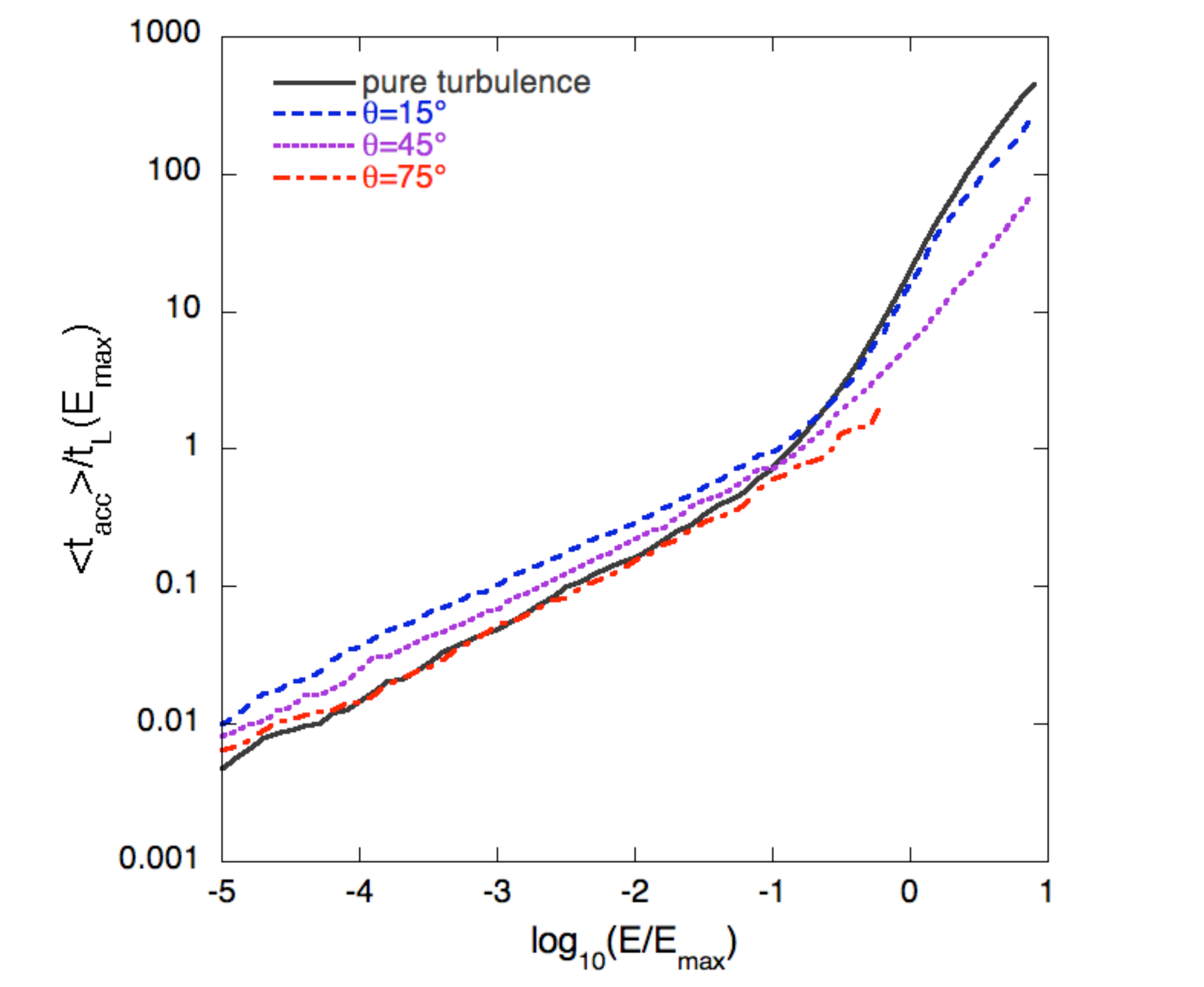}}}
\caption{Energy evolution of the average acceleration time for shocks with Lorentz factor 1.5. The case of a purely turbulent magnetic field is compared with cases including regular magnetic fields ($\rm \delta B/B_0=1$) with different obliquities ($\rm \theta=15^\circ$, $\rm \theta=45^\circ$ and $\rm \theta=75^\circ$)}
\label{fig:acc_time_angle}
\end{figure}

Shorter acceleration times can be obtained at higher energies for oblic ($\rm \theta=45^\circ$) or quasi-perpendicular ($\rm \theta=75^\circ$) shock configurations, as shown in Fig.~\ref{fig:acc_time_angle}. This is due to the fact that an oblic regular field leads, in particular, to a shorter time spent in the upstream frame (the opposite is true for the quasi-parallel shock $\rm \theta=15^\circ$). The price to pay for these shorter acceleration times is however to get much steeper spectra at high energy (as seen in the previous paragraph and in Figs.~\ref{fig:spectrum_angle} and  \ref{fig:Breg}).

\subsubsection{Comments on the particles escape} 

\begin{figure}
\centering{\rotatebox{0}{\includegraphics[scale=0.32]{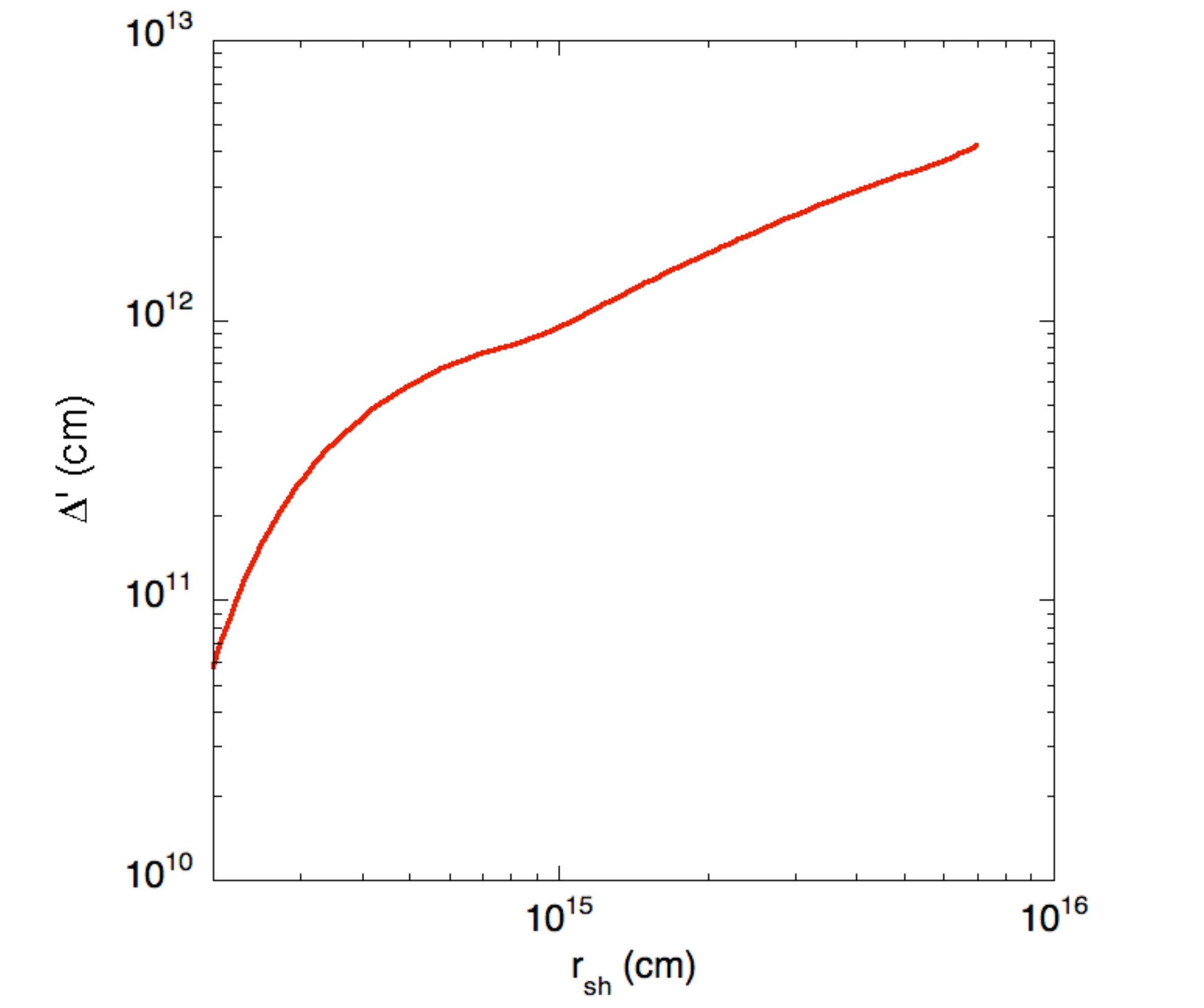}}}
\caption{Comoving thickness of the downstream medium as a function of the shock radius estimated in the central object rest frame (courtesy of Fr\'ed\'eric Daigne).}
\label{fig:epaisseur}
\end{figure}

Particles escape is a very important ingredient of cosmic-ray acceleration; in the previous paragraphs we treated the escape of particles upstream and downstream of the shock in different ways. For the escape upstream, a free boundary located at a distance from the shock front proportional to the maximum turbulence scale was assumed. We will still use this description in the following, assuming that the turbulent field upstream is produced by the accelerated cosmic-rays themselves and does not extend further away than a few times the Larmor radius of the highest energy accelerated particles from the shock at a given time. As seen in the previous paragraph, a free boundary escape upstream of the shock acts naturally as a high pass filter, since the particles must outrun the shock long enough to reach the boundary, which implies that particles must be in the weak scattering regime ($E\simeq E_{\rm max}$ and above).

In the downstream frame, in the absence of energy losses (and under the assumption of stationarity), the escape of cosmic-rays has no reason to behave as a high pass filter since particles are able to drift away from the shock at all energies and should, in principle, be able to reach a free boundary at an arbitrary location in the downstream medium after some time (that can be very long for low energy particles). To treat the escape downstream of the shock in the previous calculations, we used a residence time criterion (see above) instead of a free boundary placed at a given point of space in order to save computational time. In the absence of energy losses the two approaches should lead to identical results as soon as the residence time criterion is well chosen. This is no longer true, however, in the presence of energy losses. In this case, the location of the free boundary needs to be specified explicitly as the time needed to reach the boundary must be shorter than the energy loss time scale in order 
for a particle at a given energy to escape before being cooled. As we will see in the following, the energy loss processes at GRB internal shocks are expected to affect particles at all energies, so it is likely that the escape of particles downstream of the shock also behaves as a high pass filter.
The modeling we used in Sect.~2 to describe GRBs internal shocks and calculate SEDs does not allow to compute the thickness of the shocked region (i.e, the downstream medium) and its time evolution. A hydrodynamical treatment is required to compute this quantity. For that purpose, we used the hydrodynamic code implemented by Daigne \& Mochkovitch (2000) using the same Lorentz factor distribution as in Sect.~2. The comoving thickness of the shocked medium (i.e, the distance between the shock front and the boundary of the shocked medium) as a function of the shock radius $r_{\rm sh}$ (in the central source frame) is displayed in Fig.~\ref{fig:epaisseur}, we will use this estimate of the downstream medium thickness for our calculations of CR acceleration in the presence of energy losses in the next sections.

\section{Modeling of the energy losses and preliminary estimate of the maximum energy reachable}

\subsection{Cosmic-rays energy losses during the prompt phase}

We now turn to the estimate of cosmic-ray energy losses during the GRB prompt emission phase which we assume to be related to internal shocks. As mentioned earlier, the injection and survival of heavy nuclei (iron) in GRB jets has been studied in detail by Horiuchi et al. (2012). Nuclei can be injected at the base of the jet directly from the disk surrounding the central newly formed black hole. They can also be entrained via instabilities at the boundary of the jet during its propagation throughout the star. They can finally be synthesised within the jet itself if nucleons can condensate, forming first $\alpha$ particles and then heavier nuclei (Metzger et al. 2011).
Nuclei can be destroyed by spallation or photodisintegration reactions. Spallation occurs in regions where a strong gradient exists between the jet and its surrounding so that collisions between nuclei and energetic protons are possible. Photodisintegration can take place in regions where the photons density above a few MeV (in the nucleus rest frame) is sufficiently large. It can be the case in the inner jet or conversely at large distance where dissipation occurs and the GRB prompt emission photons are produced. Photodisintegration at the origin of the jet can be strongly suppressed in the case of magnetic acceleration since the entropy is much smaller than for thermal acceleration (the internal shock model actually favours magnetic acceleration in order to limit the brightness of the photospheric emission). Photodisintegration by the GRB prompt emission photons is discussed in detail in the following sections.\\ We will simply do the following :\\
({\it i}) assume that nuclei can be present at a significant level in the relativistic wind composition at the beginning of the internal shock phase;\\
({\it ii}) study their survival during their acceleration at internal shocks;\\
({\it iii}) study their capability to escape from the acceleration site.\\

\begin{figure}
\centering{\rotatebox{0}{\includegraphics[scale=0.32]{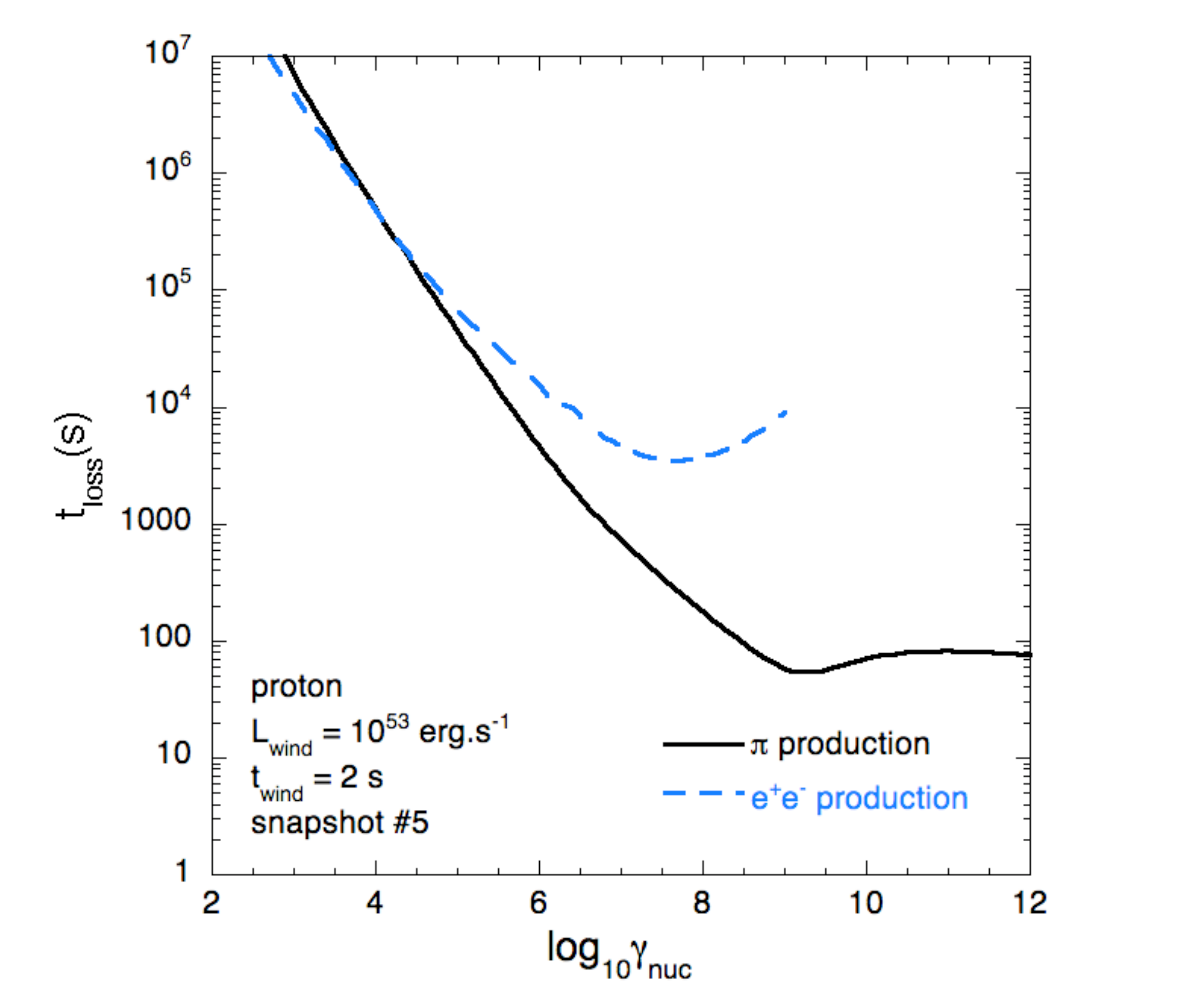}}}
\caption{Lorentz factor evolution of protons energy loss times for the pair production and pion production processes, the case of the snapshot 5 ($r_{\rm sh}=6.9\,10^{14}\,\rm cm$) for model A and $L_{\rm wind}=L_{\rm wind}^{\rm eq}=10^{53}\,\rm erg\,s^{-1}$ is shown.}
\label{fig:protloss}
\end{figure}

There are several mechanisms responsible for the energy losses of cosmic-ray protons and nuclei during their acceleration at GRBs internal shocks. Cosmic-rays can loose energy through adiabatic processes due to the expansion of the wind, by synchrotron radiation in the very intense magnetic fields that may be at work during internal shocks (see Sect.~2), by photo-interactions with the photons of the prompt emission or by hadronic interactions with the baryonic component of the wind. Concerning photo-interactions of protons and nuclei, the main processes are the same as those encountered in UHECR intergalactic propagation. The $\rm e^+e^-$ pair production is the significant photo-interaction process with the lowest photon energy threshold ($E_\gamma \sim 1$ MeV) in the proton or nucleus rest frame (PRF or NRF). We treat this process as a continuous energy loss mechanism following Rachen (1996) (in particular for the scaling with the mass number $A$ and the charge $Z$ of the 
nucleus). The spectra of the produced $\rm e^+e^-$ pairs are estimated following Kelner \& Aharonian (2008).  In the case of protons, photomeson production (with the dominant contribution of pion production) starts to dominate at $E_\gamma \sim 140$ MeV in the PRF. In the following, to model this process, we use the SOPHIA event generator presented in M\"{u}cke et al. (2000). This generator allows a precise Monte-Carlo treatment of the proton energy losses as well as the production of secondary particles for each interaction. The energy loss (or attenuation) time of protons as a function of their Lorentz factor in the shocked medium is given in Fig.~\ref{fig:protloss}, for the pair production and photomeson processes for a GRB of wind luminosity $L_{\rm wind}^{\rm eq}=10^{53}\,\rm erg\,s^{-1}$ for the snapshot \#5 ($r_{\rm sh}\simeq 6.9\times 10^{14}$ cm, see Sect.~2).

\begin{figure}
\centering{\rotatebox{0}{\includegraphics[scale=0.32]{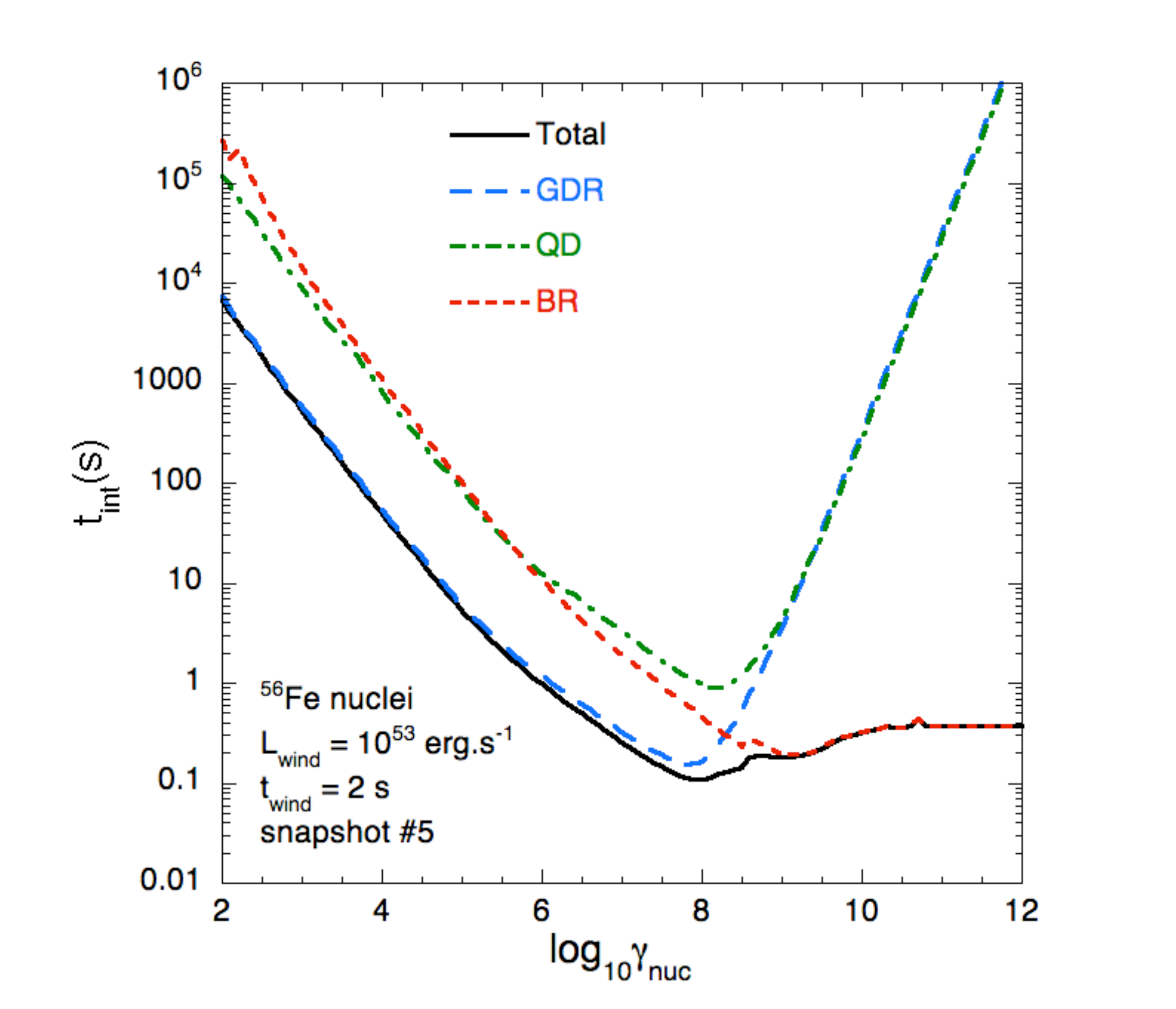}}}
\centering{\rotatebox{0}{\includegraphics[scale=0.32]{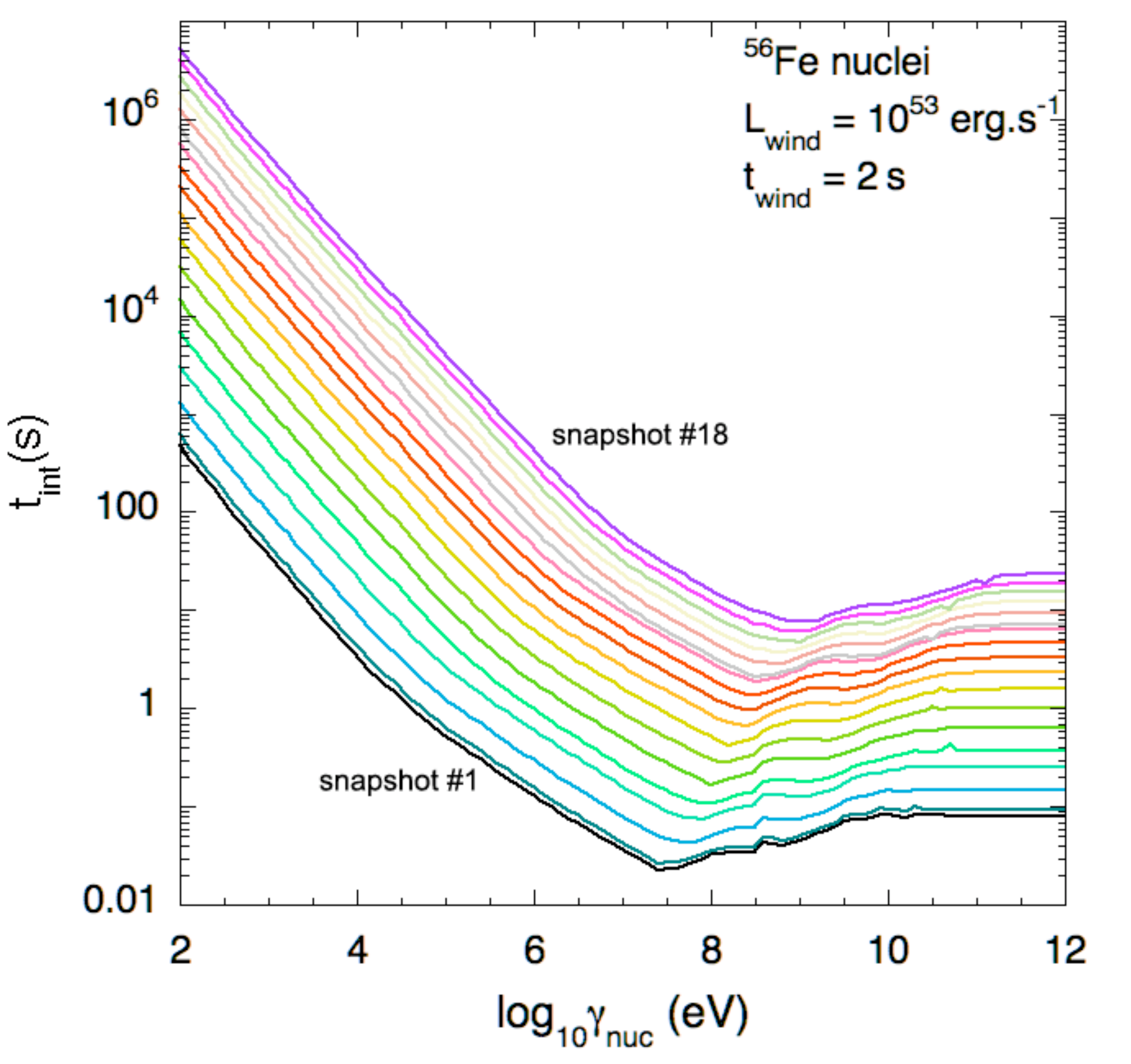}}}
\caption{Top: Lorentz factor evolution of Fe nuclei interaction times for the main photodisintegration  processes (GRD, QD and BR); the case of the snapshot 5 ($r_{\rm sh}=6.9\,10^{14}\,\rm cm$ for $L_{\rm wind}=L_{\rm wind}^{\rm eq}=10^{53}\,\rm erg\,s^{-1}$ and model A is shown. Bottom: Lorentz factor evolution of Fe nuclei total interaction time for photodisintegration for the 18 snapshots discretizing the shock propagation ($L_{\rm wind}\equiv L_{\rm wind}^{\rm eq}=10^{53}\,\rm erg\,s^{-1}$, model A).}
\label{fig:ironphotoprocess}
\end{figure}

In the case of nuclei, above the pair production threshold, different photodisintegration processes become dominant at different photon energies in the NRF. The lowest energy and highest cross section process is the giant dipole resonance (GDR). The GDR is a collective excitation of the nucleus in response to electromagnetic radiation between $\sim$ 10 and 50 MeV where a strong resonance can be seen in the photoabsorption cross section (e.g., Khan et al. 2005). The GDR mostly triggers the emission of one nucleon (most of the time a neutron but depending on the structure of the parent nucleus, $\rm \alpha$ emission can also be strong for some nuclei), two, three and four nucleons channels can also contribute significantly though their energy threshold is higher. In the following, the different channels of the GDR are modeled using theoretical calculations from Khan et al. (2005) and parametrizations for nuclei with mass $\rm A \leq 9$ from Rachen (1996). Around 30 MeV in the nucleus 
rest frame and up to the photopion production threshold, the quasi-deuteron (QD) process becomes comparable to the GDR (but much lower than at the peak of the resonance) and its contribution dominates the total cross section at higher energies. Unlike GDR, QD is predominantly a multi-nucleon emission process. The photopion production (or baryonic resonances (BR)) of nuclei becomes relevant above $\sim$ 150 MeV in the nuclei rest frame. The large excitation energy usually triggers the emission of several nucleons in addition to a pion that might be reabsorbed before leaving the nucleus. Below 1 GeV, the cross section is in good approximation proportional to the mass (this scaling slowly breaks above 1 GeV due to nuclear shadowing). The quasi-deuteron and pion production (baryonic resonances) processes are calculated following Rachen (1996), for the cross section inelasticities (nucleon yields) and secondary particle emission. Let us note that, as in the case of photomeson production for protons, the three 
main photodisintegration processes for nuclei are treated stochastically in the following sections. The contribution of the different processes to the photodisintegration interaction time, as a function of the Lorentz factor of an iron nucleus, is displayed in the upper panel of Fig.~\ref{fig:ironphotoprocess} (still for $L_{\rm wind}^{\rm eq}=10^{53}\,\rm erg\,s^{-1}$, snapshot \#5). One can see that the GDR process dominates up to relatively large Lorentz factors before the BR process takes over. In the following, we will actually never reach Lorentz factors large enough for the BR process to have a dominant contribution to nuclei photodisintegration and secondary particle production.  The Lorentz factor evolution of the total interaction time, as the shock propagates in the wind (i.e, for the 18 snapshots we consider), is displayed in the lower panel.  

Although not as significant as photo-interactions, hadronic interactions cannot be neglected in the case of large wind luminosities (namely model B and C in Sect.~2.) To model hadronic interactions we use the EPOS 1.99 event generator (Werner et al.  2006; Pierog \& Werner 2009) and assume that protons and nuclei interact with a baryon background essentially made of protons (this assumption is however not critical for the results discussed in the following sections). As the EPOS model does not treat the fragmentation/evaporation of the residual nucleus after the interaction (but instead divides the parent nucleus into participant and spectator nucleons) an addition algorithm must be added to estimate the mass and charge of the daughter particle after interaction. For that purpose, we reproduced the algorithm used in the air shower simulator CORSIKA (Heck et al., 1998) based on the works on fragmentation/evaporation of nuclei in high energy collisions by Campi \& Hufner (1981) and  Gaimard (1990). 

As seen in Sect.~2, the magnetic fields invoked in the framework of the internal shock model can be extremely strong (sometimes several tens of kilogauss) and become important source of energy losses during the acceleration of protons and nuclei through  synchrotron radiation. The synchrotron loss time of protons and nuclei is given by

\begin{equation}
t_{\rm syn} (E, M, Z)=\left(\frac{M}{m_{\rm e}}\right)^4\frac{1}{Z^4}\frac{6\pi m_{\rm e}^2 c^3}{\sigma_T B^2 E}=\left(\frac{M}{m_{\rm e}}\right)^4\frac{1}{Z^4}t_{\rm syn}({\rm e^-}, E)\,\,\rm s,
\label{eq:tsyn}
\end{equation}
where $E$, $M$ and $Z$ are the energy, mass and charge of the nucleus, $\sigma_T$ the Thomson cross section, $m_{\rm e}$ the mass of the electron and $t_{\rm syn}({\rm e^-}, E)$ the synchrotron loss time of an electron at the same energy. Moreover, the typical energy of the emitted photons by a nucleus of mass $M$, charge $Z$ and Lorentz factor $\rm \gamma$ is
\begin{equation}
E_{\rm syn}(\gamma, M, Z)\simeq 1.75\times 10^{-8}\gamma^2Z\left(\frac{B}{G}\right)\left(\frac{M}{m_{\rm e}}\right)^{-1}\,\rm eV.
\label{eq:tsyne}
\end{equation}

\subsection{Preliminary estimates of the maximum energy reachable}

\begin{figure}
{\rotatebox{0}{\includegraphics[scale=0.40]{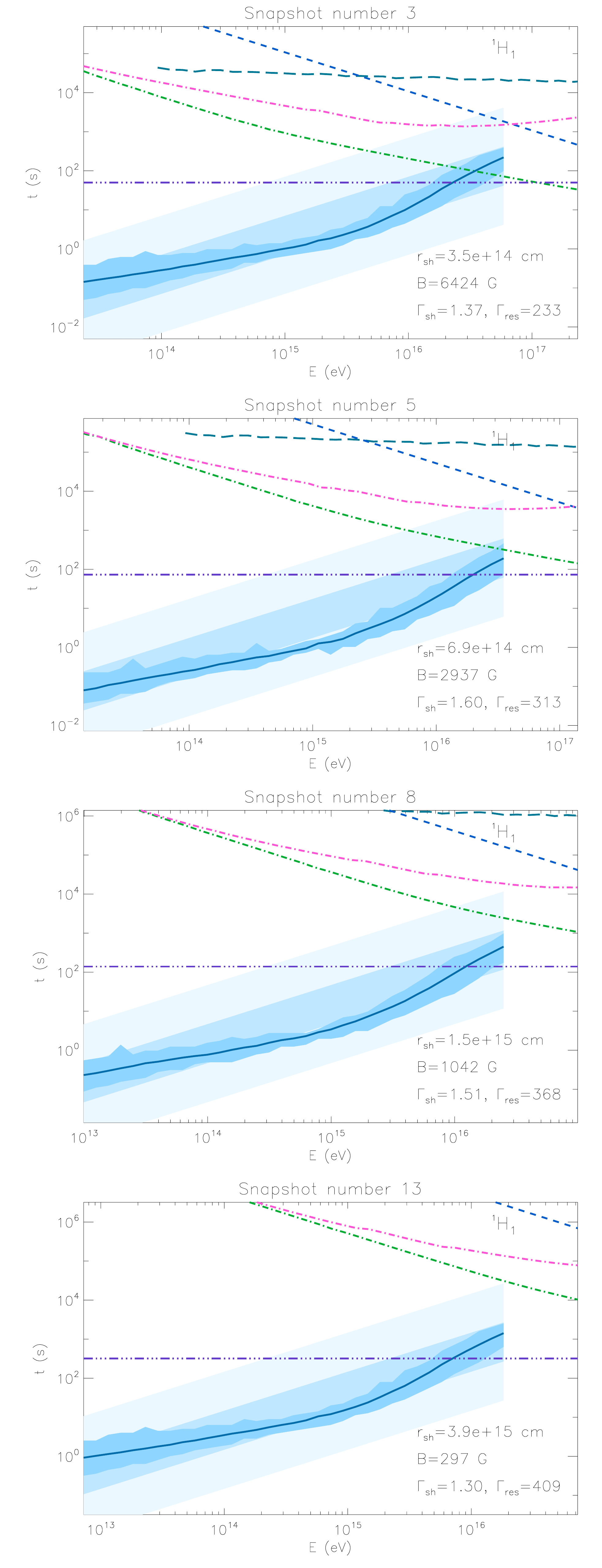}}}
\caption{Energy evolution of the competition between the acceleration time and the energy loss times for the most significant energy loss processes (see caption of Fig.~\ref{fig:legend} for more details) at different stages of the shock propapagation (snapshots 3, 5, 8 and 13) for protons,  $L_{\rm wind}\equiv L_{\rm wind}^{\rm eq}=10^{53}\,\rm erg\,s^{-1}$, model A.}
\label{fig:snapP}
\end{figure}

\begin{figure}
{\rotatebox{0}{\includegraphics[scale=0.40]{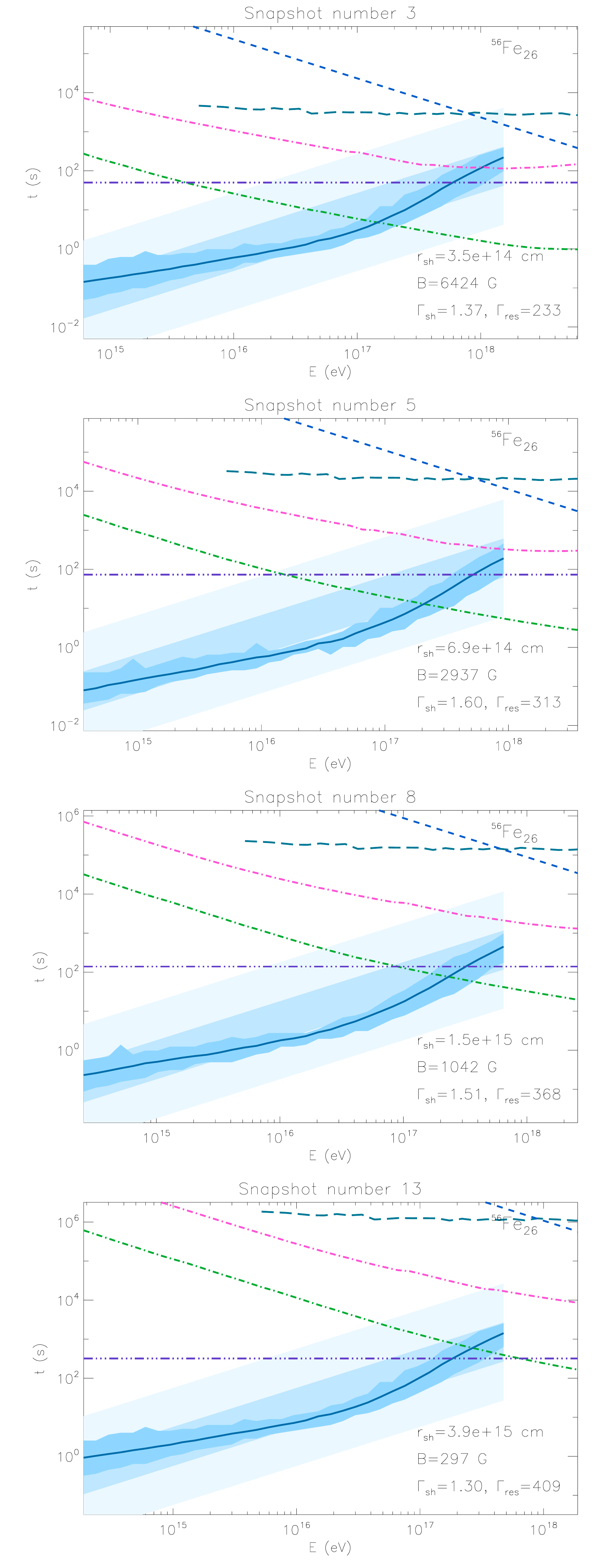}}}
\caption{Same as Fig.~\ref{fig:snapP} in the case of Fe nuclei.}
\label{fig:snapFe}
\end{figure}

\begin{figure}
{\rotatebox{0}{\includegraphics[scale=0.3]{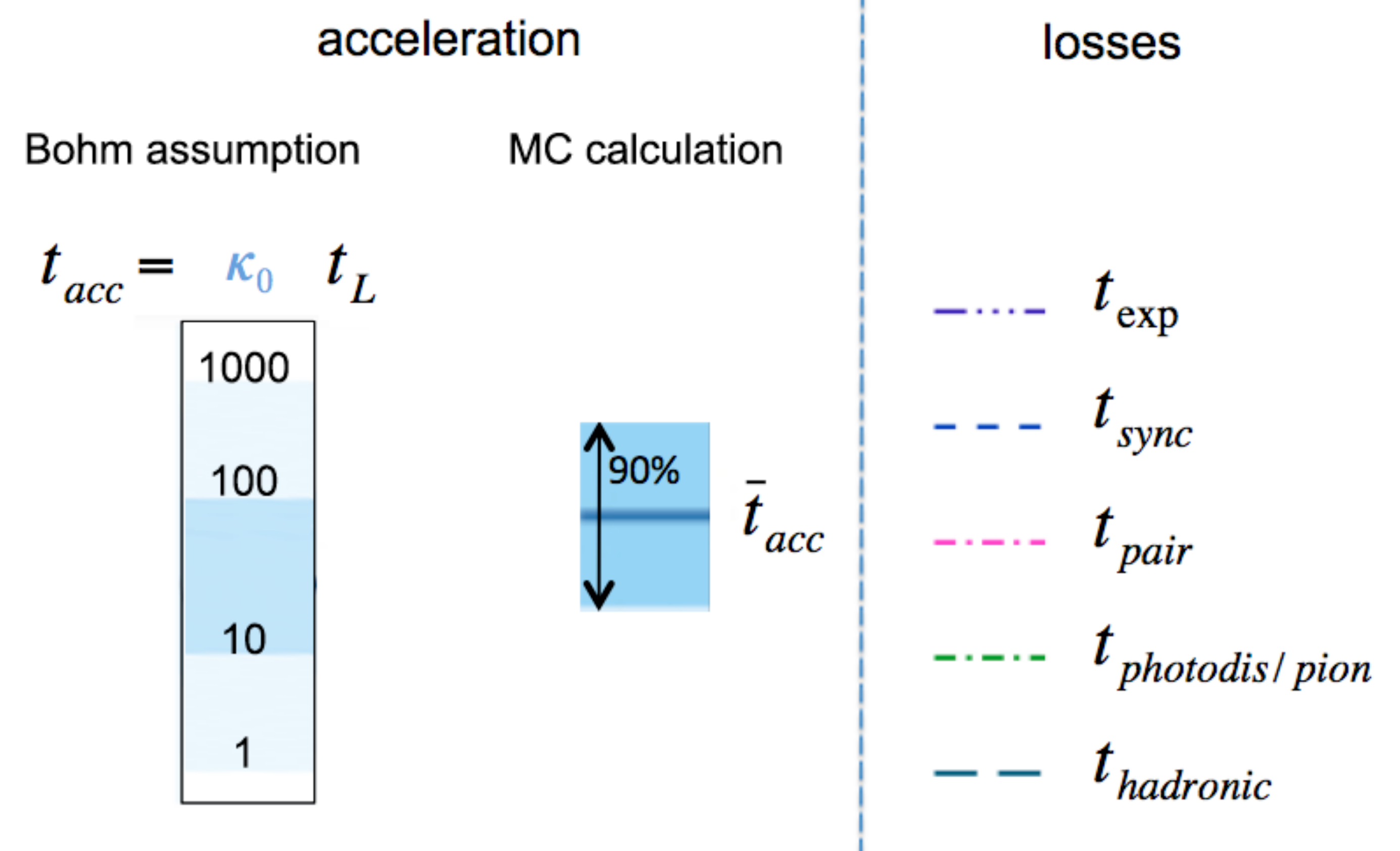}}}
\caption{Legend of Figs.~\ref{fig:snapP}, \ref{fig:snapFe} and \ref{fig:snapB}. The different broken lines represent the energy evolution of the loss time for the main energy loss processes : adiabatic losses, synchroton, pair production, pion production (for protons) or photodisintegration (for nuclei) and hadronic interactions. The heavy line shows the energy evolution of the acceleration time deduced from Sect.~3 (up to $E=3\times E_{max}$) and the dark shaded area the associated 90\% spread (see Fig.~\ref{fig:tacc_distrib}). Lighter shaded areas delimited by the assumptions $t_{\rm acc}({E})=\kappa_0\times t_L(E)$ with $\rm \kappa_0=$1, 10, 100 and 1000 (from bottom to top) are also shown for comparison.}
\label{fig:legend}
\end{figure}

\subsubsection{Competition between acceleration and energy losses}
Using our modeling of GRBs internal shocks introduced in Sect.~2, we are able to estimate the energy loss times for each of the above-mentioned processes and their evolution during the shock propagation (i.e, for each snapshot) for the different energy redistribution models and the different burst luminosities. In addition, estimating the important physical quantities of internal shocks (such as the magnetic field $B$ or the shock Lorentz factor $\rm \Gamma_{sh}$) allows us to normalize our results
on cosmic-ray acceleration at mildly relativistic shocks, given in units of $E_{\rm max}$ in Sect.~3, in terms of energy units (eV). In particular, the acceleration times we obtained can be compared to energy loss times in order to estimate the maximum energy reachable during the shock propagation for the different models and luminosities considered in Sect.~2. A supplementary hypothesis about the maximum length scale of the turbulent field, $\rm \lambda_{max}$, is needed in order to translate the acceleration times we obtained in Sect.~3 (in unit of $t_{\rm L}({E_{\rm max}})$) into energy units. We will first arbitrarily assume that $\rm \lambda_{max}$ is a small fraction of $r_{\rm sh}/\Gamma_{\rm res}$ (we remind that $\rm \Gamma_{\rm res}$ is an estimate of the Lorentz factor of the shocked medium in the central object frame, see Eq.~\ref{eq:gamma_res} and the lower panel of Fig.~\ref{fig:profil}): $ \lambda_{\rm max}=r_{\rm sh}/30\Gamma_{\rm res}$. We will however discuss later in this paragraph the influence of the choice of $\rm \lambda_{max}$ on our results and how it 
will be implemented in the Monte-Carlo calculations in the next section. Let us note that for each snapshot (which corresponds to a given value of $r_{\rm sh}$ during the shock propagation) we obtained a value of  $ \Gamma_{\rm sh}$ that does not strictly correspond to the shock Lorentz factors simulated in Sect.~3 ($\rm \Gamma_{sh}=$1.1, 1.3, 1.5, 1.7 and 1.9). In order to obtain acceleration times for a given snapshot, we simply use the results obtained for the closest Lorentz for which shock acceleration simulations where performed.

\begin{figure}
{\rotatebox{0}{\includegraphics[scale=0.4]{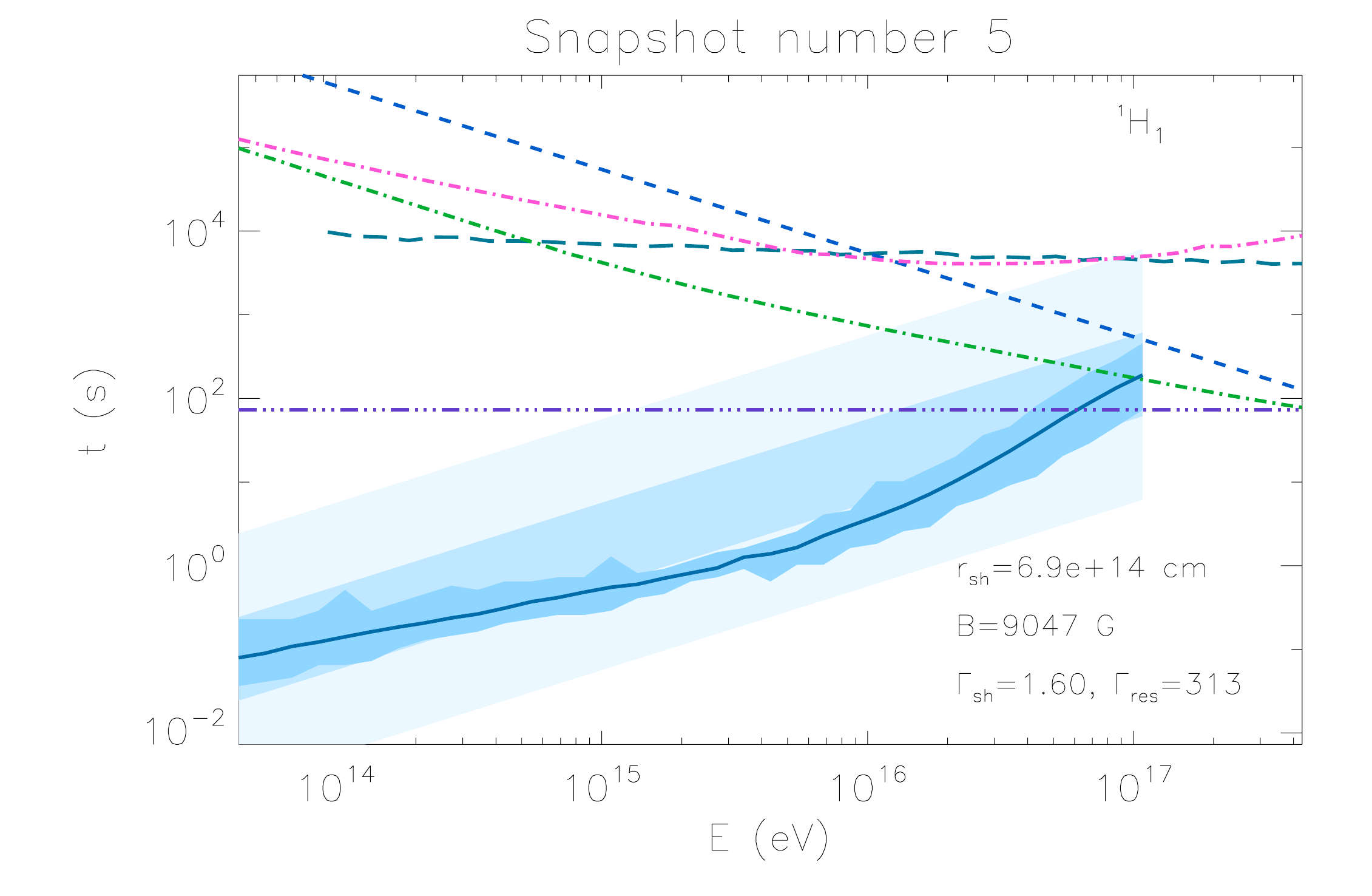}}}
{\rotatebox{0}{\includegraphics[scale=0.4]{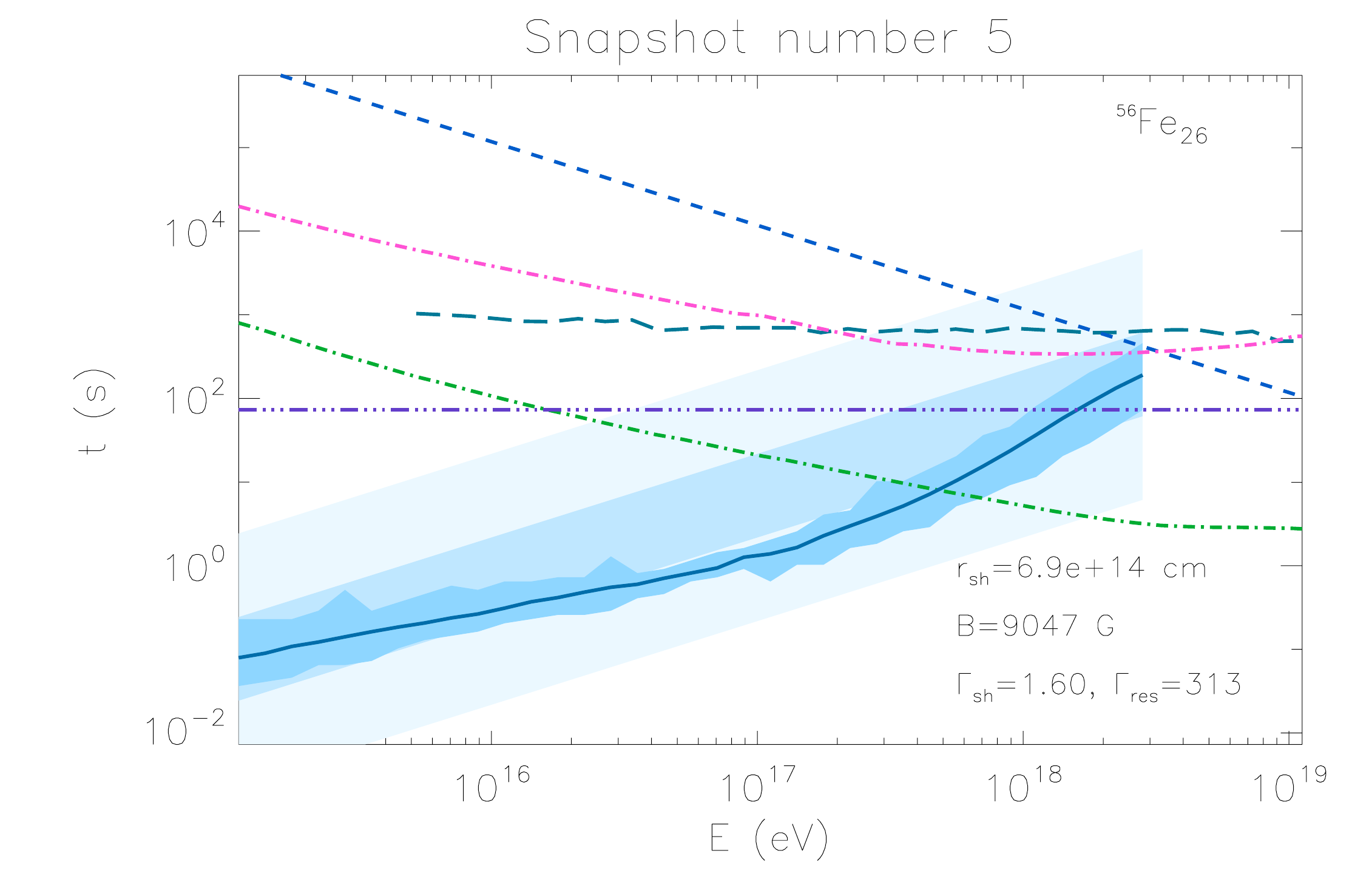}}}
\caption{Same as Figs.~\ref{fig:snapP} and \ref{fig:snapFe} for snapshot 5, $L_{\rm wind}^{\rm eq}=10^{53}\,\rm erg\,s^{-1}$, model B for protons (top) and Fe nuclei (bottom).}
\label{fig:snapB}
\end{figure}

Examples of the competition between cosmic-ray acceleration and energy loss processes are shown in Fig.~\ref{fig:snapP} for protons and Fig.~\ref{fig:snapFe} for Fe nuclei (the signification of the different lines and shaded areas is clarified in Fig.~\ref{fig:legend}) for different snapshots, for model A (equipartition) and $L_{\rm wind}\equiv L_{\rm wind}^{\rm eq}=10^{53}\,\rm erg\,s^{-1}$. In the case of protons, one sees that the acceleration is limited at all radii by adiabatic losses. Pion production and other energy loss processes are subdominant during the whole shock propagation (at least for this wind luminosity). In the case of Fe nuclei (and in general in the case of nuclei heavier than protons), the situation is slightly more complicated. During the first few snapshots, energy losses are dominated by photodisintegration processes which become progressively subdominant as the shock propagates away from the central object. The radius at which photodisintegration become subdominant obviously 
depends on physical assumptions such as the wind luminosity. By comparing the expected acceleration time and the different energy loss times, one can define the maximum reachable energy $E_{\rm max}^{\rm loss}$ as the energy at which $t_{\rm acc}(E)=t_{\rm loss}(E)$, $t_{ \rm loss}(E)$ being the total energy loss time. 
In the case of model B, the competition between acceleration and energy losses is shown in Fig.~\ref{fig:snapB}, for protons (top) and Fe nuclei (bottom) for the snapshot \#5 and $L_{\rm wind}^{\rm eq}=10^{53}\,\rm erg\,s^{-1}$. This figure can directly be compared to the same snapshot for model A in Figs.~\ref{fig:snapP} \& \ref{fig:snapFe}. In model B, the acceleration time and the energy loss time curves cross at higher energy due to the larger magnetic field which allows a faster acceleration. For protons, the difference for $E_{\rm max}^{\rm loss}$ between models A and B is proportional to the difference between the values of the magnetic field, because the main energy loss process (adiabatic losses) does not depend on the particle energy. It is not the case for nuclei heavier than protons, as photodisintegration (which is in that case the dominant energy loss process) loss time decreases as the energy increases. As a result, $E_{\rm max}^{\rm loss}$ is larger in the case of model B but the gain is not proportional to the increase of the magnetic field. One can also notice that synchrotron loss time as well as hadronic loss time are shorter in the case of model B, due to the larger value of the magnetic field and the larger wind luminosity $L_{\rm wind}$ assumed.

\subsubsection{Evolution of $E_{\rm max}^{\rm loss}$ with the shock radius and wind luminosity}
\begin{figure}
{\rotatebox{0}{\includegraphics[scale=0.3]{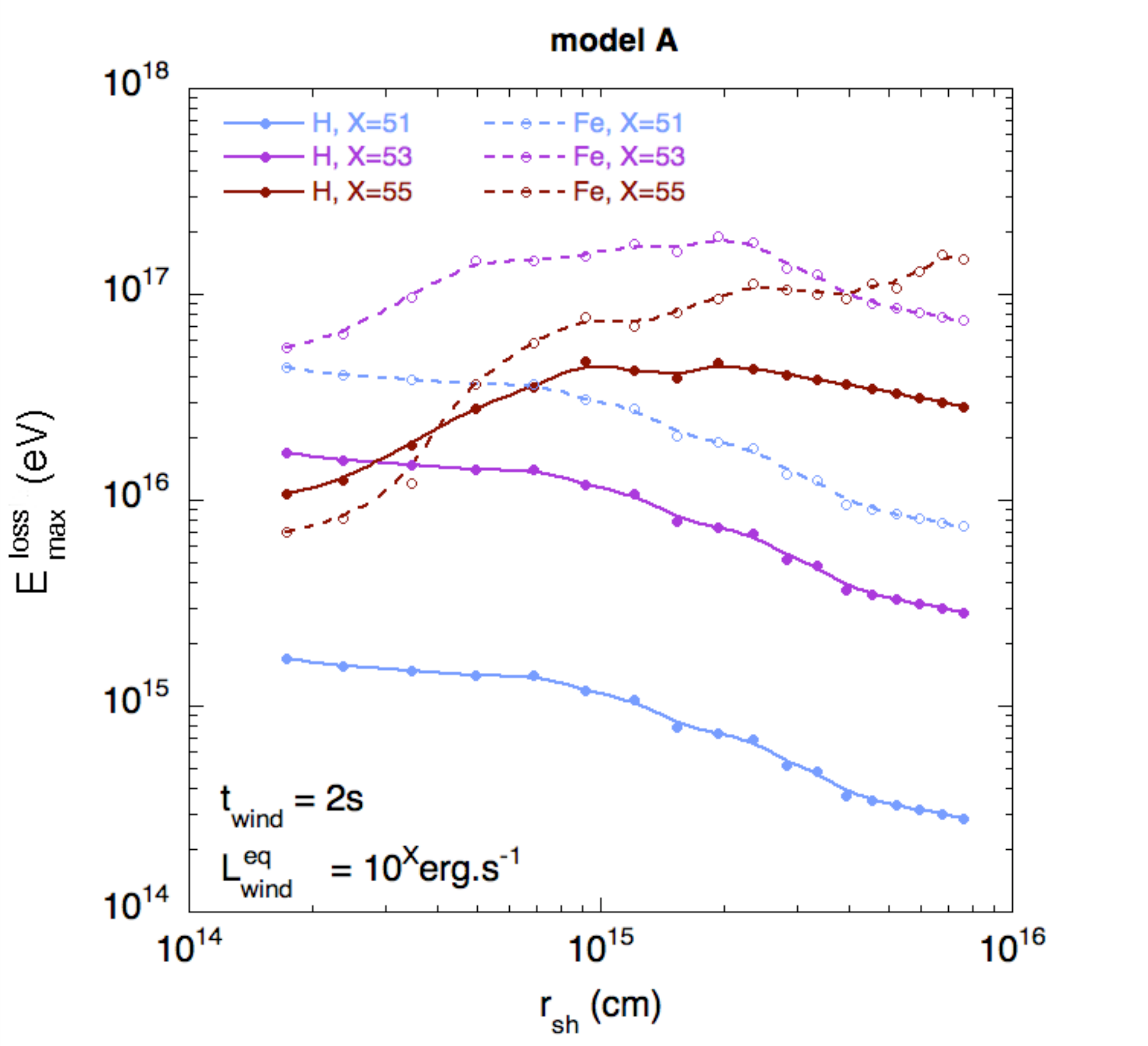}}}
{\rotatebox{0}{\includegraphics[scale=0.3]{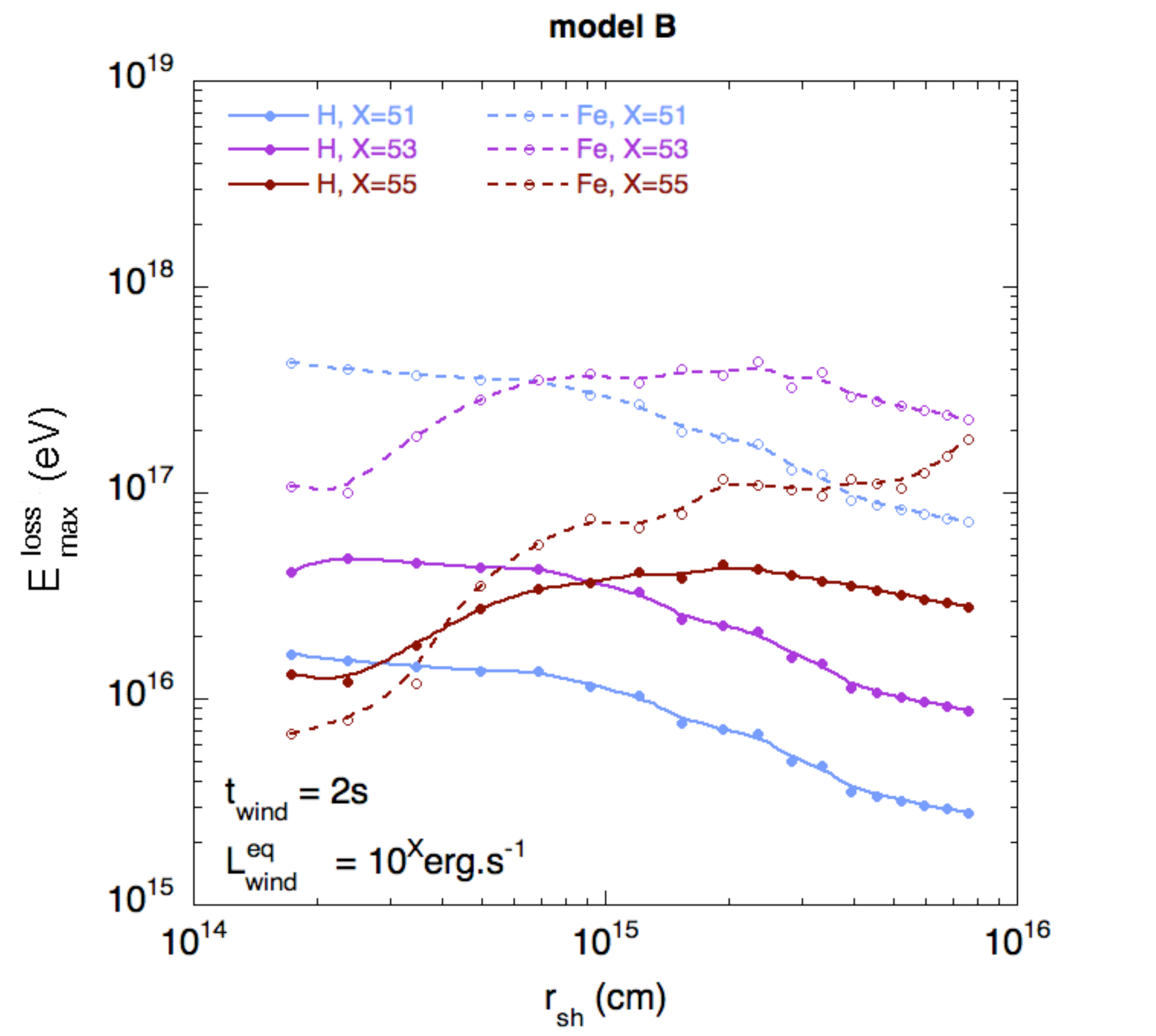}}}
{\rotatebox{0}{\includegraphics[scale=0.3]{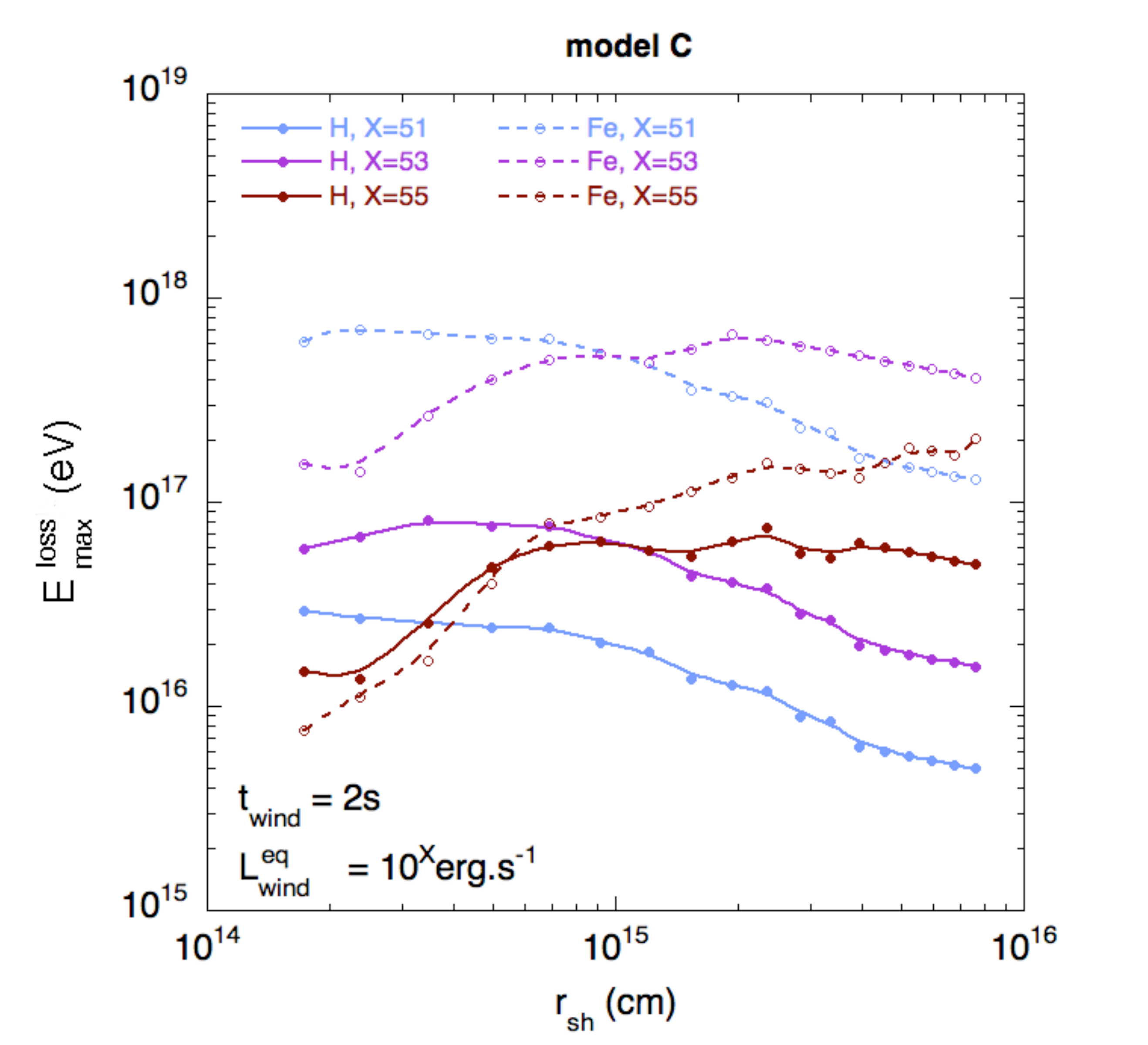}}}
\caption{$r_{\rm sh}$ evolution of $E_{\rm max}^{\rm loss}$ (see text) for different values of $L_{\rm wind}^{\rm eq}$ ($10^{51}$, $10^{53}$ and $\rm10^{55}\,\rm erg\,s^{-1}$) for protons (full lines) and Fe nuclei (dashed lines), the markers (closed circles for protons and open circles for Fe) show the value of $E_{\rm max}^{\rm loss}$ at each snapshot. These evolutions are shown for models A (top), B (center) and C (bottom).}
\label{fig:emaxmodel}
\end{figure}
Comparing acceleration times and the different energy loss times for each snapshot allows one to obtain an estimate of the evolution of $E_{\rm max}^{\rm loss}$ with the shock radius, for a internal shock modeled with a given set of physical parameters. Such an evolution is shown in Fig.~\ref{fig:emaxmodel}, for the models A (top), B (center) and C (bottom), in each case for protons (full lines), iron nuclei (dashed lines) and three different values of $L_{\rm wind}^{\rm eq}$ ($10^{51}$, $10^{53}$ and $\rm10^{55}\,$$\rm \rm erg\,s^{-1}$). Let us note that the values of $E_{\rm max}^{\rm loss}$ shown in Fig.~\ref{fig:emaxmodel} are given in the (comoving) frame of the shocked medium, they can be roughly estimated in the central source rest frame by multiplying the values of $E_{\rm max}^{\rm loss}$ obtained for each snapshot by the corresponding value of $\rm \Gamma_{\rm res}$ (see Sect.~2 and the lower panel of Fig.~\ref{fig:profil}).
Starting with model A in the case of protons, one can see that the evolution with the shock radius (hereafter, "$r_{\rm sh}$ evolution") of $E_{\rm max}^{\rm loss}$, for $L_{\rm wind}^{\rm eq}=10^{51}$ and $\rm10^{53}\,\rm erg\,s^{-1}$, have the same trend during the whole shock propagation. This is due to the fact that, for both wind luminosities, the acceleration is limited by adiabatic losses at all $r_{\rm sh}$. In that case, $E_{\rm max}^{\rm loss}$ is simply proportional to the magnetic field which evolves as $B\propto(L_{\rm wind}^{\rm e q})^{1/2}$ for model A. The values of $E_{\rm max}^{\rm loss}$ are then simply shifted upward by a factor of 10 when increasing $L_{\rm wind}$ by a factor of 100. For the largest wind luminosity considered ($ L_{\rm wind}^{\rm eq}=10^{55}\,\rm erg\,s^{-1}$), the evolution is quite different, due to the fact that photomeson interactions are dominant during the initial phase of the shock propagation ($r_{\rm sh}\lesssim 10^{15}\,\rm cm$), while adiabatic losses dominate at 
larger radii.  As a result, the maximum energy reachable for $L_{\rm wind}^{\rm eq}=10^{55}\,\rm erg\,s^{-1}$ is initially lower than the value obtained for $L_{\rm wind}^{\rm eq}=10^{53}\,\rm erg\,s^{-1}$ before becoming proportional to $(L_{\rm wind}^{\rm e q})^{1/2}$ when adiabatic losses become dominant. Similar considerations apply to model B and C with a few significant differences. First, the values of $E_{\rm max}^{\rm loss}$ obtained for each value of $L_{\rm wind}^{\rm eq}$ are larger than in the case of model A. This is again due to the fact that the magnetic field $B$ is proportional to $(L_{\rm wind}^{\rm e q})^{1/2}$ and the wind luminosities assumed for model B and C are larger than for model A for a given value of $L_{\rm wind}^{\rm eq}$ (moreover the larger value of the equipartition factor for the magnetic field $\epsilon_B$ for model C explains the higher $E_{\rm max}^{\rm loss}$ compared with model B). Second, the evolution of $E_{\rm max}^{\rm loss}$ with $L_{\rm wind}^{\rm eq}$ is slower than in the 
case of model A. This evolution is implied by the relation we inferred between $L_{\rm wind}$ and  $L_{\rm wind}^{\rm eq}$ for model B and C (see Eq.~\ref{eq:lwind}): $L_{\rm wind}\propto (L_{\rm wind}^{\rm eq})^{1/2}$, which leads to $B \propto (L_{\rm wind}^{\rm eq})^{1/4}$ and then $E_{\rm max}^{\rm loss}\propto (L_{\rm wind}^{\rm eq})^{1/4}$ (if energy losses are dominated by adiabatic losses). Let us note that, although $L_{\rm wind}^{\rm eq}$ is not a physical observable (and has been introduced in Sect.~2 to compare models with different sets of equipartition parameters), it is related to the $\rm \gamma -ray$ luminosity of the prompt emission ($L_\gamma \simeq 0.05\times L_{\rm wind}^{\rm eq}$ as discussed in Sect.~2). The difference in the evolution of $E_{\rm max}^{\rm loss}$ with $L_{\rm wind}^{\rm eq}$ at a given radius between models A, B and C illustrates the fact that the $\rm \gamma -ray$ prompt emission and the accelerated  cosmic-ray spectrum can be related in a non-trivial way depending on the physical parameters of the internal 
shocks (mostly encoded into the values of the energy redistribution factors $\epsilon_B$, $\epsilon_{\rm e}$ and $\epsilon_{\rm cr}$ in our modeling). 

\begin{figure}
{\rotatebox{0}{\includegraphics[scale=0.35]{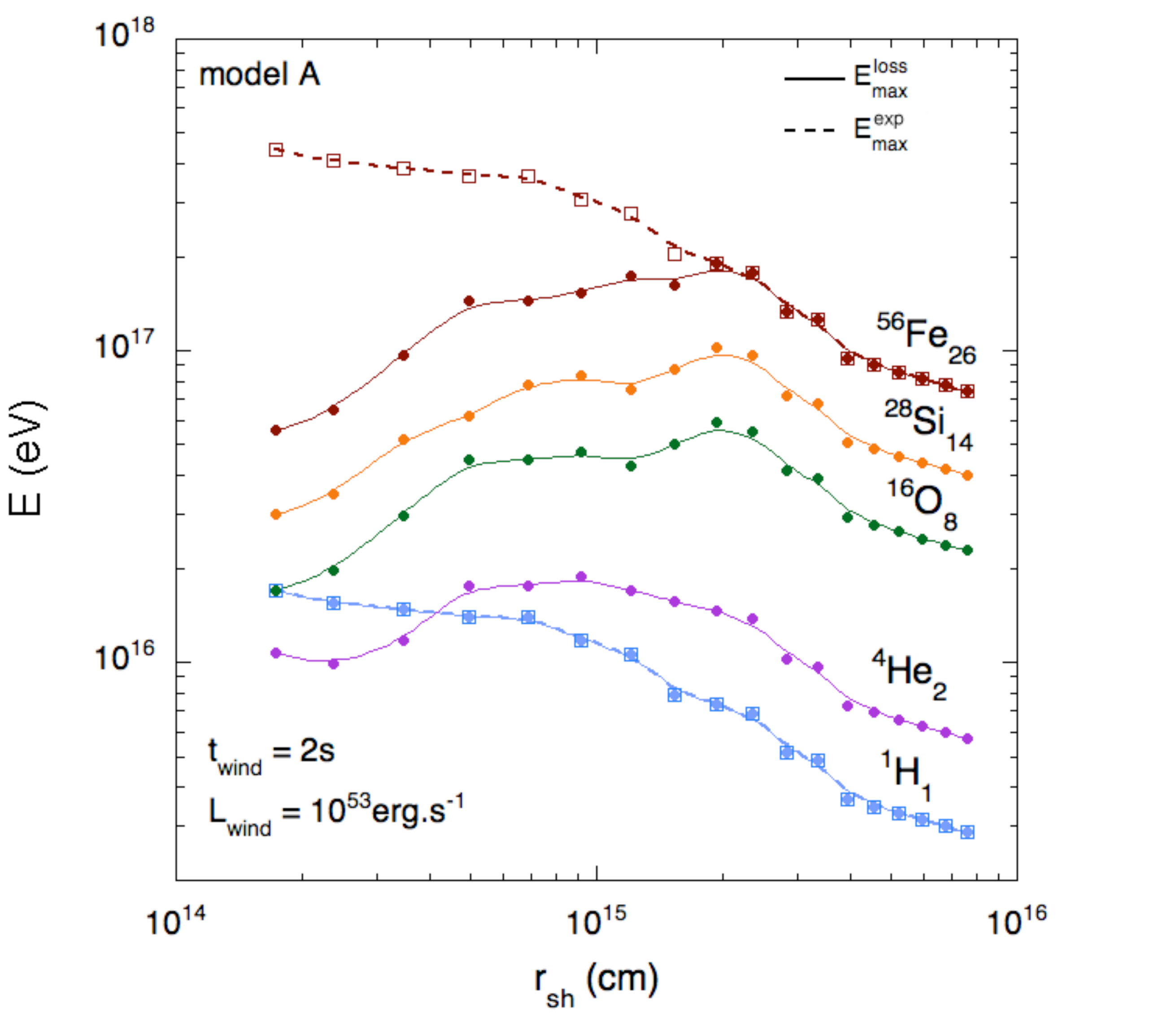}}}
\caption{$r_{\rm sh}$ evolution of $E_{\rm max}^{\rm loss}$  for different nuclear species (protons, He, O, Si, Fe) assuming model A, $L_{\rm wind}^{\rm eq}=10^{53}\,\rm erg\,s^{-1}$. The $r_{\rm sh}$ evolution of $E_{\rm max}^{\rm exp}$ (see text) is also shown (dashed lines and open squares for the value at each snapshot) for protons and Fe nuclei for comparison.}
\label{fig:emaxcompo}
\end{figure}

In the case of iron nuclei, as mentioned earlier, photo-interactions have a stronger impact than in the case of protons. Except for the lowest value of $L_{\rm wind}^{\rm eq}$, the acceleration is limited by photodisintegration processes on a significant portion of the shock propagation (actually on the whole range of $r_{\rm sh}$ for $L_{\rm wind}^{\rm eq}=10^{55}\,\rm erg\,s^{-1}$). Let us note that $E_{\rm max}^{\rm loss}$ (for a given $L_{\rm wind}^{\rm eq}$ at a given $r_{\rm sh}$) can only be proportional to $Z$ if the acceleration for each nuclear species is limited by deconfinement (see below) or adiabatic losses. The scaling $E_{\rm max}^{\rm loss}(Z)=Z\times E_{\rm max}^{\rm loss}(Z=1)$ should then only hold for certain values of $L_{\rm wind}^{\rm eq}$ or some portions of the shock propagation phase, while one expects $E_{\rm max}^{\rm loss}(Z) < Z\times E_{\rm max}^{\rm loss}(Z=1)$ when photodisintegration is the limiting process. This 
is illustrated in Fig.~\ref{fig:emaxcompo} where the $r_{\rm sh}$ evolution of $E_{\rm max}^{\rm loss}$ is shown for different nuclear species (H, He, O, Si, Fe) for model A and  $L_{\rm wind}^{\rm eq}=10^{53}\,\rm erg\,s^{-1}$. In addition, the $r_{\rm sh}$ evolution of $E_{\rm max}^{\rm exp}$ (the energy for which $t_{\rm acc}({E})=t_{\rm exp}$, where $t_{\rm exp}$ is the energy loss time due the adiabatic expansion, see Sect.~2) is represented in dashed lines for protons and Fe. As can be seen, the curves figuring $E_{\rm max}^{\rm loss}$ and $E_{\rm max}^{\rm exp}$ are superimposed for protons during the whole shock propagation whereas the two curves converge only at large radius ($r_{\rm sh} \simeq 2\,10^{15}$ cm) for Fe (and similarly for nuclei heavier than protons). The initial value of  $E_{\rm max}^{\rm loss}$ at low radius is very close to the one obtained for oxygen nuclei ($\sim$3 times lower than Fe), the gap between protons and the other species progressively increases as the shock 
propagates and the medium becomes more transparent to photodisintegration interactions. The value of $E_{\rm max}^{\rm loss}$ finally scales as the charge of the nucleus once $E_{\rm max}^{\rm loss}\simeq E_{\rm max}^{\rm exp}$ as explained earlier.

Finally, let us note that the comoving values of $E_{\rm max}^{\rm loss}$ obtained for protons, remain below $10^{17}$ eV for all models and all luminosities, meaning that they reach at most a few $10^{19}$ eV in the central object frame, for the most optimistic cases. Only nuclei, say CNO and heavier, should be able to reach $10^{20}$ eV and slightly above according to our modeling. To increase the maximum energy of protons, a significantly faster acceleration (i.e, $t_{\rm acc}\simeq t_{\rm L}$) than what we were able to obtain in Sect.~3 would be required. Alternatively, more extreme physical assumptions implying larger magnetic fields could be invoked for the same purpose. The values of magnetic fields we obtained (in particular for model B and C) are already large and synchrotron losses would become a strong limitation for proton acceleration for larger values. 

\subsubsection{Influence of $\rm \lambda_{max}$ and particles escape}

\begin{figure}
{\rotatebox{0}{\includegraphics[scale=0.35]{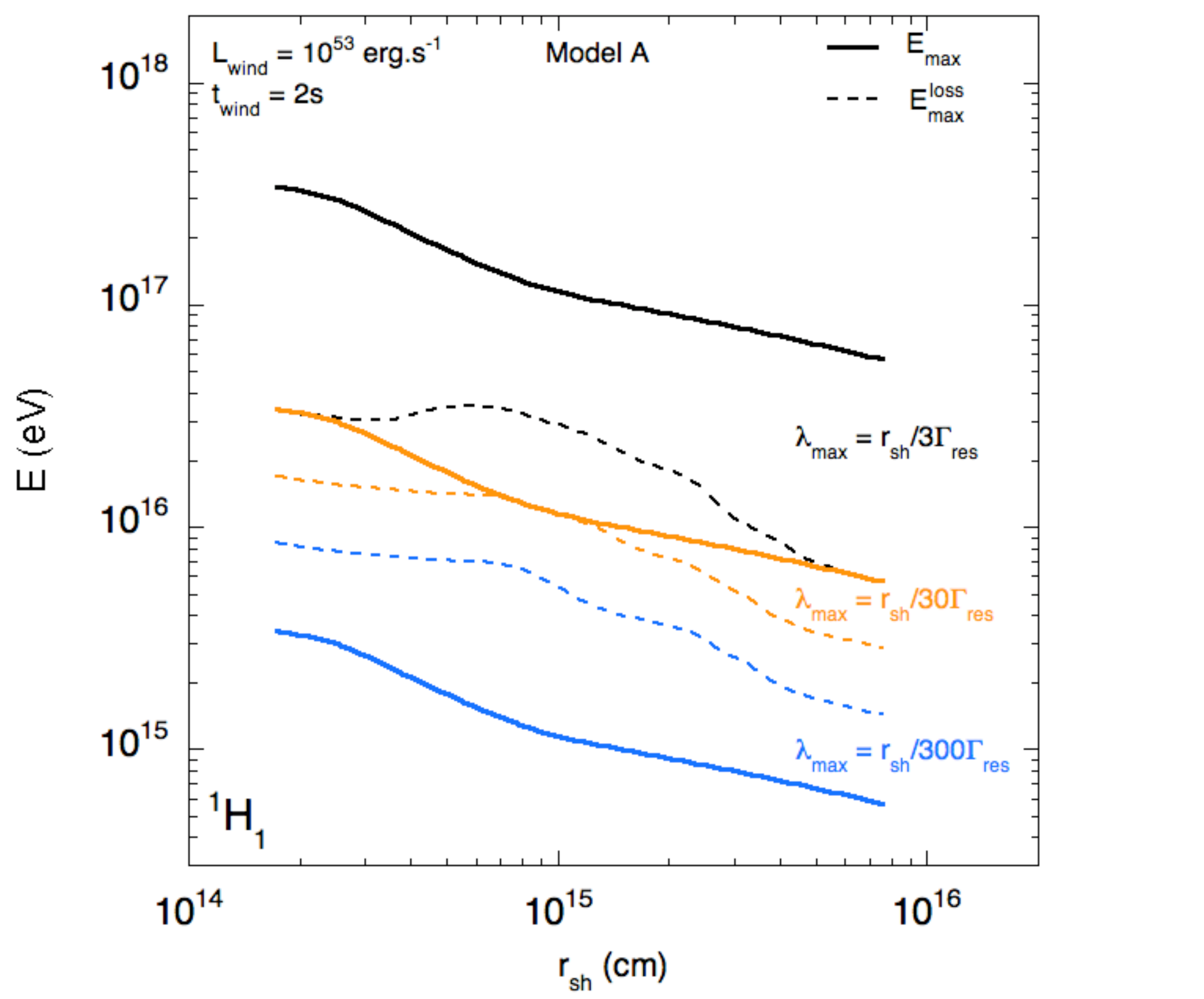}}}
\caption{$r_{\rm sh}$ evolution of $E_{\rm max}^{\rm loss}$ (dashed lines) and $E_{\rm max}$ (full lines) for different assumptions for the value of $\rm \lambda_{max}$ ($r_{\rm sh}/3\Gamma_{\rm res}$,  $r_{\rm sh}/30\Gamma_{\rm res}$ and $r_{\rm sh}/300\Gamma_{\rm res}$), for model A, $L_{\rm wind}^{\rm eq}=10^{53}\,\rm erg\,s^{-1}$, in the case of protons.}
\label{fig:lambda}
\end{figure}

As mentioned at the beginning of this section, a value for the magnetic field $B$ and the maximum turbulence length scale $\rm \lambda_{max}$ are required to scale the acceleration times and energies obtained in Sect.~3 to energy units. The values of the magnetic field were obtained using our modeling of GRBs internal shocks discussed in Sect.~2. However, this modeling does not give any prescription for the maximum turbulence length scale $\lambda_{\rm max}$ of the field which we arbitrarily set to $\lambda_{\rm max}=r_{\rm sh}/30\Gamma_{\rm res}$ in our previous calculations. It is obviously useful to study how the quantities we obtained before, such as $E_{\rm max}^{\rm loss}$, are modified when setting $\lambda_{\rm max}$ to a different value. We will test here other values of $\lambda_{\rm max}$ to understand how this physical parameter can influence the value of $E_{\rm max}^{\rm loss}$ or the escape of the accelerated particles from the acceleration region. We will then introduce the hypothesis on $\lambda_{\rm max}$ we apply in the next sections.

Fig.~\ref{fig:lambda} shows the evolution of $E_{\rm max}^{\rm loss}$ with $r_{\rm sh}$ (dashed lines) for three different choices for the value of $\rm \lambda_{max}$:  $\rm \lambda_{max} = r_{\rm sh}/3\Gamma_{\rm res}$,  $r_{\rm sh}/30\Gamma_{\rm res}$ and $r_{\rm sh}/300\Gamma_{\rm res}$, for model A, $L_{\rm wind}^{\rm eq}=10^{53}\,\rm erg\,s^{-1}$, in the case of protons. In addition, the $r_{\rm sh}$ evolution of $E_{\rm max}$ (defined in Sect.~3 as the energy for which $\rm r_L(E)=\lambda_{max}$) for each value of $\rm \lambda_{max}$ is also plotted (full lines). One sees that the values of $E_{\rm max}^{\rm loss}$ obtained, over a wide range of values for $\rm \lambda_{max}$, are similar. They differ only by a factor of $\sim$4 while the value for $\rm \lambda_{max}$ is moved over 2 orders of magnitude. Similar (or smaller) differences are obtained for all the nuclear species,  wind luminosities or energy redistribution models assumed. The slight dependence of $E_{\rm max}^{\rm loss}$ with the value 
of $\rm \lambda_{max}$ is actually due to the non monotonic energy 
evolution of $t_{\rm  acc}$ we found in Sect.~3. This dependence would not exist under the more naive assumption that the acceleration time is proportional to Larmor radius of the particle, $t_{\rm acc}({E})\propto t_{\rm L}(E)$ (or any monotonic evolution). 

On the other hand, although the chosen values of $\rm \lambda_{max}$ lead to very similar values of $E_{\rm max}^{\rm loss}$, the three cases would lead to different situations in terms of particles escape upstream or downstream of the shock. For the largest value of $\rm \lambda_{max}$ ($r_{\rm sh}/3\Gamma_{\rm res}$), one sees that we get $E_{\rm max}^{\rm loss}\ll E_{max}$ which means that the highest energy accelerated particles would hardly reach the weak scattering regime. In this case, the escape upstream or downstream of the shock would be either inefficient (upstream) or take too long compared with the energy loss time scale for particles of high energy to escape (downstream). If photo-interactions have a significant contribution to energy losses, neutrons (produced either by proton pion production or nuclei photodisintegration) may escape from the magnetized region (which would result in a 
purely protonic cosmic output at large distance from the burst) and possibly large quantities of secondary neutrinos and photons could be emitted. On the contrary, if adiabatic losses are dominant, high energy particles (either protons or heavier nuclei) should remain trapped and cool down to low energies.

An opposite situation is met for the lowest value of $\lambda_{\rm max}$ ($r_{\rm sh}/300\Gamma_{\rm res}$) for which one gets $E_{\rm max}< E_{\rm max}^{\rm loss}$. In this case, particles can reach the weak scattering regime and should manage to escape efficiently before reaching $E_{\rm max}^{\rm loss}$. If the condition $E_{\rm max}< E_{\rm max}^{\rm loss}$ is met for all nuclear species, the acceleration process is limited by deconfinement of particles from the acceleration region and the maximum energy of the escaping cosmic-rays should be proportional to their charge. This case is obviously less favorable than the previous one for secondary neutrino, photon or neutron emission.

The third case ($\lambda_{max}=r_{\rm sh}/30\Gamma_{\rm res}$, used in our previous calculations) represents an intermediate case with $E_{\rm max} \simeq E_{\rm max}^{\rm loss}$ (at least for this value of the wind luminosity\footnote{For $L_{\rm wind}^{\rm eq}=10^{55}\,\rm erg\,s^{-1}$ photo-interactions are important at the beginning of the shock propagation even for protons and we have $E_{\rm max}^{\rm loss}\ll E_{\rm max}$ at small $r_{\rm sh}$. The $E_{\rm max} \simeq E_{\rm max}^{\rm loss}$ regime is reached at larger distances.}). This represents a kind of optimal case for particle acceleration, in which particles can escape efficiently while reaching the maximum energy allowed by energy losses. As seen in Fig.~\ref{fig:emaxcompo}, this case (i.e, for the same wind luminosity) would be less favorable for Fe nuclei at low shock radius where photodisintegration is the dominant source of energy losses. In this case  $E_{\rm max}^{\rm loss}$ is quite lower than $E_{\rm max}$ and the escape for Fe nuclei 
should be less efficient than for protons. At larger distances from the shock, the 
impact of photodisingration decreases and $E_{\rm max} \simeq E_{\rm max}^{\rm loss}$ also for Fe nuclei. Let us note that if the condition $E_{\rm max} \simeq E_{\rm max}^{\rm loss}$ is met while photo-interactions are the dominant source of energy losses, an efficient cosmic-ray escape and efficient secondary particle production can occur simultaneously.

In the following Monte-Carlo calculations, we will make a different hypothesis to choose $\rm \lambda_{max}$ based on the assumption we made for models A, B and C, that cosmic-rays carry a significant fraction of the energy dissipated at internal shocks. We will assume that the maximum length scale of the turbulence $\rm \lambda_{max}$ at a given $r_{\rm sh}$ (for a given $L_{\rm wind}^{\rm eq}$ and a given energy redistribution hypothesis) is equal to the Larmor radius of the highest rigidity accelerated pariticles, which in all our cases correspond to the highest energy accelerated protons,  placing ourselves in the case $E_{\rm max} = E_{\rm max}^{\rm loss}$ for protons. As mentioned above, this represents an optimum case for the escape of protons at the maximum reachable energy while it will not necessarily be the case for heavier nuclei, for instance in a configuration for which photodisintegration is a dominant process of energy losses. The value of $\rm \lambda_{max}$, based on the value of $E_{\rm max}^{\rm loss}(Z=1)$, slightly depends on the prior hypothesis made on $\rm \lambda_{max}$ to estimate $E_{\rm max}^{\rm loss}(Z=1)$. The prior hypothesis we use is $\lambda_{\rm max}=r_{\rm sh}/30\Gamma_{\rm res}$, as in our previous calculations. As can be easily understood from our discussions, even a significant change of the prior value of $\rm \lambda_{max}$ would not strongly impact the results discussed in the next sections.

\subsubsection{Influence of $t_{\rm wind}$}

\begin{figure}
{\rotatebox{0}{\includegraphics[scale=0.35]{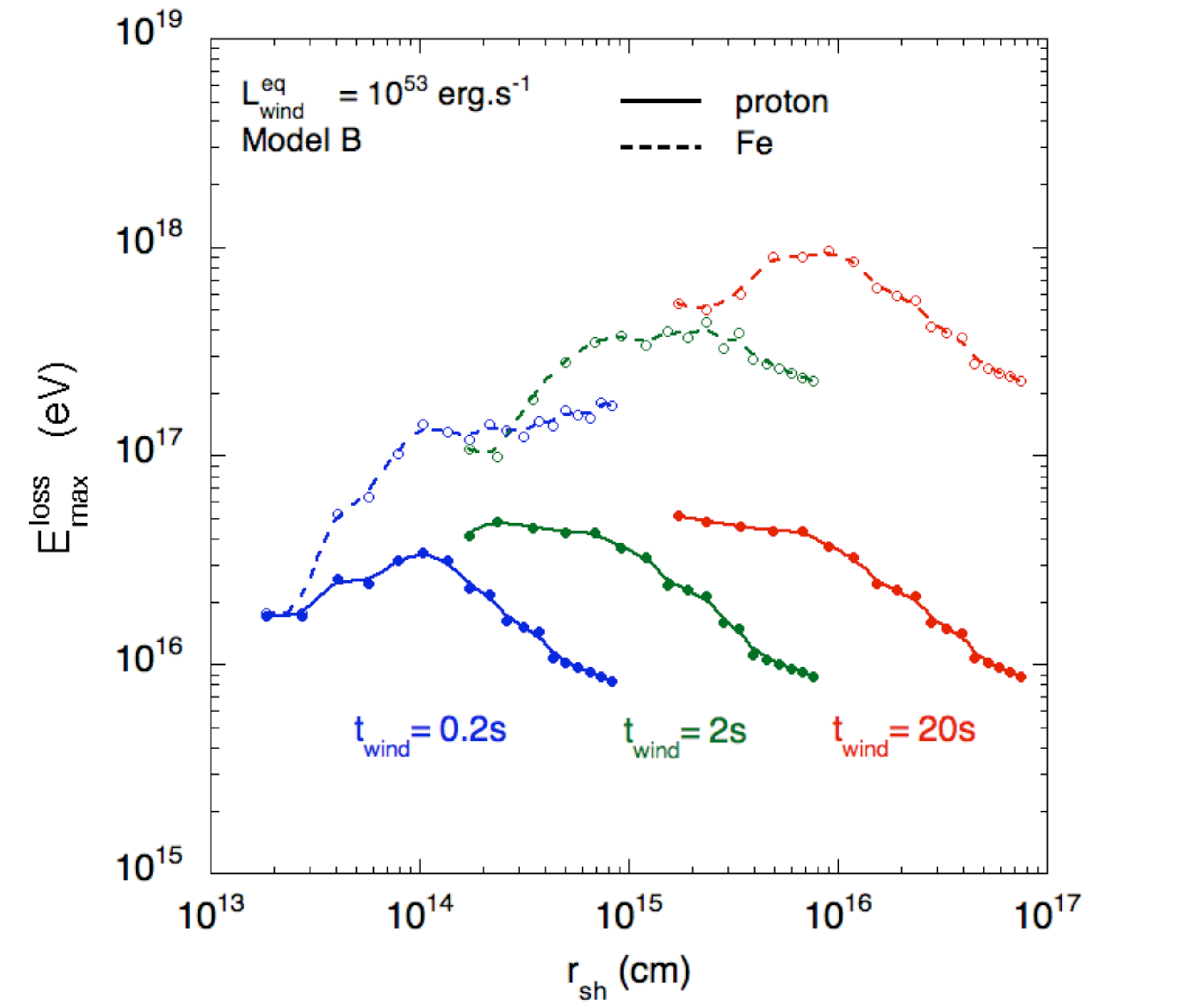}}}
\caption{$r_{\rm sh}$ evolution of $E_{\rm max}^{\rm loss}$ for protons (full lines) and Fe nuclei (dashed lines)  for model B, $L_{\rm wind}^{\rm eq}=10^{53}\,\rm erg\,s^{-1}$ for different assumptions for the value of $t_{\rm wind}$ (from left to right 0.2, 2 and 20s).}
\label{fig:twind}
\end{figure}

Before turning to the Monte-Carlo calculations, we briefly discuss the influence of the wind duration (which closely correspond to the time width of a light pulse in the central source frame) by compressing the Lorentz factor distribution we use in Sect.~2 ($t_{\rm wind}=2$s) to 0.2 or expanding it to 20s. As long GRBs have durations ranging from a few seconds to a few hundreds of seconds (regarding the prompt emission), one can observe shorter or longer light pulses, than the typical one second half width we considered so far. To compare the $E_{\rm max}^{\rm loss}$ expectation for $t_{\rm wind}=0.2$s and $t_{\rm wind}=20$s with those obtained for $t_{\rm wind}=2$s, we applied the internal shock modeling detailed in Sect.~2 to the two additional cases, tuning the $\rm \zeta_e$ parameter to obtain SEDs with similar peak energies as the $t_{\rm wind}=2$s case. The same calculations as in the previous section can be done to estimate the $r_{\rm sh}$ evolution of $E_{\rm max}^{\rm loss}$. Results of these calculations are shown in Fig.\ref{fig:twind} and compared to the $t_{\rm wind}=2$s case, assuming $L_{\rm wind}^{\rm eq}=10^{53}\,\rm erg\,s^{-1}$, model B, protons (full lines) and Fe nuclei (dashed lines), for illustration (results for models A and C show the same trends). As can be intuitively anticipated, for a given Lorentz factor distribution, the radius at which the shock is formed as well as the range of distances over which the shock propagation takes place are, in good approximation  proportional to $t_{\rm wind}$. As a result (see Sect.~2), for a given wind luminosity, baryons and photon densities are expected to scale roughly as $t_{\rm wind}^{-2}$, the magnetic field as well as the dynamical time scale roughly as $t_{\rm wind}^{-1}$. It means that the energy loss rates by hadronic interactions, photo-interactions or synchrotron are expected to scale as well as $t_{\rm wind}^{-2}$ while the adiabatic loss rate and the acceleration should scale as $t_{\rm wind}^{-1}$. The competition between acceleration and the different sources of energy losses and acceleration is then expected to depend on $t_{\rm wind}$ (all the other parameters being fixed). Passing from $t_{\rm wind}=2$s to $t_{\rm wind}=0.2$s, the synchrotron, photo-interactions and hadronic interactions compete more efficiently with acceleration and adiabatic losses. The acceleration is then limited by photodisintegration during the whole shock propagation for Fe nuclei and the values of $E_{\rm max}^{\rm loss}$ we obtain are lower than for the $t_{\rm wind}=2$s case. For protons, acceleration is initially limited by photo-interactions which results in lower values of $E_{\rm max}^{\rm loss}$, compared with the $t_{\rm wind}=2$s case. At larger distance (above $\sim10^{14}$cm in Fig.~\ref{fig:twind}), adiabatic losses become dominant and the values $E_{\rm max}^{\rm loss}$ is equivalent to the one obtained in the $t_{\rm wind}=2$s case, as the adiabatic loss time and the acceleration time scale in approximately the same way with $t_{\rm wind}$. For the longest wind duration ($t_{\rm wind}=20$s), on the contrary, the acceleration region is more transparent, the acceleration rate and 
adiabatic loss rates decrease more slowly than the interaction and synchrotron loss rates. Photodisintegration as a weaker impact on Fe nuclei acceleration at the beginning of the shock propagation, and the values of $E_{\rm max}^{\rm loss}$ obtained are larger than in the 2s case. At larger distances (or for the whole shock propagation in the case of protons), the acceleration is limited by adiabatic losses as in the 2s case. Thus $E_{\rm max}^{\rm loss}$ is equivalent to the $t_{\rm wind}=2$s case, for the reasons explained above. For a given Lorentz factor distribution and a given wind luminosity, taking into account the results shown in Fig.~\ref{fig:twind} and our discussion in the previous paragraph, the escape of nuclei would be favored in the case of the longest wind duration while the emission of secondary particles (neutrons, neutrinos, photons) would be enhanced for shorter durations.

Finally, let us note that changing the Lorentz factor profile of the wind (which goes from $\rm \Gamma_{min}=100$  to $\rm \Gamma_{max}=600$ for our generic case) could also slightly modify the results, but to a lower extent. A larger (resp. lower) Lorentz factor contrast would result in more (resp. less) energy dissipated in internal shocks, slightly larger (resp. lower) shock Lorentz factors ($\rm \Gamma_{sh}$) and shocked medium Lorentz factors ($\Gamma_{\rm res}$) resulting in a slightly faster (resp. slower) cosmic-ray acceleration and slightly larger (resp. lower) cosmic-ray energies in the central object frame. These quantities evolve however quite slowly with the Lorentz factor contrasts and we checked that our results would not be significantly affected by assuming $\Gamma_{\rm max}=400$ or $\Gamma_{\rm max}=800$ for instance.

\section{Monte-Carlo calculations in the presence of energy losses}

\subsection{Brief description of the code and physical hypotheses}

We will now explicitly calculate the expected cosmic-ray output from GRB internal shocks, implementing the energy loss processes introduced in Sect.~4 into the relativistic Fermi acceleration code described in Sect.~3. In the following, as mentioned earlier, adiabatic, synchrotron and pair production losses are modeled as continuous energy loss processes while photomeson production (for protons), photodisintegration (for nuclei) and hadronic interactions are treated stochastically. Besides the treatment of energy losses and secondary particle production, the principles of the calculation remain the same as what was done in Sect.~3.

Practically, the integration of the trajectory of a given particle is divided in small time steps corresponding to either a tenth of the particle Larmor time, a thirtieth of the particle interaction time (summed over all the processes treated stochastically) or a hundredth of the particle energy loss time (summed over all the continuous energy loss processes), whichever the shortest. Once the trajectory integration has been performed, the energy of the particle is corrected to take into account continuous energy losses and the emitted synchrotron photons and $\rm e^+e^-$ pairs are stored as attributes of the particle. The occurrence of stochastic interactions is then checked according to their probability on the time step. Whenever an interaction occurs, the process involved is chosen randomly (according to the relative probability of the different processes) along with its inelasticity (energy and/or nucleon losses) and the corresponding secondary particle emission (mainly charged mesons and photons). The secondary particles are stored (as well as the synchrotron photons and $\rm e^+e^-$ pairs from continuous processes) and treated in post analysis, to calculate their energy losses (mainly synchrotron for charged particles) and decay, taking into account the physical conditions at the time and place of their production. This ultimately results in the emission of (mostly) very high energy neutrinos and photons. Secondary nucleons are treated as new particles created at the time and place of their production, with the weight of their parent nucleus. They can be further accelerated by performing Fermi cycles once they are produced (in the case of neutrons, they have to decay or interact before leaving the acceleration region). Their trajectories and energy losses are calculated from their production point in the same way as their parent nucleus.

For a given burst defined by a set of physical parameters particles are injected at 18 different times/radii corresponding to the 18 snapshots used in the previous sections to discretize the shock propagation. The physical conditions relevant for a given snapshot (i.e, the variance of the turbulent field, the shock Lorentz factor, the baryon density, the prompt photon density, the dissipated energy available for cosmic-rays) are derived in the framework of the GRB internal shock model described in Sect.~2. The magnetic field is assumed to be purely turbulent and to follow a Kolmogorov power spectrum. The maximum turbulence length scale $\rm \lambda_{max}$ is set to be equal to the Larmor radius of protons at the energy  $E_{{\rm max,}i}^{ \rm loss}(Z=1)$, where $i$ is the snapshot number, as discussed in Sect.~4. For each snapshot, 7500 particles (protons, He, O, Si and Fe nuclei in equal number) are initially injected upstream of the shock with 
an energy $E_0=Z\times10^{-6}E_{{\rm max,}i}^{ \rm loss}(Z=1)$ (where $Z$ is the charge of the nucleus). The different nuclear species are given a weight in order to reproduce an effective composition equivalent to the galactic cosmic-ray source composition (Du Vernois \& Thayer, 1996) with a metallicity (i.e, the relative abundance of nuclei heavier than He) enhanced by a factor 10. To normalize the spectrum, obtained for a given snapshot, we take advantage of the fact that low energy particles, say below $10^{-2.5}\times E_{\rm max}$, loose very little energy during the time elapsed between their injection and their last shock crossing before being advected downstream. We then assume that these low energy particles carry the same fraction of the total energy communicated to cosmic-ray ($\epsilon_{\rm cr}\times e_{{\rm diss,}i}$ for a given snapshot) as in the absence of energy losses (i.e, the spectra obtained in Sect.~3). This assumption allows us to estimate the correspondence between the particles weights and the actual number of particles.

Concerning particles escape, we assume, as discussed in Sects.~3.3.3 and 4.2.3, that particles might escape through boundaries upstream and downstream of the shock. The boundary downstream is determined by the hydrodynamic simulation of Daigne \& Mochkovitch (2000) (see Fig. \ref{fig:epaisseur}) which gives us, at any time of the shock propagation, the comoving distance $\rm \Delta^{\prime}$ between the shock front and the extremity of the shocked medium. The relation between $\rm \Delta^{\prime}$ and $\rm \lambda_{max}$ (under our assumption $\lambda_{\rm max}=r_{\rm L}(E_{\rm max}^{\rm loss}(Z=1)$)) depends on the GRB physical parameters and on the distance to the central source and is dermined at each step in a consistent way. The width of the shocked region $\rm \Delta^{\prime}$ goes from  a few $\rm \lambda_{max}$ (relatively large distance from the central source) to a few tens of $\rm \lambda_{max}$ (for very luminous bursts and early times of the shock propagation). For the boundary upstream, we keep 
the hypothesis used in Sect.~3, that the magnetic 
turbulence only extends up to a distance $\rm \Delta_{up}^{\prime}\simeq\lambda_{max}$ away from the shock. In the following, we use $\rm \Delta_{up}^{\prime}=3\times \lambda_{max}$. As discussed in Sect.~3.3.3, both boundaries should behave like high pass filters in the presence of energy losses. Indeed, in the downstream region, particles have to reach the boundary fast enough not to be cooled down to low energies by adiabatic losses (or any other significant energy loss mechanism). Once particles have reached a boundary, either upstream or downstream, we assume they are decoupled from the magnetic field and the wind expansion and can freely propagate away from the source environment. Although particles might still interact with photons created in other parts of the wind, once they have escaped from the acceleration site, it is likely that the photon fields encountered will be much more tenuous than those experienced during the acceleration, and that head-on photon-nucleus collisions will be highly 
suppressed (see discussion in Sect.~2). 

In the following, we simulate the acceleration of cosmic-rays at GRB internal shocks for a wide range of wind luminosities, from $L_{\rm wind}^{\rm eq}=10^{51}\,\rm erg\,s^{-1}$ to $\rm10^{55}\,\rm erg\,s^{-1}$ (which corresponds to $L_\gamma \simeq 5\,10^{49}\, \rm erg.s^{-1}$ to $\rm 5\,10^{53}\,erg.s^{-1}$). We consider the three energy redistribution models, A, B and C introduced in Sect.~2. To take into account the evolution of the cosmic-ray and secondary particle outputs during the shock propagation, the 18 snapshots corresponding to different stages of the shock evolution and the corresponding set of physical parameters are considered for each simulated burst. Let us note that, for a given snapshot, we take into account the evolution with time of the physical quantities, such as photon and baryon densities, magnetic fields. This is especially important not to overestimate energy losses and secondary particles production of low energy nuclei advected downstream of the shock (i.e, below the weak 
scattering 
regime). These particles may, indeed, be confined during a significant fraction of the whole shock propagation time, much longer than the time a given snapshot remains representative of the physical conditions at work.

\subsection{Cosmic-ray spectra}

\begin{figure}
{\rotatebox{0}{\includegraphics[scale=0.32]{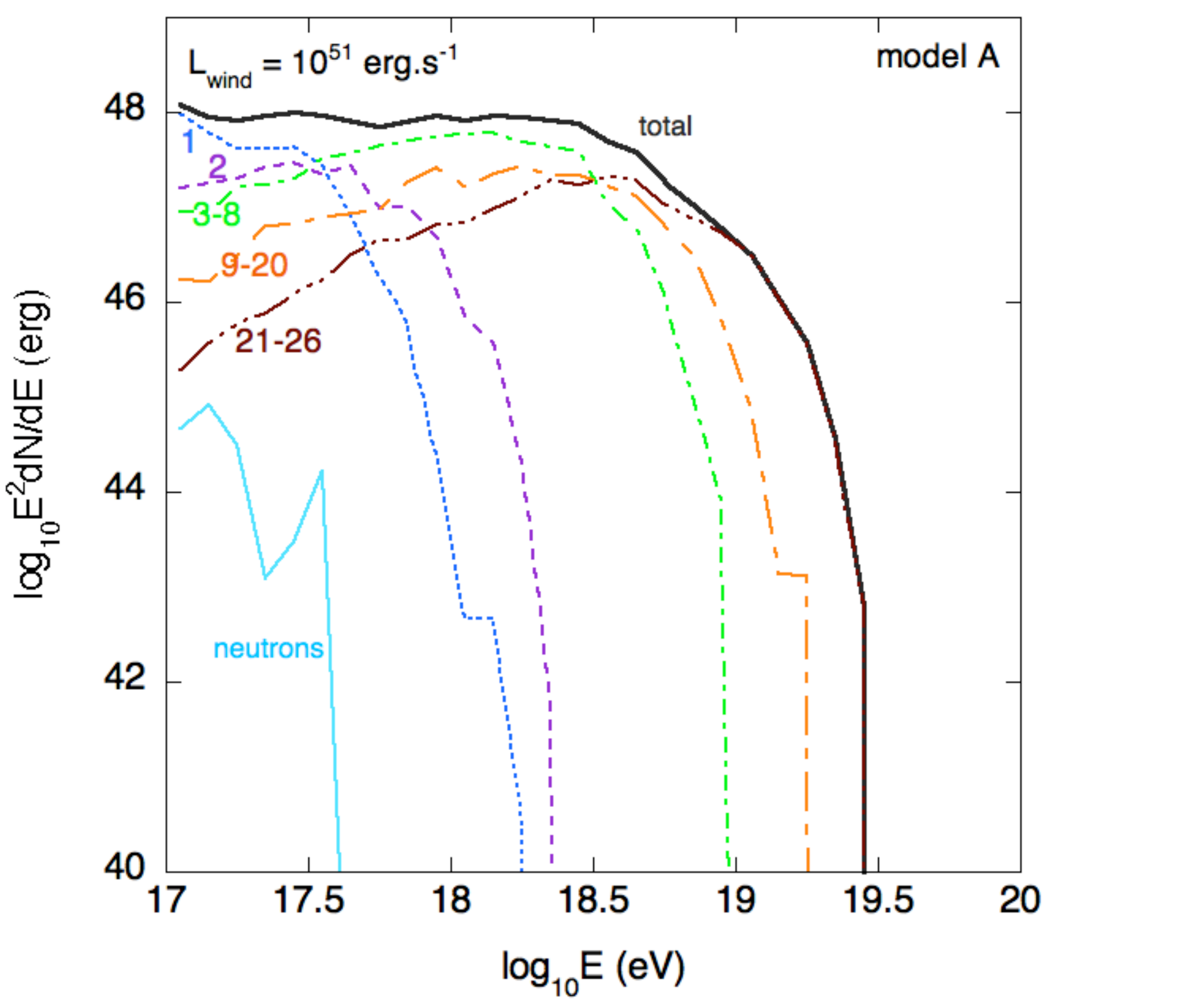}}}
{\rotatebox{0}{\includegraphics[scale=0.32]{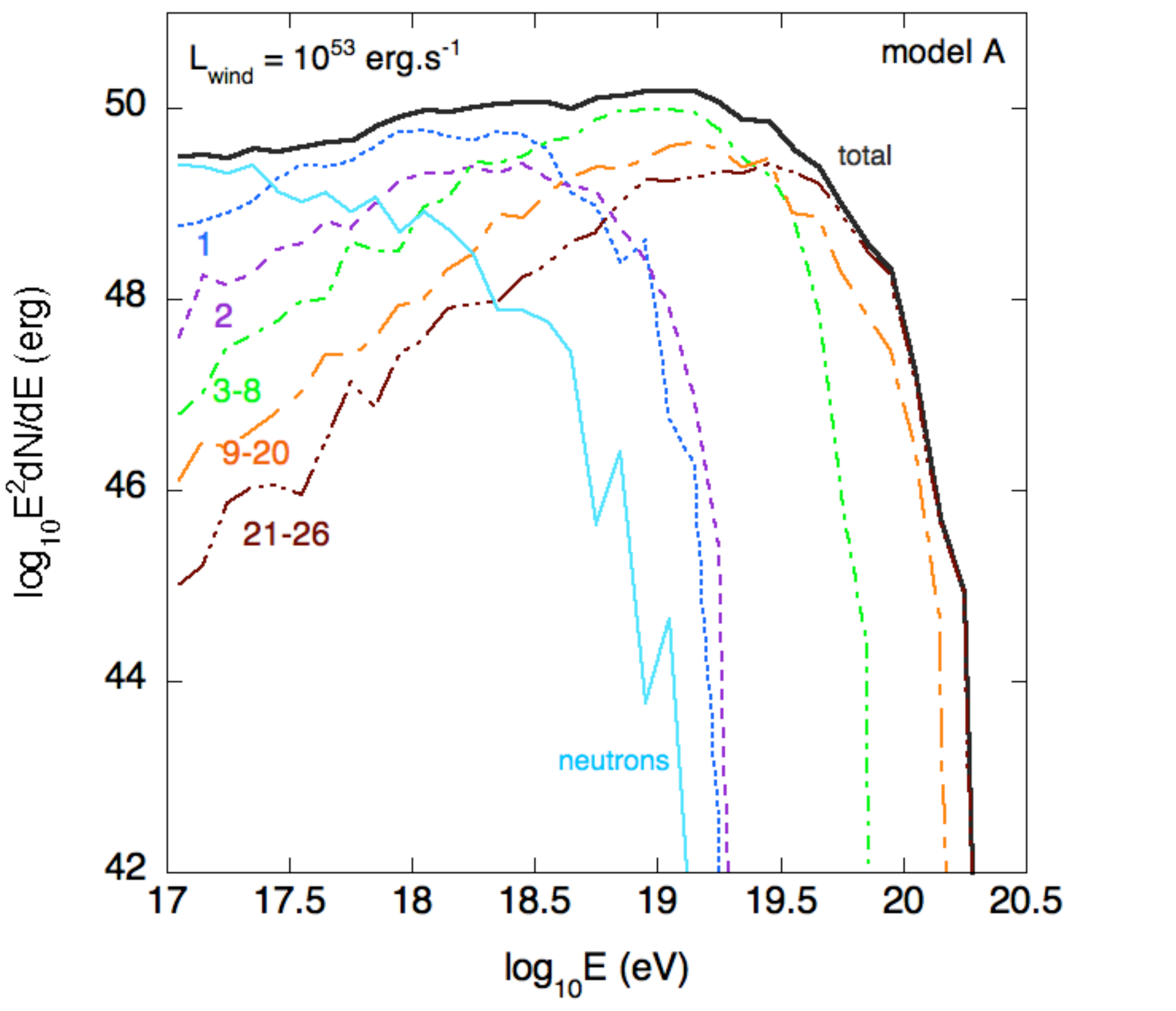}}}
{\rotatebox{0}{\includegraphics[scale=0.32]{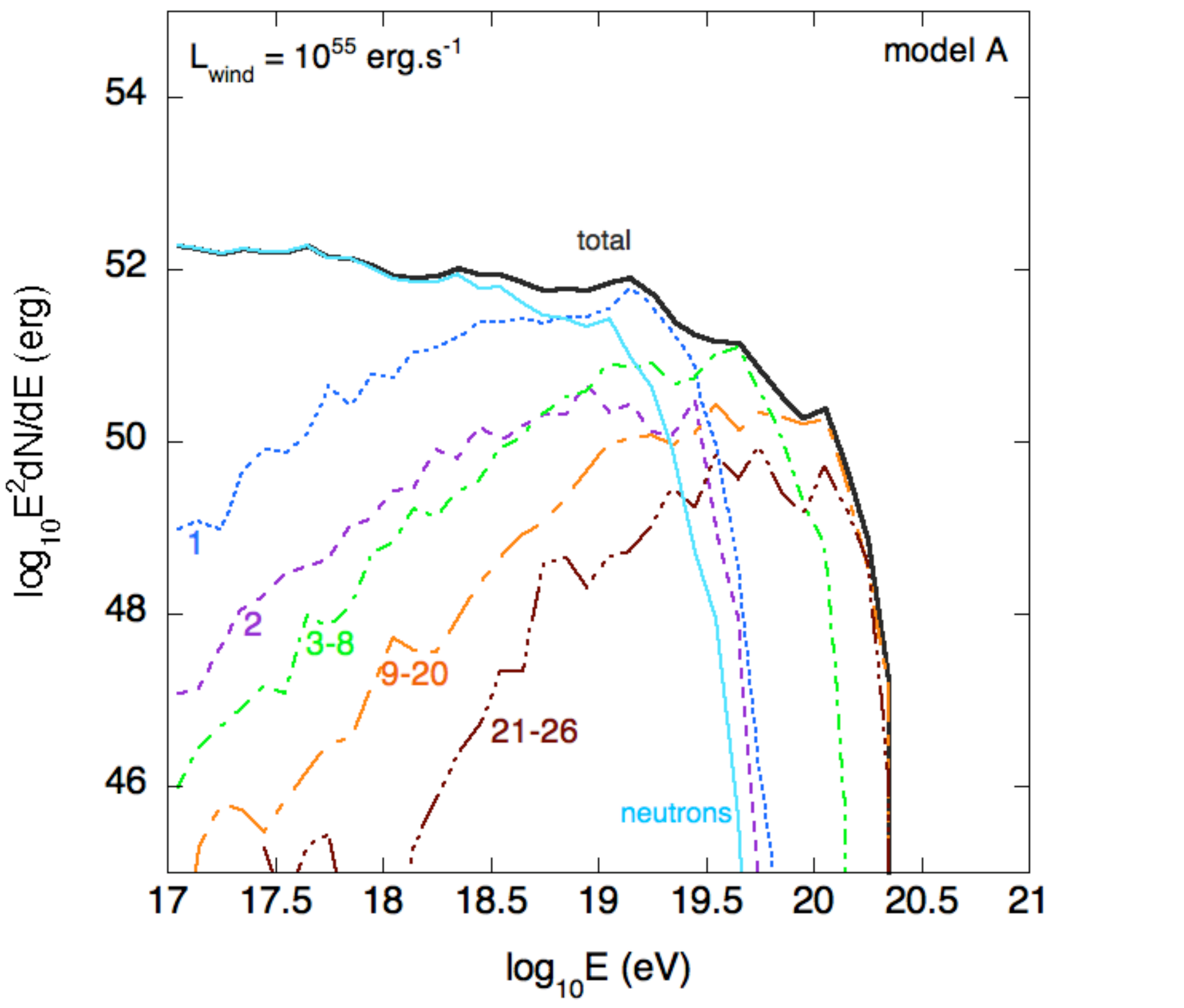}}}
\caption{Cosmic-ray spectra, multiplied by $E^2$, emiited by GRBs, in the central source frame, for different values of $L_{\rm wind}^{\rm eq}$ ($10^{51}$ (upper panel), $10^{53}$ (central panel) and $10^{55}\, \rm erg\,s^{-1}$ (lower panel)), in the case of model A. The contribution of different groups of nuclear species is shown (the labels refer to the nuclei atomic number $Z$). The normalization is obtained by integrating over the whole shock propagation.}
\label{fig:spectra_a}
\end{figure}

\begin{figure}
{\rotatebox{0}{\includegraphics[scale=0.32]{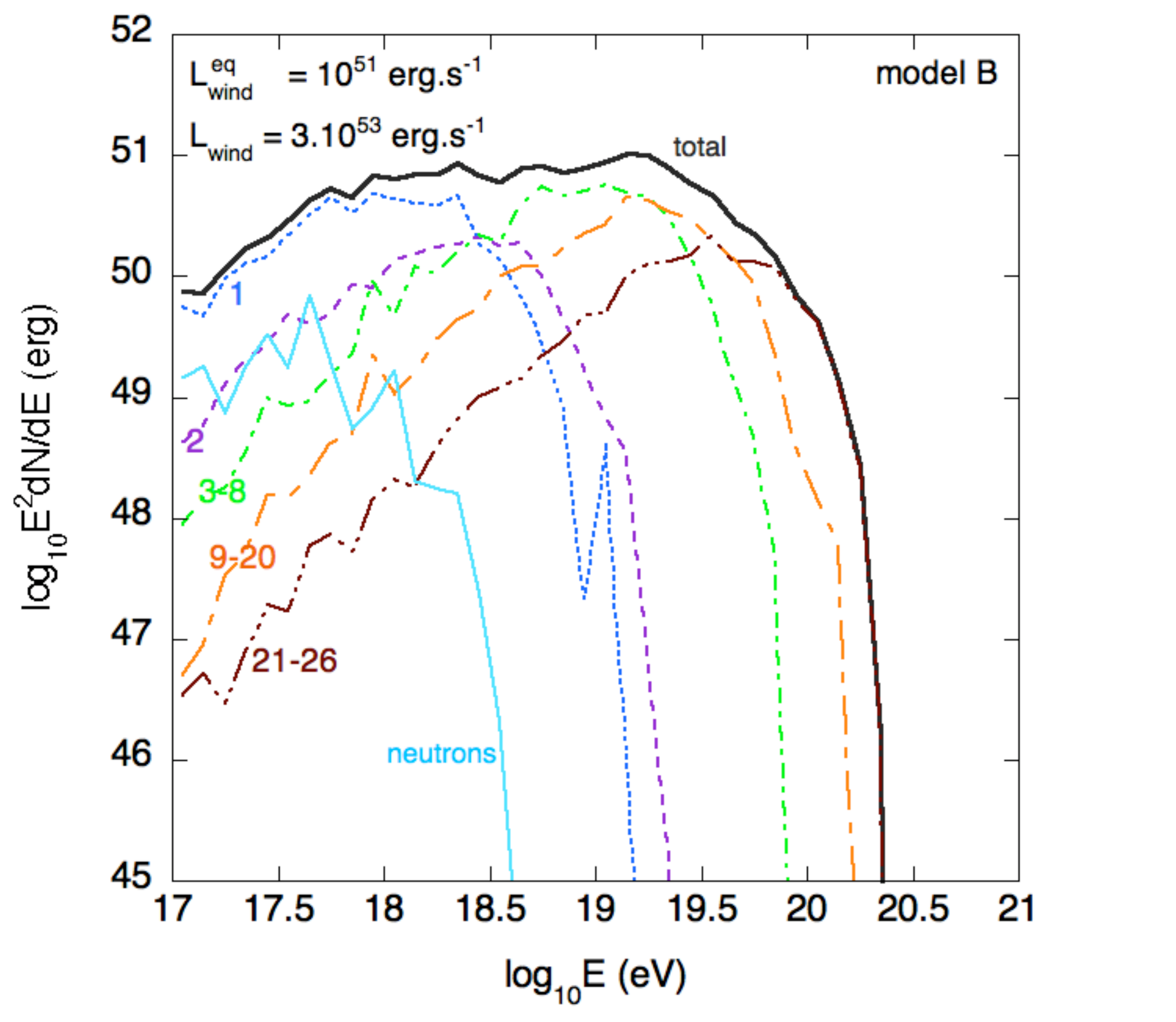}}}
{\rotatebox{0}{\includegraphics[scale=0.32]{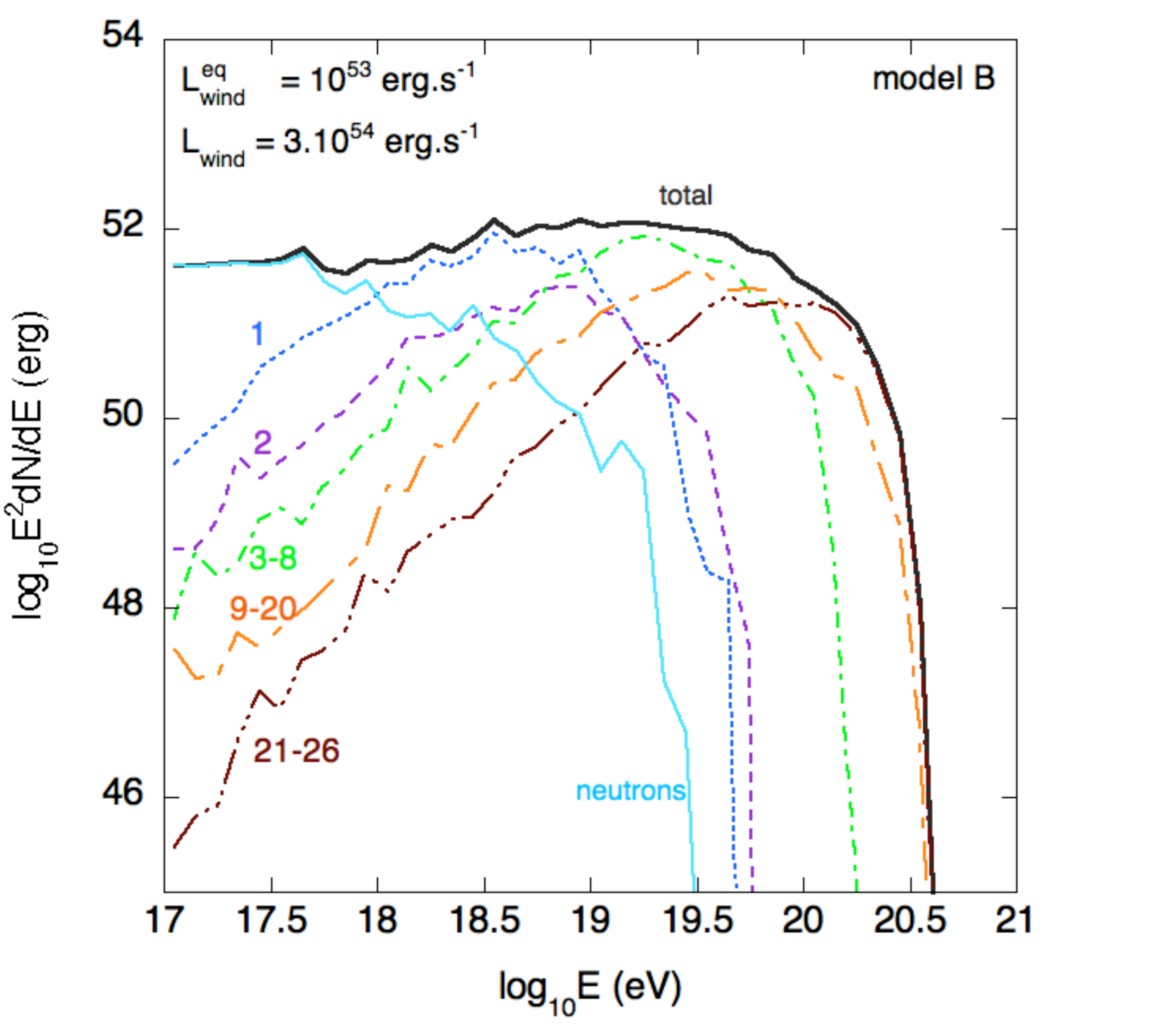}}}
{\rotatebox{0}{\includegraphics[scale=0.32]{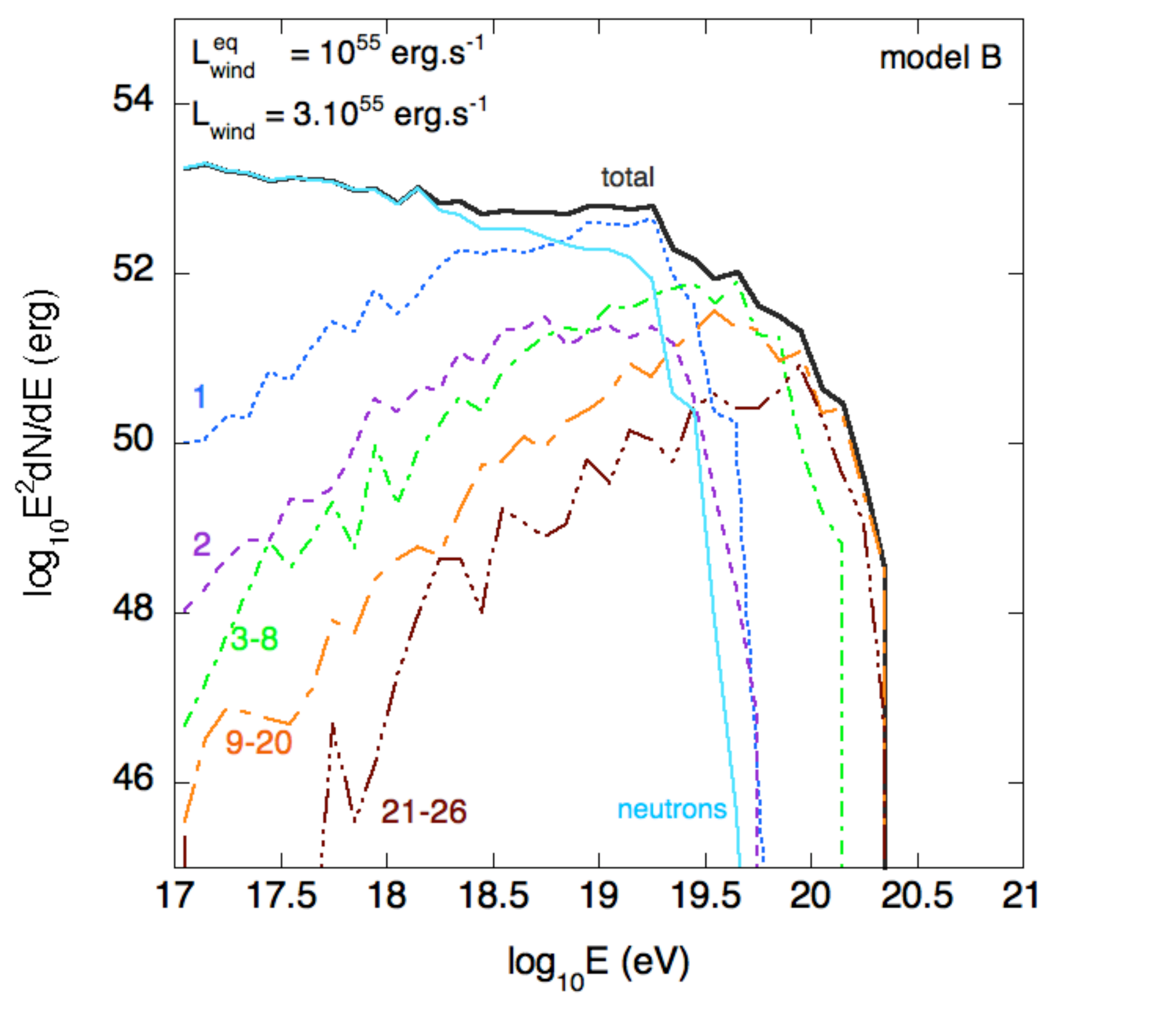}}}
\caption{Same as Fig.~\ref{fig:spectra_a}, in the case of model B.}
\label{fig:spectra_b}
\end{figure}

The cosmic-ray output expected for our modeling of GRB internal shocks is displayed in Figs.~\ref{fig:spectra_a} and \ref{fig:spectra_b}, for different values of $L_{\rm wind}^{\rm eq}$ between $10^{51}$ and $\rm 10^{55}\,\rm erg\,s^{-1}$ (corresponding to a prompt emission luminosity $\rm L_{\gamma}$ between $\sim 5\,10^{49}$ and $\rm \sim 5\,10^{53}\,\rm erg\,s^{-1}$) for model A (Fig.~\ref{fig:spectra_a}) and model B (Fig.~\ref{fig:spectra_b}). In each case, the cosmic-ray output is given in the central source frame, summed over the 18 snapshots. The normalization represents the energy released in cosmic-rays, in a given energy bin, integrated over the whole shock evolution for the assumptions we made in Sect.~2 about the wind Lorentz factor distribution and wind duration ($t_{\rm wind}=2$s). 

For low values of $L_{\rm wind}^{\rm eq}$ ($L_{\rm wind}^{\rm eq}=10^{51}\,\rm erg\,s^{-1}$ on the upper panels of Figs.~\ref{fig:spectra_a} and \ref{fig:spectra_b}, which correspond to $\rm L_{\gamma}\sim5\,10^{49}\,\rm erg\,s^{-1}$) photo-interactions are not efficient and do not limit cosmic-rays acceleration whatever the nuclear species considered. The maximum energy reached is in good approximation proportional to the charge of the nucleus, and the spectra of the different species are very hard due to the high pass filter effect of the cosmic-ray escape through the assumed boundaries downstream and upstream of the shock. A faint neutron component is present. These neutrons are produced in vast majority by the photodisintegration of nuclei during the early stage of the shock propagation.

At intermediate values of $L_{\rm wind}^{\rm eq}$, represented by the case $L_{\rm wind}^{\rm eq}=10^{53}\,\rm erg\,s^{-1}$ ($\rm L_{\gamma}\sim5\,10^{51}\,\rm erg\,s^{-1}$) on the central panels of  Figs.~\ref{fig:spectra_a} and \ref{fig:spectra_b}, photo-interactions start to have a significant impact on the cosmic-ray output. As seen in Sect.~4, the acceleration of nuclei is limited by photodisintegration during the early stages of the shock propagation. As a result, the gap in maximum energy between protons and nuclei is slightly lower than in the $L_{\rm wind}^{\rm eq}=10^{51}\,\rm erg\,s^{-1}$ case.  Neutron emission from nuclei photodisintegration are also more intense. As the cross sections for nuclei photodisintegration is larger than the proton photomeson production cross section (and the interaction threshold is lower in the nucleus rest frame), nuclei start to interact significantly at lower Lorentz factors than the protons. The neutrons are then produced at energies lower than the highest energy 
reached by accelerated protons, where the probability to interact before leaving the acceleration region is very low. Their escape is then very efficient. 

At high luminosities, (case $L_{\rm wind}^{\rm eq}=10^{55}\,\rm erg\,s^{-1}$, $L_\gamma\sim 5\,10^{53}\,\rm erg\,s^{-1}$, on the lower panel of Figs.~\ref{fig:spectra_a} and \ref{fig:spectra_b}), the acceleration of nuclei is severely limited by photodisintegration. During a significant portion of the shock propagation, we have $E_{\rm max}^{\rm loss}(Z>1)\ll Z\times E_{\rm max}^{\rm loss}(Z=1)$ meaning that nuclei acceleration is limited by photodisintegration before reaching the weak scattering regime. As a consequence nuclei cannot escape from the shock environment before the last stages of the shock propagation (see Sect.~4). This is not the case for protons since we assumed $\lambda_{\rm max}=E_{\rm max}^{\rm loss}(Z=1)$ for our calculations. This results in a lower relative abundance of nuclei compared to protons and a more reduced difference between the maximum energies reached by the different species. The neutron component is in this case very intense as the photon density is so large that nuclei are 
photodisintegrated efficiently even at lower energies. The gap in energy between the escaping protons and neutrons is also reduced due to the fact that proton acceleration is limited by photomeson production during the early stage of the shock propagation, which implies that the accelerated protons also contribute to neutron production at high luminosity (although the contribution of nuclei photodisintegration is still largely dominant). These neutrons should ultimately decay back to protons outside of the source environment and contribute to the proton component far away from the central source. As a consequence, we see that for intermediate and high luminosity bursts, one can expect the spectrum of escaping protons to be significantly softer than the spectrum of escaping nuclei due to the contribution of escaping neutrons which are mostly produced by nuclei photodisintegration.

These qualitative discussions on the evolution of the cosmic-ray output as a function of $L_{\rm wind}^{\rm eq}$ and the corresponding $L_\gamma$ hold for the three energy redistribution models A, B and C which, by assumption, predict similar photon densities within the acceleration region for a given value of $L_{\rm wind}^{\rm eq}$. There are however important differences concerning the cosmic-ray output between the three models. In the cases of models B and C (not displayed), the most important difference is of course the amount of energy released in very high and ultra-high energy cosmic-rays. The larger values obtained for models B and C are implied by the assumption of a larger wind luminosity for a given prompt emission luminosity. Concerning the maximum energy, the values obtained for models B and C are larger than for model A, due to the larger wind luminosity assumed. For instance, in the case of $L_{\rm wind}^{\rm eq}=10^{51}\,\rm erg\,s^{-1}$, the wind luminosity for model B is $L_{\rm wind}=3\,
10^{53}\,\rm erg\,s^{-1}$ and the energy redistribution factor for the magnetic field $\epsilon_B=0.1$. The magnetic field is expected to be proportional to $(\epsilon_B\times L_{\rm wind})^{1/2}$ and the maximum energy achievable is proportional to the magnetic field (in a case not limited by photo-interactions). The maximum energy is $\sim$10 times larger for model B, as observed when comparing the upper panels of Figs.~\ref{fig:spectra_a} and \ref{fig:spectra_b}. In the case of model C, the maximum energy is even $\sim$ twice larger than for model B, due to the larger assumed value of $\epsilon_B=0.33$. The gap in maximum reachable energy between model A and models B and C is lower for the higher values of $L_{\rm wind}^{\rm eq}$, due to the slower evolution of the maximum energy with $L_{\rm wind}^{\rm eq}$ for models B and C ($E_{\rm max}\propto L_{\rm wind}^{\rm eq\,1/4}$, see Sect.~4) and also to the increasing importance of photo-interactions with increasing values of $L_{\rm wind}^{\rm eq}$. Part of 
the increase of the maximum energy due to the larger value of the magnetic fields for models B and C is indeed counterbalanced to some extent as particles are accelerated to higher energies where photo-interaction rates are higher. Note that this latter consideration is also relevant for neutron and secondary particle emission. 

Overall, one can see that according to our modeling of mildly relativistic acceleration at GRBs internal shock, relatively heavy nuclei (say, Si and above) can reach energies above $10^{20}$ eV in most cases, except for the lowest values of $L_{\rm wind}^{\rm eq}$ for model A. Intermediate nuclei such as CNO can reach energies above $10^{19.5}$ eV and approach $10^{20}$ eV for most values of $L_{\rm wind}^{\rm eq}$ for models B and C, while protons can only approach at most $10^{19.5}$ eV for the highest values of $L_{\rm wind}^{\rm eq}$ and remain below $10^{19}$ eV in most cases. As already mentioned in Sect.~4, these low maximum energies for the proton component can only be increased, in principle, by invoking a faster acceleration mechanism, a magnetic field configuration allowing acceleration rates close to the so-called Bohm rate or larger magnetic fields than those implied by our hypotheses\footnote{Invoking significantly larger magnetic fields, rather than a faster acceleration process, would imply more extreme assumptions on the wind luminosities than those of models B and C. In this case, relying only on stronger magnetic fields, synchrotron losses could prevent protons from being accelerated above $10^{20}$ eV.}. 

\subsection{Neutrino spectra}

Neutrinos should be natural by-products of cosmic-rays acceleration at GRB internal shocks. Interaction of very-high and ultra-high energy cosmic-rays through photomeson production (and to a lower extent through hadronic interactions) during their acceleration or their escape is expected to ultimately trigger ample neutrino emission.
In our Monte-Carlo simulation, the production of neutral and charged mesons by protons and neutrons is handled by the SOPHIA event generator (M\"{u}cke et al. 2000), while pion production for nuclei is treated following Rachen (1996) in the first delta resonance approximation, taking into account pion reabsorption and energy losses in the residual nucleus (see also Allard et al., 2006, for more details on the implementation). Let us note that the latter treatment  is not necessarily a good approximation for interactions with photons above 1 GeV (in the nucleus rest frame), but this energy range for interactions does not have a dominant contribution (at most a few percent for nuclei and $\sim 10-15\%$ for protons) for the range  of Lorentz factors protons and nuclei manage to reach in our calculations ($\gamma_{\rm nuc}<10^8$ in the wind frame, in all the cases we previously considered). Once produced, the charged mesons and muons may suffer severe synchrotron losses before decaying, depending on their mass 
and on the ambient magnetic fields at the time of their production (see eq.~\ref{eq:tsyn}). Synchrotron losses of charged mesons and muons significantly affect the energy of the produced resulting at the end of the decay chain.

\begin{figure}
{\rotatebox{0}{\includegraphics[scale=0.3]{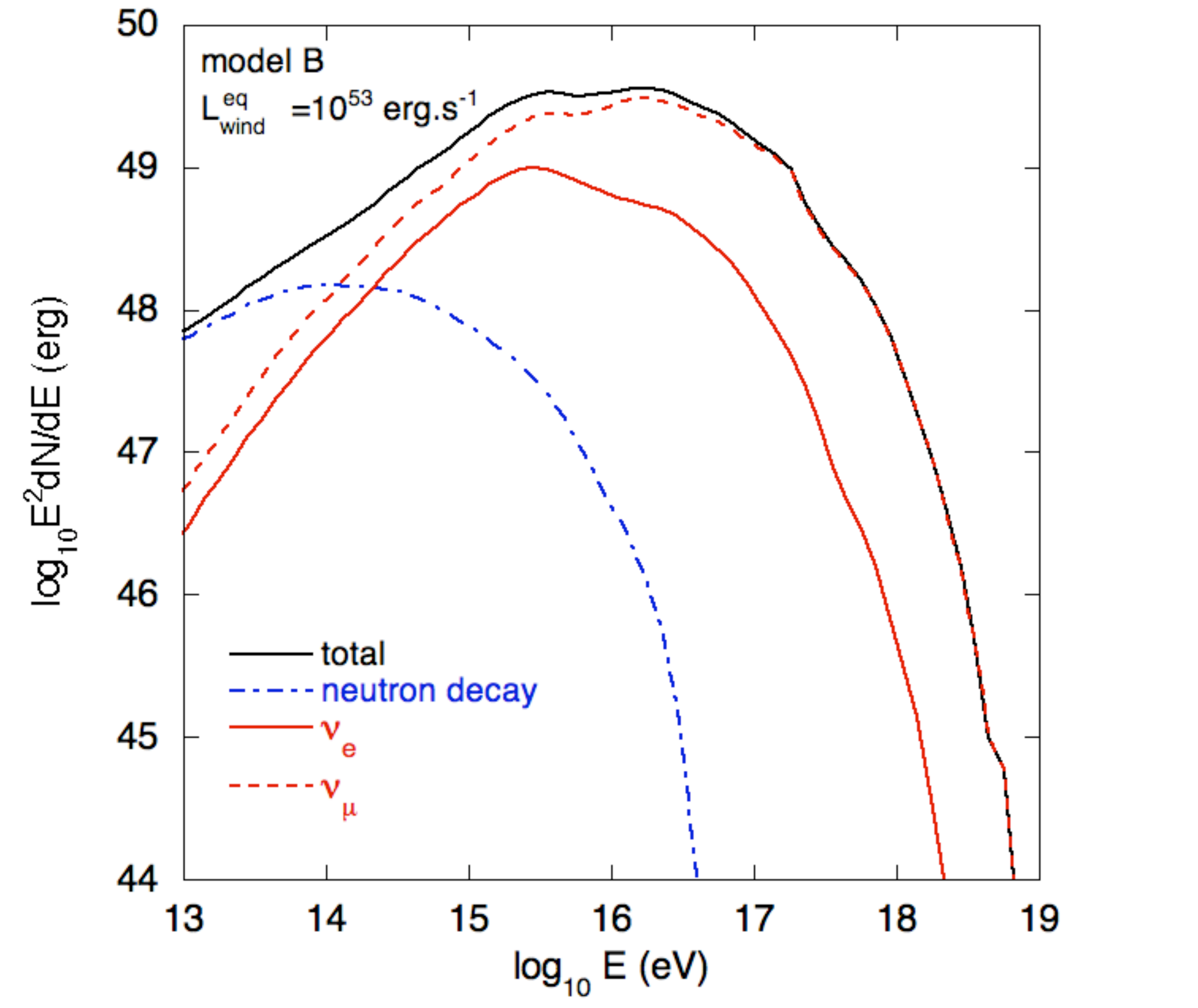}}}
{\rotatebox{0}{\includegraphics[scale=0.3]{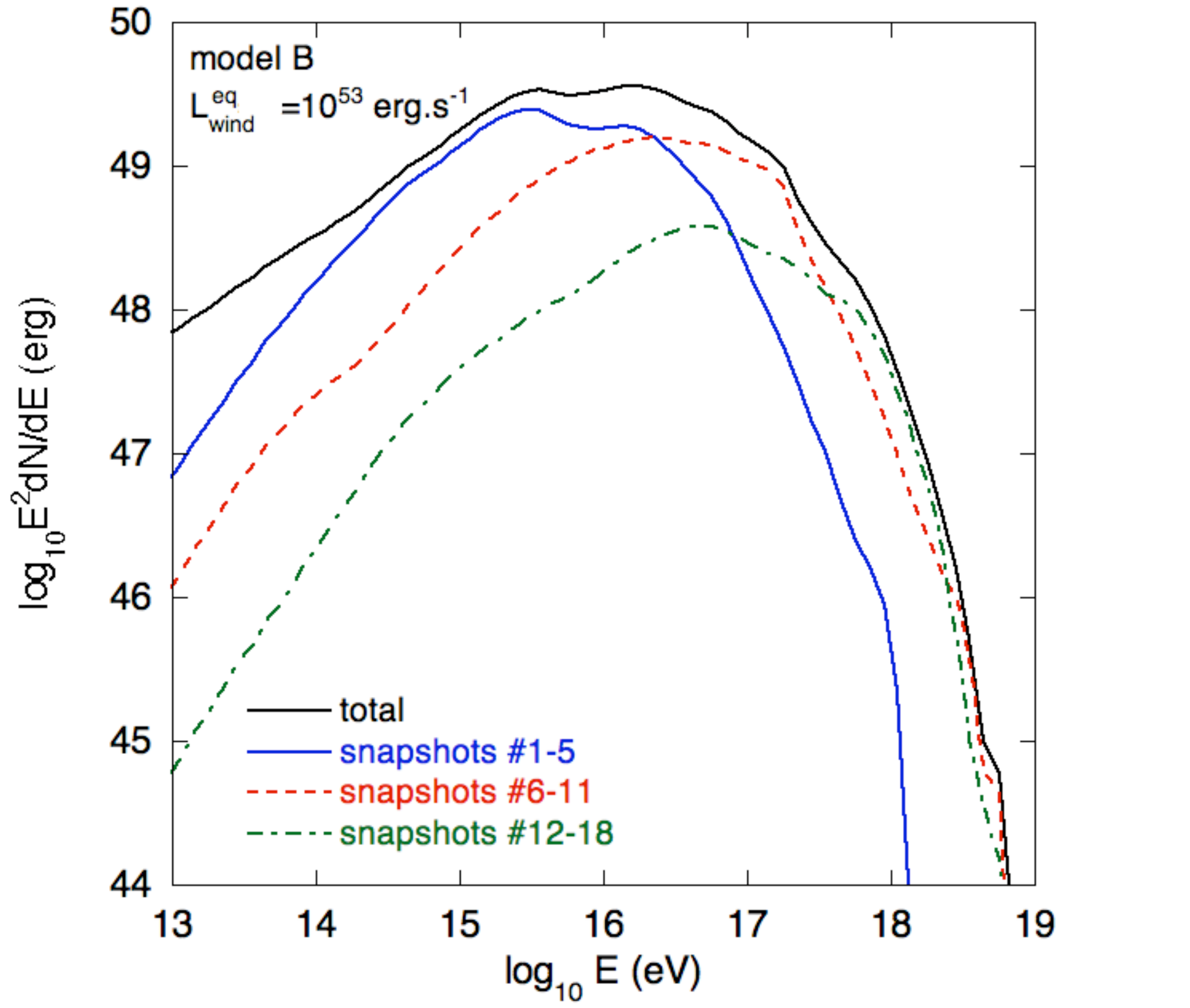}}}
\caption{Neutrino spectrum, multiplied by $E^2$, emitted by a GRB (as seen in the central source frame), assuming for illustration $L_{\rm wind}^{\rm eq}=\rm 10^{53}\,erg\,s^{-1}$, in the case of model B. The normalization is obtained by integrating over the whole shock propagation. The upper panel shows the contribution of $\nu_e$ for neutron decay, $\nu_e$ and $\nu_\mu$ for photomeson production and hadronic interactions. The lower panel shows the contribution of different phases of the shock propagation (see text).}
\label{fig:neut53}
\end{figure}

For the time being, we only consider neutrinos produced at the sources, associated with the cosmic-ray output discussed in the previous section. Let us note that we add the contribution of neutrinos from neutron decay to the neutrinos produced during cosmic-ray acceleration and escape, although these neutrinos are not produced \emph{stricto sensu} in the source. The spectrum of neutrinos emitted during cosmic-ray acceleration and escape (integrated on the whole shock propagation) for the case $L_{\rm wind}^{\rm eq}=10^{53}\,\rm erg\,s^{-1}$ and model B is displayed in Fig.~\ref{fig:neut53}. On the upper panel, we show the contribution of electron neutrinos from neutron decay and the electron and muon neutrinos from photomeson production. Let us note that we do not separate the contribution of the neutrinos from that of the anti-neutrinos or the contribution of the different charged mesons, detailed studies for the pure proton case can for instance be found in Hummer et al. (2010). At low energy, below $\sim 
10^{14}$ eV, the neutrino spectrum is dominated by electronic (anti-)neutrinos (originating mainly from nuclei photodisintegration, see the previous section). Above this energy, the contribution of photomeson production becomes dominant. As expected in the case of highly magnetized astrophysical sources (see for instance M\"{u}cke \& Protheroe 2001; M\"{u}cke et al. 2003,  for the case of blazars and BL Lacs, Murase \& Nagataki 2006; H\"{u}mmer, Baerwald \& Winter 2012; Baerwald et al. 2014 for the case of GRBs), the spectrum of electron neutrinos (whose energy is affected by the successive synchrotron cooling of pions and muons) cuts at lower energy than the spectrum of muon neutrinos. We emphasize that, except for the low energy part of the spectrum dominated by the contribution of neutron decay, most of the neutrinos displayed in Fig.~\ref{fig:neut53} are emitted by primary protons (at a level of $\sim 60\%$ whatever the assumed luminosity) due to the larger gap in energy between the primary cosmic-ray 
and the produced neutrino in the case of nuclei and the fact that the pion production process is significantly reduced by the dominant contribution of giant dipole resonance to nuclei photodisintegration  for the Lorentz factors, which nuclei are able to reach during their acceleration.

In the lower panel of Fig.~\ref{fig:neut53}, the total neutrino spectrum is decomposed between different stages of the shock propagation, namely between the snapshots \#1 to \#5 ($r_{\rm sh}$ between $\sim10^{14}$ and  $\sim 6\,10^{14}$ cm), \#6 to \#11 ($r_{\rm sh}$ between $\sim 6\,10^{14}$ and  $\sim 3\,10^{15}$ cm) and \#12 to \#18 ($r_{\rm sh}$ between $\sim 3\,10^{15}$ and  $\sim 10^{16}$ cm). Neutrinos from neutron decay are not included in the decomposition. As can be seen the neutrino emission is more intense and peaks at lower energy during the early stages of the shock propagation, while neutrino emission becomes fainter and shifted to higher energy at later stages. This trend (which is met for all the luminosity cases we considered) is due to the combined evolution of the comoving photon background energy and density and the magnetic field during the shock propagation. At the beginning of the shock propagation, the photon density as well as the comoving peak energy are higher, cosmic-rays have 
larger interaction probabilities and interactions mostly occur at lower energies than at later stages of the shock propagation. Moreover, the magnetic field is also initially more intense which results in a faster synchrotron cooling of the produced charged mesons and muons and also contribute to the emission of lower energy neutrinos. Unlike photon prompt emission, neutrino emission from GRBs internal shocks would then be expected to evolve from soft to hard spectra. The neutrinos "pulse shape" is also expected to be slightly different from the photon pulse shape as a the fraction of the total energy dissipated in neutrinos emitted during the early phase of the shock propagation is expected to be larger than the corresponding fraction for photons : $\sim 55\%$ of the total energy dissipated in neutrinos is emitted on the distance range corresponding to snapshots \#1 to \#5 while less than $\sim30\%$ of the total energy dissipated in prompt emission photons is emitted on the same distance range, as can be deduced from the bottom-
right panel of Fig.~\ref{fig:param}.

\begin{figure}
{\rotatebox{0}{\includegraphics[scale=0.33]{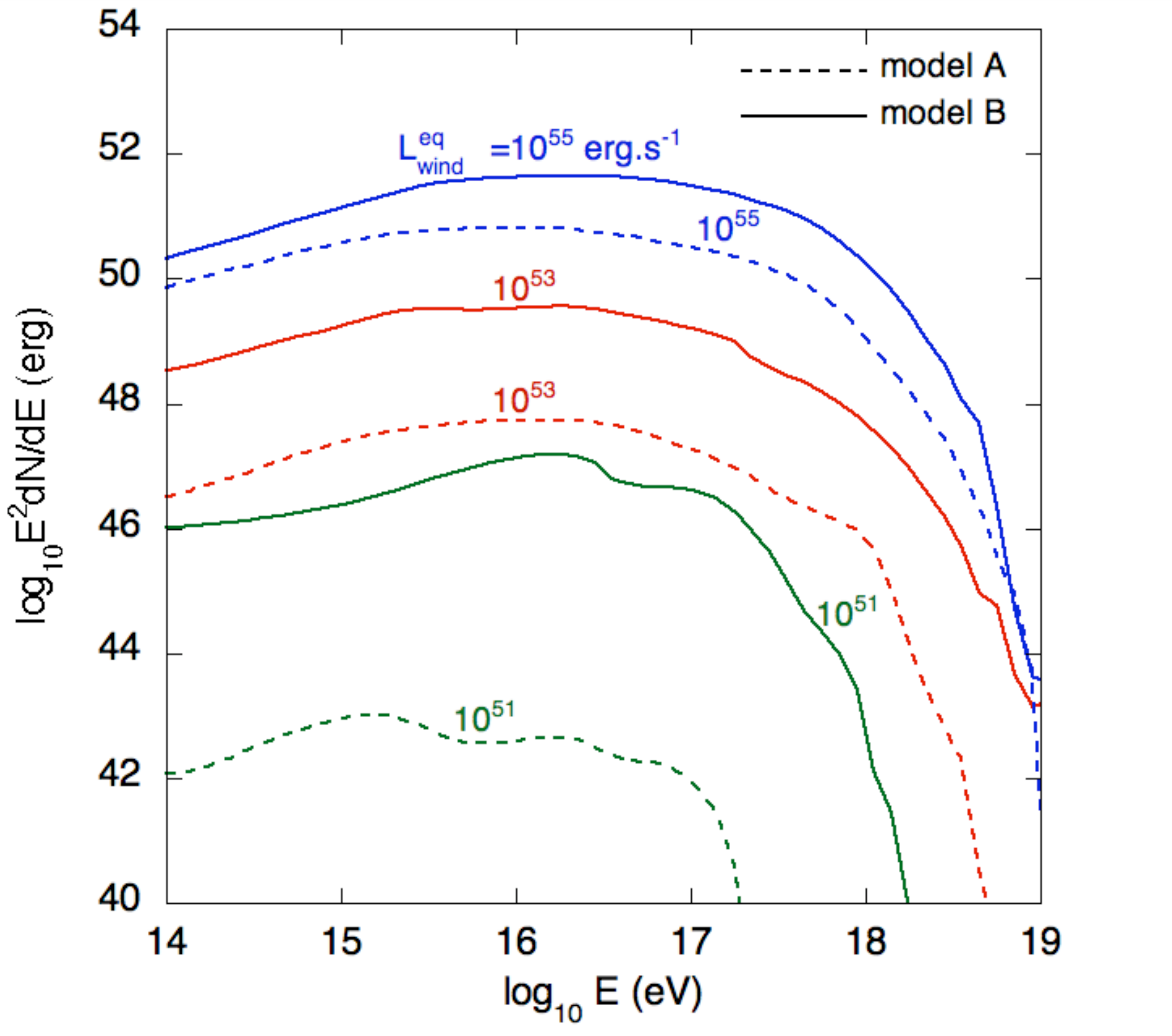}}}
\caption{Neutrino spectra emitted by GRBs for different values of $L_{\rm wind}^{\rm eq}$: $10^{51}$, $10^{53}$ and $10^{55}\rm \,erg\,s^{-1}$, for models A (dashed lines) and B (full lines).}
\label{fig:neutalllum}
\end{figure}

We now come to the calculations of the neutrino emission for different values of $L_{\rm wind}^{\rm eq}$.  The emitted neutrino spectrum for the cases $L_{\rm wind}^{\rm eq}=10^{51}$, $10^{53}$ and $\rm 10^{55}\,\rm erg\,s^{-1}$ is displayed in Fig.~\ref{fig:neutalllum} for models A (dashed lines) and B (plain lines). One expects in principle the neutrino output of a given GRB to be more or less proportional to the energy communicated to cosmic-rays and to the ambient photon density. In the case of model A, both quantities are proportional to $L_{\rm wind}^{\rm eq}=L_{\rm wind}$ so the amount of energy emitted in neutrinos is approximately proportional to $(L_{\rm wind}^{\rm eq})^2$, as can be seen in Fig.~\ref{fig:neutalllum}. As a result, while the wind luminosity and the prompt emission luminosity spread over 4 orders of magnitude, the corresponding neutrino luminosity spreads over $\sim$8 orders of magnitude. For model B, the expected range of neutrino luminosities is reduced, since the energy released in 
cosmic-rays is proportional to $L_{\rm wind}$, i.e proportional to $L_{\rm wind}^{\rm eq\,1/2}$, while the photon density is still proportional to $L_{\rm wind}^{\rm eq}$. The difference  between models A and B in terms of neutrino output, for a given value of $L_{\rm wind}^{\rm eq}$,  mostly relies on the difference in the energy communicated to cosmic-rays, which is expected to be proportional to the product $\epsilon_{\rm cr}\times L_{\rm wind}$ (the photon densities and energy distribution being in good approximation identical for both models). As an example, for the case $L_{\rm wind}^{\rm eq}=10^{51}\,\rm erg\,s^{-1}$, we assumed $L_{\rm wind}=10^{51}\,\rm erg\,s^{-1}$ and $\rm \epsilon_{\rm cr}=1/3$ for model A, while  $L_{\rm wind}=3\,10^{53}\,\rm erg\,s^{-1}$ and $\rm \epsilon_{\rm cr}\simeq0.9$ for model B. These different assumptions imply a neutrino output three orders of magnitude larger for model B, following our simple reasoning, which is more or less what can be observed in Fig.~\ref{fig:neutalllum}, for 
this particular value of  $L_{\rm wind}^{\rm eq}$. For more detailed comparisons of the neutrino output, between model A and B or between different values of $L_{\rm wind}^{\rm eq}$, one has to take into account the value of the ambient magnetic field and the energy distribution of the ambient photon background, which both affect the energy reached by the cosmic-rays, the energy of the photoproduced charged mesons, their cooling and finally the energy of the emitted neutrinos and  the energy transferred from cosmic-rays to neutrinos. Fuller discussions of these aspects can be found, for instance, in the recent paper by He et al. (2012) or in Murase \& Nagataki (2006).

It is important to note that high energy photons are also emitted as by-products of cosmic-ray acceleration. The mechanisms at the origin of this emission, besides the cooling of accelerated electrons considered in Sect.~2, are the decay of neutral mesons, synchrotron cooling of photoproduced $\rm e^+e^-$ pairs and charged mesons and muons. The high energy photons emitted by these processes can either escape from the GRB environment or induce electromagnetic cascades that would contribute to the prompt emission at lower energy. Although we did not fully calculate the cascading of these high energy photons and their final contribution to the prompt emission, we however estimated the energy transferred from cosmic-rays to high energy photons and compared it with the prompt emission energy. It turns out that the amount of energy transferred from cosmic-ray to photons does not exceed a few percents of the prompt emission calculated in Sect.~2 in the case of model A (it is even almost negligible at low luminosity)
 while it represent 10 to 20\% of the prompt emission in the case of models B and C. Although non negligible, this supplementary contribution to the prompt emission does not invalidate our calculations of cosmic-ray photo-interactions. The full calculations of the contribution of high energy photons emitted by accelerated cosmic-rays to the prompt emission will be considered in a forthcoming paper.

\section{Diffuse fluxes}

\subsection{GRB luminosity function and cosmic-ray effective spectrum}

\begin{figure}
{\rotatebox{0}{\includegraphics[scale=0.32]{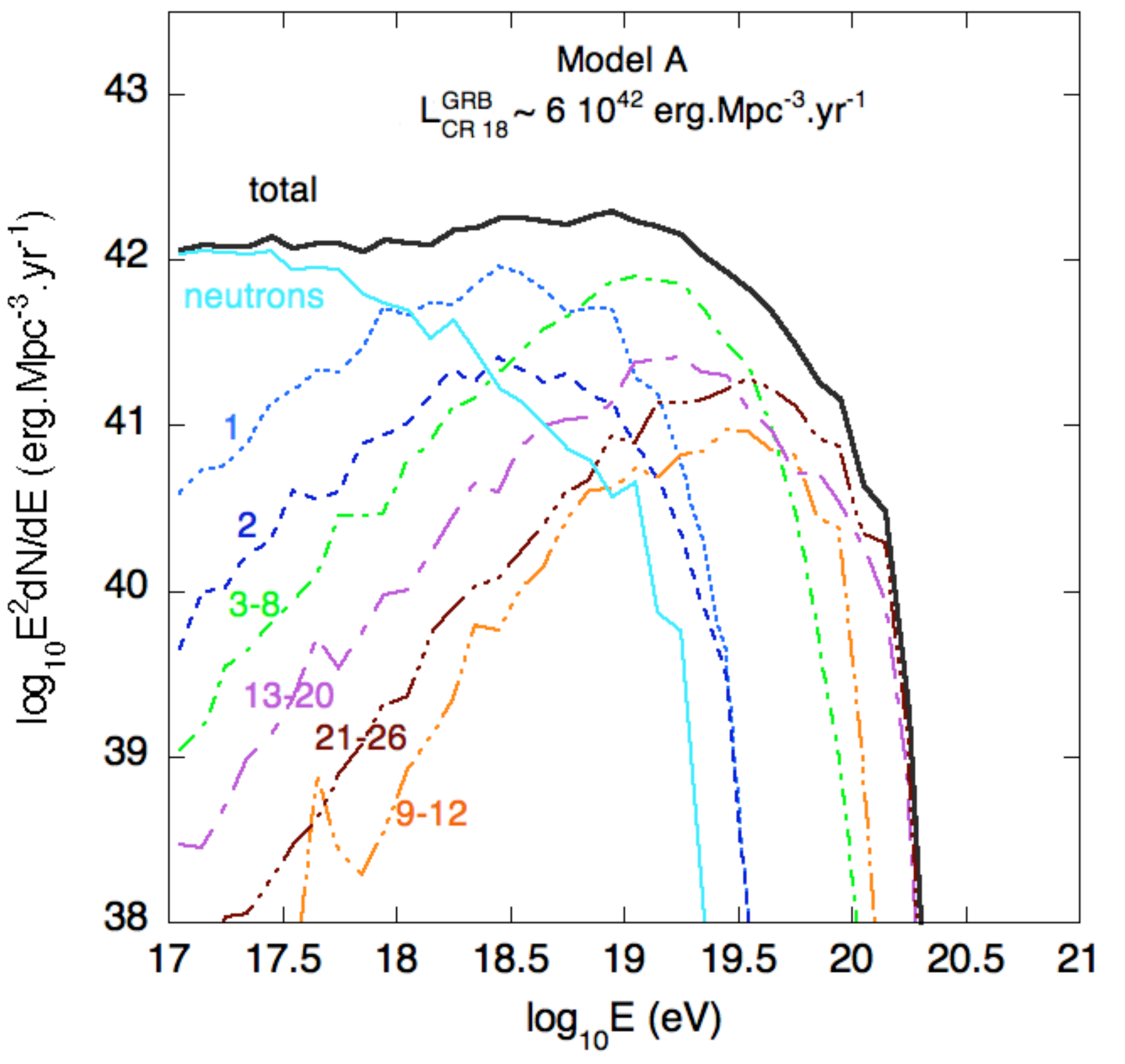}}}
{\rotatebox{0}{\includegraphics[scale=0.32]{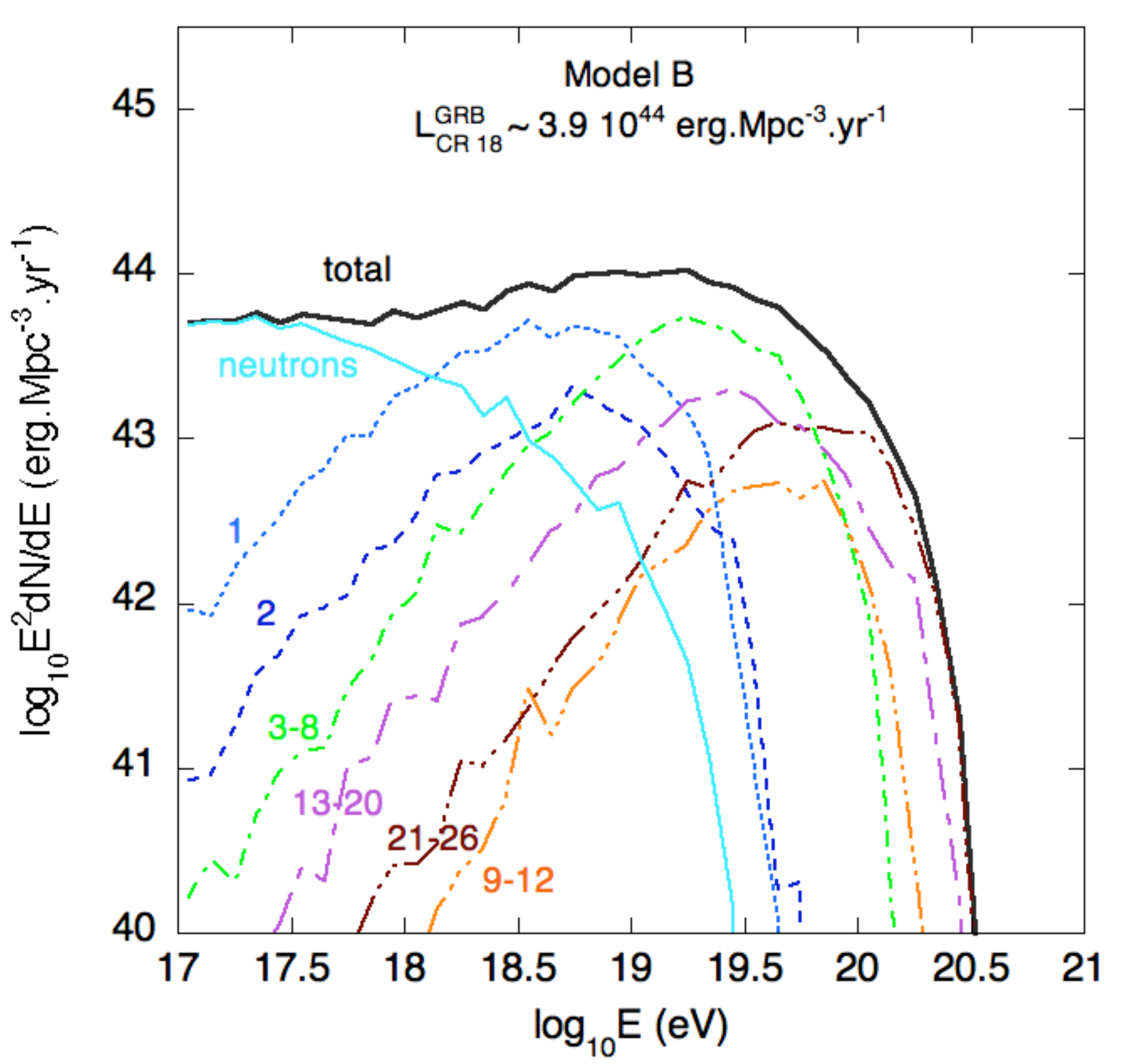}}}
{\rotatebox{0}{\includegraphics[scale=0.32]{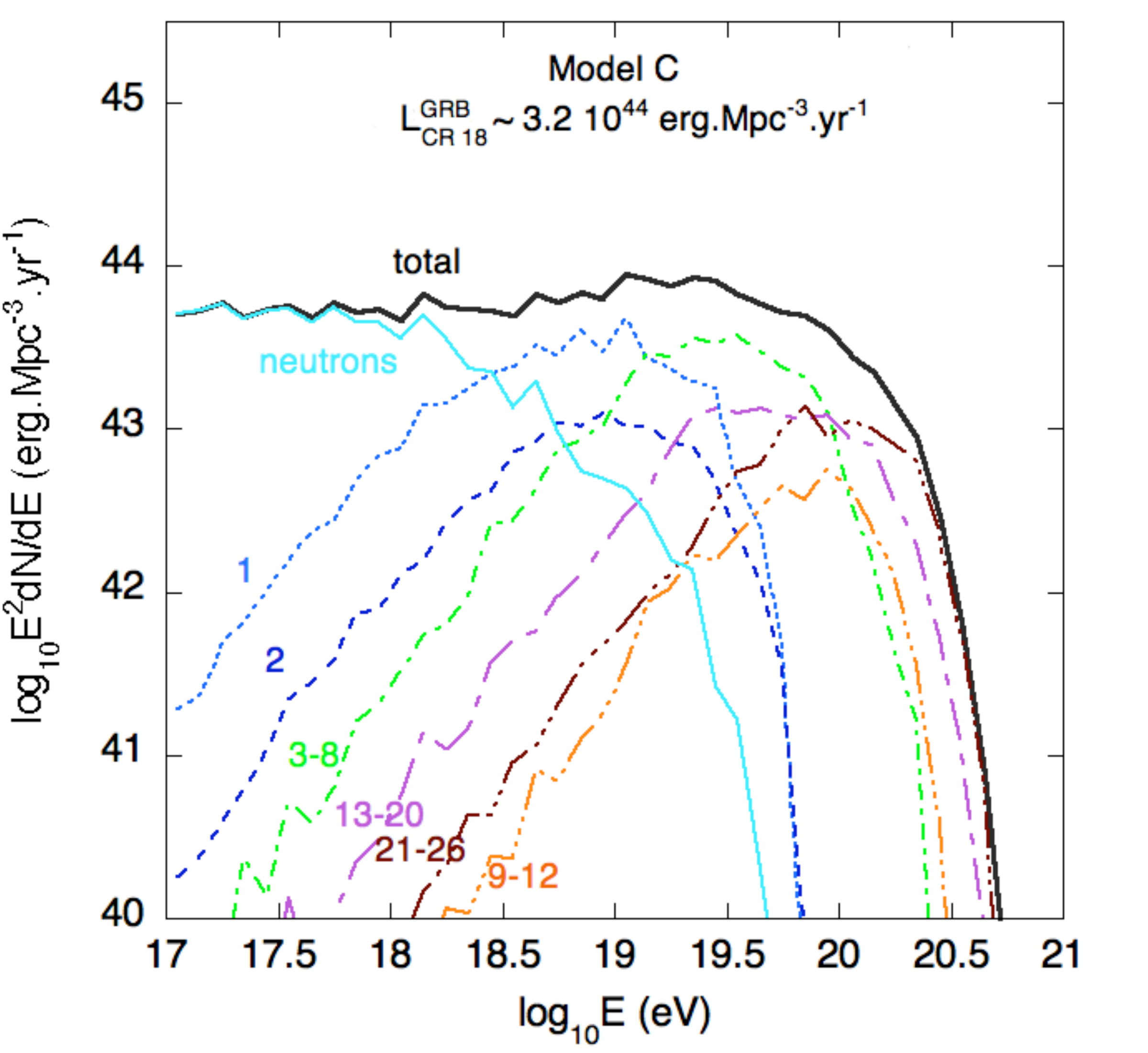}}}
\caption{Differential luminosity density (also called effective injection spectrum in the text) of UHECRs emitted by GRBs, calculated for models A (top), B (center) and C (bottom). The contributions of different groups of nuclear species is shown.}
\label{fig:eff}
\end{figure}

In the previous section, we calculated the emitted cosmic-ray and neutrino spectra for synthetic single pulse GRBs assuming values of $L_{\rm wind}^{\rm eq}$ between $10^{51}$ and $\rm 10^{55}\,\rm erg\,s^{-1}$ for models A, B and C. To calculate diffuse fluxes observable on Earth, which is the main goal of our study, one needs to make additional assumptions on the GRB luminosity function, the local rate density and its cosmological evolution. Numerous works have been recently dedicated to estimate the GRB prompt emission luminosity function, mostly based on the analysis of BATSE, HETE2 and/or \emph{Swift} data (see for instance Daigne et al. 2006; Guetta \& Piran 2007; Liang et al. 2007; Le \& Dermer 2007;  Guetta \& Della Valle 2007; Zitouni et al. 2008; Wanderman \& Piran 2010). In the following, we use one of the most recent estimates of the GRB prompt emission luminosity function and rate density derived from the analysis of \emph{Swift} GRBs by Wanderman and Piran (2010). The authors found that the 
luminosity function of GRBs with $L_\gamma \geq10^{50}\,\rm erg\,s^{-1}$ is well reproduced by a broken power law:
\begin{equation}
\frac{dN_{\rm GRB}}{dL_\gamma}(L_\gamma) \propto \left\lbrace\begin{array}{cl}
& L_\gamma^{-\alpha} \ \ \ {\rm for}\ \ L_\gamma\leq L_\star \\
& L_\gamma^{-\beta}\ \ \ {\rm for}\ \  L_\gamma > L_\star\\
\end{array}\right.
\label{eq:lf}
\end{equation}
where the power indices are $\rm \alpha=1.2$, $\rm \beta=2.4$ and the break luminosity $L_\star=10^{52.5}\,\rm erg\,s^{-1}$. To normalize the luminosity function, the authors estimated the local (at redshift $z$=0) rate density of GRBs with $L_\gamma \geq10^{50}\,\rm erg\,s^{-1}$,  $\rm \rho_{GRB}(0)=1.3\,Gpc^{-3}\,yr^{-1}$.\\
For the redshift evolution of the (comoving) rate density, $\rho_{GRB}(z)$, a broken power law behaviour was also found by the same authors,
\begin{equation}
\rho_{GRB}(z)=\rho_{GRB}(0) \times \left\lbrace\begin{array}{cl}
& (1+z)^{n_1} \ \ \ {\rm for}\ \  z\leq z_\star\\
& (1+z_\star)^{n_1-n_2}\times (1+z)^{n_2}\ \ \ {\rm for}\ \  z>z_\star\\
\end{array}\right.
\label{eq:evol}
\end{equation}
where $n_1=2.1$, $n_2=-1.4$ and $z_\star=3$.\\
For our assumption on the Lorentz factor distribution (see Sect.~2), we found that $L_\gamma$ is approximately related to $L_{\rm wind}^{\rm eq}$ by the relation $L_\gamma\simeq 0.05 \times L_{\rm wind}^{\rm eq}$. It is then easy to translate the prompt emission luminosity function into a cosmic-ray luminosity function for each energy redistribution model. In the previous sections, we considered single pulse synthetic GRBs with a wind duration $t_{\rm wind}=2$s, which results, in the central source frame, in a photon emission of $\sim$2s (a pulse with a fast rise and exponential decay  with a half width of $\sim$1s). Although this kind of pulse is quite typical of long GRB light curves, the total photon emission lasts in general much longer than 2s, which means that the central source is active on timescales larger than 2s. 
To calculate the GRB total luminosity density in cosmic-ray, we treat this single synthetic pulse as an elementary pattern repeated several times during the total duration of the central source activity.
We assume that the Lorentz factor distribution (see Fig.~\ref{fig:lwind}) and wind duration we considered so far is repeated 10 times during the total source activity, which give a total duration of 20s
which closely corresponds to the peak of the $\rm T_{90}$ distribution (estimated in the central source frame), deduced from \emph{Swift} observations for GRBs with known redshift (Sakamoto et al. 2011). Of course, these assumptions represent an oversimplification that shortcuts the great diversity of light curves and other properties seen in GRBs observations. These choices are however dictated by our current lack of understanding of the origin of the diversity of the observed properties of GRBs (see for instance Bosnjak \& Daigne 2014) and also by the large computational time required for our calculation. However, our previous results and discussions,
 in particular in sections 2 and 4,  give some hints on how the cosmic-ray output could vary assuming longer or shorter light pulses or different Lorentz factor contrasts. We believe that, if we had modeled the burst to burst variations of these parameters, the overall UHECR emission from GRBs in the extragalactic medium  would not be very different from what we obtain here with the above simplifications.  

As mentioned above, the prompt emission luminosity function given in Eq.~\ref{eq:lf} can easily be translated into a wind luminosity function and then into a cosmic-ray luminosity density (expressed in $\rm erg\,Mpc^{-3}\,yr^{-1}$) by convoluting the luminosity function with the cosmic-ray spectra emitted by single GRBs for various values of $L_{\rm wind}^{\rm eq}$ calculated in the previous section. We use the spectra (see Figs.~\ref{fig:spectra_a} and \ref{fig:spectra_b}) calculated for 21 values of $L_{\rm wind}^{\rm eq}$ between $10^{51}$ and $\rm 10^{55}\,\rm erg\,s^{-1}$ (with a step of 0.2 in log of $L_{\rm wind}^{\rm eq}$). The result is shown in Fig.~\ref{fig:eff}, where our estimates for the differential luminosity density of cosmic-rays emitted by GRBs are shown for models A, B and C. These \emph{effective} spectra of escaping cosmic-rays, which represent a weighted sum of the spectra obtained in Sect.~5, retain the main features we already discussed. This includes very hard spectra for nuclei 
heavier than protons and  much softer spectra for protons if one includes the contribution of neutrons (mostly produced by photodisintegration of nuclei in high luminosity GRBs). The maximum energy reached by nuclei (CNO and above) is significantly larger than that of protons, but the difference is smaller than if it were proportional to the charge of the nucleus. This behaviour was already visible for individual GRBs of intermediate luminosities ($L_{\rm wind}^{\rm eq}\simeq10^{53}\,\rm erg\,s^{-1}$) for which photodisintegration already limits the maximum energy reachable by nuclei. The difference in maximum energy for the different energy redistribution models is also visible on these \emph{effective} spectra. As already discussed, the energies reached in the case of model C are the highest since the physical parameters assumed for this model imply the largest magnetic fields for a given value of $L_{\rm wind}^{\rm eq}$. Conversely, model A, with the lowest magnetic fields, predicts the lowest fraction of 
cosmic-rays accelerated above $10^{20}$~eV.\\
Perhaps the most interesting information is obtained by integrating the differential luminosity density in order to estimate the total luminosity density emitted in cosmic-rays above $10^{18}$~eV, $ L_{\rm CR\,18}^{\rm GRB}$, for each energy redistribution model. For models A, B and C, we obtain respectively $ L_{\rm CR\,18}^{\rm GRB}\simeq6\,10^{42}$, $3.9\,10^{44}$ and $\rm 3.2\,10^{44}\,erg\,Mpc^{-3}\,yr^{-1}$. These values can be compared with the estimates of the luminosity density, $L_{\rm CR\,18}$, required from cosmic-ray sources above $10^{18}$~eV to reproduce the UHECR flux observed on Earth. This luminosity is of the order of a few times $\rm 10^{44}\,erg\,Mpc^{-3}\,yr^{-1}$, depending on the  assumed UHECR phenomenology  (see for instance Berezinsky et al. 2006; Katz, Budnik \& Waxman 2009; Decerprit \& Allard 2011, the latter includes estimates for various compositions at the sources).\\ A rapid comparison between the values of $ L_{\rm CR\,18}^{\rm GRB}$ found for each model and $L_{\rm CR\,18}
$ shows that the cosmic-ray luminosity density obtained with model A, assuming equipatition, is well below what is required to reproduce the UHECR flux on Earth. This should not come as a surprise since the luminosity function we use, together with the assumed 20s average duration, implies a prompt emission luminosity density of the order of $\sim 2\,10^{44}\rm \,erg\,Mpc^{-3}\,yr^{-1}$, which is approximately the luminosity density required for cosmic-ray above $10^{18}$ eV. This means that, in the case of GRBs, the luminosity density in cosmic-ray above $10^{18}$ eV has to be of the order or even larger than the whole prompt emission luminosity density. This, of course, cannot be achieved if the energy dissipated at internal shocks is equally shared between electrons and cosmic-rays, unless for some reason most of this energy goes to particles accelerated above $10^{18}$ eV. In the case of our modeling, however, escaping cosmic-rays above $10^{18}$ eV only carry a few percent of the total energy communicated to cosmic-rays.\\
On the other hand, in the case of models B and C, for which we assume a much larger fraction of the dissipated energy communicated to cosmic-rays together with significantly larger wind luminosities (at least for low luminosity GRBs), the obtained values of $ L_{\rm CR\,18}^{\rm GRB}$  are of the same order as the required cosmic-ray luminosity density above $10^{18}$ eV. \\
To further discuss the compatibility of our predictions with UHECR observations, one needs to propagate these cosmic-rays accelerated at GRBs from their sources to the Earth, through the extragalactic medium.

\subsection{Cosmic-ray diffuse flux}

\subsubsection{Calculation method and hypotheses}

 To calculate the diffuse UHECR flux received from GRBs on Earth, the simplest approach would be to use the "effective" injection spectra, calculated in the previous section, in the continuous injection and continuous source distribution approximations. Such an approach would, however, ignore some major subtleties implied by the fact that GRBs are very rare events, within the (spacetime) horizon\footnote{The term "horizon" refers to the largest distance from which a cosmic-ray of a given mass and  energy can come, due to the energy losses experienced during its propagation. Equivalently the "horizon" can refer to the maximum possible propagation time. Although the latter is more relevant to the propagation of cosmic-rays in the presence of magnetic fields, we will mostly discuss the horizon as a (curvilinear) distance in the following.} of the highest energy cosmic-rays, and that, in very good approximation, UHECRs are emitted instantaneously. In that case, even if our models predict the emission of 
cosmic-rays 
above $10^{20}$ eV, it is not guaranteed, in principle, that the UHECR emission from GRBs should result in a roughly steady UHECR flux on Earth.
It would not be the case for instance, in the absence of intervening magnetic fields, in which case the duration of the cosmic-ray signal from a single GRB would be of the same order as that of the prompt photon emission. Within a radius of 100 Mpc from the Earth, which approximately corresponds to the cosmic-ray horizon above 80 EeV, the GRB rate should be $\sim10^{-3}\,\rm yr^{-1}$ (according to the estimate of the local rate by Wanderman \& Piran 2010). Thus, if there were no magnetic field to spread the cosmic-ray signal in time, it would be very unlikely to observe any cosmic-ray signal above a few $10^{19}$ eV, even after several years of observations. If intervening magnetic fields are strong enough to spread cosmic-ray arrival times, from a single GRB, by a few thousand years or more, the overlap  between cosmic-ray from several GRBs can lead to a relatively steady cosmic-ray flux on Earth, even at the highest energies.
To estimate quantitatively  the diffuse cosmic-ray flux expected on Earth, one must then take into account the discreteness and instantaneity of GRB events as well as the effects of extragalactic magnetic fields on UHECR propagation. 

The extragalactic magnetic field (EGMF) is poorly known, and its spatial distribution, intensity, coherence length, time evolution, and origin are all uncertain. Observations imply the presence of $\mu$G fields in the core of large galaxy clusters. However, the spatial extension of these large field regions and their volume filling factor in the universe are difficult to evaluate. Efforts have been made to model local magnetic fields using simulations of large-scale structure formation that include an MHD treatment of the magnetic field evolution (see the pioneering studies by Dolag et al. 2002; Sigl et al. 2004; or more recent calculations by Das et al. 2008; Ryu et al. 2008, 2010; Donnert et al. 2009). Some of these simulations are constrained by the local density and velocity field to provide more realistic field configurations in the local universe. These simulations rely on different assumptions regarding the origin of the fields and the mechanisms involved in their growth. They are ultimately 
normalized to 
the values observed at the present epoch in the central regions of galaxy clusters (see the discussion in Kotera \& Lemoine 2008). The outcome of the different simulations strongly differ. In particular, the volume filling factors for strong fields (above 1 nG, say) vary by several orders of magnitude from one simulation to the other. In contrast, an interesting simple alternative to complex hydrodynamical simulations, offering more freedom to test different models of the magnetic field evolution with local density, has been proposed by Kotera \& Lemoine (2008).\\
In view of the above-mentioned uncertainties, and since we are only interested in the spread in time of the propagated UHECR, we use a simplified approach, assuming that the universe is filled with a purely turbulent, homogeneous magnetic field, and will study the dependence of the high-energy cosmic-ray spectrum and composition on the variance of the turbulent field. Following Globus, Allard \& Parizot (2008), we model the extragalactic turbulent field using the algorithm proposed by Giacalone and Jokipii (1999) (see Sect.~3) normalized to different variances. We set the value of the maximum turbulent scale $\lambda_{\rm max}^{\rm EGMF}$ to 1 Mpc, which corresponds to a coherence length $\lambda_{\rm c}^{\rm EGMF}\simeq \lambda_{\rm max}^{\rm EGMF}/5 \simeq 0.2\,\rm Mpc$.

Using the GRB luminosity function (Eq.~\ref{eq:lf}), the GRB local rate, $\rm \rho_{\rm GRB}(0)$, and the associated cosmological evolution of the comoving GRB rate (Eq.~\ref{eq:evol}), estimated by Wanderman \& Piran (2010), one can simulate realizations of the history of GRB explosions in the universe. In practice, we discretize the universe in space and redshift bins. We also discretize the GRB luminosity range in 21 bins between $L_{\rm wind}^{\rm eq}=10^{51}$ and $10^{55}\rm\,erg\,s^{-1}$ (which corresponds to $L_\gamma$ between $5\,10^{49}$ and $5\,10^{53}\rm\,erg\,s^{-1}$). For each luminosity bin, we compute the number of GRBs occurring in each bin of spacetime, according to the mean values expected (according to Eqs.~\ref{eq:lf} and \ref{eq:evol}) and using Poisson statistics. Since the average number of GRBs at low distance and low redshift is small, we compute 300 different realizations in order to estimate the effect of the fluctuations of the number of GRBs in the very local universe on the 
expected UHECR diffuse flux.\\
For a given realization of the history of GRB explosions, the diffuse cosmic-ray flux, observed \emph{now} on Earth, can be calculated by summing the contributions of each individual GRBs. A given GRB is characterized by its injection spectrum and composition (calculated in Sect.~5), $Q_i(E; L_{\rm \gamma}, \epsilon)$, where $i$ refers to a given nuclear species, $L_\gamma$ is related to $L_{\rm wind}^{\rm eq}$ and $\epsilon$ refers to a given energy redistribution model (either model A, B or C), emitted instantaneously at the look-back time $t$ (or equivalently at the redshift $z(t)$) and at a proper comoving distance $r$, which does not have to correspond to the retarded time of the photon emission because the trajectory is not rectilinear. \\
One can define the function $F(A, E, r; t, L_\gamma, \epsilon)$ (hereafter, contribution function) representing the contribution to the observed UHECR flux of the particles emitted by a GRB at a proper comoving distance $r$ exploding at a given time $t$ with a photon luminosity $L_\gamma$. To compute this contribution function, we use a Monte-Carlo procedure. We inject particles according to a given injection spectrum, $Q_i(E; L_{\rm \gamma}, \epsilon)$. Then we propagate particles from the injection time $t$ to the present time (i.e, from redshift $z(t)$ to $z=0$). After that propagation time, particles are spread over a large volume around the source, with a variety of masses and energies as a result of their interactions with the photon backgrounds. We then simply register their number density, energy spectrum and composition in different bins of proper comoving distance from the GRB, $r_i$ (where isotropy is assumed according to our simplifying assumption on the EGMF structure). Specifically, we count 
the 
number of particles $N(A_i, E_j, r_k; t, L_\gamma, \epsilon)$ found with mass $A_i$, in the $j$th energy bin, i.e between $E_j$ and $E_{j+1}=E_j+\Delta E_j$ and in the $k$th distance bin, i.e between $r_k$ and $r_{k+1}=r_k+\Delta r_k$, from which we derive the contribution function :
\begin{equation}
F(A, E, r; t, L_\gamma, \epsilon)=\frac{N(A_i, E_j, r_k; t, L_\gamma, \epsilon)}{ \Delta E_j \times 4 \pi r_k^2\Delta r_k}
\label{eq:contrib}
\end{equation}
This is a step function, constant for $E_j \leq E \leq E_j+\Delta E_j$ and $r_k \leq r \leq r_k + \Delta r_k$. $F(A, E, r; t, L_\gamma, \epsilon)$ is expressed in $\rm eV^{-1}\,cm^{-3}$ and depends of course on the injection spectrum which for our modeling depends on $L_\gamma(L_{\rm wind}^{\rm eq})$ and the assumed energy redistribution model. The cosmic-ray flux of a single GRB exploding at a look-back time $t$ and a proper comoving distance $r$ that contributes \emph{now} to the total flux received on Earth is finally given by 
\begin{equation}
\phi(A, E)=\frac{c}{ 4 \pi }\times F(A, E, r; t, L_\gamma, \epsilon)\,,
\label{eq:flux}
\end{equation}
where the factor $c/4\pi$ converts the cosmic-ray density into a flux (in $\rm cm^{-2}\,s^{-1}\,eV^{-1}\,sr^{-1}$).\\
Practically, in order to calculate these "contribution functions" and fluxes from single GRBs, we used the Monte Carlo code presented in Allard et al. (2005) to compute the energy losses and photodisintegration processes in the extragalactic photon backgrounds (see Decerprit \& Allard 2011 for a more detailed description of the latest version of the code). We also computed the 3D geometrical trajectories as influenced by the magnetic fields using the fast integration method described in detail in Globus, Allard \& Parizot (2008), where a comparison with a full numerical integration is given. This numerical treatment permits, in particular, to keep track of the redshift dependence of the energy losses without having to assume a rectilinear transport, and to take into account the expansion of the universe at each step of the trajectory integration\footnote{For all our calculations, we assume a standard universe with $\Omega_{\rm m} = 0.3$, $\Omega_{\rm \Lambda} = 0.7$ and $H_{\rm 0} 
=71 \rm km\,s^{-1}\, Mpc^{-1}$. Moreover we use the most up-to date estimate of the density of the infrared, optical and ultra-violet backgrounds (and their redshift evolution) described in Kneiske et al. (2004). }.\\
We simulate the propagation of UHECRs above $10^{17.5}$ eV, considering various nuclear species, following a power law spectrum injection at 156 different redshifts (from a few $10^{-4}$ to $\sim$5), for three different hypotheses about the EGMF variance : 0.01, 0.1 and 1 nG. In total, more than $10^9$ trajectories from the injection redshift to $z=0$ were simulated in order to calculate the above-mentioned contribution functions and fluxes with large statistics. Once the simulations have been completed, the distributions of proper comoving distances from the GRB, of masses and of  energies, after propagation, can be calculated for each injection redshift and EGMF variance. The particles are reweighted in order to simulate the injection and composition calculated in Sect.~5 at 21 different luminosities, for each of the three models. This allows us to estimate the contribution functions $F(A, E, r; t, L_\gamma, \epsilon)$ and fluxes expected from single GRBs, for all the luminosities and energy redistribution 
models, at each redshift and each different EGMF variance. For each case of redshift and EGMF variance, we divide the proper comoving range obtained into 280 bins\footnote{The bin size is chosen so that particles are spread over more than two bins at all energies.} and calculate the contribution functions and fluxes in each bin. We are thus able to estimate the flux on Earth contributed now by a GRB of a given luminosity, exploding at a given proper comoving distance at a given look-back time/redshift. Finally, the total UHECR diffuse flux from GRBs can be estimated by summing the contribution of all single GRBs (at all luminosities) for each of the 300 realizations of the GRB history in the universe.\\
In what follows, we restrict ourselves to a qualitative discussion of the impact of the EGMF on the expected UHECR diffuse flux from GRBs. More quantitative aspects, such as the number of GRBs contributing as a function of the energy, time delays, anisotropy expectation (which requires to include the galactic magnetic field) as well as the study of inhomogenous magnetic fields configurations (see for instance Takami \& Murase 2012; Kalli, Lemoine \& Kotera 2011) will be considered in a forthcoming paper.

\subsubsection{Results}

\begin{figure}
{\rotatebox{0}{\includegraphics[scale=0.32]{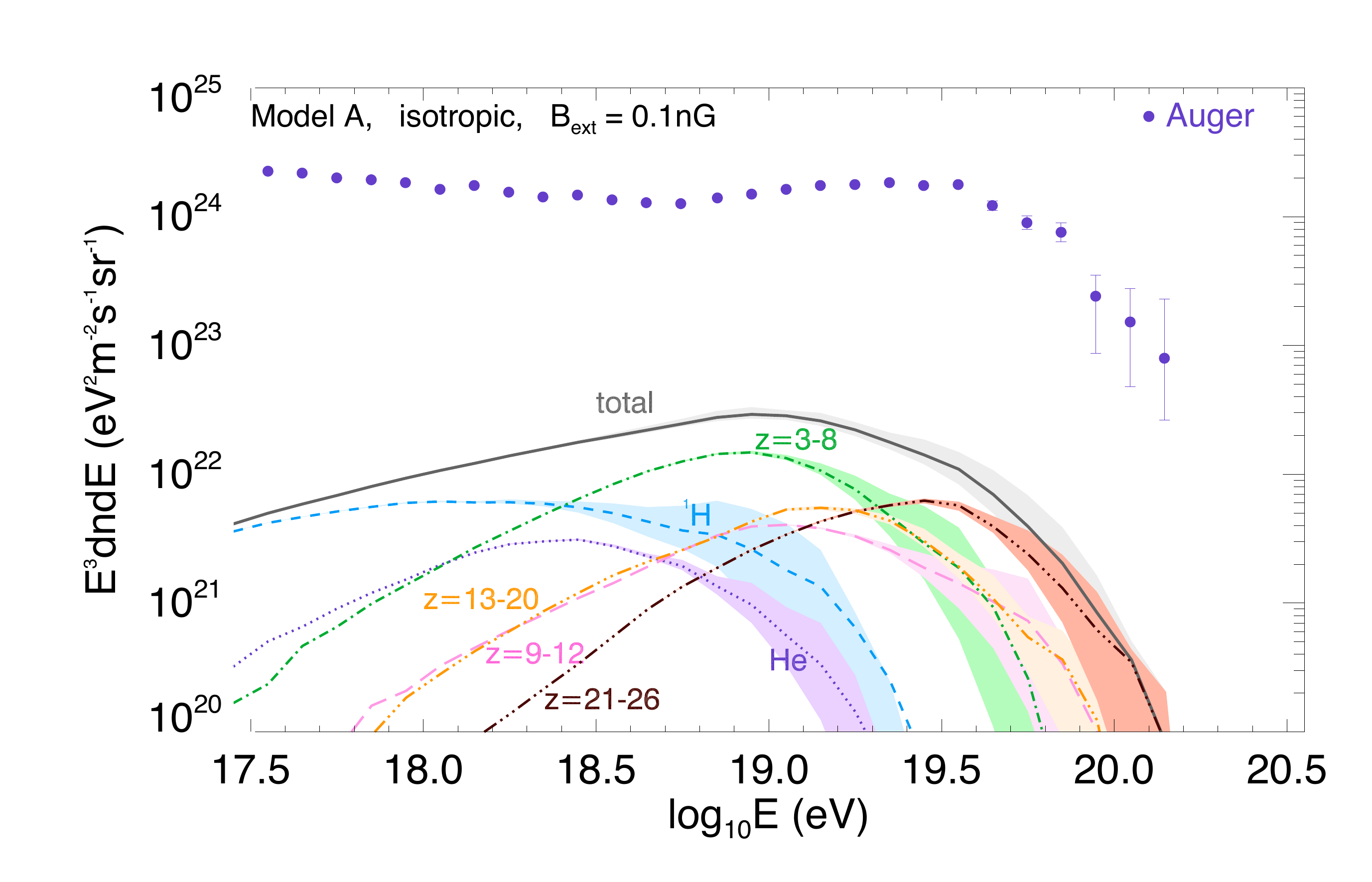}}}
{\rotatebox{0}{\includegraphics[scale=0.32]{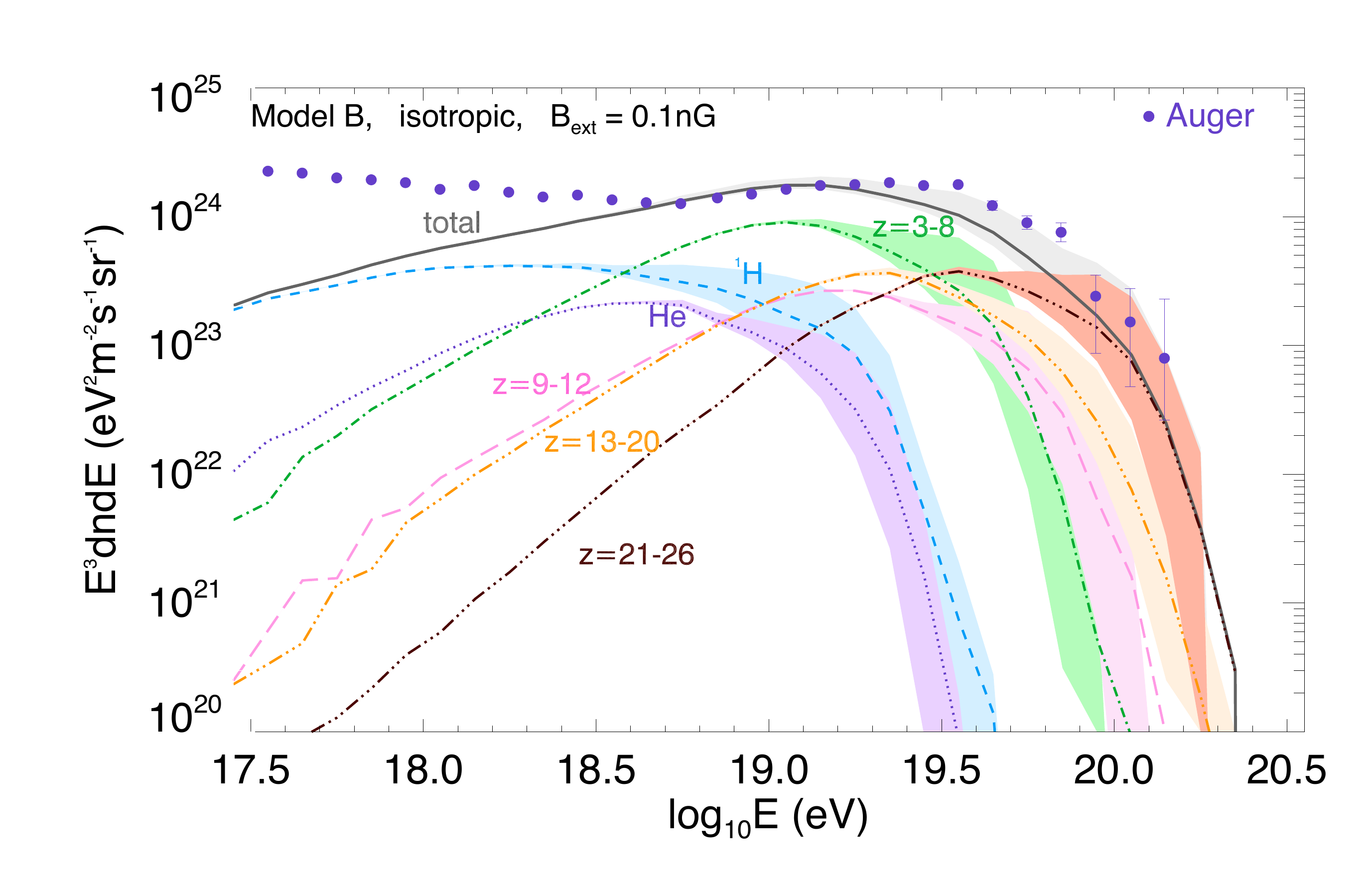}}}
{\rotatebox{0}{\includegraphics[scale=0.32]{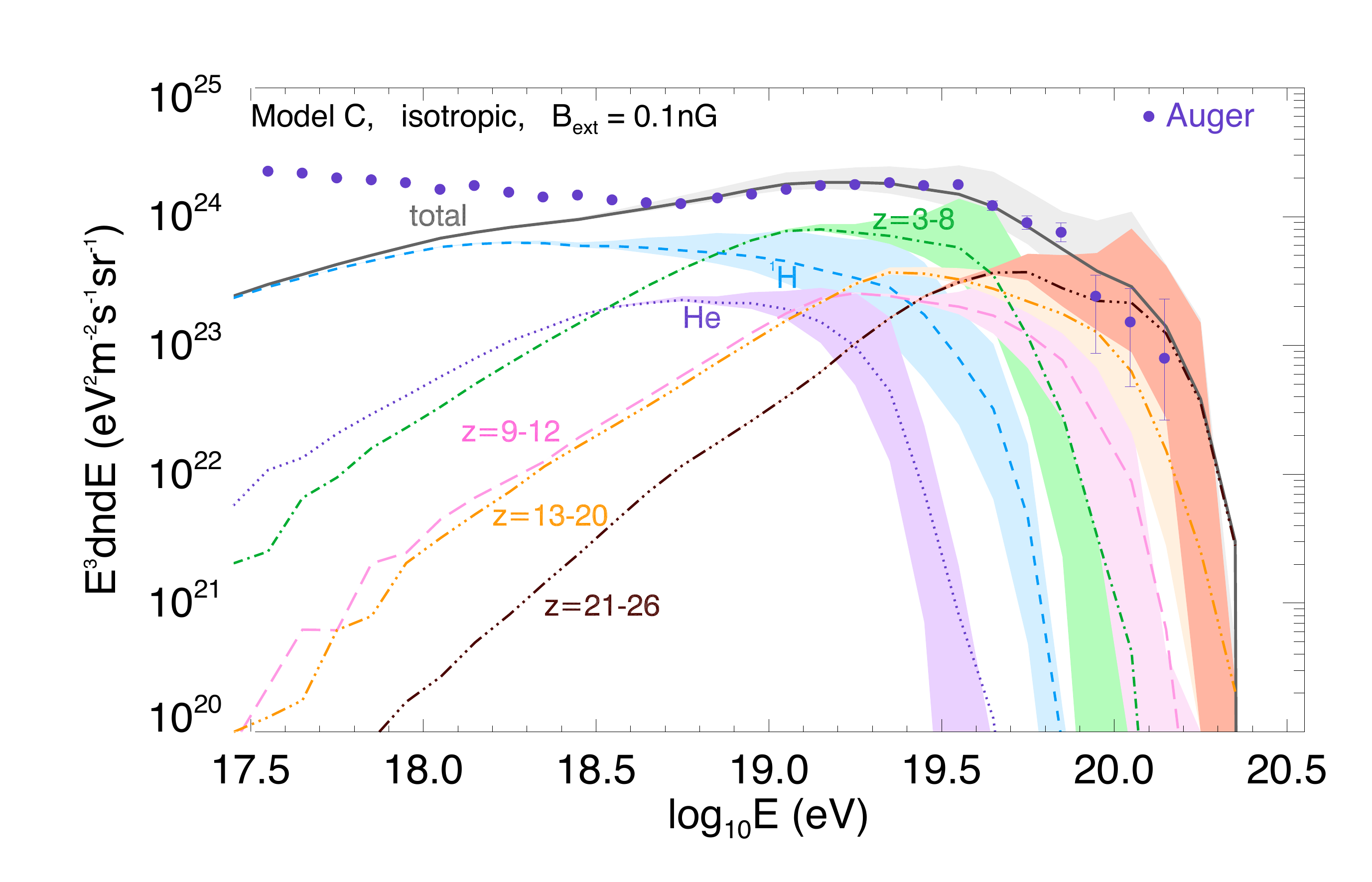}}}
\caption{Diffuse cosmique ray flux spectrum, expected on Earth, assuming an EGMF variance of 0.1 nG, in the cases of model A (top), B (center) and C (bottom). The contributions of different groups of nuclei are shown, the lines (plain lines for the total spectrum) represent the mean value calculated over 300 realizations of GRB history in the universe, the shaded areas represent the 90\% intervals (excluding the 5\% highest and the 5\% lowest realizations) of the 300 realizations.  These fluxes are compared with the latest Auger estimate of the UHECR flux.}
\label{fig:diff_iso}
\end{figure}
\begin{figure}
{\rotatebox{0}{\includegraphics[scale=0.32]{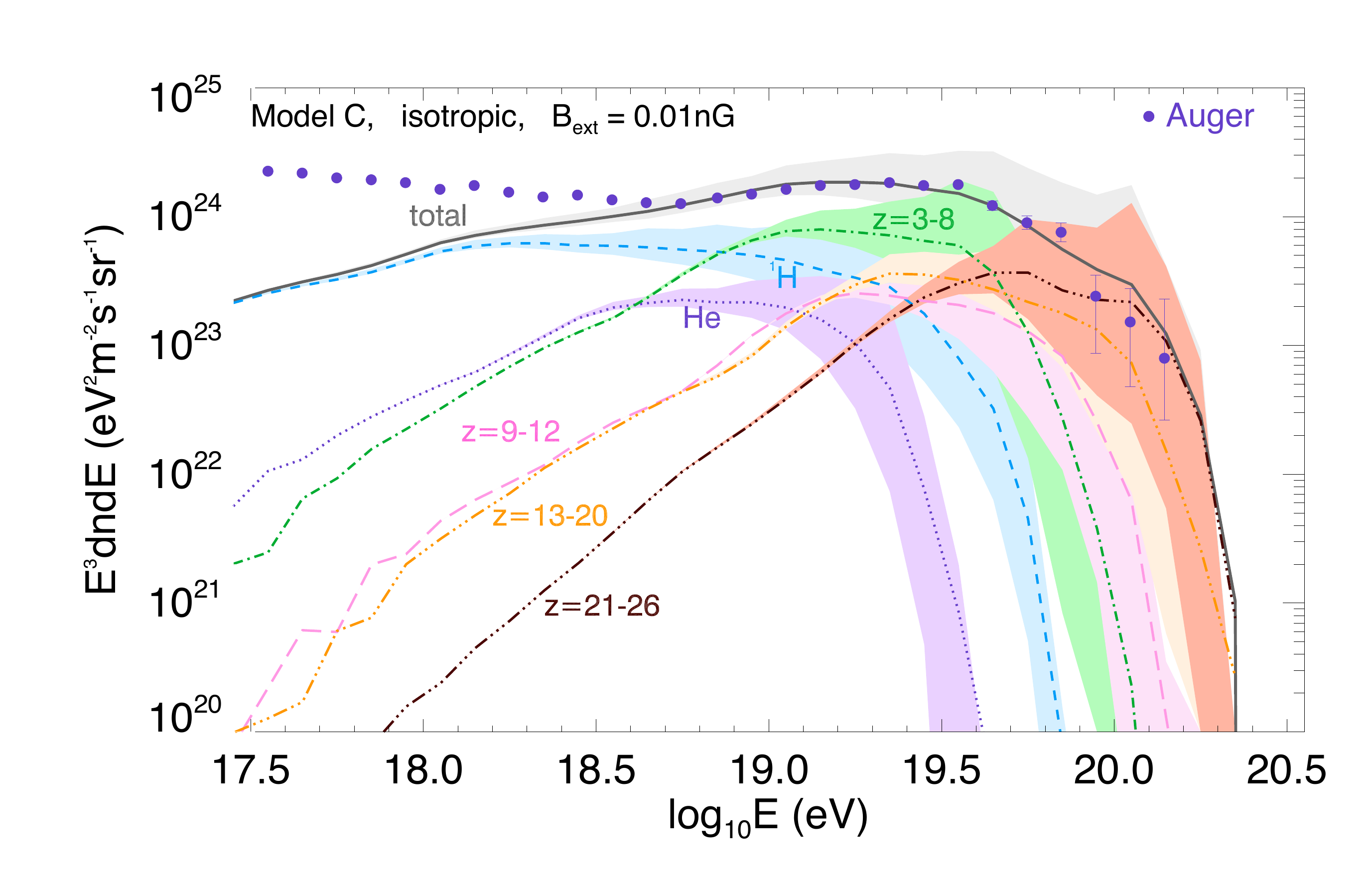}}}
{\rotatebox{0}{\includegraphics[scale=0.32]{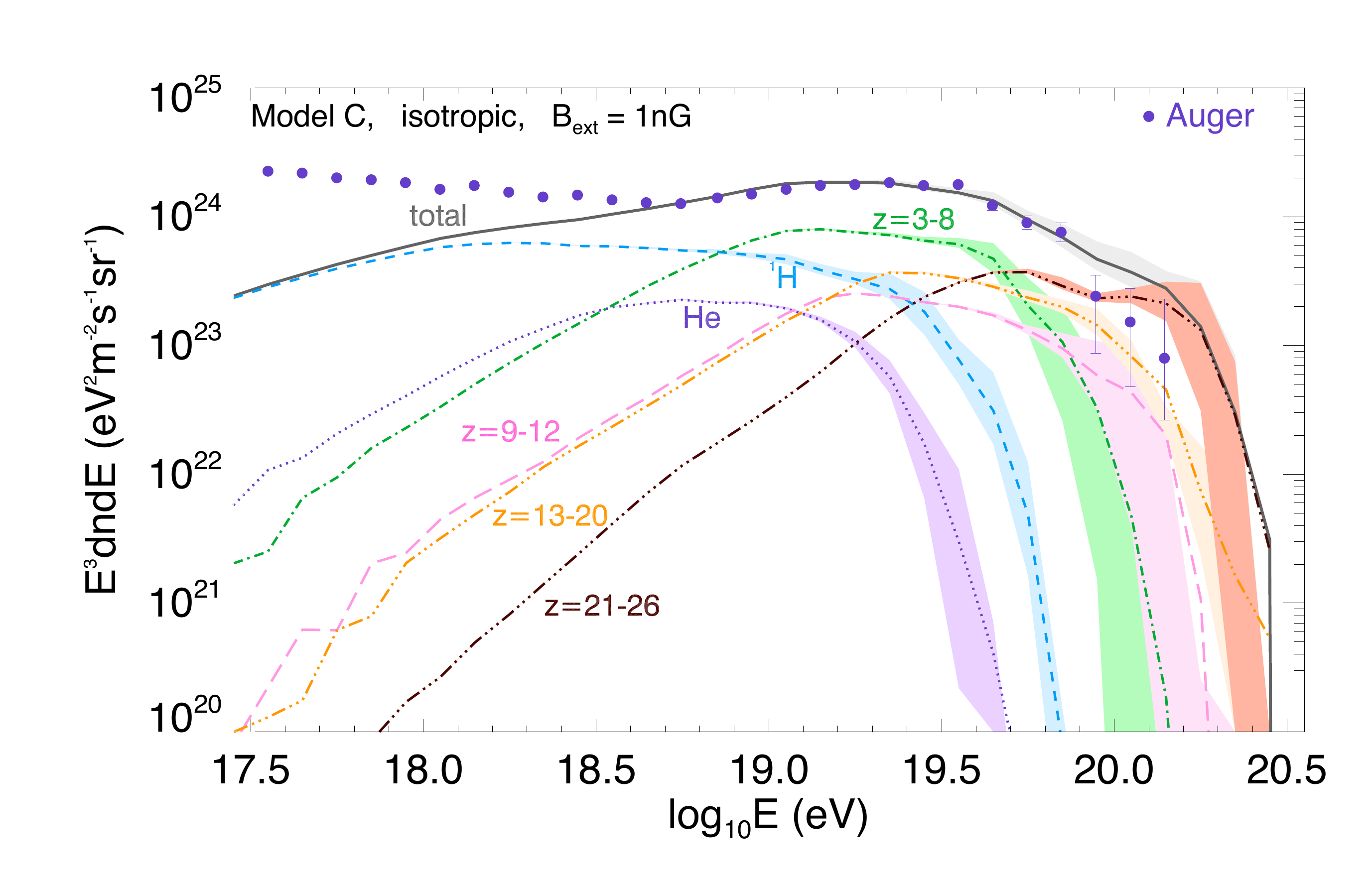}}}
{\rotatebox{0}{\includegraphics[scale=0.32]{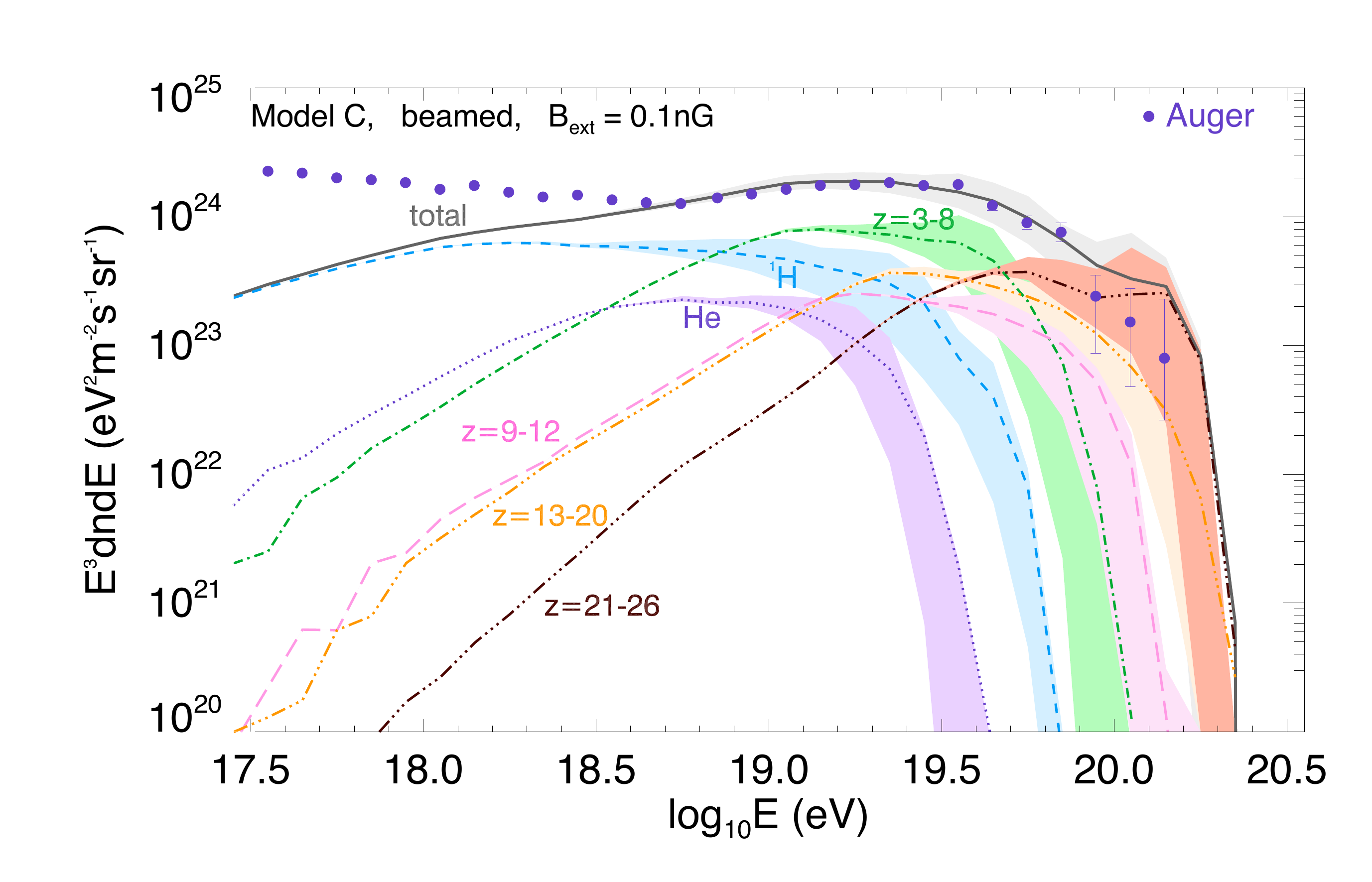}}}
\caption{Same as Fig.~\ref{fig:diff_iso}, predictions for model C are shown assuming EGMF variances of 0.01 nG (top) and 1 nG (center). The lower panel shows the case of beamed GRBs, assuming an opening angle $\theta_{\rm GRB}=5^\circ$, for an EGMF variance of 0.1 nG (to be compared to the bottom panel of Fig.~\ref{fig:diff_iso}).}
\label{fig:diff_compare}
\end{figure}

UHECR diffuse fluxes, calculated for models A, B and C and three hypotheses about the EGMF variance are shown in Fig.~\ref{fig:diff_iso} and \ref{fig:diff_compare}. Fig.~\ref{fig:diff_iso} displays the spectra obtained for the three models assuming an EGMF variance of 0.1 nG. The lines show the mean values obtained averaging over 300 realizations, for the total spectrum (plain line) and for different nuclear species. The shaded areas show the 90\% intervals of the 300 realizations, excluding the 5\% highest and lowest ones.\\
Concerning model A, one sees, as anticipated in the previous section, that the power emitted in UHE cosmic-ray is too weak to contribute significantly to the flux observed on Earth. It is almost two orders of magnitude (at least a factor $\sim$70) below the flux recently reporter by Auger (Abraham et al. 2010a, Aab et al. 2013).\\
In the cases of models B and C,  the predicted fluxes are in very good agreement with what is observed : the absolute scale of the fluxes have only been slightly rescaled, by $\sim$5\% downward for model B and $\sim25$\% upward for model C, to fall on top of Auger data.\\ 
\begin{figure}
{\rotatebox{0}{\includegraphics[scale=0.32]{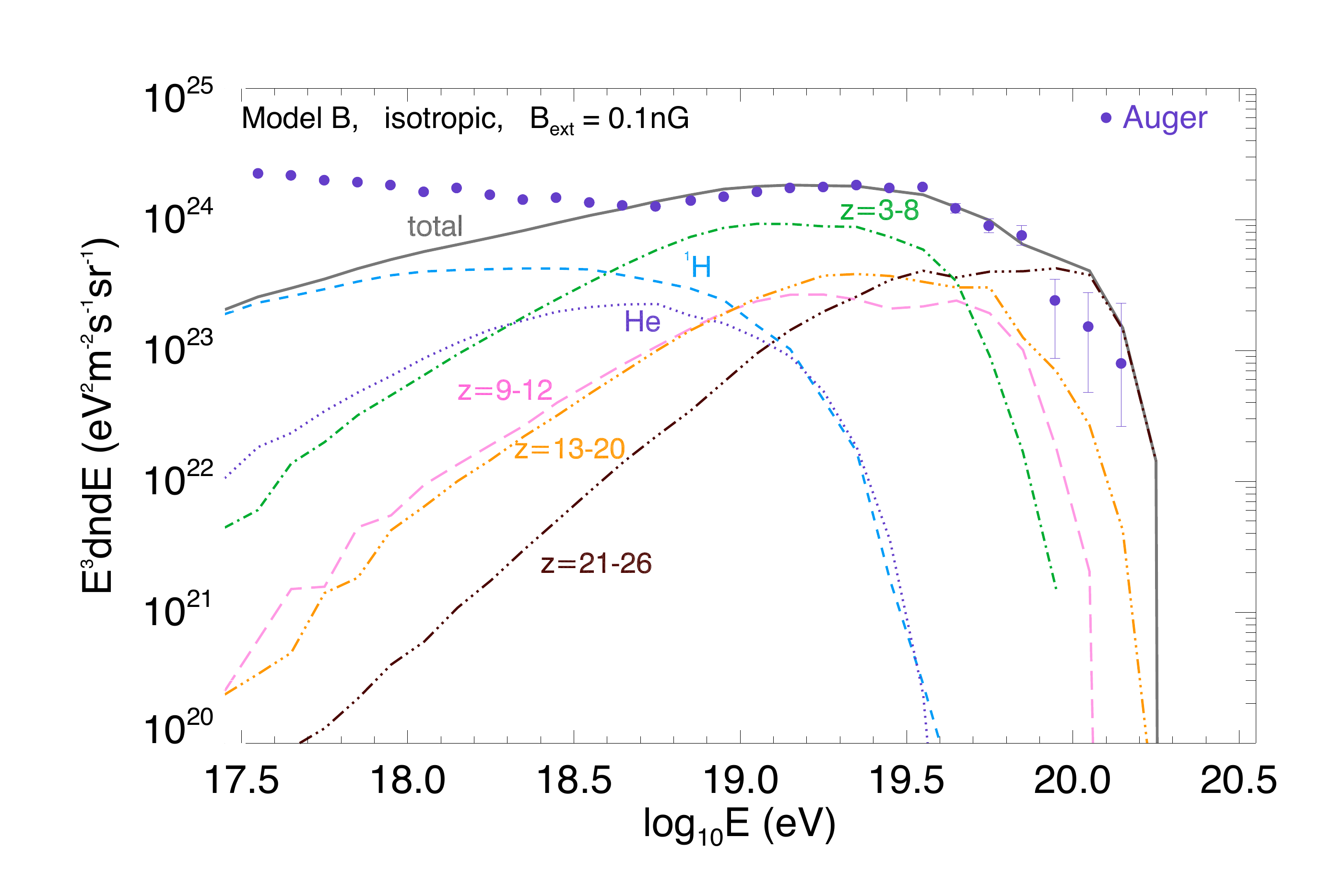}}}
\caption{UHECR spectrum predicted for one of the 300 realizations of the history of GRB explosion in the universe, in the case of model B, assuming an EGMF variance of 0.1 nG.}
\label{fig:real44}
\end{figure}
In the case of model B, the mean value obtained for the total spectrum is slightly too soft to reproduce the Auger data at the highest energies. For the 0.1 nG EGMF variance, displayed in the central panel of Fig.~\ref{fig:diff_iso}, only a few realizations (among the 5\% highest) allow to reproduce correctly the shape of the Auger spectrum at the highest energies. For instance, Fig.~\ref{fig:real44} shows one of the 300 realizations, which appears to give a relatively good fit to Auger data. Let us note that for lower values of the EGMF variance, 0.01 nG for instance, the spread of the predictions becomes larger (see below) and the Auger data lie comfortably within the 90\% interval.\\ 
Model C shows the best compatibility with the Auger data, the shape of the measured spectrum being very well reproduced above the ankle. The top and central panels of Fig.~\ref{fig:diff_compare} show the predictions, in the case of model C, for two other values of the EGMF variance, respectively 0.01 and 1 nG. As can be seen, the mean values are not significantly modified by changing the EGMF variance. The spread of the 300 realizations, however, increases with decreasing EGMF variance. This is due to the fact that larger magnetic fields result in  larger spreads in rectilinear distances from the source reached after a given propagation time, or equivalently of the cosmic-rays arrival times at a given distance from the source. Even at the highest energies, if the spread of the cosmic-ray signal is large enough, several GRBs can contribute to the UHECR flux at a given time. The fluctuations of the number of GRBs in the local universe have thus less impact on the spread of the UHECR spectrum. The relevance of 
this discussion, of course, depends on the cosmic-ray energy. Low energy particles (say, below $10^{19}$ eV) loose their energy quite slowly and thus have a very large horizon (either in space or time), within which the number of contributing GRBs  is large enough to limit the fluctuations of the UHECR spectrum even for the lowest magnetic field variance and even for the proton component. At higher energy, the discussion is complicated by the fact that the composition is mixed and that the proton component does not reach the highest energies unlike heavier nuclei. The highest rigidity particles in the spectrum are actually protons of a few $10^{19}$ eV, accelerated by intermediate or high luminosity GRBs (see Sect.~5). At these energies, say around $2\,10^{19}$ eV, the proton horizon is still large, keeping the spread of the spectrum predictions to a relatively modest level even for the lowest EGMF variance assumed. At higher energies, quite comparable spreads are observed, for instance for the Fe group around $10^{20}$ eV (or equivalently for CNO around $5\,10^{19}$ eV. From a given GRB, these particles are, however,  more spread in terms of arrival time or  distance from the source. As a result, if protons around $2\,10^{19}$ eV and Fe around $10^{20}$ eV had similar horizons, one would expect larger spreads of the spectrum predictions for the protons around $2\,10^{19}$ eV. The horizon of protons around $2\,10^{19}$ eV is, however, several times larger than that of Fe around $10^{20}$ eV (which is of the order of 100 Mpc, see for instance Globus, Allard \& Parizot 2008). The Fe component, around $10^{20}$ eV, is then more impacted by realization to realization fluctuations of the number of GRBs within its horizon, which is why the fluctuations of the spectrum predictions are of the same order as that of protons at $\sim2\,10^{19}$ eV instead of being smaller. The spread of the predictions would of course be much larger if protons were dominant, say around $10^{20}$ eV, since these protons would cumulate both characteristics of being the highest rigidity particles (i.e, the least spread by the effect of the EGMF) and having a reduced horizon because of severe interactions with CMB photons.

We also computed UHECR spectra in the case of beamed GRBs. Let us consider a GRB with a conical geometry and semi-opening angle $\theta_{\rm GRB}$, and an angle $\chi$ between the GRB axis and the vector connecting the central source to the Earth. The additional condition (still benefiting from the simplifying assumption of an homogenous and isotropic turbulent EGMF) for a given cosmic-ray to reach the Earth \emph{now}, in the case of beamed GRBs, is that $\alpha \in \left[\chi-\theta_{\rm GRB}\,,\,\chi+\theta_{\rm GRB}\right]$, where $\alpha$ is the angle between the initial cosmic-ray velocity $\bold{\overrightarrow{u_0}}$ and its position vector at the end of its propagation $\bold{\overrightarrow{r}}$. This means that a particle, with a given value of $\alpha$ and a given proper comoving distance $r$ from its source after propagating from $z(t)$ to $z=0$, can contribute to the flux from GRBs at a proper comoving distance $r$ and a value of $\chi$ such that $\chi \in \left[\alpha-\theta_{\rm GRB}\,,\,\alpha+\theta_{\rm GRB}\right]$. If the Earth lies within the emission cone, the condition becomes $\chi \in \left[0\,,\,\alpha+\theta_{\rm GRB}\right]$. To compute the expected UHECR spectra one has, of course, to take into account the fact that the GRB rate, derived assuming isotropic GRBs, has to be increased by a factor $f_{\rm GRB}\simeq (\theta_{\rm GRB}^2/4)^{-1}$ (for small values of  $\theta_{\rm GRB}$) and the total energy emitted by each single GRB divided by the same factor $f_{\rm GRB}$. From a practical point of view, to treat the case of beamed GRBs, we add bins of  $\chi$ angle to the discretization of the GRB history in bins of space and time and an additional variable to the contribution function, which becomes :
\begin{equation}
F(A, E, r, \alpha; t, L_\gamma, \epsilon)=\frac{N(A_i, E_j, r_k, \alpha_l; t, L_\gamma, \epsilon)}{ \Delta E_j \times 2 \pi r_k^2\Delta r_k \sin\alpha_l\Delta\alpha_l}
\label{eq:contribalf}
\end{equation}
The result of our calculations, in the case of model C, assuming an EGMF variance of 0.1 nG, $\theta_{\rm GRB}=5^\circ$ ($f_{\rm GRB}\simeq 526$),  is shown in the lower panel of Fig.~\ref{fig:diff_compare}.\\
Comparing with the lower panel of Fig.~\ref{fig:diff_iso}, which shows the predictions for model C with the same EGMF variance and isotropic GRBs, one sees that the means values obtained for the total spectrum and its components are not significantly modified by considering the beaming of GRBs. On the other hand, the spread of the predictions is significantly smaller than in the isotropic case. This is due to the much larger GRB rate implied by the assumed beaming angle $\theta_{\rm GRB}=5^\circ$. Unlike prompt emission photons, cosmic-rays are deflected by the EGMF during their propagation and as a result, off-axis GRBs can contribute to the UHECR flux on Earth. The decrease of the spread of our predictions remains, however, relatively moderate since, at the highest energies (where the horizon is of the order of 100 Mpc), cosmic-rays do not propagate long enough to be isotropized by the EGMF. The number of GRBs that actually contribute to the UHECR flux on Earth is not as much larger, compared to the isotropic case, as what the beaming factor $f_{\rm GRB}$ would suggest.   

\subsubsection{Evolution of the composition and implications for the transition from Galactic to extragalactic cosmic-ray}

From the point of view of the UHECR composition on Earth, all the models predict an increase of the cosmic-ray mean mass with energy, as could be anticipated form the results shown in Sect.~5. This trend is indeed entirely due to the fact that, according to our modeling, protons are unable to reach energies above a few $10^{19}$ eV during the acceleration process, while intermediate and heavy nuclei could reach energies above $10^{20}$ eV. The composition trend at the source is then not very much affected by the extragalactic propagation. In the case of models B and C, however, the maximum energy reached by nuclei heavier than protons is large enough to allow GDR photo-nuclear interactions with CMB photons. The spectra of the different nuclear species heavier than protons,  after extragalactic propagation, is then cut at a lower energy than at the source. In the case of protons, the source spectrum cuts-off below the photomeson production threshold with CMB photons in all cases. As a result, the highest energy protons are less affected by energy losses in the intergalactic medium than the highest energy nuclei. The gap in energy observed on Earth between the highest energy protons and the highest energy nuclei is then smaller after propagation than at the sources (this is true in particular for model C, as can be seen comparing Figs.~\ref{fig:eff} and \ref{fig:diff_iso}). 

Overall, the evolution of the UHECR composition implied by models B and C is in very good qualitative agreement with the trend suggested by Auger composition analyses, based on the measurements of the maximum shower depth (Abraham et al 2010b; Aab et al. 2013). For both models, the composition above the ankle is becoming increasingly heavier, with fewer and fewer protons. It is dominated first by protons, then by intermediate and finally by heavy nuclei. The energies of the transitions between the successive dominant nuclear species are lower in the case of model B, since the maximum energies reached at the sources are lower. Concerning proton abundances above the ankle, they are on average of the order of $\sim$25\% (resp. $\sim15\%$) at $10^{19}$ eV, and $\sim$10-12\% (resp. very low) at $10^{19.5}$ eV (but with larger fluctuations), respectively for models B and C.

These considerations, of course, depend to some extent on the assumed composition in the acceleration site. As mentioned earlier, we assumed for our calculations a metallicity ten times larger than in the estimate of the Galactic cosmic-ray source composition. Good fits to the data can still be found with a metallicity twice lower than in our assumption but not with the typical Galactic abundances (in which case the predicted spectra are too soft at the highest energies). Larger assumed metallicities can also provide good fits to the data. We emphasize that secondary nucleons, emitted by nuclei photodisintegration during the acceleration process, actually guarantee a significant proton component even if the proton abundance in the acceleration site is initially small.

Perhaps the most interesting composition feature, visible in Figs.~\ref{fig:eff}, \ref{fig:diff_iso} and \ref{fig:diff_compare}, is the softer spectrum found for the proton component, as a result of the efficient escape of secondary neutrons produced by the photodisintegration of accelerated nuclei (see Sect.~5). This prediction has interesting consequences for the relative abundance of extragalactic protons below the ankle energy and for the phenomenology of the transition from Galactic to extragalactic cosmic-ray (GCR-to-EGCR transition). If the proton component was as hard as that of the other nuclear species (which evolve approximately as $E^{-1}$ in Figs .~\ref{fig:diff_iso} and \ref{fig:diff_compare}), the relative abundance of extragalactic protons would fall rapidly below the ankle. With the much softer energy spectrum that we found, however, the relative abundance of extragalactic protons below the ankle is of the order of $\sim$15\% (resp. $\sim$10\%) at $10^{17.5}$ eV and $\sim$50\% (resp. $\sim$40\%) at $10^{18}$ eV for model C (resp. model B). These relative abundances are very similar to those obtained for the extragalactic mixed composition model studied in Allard et al. (2005, 2007a, 2007b) which means that all these models have basically identical implications for the GCR-to-EGCR transition (see, for instance, Fig.~2 of Allard et al. 2007b).  For all these models, the GCR-to-EGCR transition starts around $10^{17}$ eV and is completed at the ankle. The evolution of the composition implied by the extragalactic mixed composition model between $\sim10^{17}$ eV and the ankle was shown to be in very good agreement with the data of Fly's eye (Bird et al. 1993) and HiRes-Mia (Abu-Zayyad et al. 2000), as shown by Allard et al. (2007b). However, at the highest energies, this scenario, unlike what we found for models B and C, is in conflict  with the composition trend suggested by Auger composition analyses (Abraham et al. 2010), since the extragalactic mixed composition model predicted a light composition above $10^{19}$ eV (because the acceleration of protons at the sources above $10^{20}$ eV was assumed in this model). The spectra we obtained for models B and C, on the other hand, predict a composition increasingly heavy at the highest energies while preserving the GCR to EGCR transition scenario of the extragalactic mixed composition model.\\
Recently, the KASCADE-Grande collaboration released two successive composition analyses (Apel et al. 2011, 2013) showing :\\
({\it i}) a knee-like structure in the spectrum of the heavy component of cosmic rays, with a break located slightly below $10^{17}$ eV. This feature is compatible with the presence of a "Fe-knee" located at an energy $\sim26$ times larger than the proton-knee at $\sim 3-4\,10^{15}$ eV (see e.g., Antoni et al. 2005).\\
({\it ii}) an ankle-like feature in the energy spectrum of light elements of cosmic rays with a break located at $\sim 10^{17}$ eV.\\
({\it iii}) a slope of the light component fitted below the ankle-like break compatible with the slope of the heavy component fitted above the "Fe-knee".\\
Put together, these observations point toward a rather natural picture. The ankle-like break in the light CR spectrum at $\sim10^{17}$ eV (Apel et al. 2013) implies that the light Galactic component (and then probably the proton component) extends at least up to this energy. It further implies that the heavy Galactic component (at least the Fe component) should extend at least up to $\sim 3\,10^{18}$ eV, which is very close to the ankle energy. This point is further supported by the fact that  the post "Fe-knee" heavy component (which again has a spectral slope compatible with that of the pre-ankle light component) does not show any evidence for a cut-off in the whole energy range covered by KASCADE-Grande, which extends up to $10^{18}$ eV (Apel et al. 2011). An extragalactic component such as that predicted by models B and C would fit rather well in this picture, emerging around $10^{17}$ eV (where it represents a few percents of the total spectrum), completely dominating the Galactic component at the ankle 
with a composition still dominated by light nuclei, and getting progressively heavier as the energy increases.\\
Although the above picture looks quite consistent, let us not for forget that a full study of the hadronic model dependence of the features claimed by the KASCADE-Grande experiment is still needed. Moreover, the current overlap between the energy ranges covered by the KASCADE-Grande and Auger experiments is very small. This situation should however improve significantly in the next few years since the low energy extensions of Auger (Abreu et al. 2011) and TA (TALE, Sagawa et al. 2009) are expected to increase the energy range covered by both experiments down to $\sim10^{17}$ eV (and even below in the case of TALE), then fully covering the energy range where the GCR-to-EGCR transition is expected to take place.

It is interesting to note that our results cannot, in principle, be obtained in standard extragalactic propagation studies, where it is usually assumed that all the nuclear species are emitted from their sources with similar injection spectra (following the same power law) and maximum energies scaling with the charge of the nuclei. In that case, one can explain the Auger composition trend within the framework of the so-called "low proton $E_{\rm max}$ models" (Allard et al. 2008), assuming that the proton component is not accelerated up to the highest energies, unlike heavier nuclei. In this case, the models that best reproduce the Auger composition trend usually need very hard spectral indices at the sources. Although these models can give a quite consistent account of the Auger data, they have more difficulties explaining the presence of a significant extragalactic component down to a few $10^{17}$ eV. In the acceleration models investigated here, these usual assumptions are broken in a completely natural way, as a direct consequence of the acceleration process (including the energy/nucleon losses during the process) and not as an {\it ad hoc} attempt to reproduce the data.\\
Finally, it is worth mentioning that the GCR-to-EGCR transition scenario implied by our results can certainly not claim to be a characteristic signature of the acceleration of UHECR at GRB internal shocks. The softer proton component we obtained seems to be a natural consequence of the acceleration of cosmic-ray nuclei in a high density radiation environment and can probably be expected in the context of other candidate sources.

\subsection{Diffuse neutrino flux}

We finally turn to the calculation of the diffuse neutrino flux expected from GRBs on Earth, for our different models. It is well known that, unlike UHECRs, neutrinos suffer only from adiabatic losses due to the expansion of the universe, and then have horizons comparable to the size of the universe. In this context, the calculation of the diffuse neutrino flux is much more straightforward  than the diffuse UHECR flux. Using the neutrino emission by single GRBs calculated in Sect.~5,  for different prompt emission luminosities and the GRB luminosity function and cosmological evolution, the diffuse neutrino flux expected on Earth is given by 
\begin{figure}
{\rotatebox{0}{\includegraphics[scale=0.33]{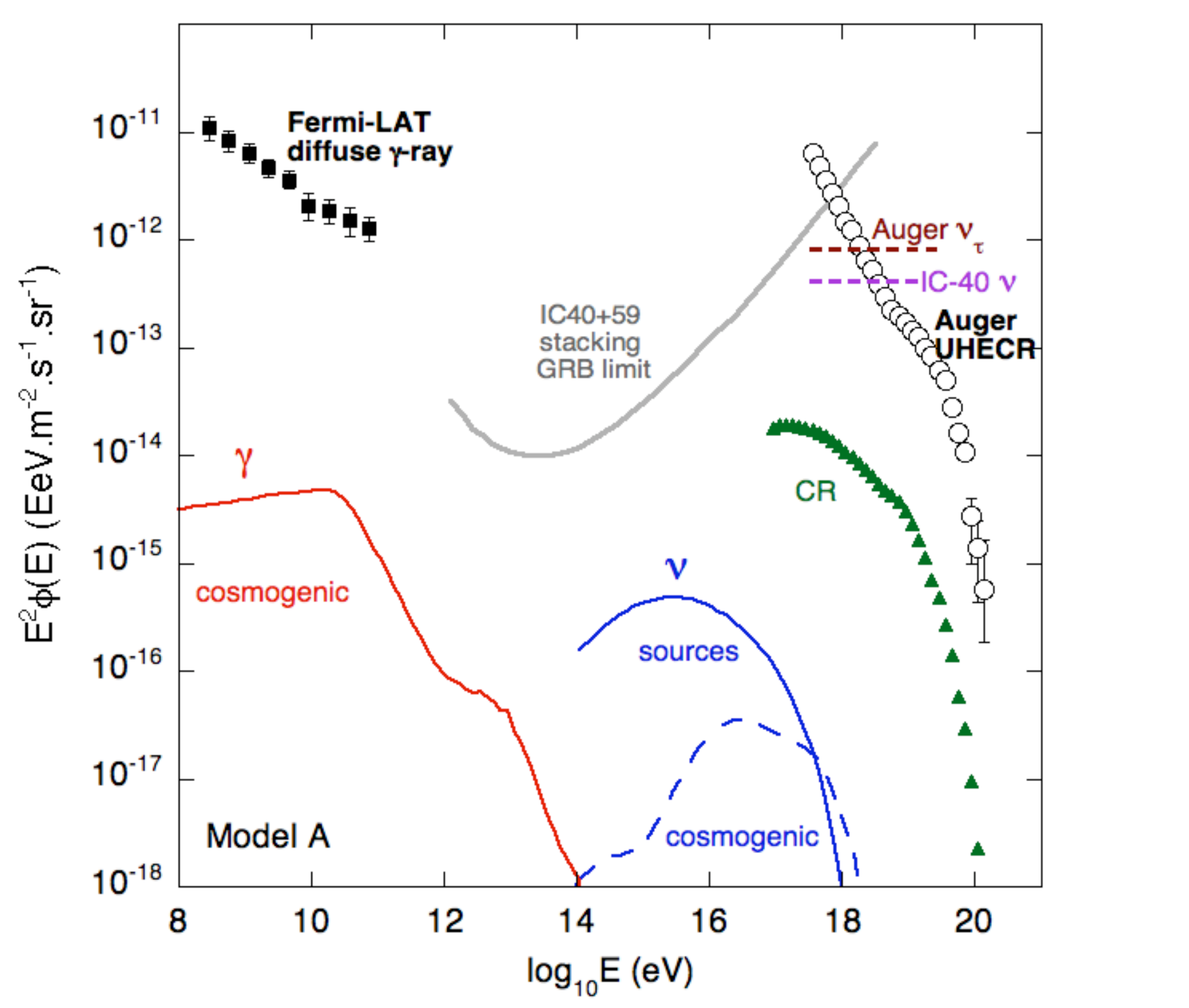}}}
{\rotatebox{0}{\includegraphics[scale=0.33]{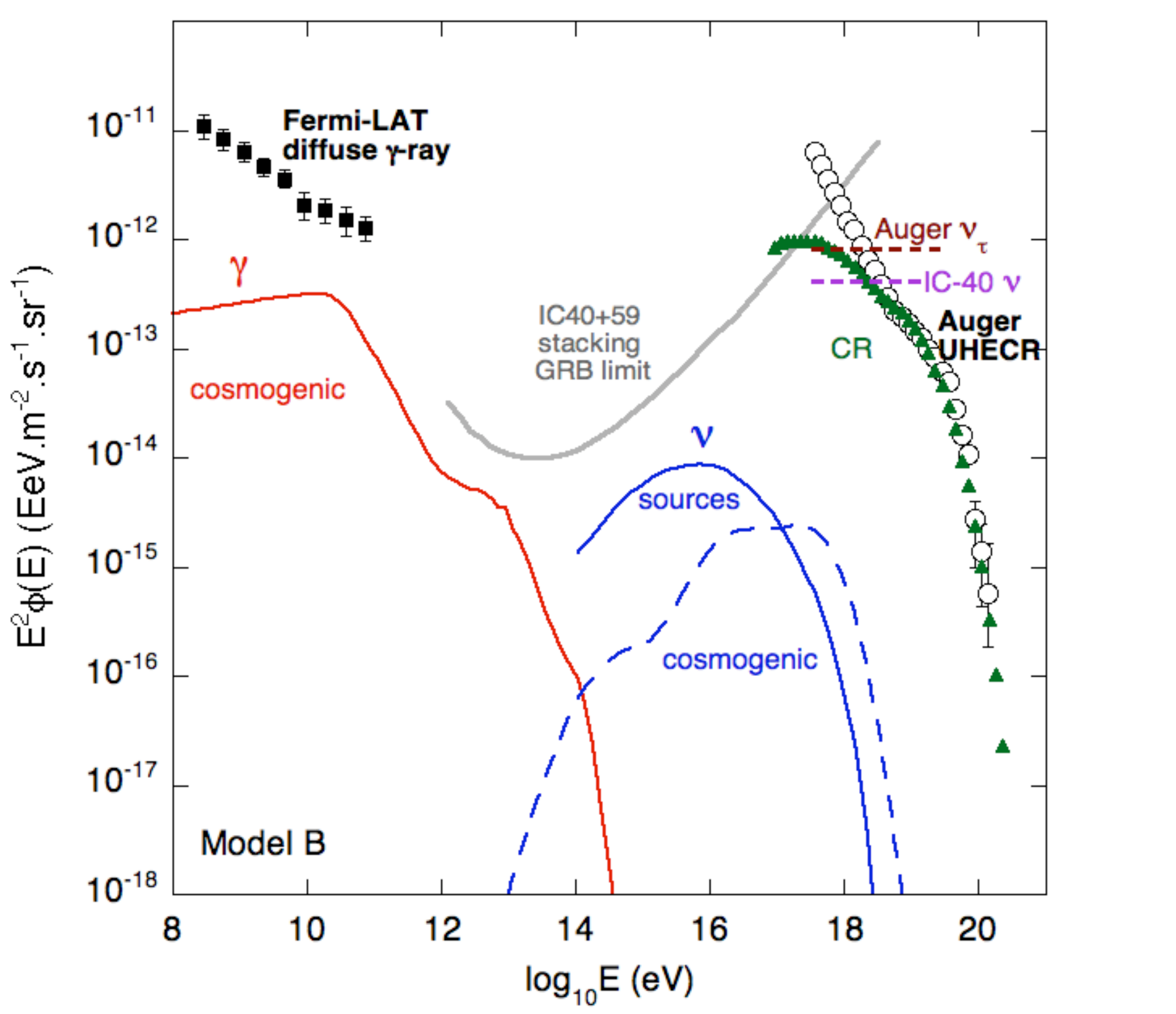}}}
{\rotatebox{0}{\includegraphics[scale=0.33]{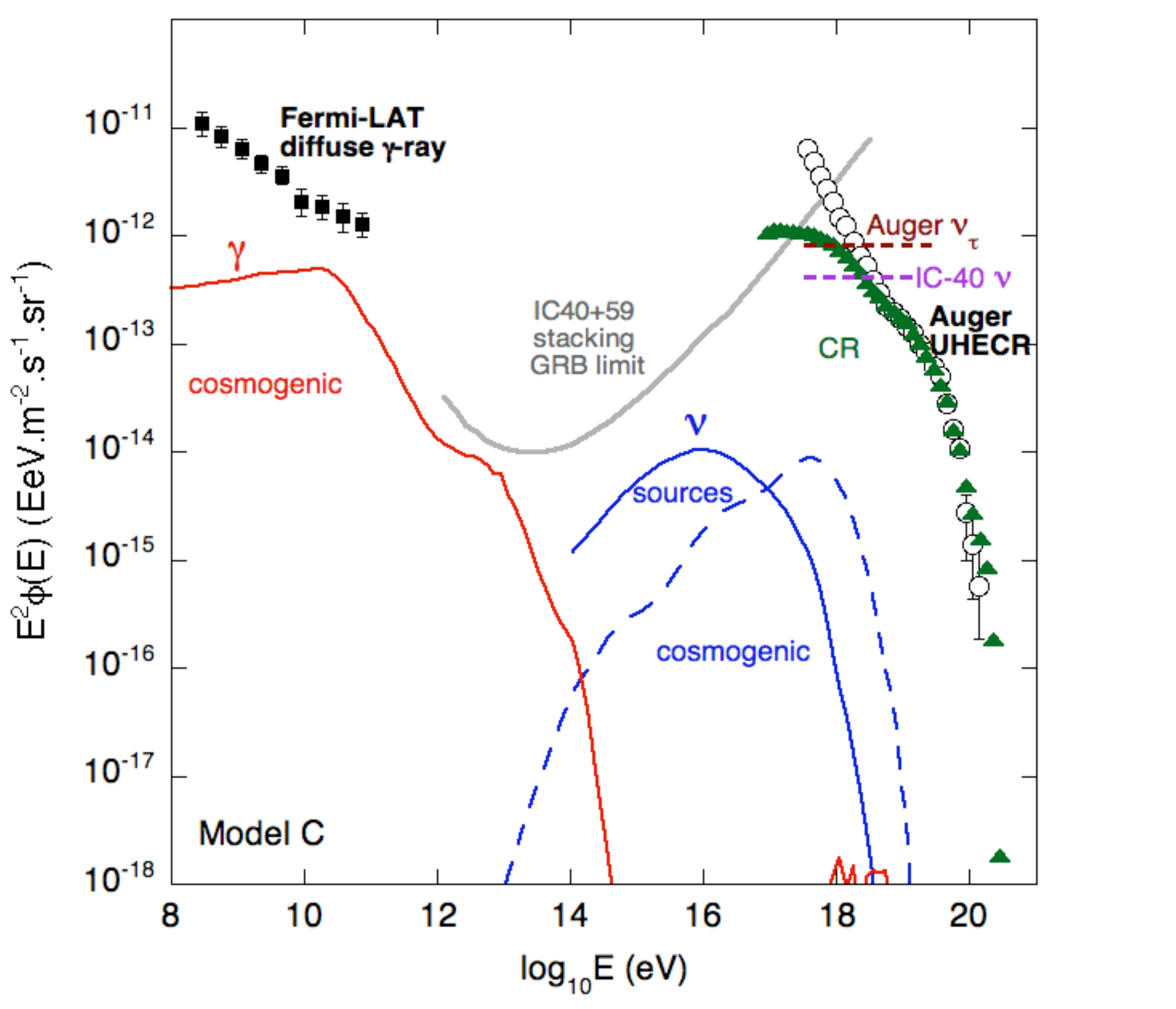}}}
\caption{UHECR spectrum, cosmogenic neutrinos and photons spectra and diffuse neutrino spectrum from GRB sources predicted for models A (top), B (center) and C (bottom), compared to various experimental limits or measurements (see text).}
\label{fig:diff_sec}
\end{figure}
\begin{eqnarray}
\Phi_\nu(E_\nu^{\rm obs})&=&\frac{1}{4\pi}\int dz \frac{dV_{\rm co}}{dz}\frac{1}{4\pi d_L^2(z)}\nonumber\\ 
&&\times\int dL_\gamma\, \frac{dN_{\rm GRB}}{dL_\gamma}(L_\gamma , z)\,L_\nu(E_\nu^{\rm em}(z), L_\gamma)\,,
\label{eq:fluxdiff}
\end{eqnarray}
where $\Phi(E_\nu^{\rm obs})$ is the neutrino differential energy flux on Earth (in $\rm eV\,m^{-2}\,s^{-1}\,sr^{-1}$), $L_\nu(E_\nu^{\rm em}, \rm L_\gamma)$ is the differential luminosity in neutrinos of a GRB with prompt emission luminosity $L_\gamma$ (calculated in Sect.~5, in $\rm erg\,s^{-1}$), $d_L(z)$ is the luminosity distance at redshift $z$, $V_{\rm co}$ is the comoving volume and $dN_{\rm GRB}/dL_\gamma(L_\gamma , z)$ is the GRB luminosity function and cosmological evolution combining Eqs.~\ref{eq:lf} and \ref{eq:evol}. The energy of a neutrino observed on Earth $E_\nu^{\rm obs}$ and its energy at the source $E_\nu^{\rm em}$ are, of course, related by $E_\nu^{\rm em}=(1+z)E_\nu^{\rm obs}$, where $z$ is the redshift of the source.\\
In addition to the diffuse flux due to neutrinos emitted at the sources, one has also to consider the \emph{cosmogenic} neutrino flux (Berezinsky \& Zatsepin 1969) emitted during the extragalactic propagation of UHECRs. This cosmogenic neutrino flux is accompanied by a flux of cosmogenic photons resulting from the electromagnetic cascades induced by very high energy photons and ${\rm e^+e^-}$ pairs, also produced during UHECRs extragalactic propagation (Strong et al. 1973, 1974; Berezinsky \& Smirnov 1975). These cosmogenic photons can give key constraints on UHECR models (Ahlers et al. 2010; Berezinsky et al. 2010; Decerprit \& Allard 2011) since they can be compared with the diffuse $\gamma$-ray flux measured by Fermi (Abdo  et al. 2010).\\
To calculate the development of extragalactic electromagnetic cascades and the cosmogenic neutrino and photon fluxes, we use the Monte-Carlo codes described in Decerprit \& Allard (2011). In the case of cosmogenic neutrinos and photons, the contribution of UHECRs emitted at low redshift (say, below $z$=0.5) is small (see for instance Allard et al. 2006). This means that one can neglect the effects due to the fluctuations of the number of GRBs in the very local universe and the discreteness of UHECR emission\footnote{It would actually not be a good approximation for UHE photons (i.e, photons produced around $10^{19}$ eV through the disintegration of neutral pions). This signal on Earth crucially depends on the source distribution in the local universe. In the cases we study, however, the production of these photons is highly suppressed due the fact that protons do not reach high enough energies to interact through photomeson production with CMB photons in the local universe.}. We thus assume for our calculations UHECR emission in the continuous injection and continuous source distribution approximation. The injection spectrum of UHECRs is, for each energy redistribution model, the effective injection spectrum calculated at the beginning of Sect.~6 and shown in Fig.~\ref{fig:eff}.\\
The results of our calculations are shown in Fig.~\ref{fig:diff_sec} for models A (top), B (center) and C (bottom). Our predictions are compared with different experimental measurements or limits :\\
({\it i}) The cosmogenic photon fluxes are compared with the diffuse $\gamma$-ray flux measured by Fermi (Abdo et al. 2010).\\ 
({\it ii}) The cosmogenic  neutrino fluxes (summed over all flavors) are compared with the IceCube limit on UHE neutrinos (Abbasi et al. 2011) and the Auger limit on UHE Earth-skimming $\nu_\tau$ (Abreu et al. 2012, the original limit is simply multiplied by three to be compared with our "all flavors" fluxes).\\ 
({\it iii}) The neutrino fluxes from the sources are compared with the IC40+59 stacking GRB limit estimated by Baerwald et al. (2014) and based on the non-observation of high energy neutrinos from 215 GRB in IceCube field of view (Abbasi et al. 2012).\\ 
({\it iv}) The UHECR flux, calculated in the continuous injection and continuous source distribution approximation (the resulting spectra follow closely the mean values obtained in the previous sections) is compared to Auger data.\\
One can see that our predictions do not upset any experimental measurement or limit. The cosmogenic photon flux lies below the diffuse $\gamma$-ray flux observed by Fermi for all the models while the predicted cosmogenic neutrino fluxes are well below the current limits given by Auger or IceCube. Even the highest flux predicted for model C would be difficult to detect within the lifetime of the IceCube experiment. Slightly higher maximum energy reached by protons (compared with what we found for model C) would, however, increase significantly the chance of a detection.\\
The diffuse neutrino flux emitted from the sources is not currently constrained by the non-observation of a neutrino signal in the stacking analyses reported by IceCube.  It is worth mentioning that the contribution of the highest luminosity GRBs to the diffuse neutrino flux from the source is much larger than their contribution to the diffuse UHECR flux. As a result, the difference between the neutrino flux predicted for model A and those predicted for models B and C is smaller than the differences we found for UHECRs. Due to the important contribution of the highest luminosity GRBs, these predictions are more sensitive to the details of the luminosity function at very high luminosities. Despite these considerations, fluxes of the order of those predicted here should be within the reach of IceCube or the future KM3Net experiment (KM3Net consortium 2011). The observation (or non observation) of neutrinos associated with GRBs, should then give key constraints on UHECR acceleration at GRB internal shocks within a few years.  
  
\section{Summary}

Throughout this paper, we investigated the acceleration of UHECR protons and nuclei at GRB internal shocks. In Sect.~2, we modeled internal shocks using the simple approach implemented by Daigne \& Mochkovitch (1998), where the relativistic outflow emitted by the central engine is represented by a large number of shells that interact with one another, discretizing the propagation of a collisionless shock within the outflow. This treatment allowed us to estimate the key physical quantities at work during the internal shock phase and their evolution during the shock propagation. Within this framework, we assumed a given initial Lorentz factor distribution of the relativistic wind and considered three different models consisting in different combinations of the dissipated energy redistribution parameters, $\epsilon_{\rm e}$, $\epsilon_{\rm cr}$, $\epsilon_B$ (which refer respectively to the fraction of the dissipated energy communicated to accelerated electrons, accelerated cosmic-ray and to the magnetic field). The first model (model A) assumed equipartition of the dissipated energy leading to an efficiency of the prompt emission of the order of $\sim5\%$ independent of the burst luminosity $L_\gamma$. For models  B and C, we made a  different assumption on the dissipated energy redistribution parameters using much lower values for $\epsilon_{\rm e}$, which implies that most of the dissipated energy is communicated to the accelerated cosmic-rays and the magnetic field (using $\epsilon_{\rm cr}\sim0.9$ (resp. 0.66) and $\epsilon_B\sim0.1$ (resp. 0.33) for model B (resp. model C)).  These assumptions imply larger values of the wind luminosity for a given prompt emission luminosity of the GRB than in the case of model A, and lower efficiencies for the prompt emission (from approximately 0.01\% for low values of $L_\gamma$ to 1\% for the highest prompt emission luminosities). On the other hand, the larger wind luminosities and large values of $\epsilon_{\rm cr}$ and $\epsilon_B$, imply larger cosmic-ray luminosities and higher magnetic fields which is, in principle, more favorable for UHECR acceleration. For each of these models, and for different prompt emission luminosities, we calculated the evolution of some important physical quantities, such as the dissipated energy, the magnetic field, the baryon density, the shock Lorentz factor, the prompt emission photon spectrum and density, at different times of the shock propagation. 

In Sect.~3, we performed numerical simulations of cosmic-ray acceleration at mildly relativistic shocks, using numerical tools inspired by the work of Niemiec \& Ostrowski (2004). We considered Lorentz factors between 1 and 2, which is the typical range expected at GRB internal shocks. We tested various magnetic field configurations in the vicinity of the shock, assuming different turbulence power spectra, turbulence levels or regular field obliquities. After setting escape conditions upstream and downstream of the shock, we calculated the spectra of particles escaping downstream and upstream as well as the energy evolution of the acceleration time for different magnetic field configurations. Most of the spectra of particles advected downstream deviated significantly from single power laws, as already pointed out by Niemiec \& Ostrowski (2004). In the cases of purely turbulent fields, we found relatively hard spectra with a sizeable fraction of the available energy communicated to high energies (i.e, $E>E_{\rm max}$, with the definition of $E_{\rm max}$ given by Eq.~\ref{eq:Emax}). However, the expected slope of the spectrum of the particles advected downstream was found to be strongly dependent on the assumed magnetic field configuration in the vicinity of the shock. Quasi parallel shocks with low turbulence level were found to give the hardest spectra (well harder than $E^{-2}$), while quasi-perpendicular "superluminal" shocks resulted in much softer spectra with sharp cut-offs below $E_{\rm max}$.\\
For the particles escaping upstream of the shock, assuming a free boundary escape at a few $\lambda_{\rm max}$ away from the shock front, we found extremely hard spectra, starting at energies slightly below $E_{\rm max}$ where the particles enter the weak scattering regime and can outrun the shock for a long enough time to reach the boundary. The very efficient escape at energies larger than $E_{\rm max}$ results in a cut-off of the spectrum and gives a bell-like shape to the spectrum of particles escaping upstream of the shock. Compared to the purely turbulent case, the escape upstream was found to be more efficient for quasi-parallel shocks and highly suppressed for quasi-perpendicular "superluminal" shocks.\\
Concerning the energy evolution of the acceleration time, we also found a broken power law shape with a significant hardening of the evolution at energies slightly below $E_{\rm max}$ where particles enter the weak scattering regime. At low energy, the shape of the evolution and the value of the acceleration time at a given energy strongly depend on the assumed magnetic field configuration (in particular on the assumed turbulence power spectrum). At higher energy, in the weak scattering regime, the shape of the evolution becomes practically independent of the assumed power spectrum in the purely turbulent case. The acceleration time at $E=E_{\rm max}$, was found to lie between $\sim$ 15 and 80 (depending on the shock Lorentz factor) Larmor times of the particles at this energy, quite far from the so-called "Bohm scaling" ($t_{\rm acc}(E)\simeq t_{\rm L}(E)$). Such an optimistic assumption on the acceleration time would require very different magnetic field configurations from those envisaged here, if not a different acceleration mechanism.        

In Sect.~4, using the physical quantities estimated in Sect.~2, we calculated the UHECR energy loss times expected at internal shocks for the most relevant energy loss processes, i.e, photo-hadronic interactions, hadron-hadron interactions, synchrotron radiation and adiabatic losses. These energy loss times were calculated as functions of energy for different nuclear species, from protons to Fe nuclei, at different times of the shock propagation and different prompt emission luminosities, for each of the above-mentioned models.  Preliminary estimates of the maximum energy achievable at GRB internal shocks could thus be obtained by comparing these energy loss times with the acceleration times obtained in Sect.~3, assuming purely turbulent magnetic fields following a Kolmogorov power spectrum.  We found that, in most cases, the maximum energy reachable by UHECRs is either limited by photo-interactions or adiabatic losses, depending on the burst luminosity and the stage of the shock propagation. Photodisintegration is the limiting process for the acceleration of UHECR nuclei at the beginning of the shock propagation in the case of intermediate luminosity GRBs ($L_{\rm wind}^{\rm eq} \simeq 10^{53}\,\rm erg\,s^{-1}$, $L_\gamma \simeq 5\,10^{51}\,\rm erg\,s^{-1}$) and during most of the shock propagation for the highest luminosity GRBs. Photomeson production is the limiting process for the acceleration of protons only at the early stages of the shock propagation for very high luminosity bursts. Both protons and nuclei are limited by adiabatic losses during the whole shock propagation for low luminosity GRBs and during a significant fraction of the shock propagation for intermediate luminosity GRBs. In these cases, the maximum energy of the different nuclear species are expected to scale with the charge of the nucleus. In all other cases, the gap in maximum energy between protons and the other nuclear species is expected to be smaller than the factor $Z$. \\
Comparing the maximum energies obtained for the different energy redistribution models, we found that those obtained for models B and C were significantly larger than those predicted for model A, especially at low and intermediate GRB luminosities, due to the larger magnetic fields implied by the physical hypotheses defining models B and C. In the case of model A, maximum energies of the order of $10^{20}$ eV (in the central source rest frame) were only reached by heavy nuclei for intermediate luminosity bursts. For models B and C, on the other hand, heavy nuclei were able to reach energies of the order of $10^{20}$ eV or above, already for low luminosity GRBs (especially in the case of model C), while intermediate mass nuclei (CNO) reached $10^{20}$ eV in the cases of intermediate luminosity GRBs. The maximum energy of protons remained significantly lower than $10^{20}$ eV, in all the cases we studied, reaching at most $\sim 10^{19.5}$ eV, in the case of model C and high luminosity bursts.

We also discussed the escape of accelerated particles from the acceleration site and the importance of the value of the maximum turbulence scale of the magnetic field, $\lambda_{\rm max}$, in particular when energy losses are considered. Accelerated particles are expected to suffer from energy losses at all energies and must leave the acceleration site before being cooled. Particles that have reached the weak scattering regime (i.e, particles with Larmor radii of the order of $\lambda_{\rm max}$) should escape more efficiently than those at lower energy. As a result, the escape process should behave as a high-pass filter, both upstream and downstream of the shock.\\ After discussing different  possible situations and their consequences on accelerated cosmic-ray escape and secondary particle production, we set a prescription on the assumed value $\lambda_{\rm max}$ for our subsequent calculations : since for all our models, accelerated cosmic-rays are assumed to carry a significant fraction of the dissipated energy, we assumed that the maximum turbulence scale is equal to the Larmor radius of the highest rigidity accelerated particles (which correspond to the highest energy accelerated protons in all the cases we considered), for a given GRB at a given stage of the shock propagation. This assumption represents an optimum situation for protons escape but not necessarily for heavier nuclei, since the rigidities reached by accelerated nuclei can be significantly lower than those of protons when the acceleration process is limited by photodisintegration. As a result, nuclei heavier than protons are not guaranteed to reach the weak scattering regime, for instance in the case of high luminosity GRBs. 

In Sect.~5, we included the energy loss mechanisms into the Monte-Carlo calculation of cosmic-ray acceleration at mildly relativistic shocks. We performed simulations reproducing the physical conditions (magnetic fields, baryon and photon densities, shock Lorentz factors) estimated in Sect.~2, at different stages of the shock propagation. This treatment allowed us to estimate the cosmic-ray and secondary neutrino emission, for a GRB of a given luminosity, and its evolution during the shock propagation. We calculated cosmic-ray and neutrino outputs for different GRB luminosities and our three energy redistribution models, assuming a composition with ten times the metallicity of Galactic cosmic-ray sources. The main trends we extracted from our calculations were very similar to those anticipated from our discussion in Sect.~4 :\\
({\it i}) We found hard escaping cosmic-ray spectra for the different nuclear species as a result of the selection of high energy/rigidity particles by the escape process.\\
({\it ii}) The maximum energies of protons did not exceed a few $10^{19}$ eV, in the most optimistic cases. Intermediate and heavy nuclei could reach energies of the order or above $10^{20}$ eV, in particular in the cases of models B and C.\\
({\it iii}) The escape of nuclei heavier than protons was strongly suppressed for high luminosity GRBs.\\
In addition to these trends, we found, in the cases of intermediate or high luminosity GRBs, significant neutron components due to the photodisintegration of nuclei during the acceleration process. These neutron spectra appear to be much softer than those of the other nuclear species and to dominate at low energy, since neutron escape is not impacted by magnetic fields and the interaction probabilities of these neutrons in the acceleration site is low. These soft neutron component is, of course, expected to ultimately decay to protons during the extragalactic propagation.\\
Concerning neutrinos, we found a very strong evolution of the neutrino emission with the GRB luminosity, which was expected since both the photon density and the energy communicated to accelerated cosmic-rays increase with the GRB luminosity (the latter, however, evolves differently for model A, on the one hand, and models B and C, on the other hand). For a given GRB luminosity, we also found an evolution from a soft to a hard neutrino output as the shock propagates. This evolution is due to the combined effect of lower cosmic-ray photomeson interaction thresholds and larger magnetic fields expected at early times of the shock propagation.

Finally, in Sect.~6, we calculated the expected UHECR and neutrino diffuse fluxes on Earth, by convoluting the cosmic-ray and neutrino release obtained for given GRB luminosities, with the distribution of GRB luminosities, taking also into account their cosmological evolution as recently derived by Wanderman \& Piran (2010). \\
In the case of model A, we obtained an integrated cosmic-ray luminosity density, above $10^{18}$ eV, of the order of $6\,10^{42}\,\rm erg\,Mpc^{-3}\,yr^{-1}$, almost two orders of magnitude below that required to reproduce the observed UHECR flux. This means that GRBs are very strongly disfavored as the dominant source of UHECRs, if the dissipated energy is equally shared between accelerated electrons and cosmic-rays, as already concluded by several authors.\\
For models B and C, we found UHECR luminosity densities of $\sim3.9\,10^{44}$ and $3.2\,10^{44}\,\rm erg\,Mpc^{-3}\,yr^{-1}$ respectively, of the same order as those required. We then simulated the extragalactic propagation of UHECR from GRB sources, considering ({\it i}) different assumptions on the EGMF variance, ({\it ii}) the cases of isotropic or beamed GRBs, ({\it iii}) a large number of realizations of the history of GRB explosions in the universe. We found that the magnitude of the observed UHECR flux is well reproduced by models B and C, providing very moderate renormalizations of the predicted fluxes by $\sim 5\%$ downward and $\sim 25\%$ upward, respectively. The shape of the observed UHECR spectrum is also well reproduced above the ankle by both models, particularly by model C.\\
Concerning the evolution of the composition, we found a good qualitative agreement with the trend suggested by the Auger data: a light composition at the ankle becoming gradually heavier with increasing energy. Because of the slightly higher maximum energies predicted, model C differs from model B by a slight shift in energy of the decrease of the proton component.\\
At lower energies, the presence of a softer proton component has interesting consequences for the phenomenology of the GCR-to-EGCR transition. The progressive increase of the contribution of the extragalactic component between $\sim10^{17}$ eV and the ankle, implied by model B or C, is very similar to that of the mixed composition model studied in Allard et al. (2007b) which was shown to be in good agreement with Fly's Eye and Hires-Mia data. Moreover, by predicting extragalactic components which are already significant at a few $10^{17}$ eV, models B and C are also in good qualitative agreement with the recent results of the KASCADE-Grande experiment. As we discussed earlier, the prediction of softer proton components is probably not a distinctive signature of GRB sources but a generic feature of strongly magnetized sources harbored in dense environments, partially opaque to cosmic-rays.\\
We finally calculated diffuse neutrino fluxes from GRB sources as well as cosmogenic neutrino and photon fluxes, and found that those fluxes were currently not constrained by any experimental measurement or limit. 
 
In conclusion, we have developed a numerical tool to compute the acceleration of particles at mildly relativistic shocks. This code includes energy losses and handles particle escape according to specific prescriptions, which can be adapted to different astrophysical environments and physical conditions. In this paper, we applied our code to the case of GRBs in the framework of the internal shock model, with different assumptions about the redistribution of the energy released by the collision of shells of different Lorentz factors. The model takes into account the dynamical evolution of the GRB, by summing the weighted contributions of successive stages of the shock propagation, with the corresponding values of the relevant physical parameters.

We have shown that particle acceleration can indeed be efficient in such environments, including for heavy nuclei, which survive photo-dissociation for a large range of GRB luminosities, during a significant fraction of the shock propagation. We also showed that, once the distribution of GRBs in the universe is taken into account (in luminosity and rate density), the overall energetic particle production of GRBs may account for the observed UHECR flux, in the case of the models B and C investigated here. This includes not only the total flux, but also the shape of the UHECR spectrum and the evolution of the composition as a function of energy. In particular, protons are found to be accelerated up to energies of the order $10^{19.5}$~eV at most, while intermediate and heavy nuclei are able to reach higher values, of the order of $10^{20}$ eV and above. On the other hand, in the case the energy redistribution parameters corresponding to model A, the GRBs would fall short of being able to power the UHECRs by a large factor.

It may seem that the physical assumptions behind models B and C are somewhat extreme, from the energetics point of view, by requiring a much larger power in cosmic rays than in the photons responsible for the GRB prompt emission. However, we draw attention on the fact that this does not increase the maximum power usually assumed for the GRBs by unreasonable factors. As a matter of fact, the highest luminosity GRBs ($L_\gamma \ge 5\,10^{53}$~erg $\rm s^{-1}$) require a wind power of only a factor of 3 for models B and C larger than for model A, while the difference in the power needed increases for lower luminosity GRBs, to reach a factor of 300 at the lowest luminosity. As a consequence, the spread in the assumed intrinsic power of GBRs is smaller in the case of models B and C than in the case of model A (2 orders of magnitude instead of 4, see Fig.~\ref{fig:lwind}).

Another key assumption in our calculation of the UHECR input from GRBs is the fact that early development of the burst is compatible with the presence of nuclei heavier than protons in the relativistic wind at the beginning of the internal shock phase. Further observational and theoretical constraints on the GRB central engine and the origin of the prompt emission will be needed to evaluate the validity of our hypotheses.

Finally, we note that the diffuse neutrino fluxes from GRBs of the order of those calculated in Sect.~6 should be within the reach of high-energy neutrino observatories such as IceCube or the future KM3Net. Within the lifetime of these observatories, neutrinos should then give key constraints on the scenarios elaborated in this study. Likewise, more constraints on the nature of the UHECR sources should be provided by accumulating larger statistics at the highest energies, with the Pierre Auger and Telescope Array observatories, and with the future JEM-EUSO experiment onboard the international space station.

\section*{Acknowledgments}
We thank Fr\'ed\'eric Daigne, Guillaume Decerprit, Glennys Farrar, Andrei Gruzinov, Elias Khan, Amir Levinson, Andrew MacFadyen, Shigehiro Nagataki for interesting discussions and useful comments.


\begin{thebibliography}{}
\bibitem[\protect\citeauthoryear{Aab et al.}{2013}]{Aab13}Aab A. et al. [the Pierre Auger collaboration], 2013, proceedings of the 33rd International Cosmic-Ray Conference, Rio (Brasil), arXiv:1307.5059
\bibitem[\protect\citeauthoryear{Abbasi et al.}{2011}]{Abbasi11}Abbasi R. U. et al. [IceCube collaboration], 2011, Phys. Rev. D, 83(11), 092003
\bibitem[\protect\citeauthoryear{Abbasi et al.}{2012}]{Abbasi12}Abbasi R. U. et al. [IceCube collaboration],  2012, Nature, 484, 351 
\bibitem[\protect\citeauthoryear{Abdo et al.}{2010}]{Abdo10}Abdo A. A. et al. [Fermi Collaboration] 2010, Phys. Rev. Let, 104, 101101
\bibitem[\protect\citeauthoryear{Abraham et al.}{2010}]{Abraham10}Abraham J. et al. [Pierre Auger Collaboration], 2010, Physics Letters B, vol. 685 pp. 239
\bibitem[\protect\citeauthoryear{Abraham et al.}{2010b}]{Abraham10b}Abraham J. et al. [Pierre Auger Collaboration], 2010, Phys. Rev. Let., vol. 104 pp. 91101
\bibitem[\protect\citeauthoryear{Abreu et al.}{2011}]{Abreu11}Abreu P. et al. [Pierre Auger Collaboration], 2011, proc. of the 32nd international cosmic-ray conference, Beijing (China), arXiv:1107.4807
\bibitem[\protect\citeauthoryear{Abreu et al.}{2012}]{Abreu12}Abreu P. et al. [Pierre Auger Collaboration], 2012, ApJ Letters, 755, 4 
\bibitem[\protect\citeauthoryear{Abu}{2000}]{Abu00}Abu-Zayyad T. et. al., [HiRes-Mia collaboration], 2000, Phys. Rev. Lett. 84, 4276
\bibitem[\protect\citeauthoryear{Achterberg et al.}{2001}]{Achterberg01}Achterberg A., Gallant Y. A., Kirk J. G., Guthmann A. W., 2001, MNRAS, 328, 393
\bibitem[\protect\citeauthoryear{Achterberg et al.}{2007}]{Achterberg01}Achterberg A., Wiersma J., Norman C. A., 2007, A\&A, 475, 19
\bibitem[\protect\citeauthoryear{Ahlers et al.}{2010}]{Ahlers10}Ahlers M., Anchordoqui L. A., Gonzalez-Garcia M. C., Halzen F., Sarkar S., 2010, Astroparticle Physics, 34, 106
\bibitem[\protect\citeauthoryear{Ahlers et al.}{2011}]{Ahlers11}Ahlers M., Gonzalez-Garcia M. C., Halzen F., 2011, Astropart. Phys., 35, 87
\bibitem[\protect\citeauthoryear{Allard et al.}{2005}]{Allard05}Allard D., Parizot E., Olinto A. V., Khan E., Goriely S., 2005, A\&A, 443, L29
\bibitem[\protect\citeauthoryear{Allard et al.}{2006}]{Allard06}Allard D., Ave M., Busca N. et al., 2006, J. Cosmo. Astro-Part. Phys., 9, 5
\bibitem[\protect\citeauthoryear{Allard et al.}{2007a}]{Allard07a}Allard D., Parizot E., Olinto A. V., 2007, Astropart. Phys. 27, 6, arXiv:astro-ph/0512345
\bibitem[\protect\citeauthoryear{Allard et al.}{2007b}]{Allard07b}Allard D., Parizot E., Olinto A. V., 2007, A\&A, 473, 59
\bibitem[\protect\citeauthoryear{Allard et al.}{2008}]{Allard08}Allard D., Busca N. G., Decerprit G., Olinto A. V., Parizot E., 2008, Journal of Cosmology and Astroparticle Physics vol.10, pp 33.
\bibitem[\protect\citeauthoryear{Antoni et al.}{2005}]{Antoni05}Antoni T. et al., [KASCADE collaboration], 2005, Astropart. Phys., 24, 1, astro-ph/0505413.
\bibitem[\protect\citeauthoryear{Apel}{2011}]{Apel11}Apel W. D. et al. [KASCADE-Grande collaboration], 2011, Phys. Rev. Lett., 107, 171104
\bibitem[\protect\citeauthoryear{Apel}{2013}]{Apel13}Apel W. D. et al. [KASCADE-Grande collaboration], 2013, Phys. Rev. D, 87, 081101
\bibitem[\protect\citeauthoryear{Asano \& Inoue}{2007}]{AsanoInoue07}Asano K., Inoue S., 2007, ApJ, 671, 645
\bibitem[\protect\citeauthoryear{Asano et al.}{2009}]{Asano09} Asano K., Inoue S., M\'esz\'aros P., 2009, ApJ, 699, 953
\bibitem[\protect\citeauthoryear{Atoyan \& Dermer}{2006}]{AtoyanDermer06}Atoyan A., Dermer C. D., 2006, New Journal of Physics, Volume 8, Issue 7, pp. 122
\bibitem[\protect\citeauthoryear{Baerwald}{2014}]{Baerwald14}Baerwald P., Bustamante M., Winter W., 2014, arXiv:1401.1820	
\bibitem[\protect\citeauthoryear{Bednarz \& Ostrowski}{1998}]{Bednarz98}Bednarz J., Ostrowski M., 1998, PhRvL, 80, 3911	
\bibitem[\protect\citeauthoryear{Begelman \& Kirk}{1990}]{BegKirk90}Begelman M. C., Kirk J. G., 1990, ApJ, 353, 66	
\bibitem[\protect\citeauthoryear{Beloborodov}{2003}]{Belob03}Beloborodov A. M., 2003, ApJ, 588, 931
\bibitem[\protect\citeauthoryear{Beloborodov}{2010}]{Beloborodov10}Beloborodov A. M., 2010, MNRAS, 407, 1033 
\bibitem[\protect\citeauthoryear{Beloborodov}{2013}]{Beloborodov13}Beloborodov A. M., 2013, ApJ, 764, 157 
\bibitem[\protect\citeauthoryear{Berezinsky et al.}{1969}]{Berez69}Berezinsky V. S., Zatsepin G. T., 1969, Phys. Lett. B, 28, 423
\bibitem[\protect\citeauthoryear{Berezinsky et al.}{1975}]{Berez06}Berezinsky V. S., Smirnov A., 1975, Astrophysics and Space Science, 32, 461
\bibitem[\protect\citeauthoryear{Berezinsky et al.}{2006}]{Berez06}Berezinsky V. S., Gazizov A. Z., Grigorieva S.I., Phys. Rev. D., 74, 04300
\bibitem[\protect\citeauthoryear{Berezinsky et al.}{2010}]{Berez10}Berezinsky V. S., Gazizov A. Z., Kachelriess M., Ostapchenko S., 2010, Phys. Lett. B, 695, 13
\bibitem[\protect\citeauthoryear{Bird et al.}{1993}]{Bird93}Bird D. et.al [Fly's Eye collaboration], 1993, Phys. Rev. Let. 71, 3401
\bibitem[\protect\citeauthoryear{Bosnjak}{2009}]{Bosnjak09}Bosnjak Z., Daigne F., Dubus G., 2009, A\&A, 498, 677
\bibitem[\protect\citeauthoryear{Bosnjak}{2014}]{Bosnjak14}Bosnjak Z., Daigne F., 2014, ArXiv e-prints arXiv:1404.4577
\bibitem[\protect\citeauthoryear{Bottcher}{1998}]{Bottcher98}B\"{o}ttcher M., Dermer C. D., 1998, ApJ, 499, 131
\bibitem[\protect\citeauthoryear{Calvez et al.}{2010}]{Calvez10}Calvez A., Kusenko A., Nagataki S., 2010, Physical Review Letters, vol. 105, Issue 9, id. 091101
\bibitem[\protect\citeauthoryear{Campi \& Hufner}{1981}]{Campi81}Campi X., Hufner J., 1981, Phys. Rev. C24 2199
\bibitem[\protect\citeauthoryear{Daigne \& Mochkovitch}{1998}]{DaigneMochko98}Daigne F., Mochkovitch R., 1998, MNRAS, 296, 275
\bibitem[\protect\citeauthoryear{Daigne \& Mochkovitch}{2000}]{DaigneMochko00}Daigne F., Mochkovitch R., 2000, A\&A, 358, 1157 
\bibitem[\protect\citeauthoryear{Daigne \& Mochkovitch}{2002}]{DaigneMochko02}Daigne F., Mochkovitch R., 2002, MNRAS, 336, 1271
\bibitem[\protect\citeauthoryear{Daigne}{2006}]{Daigne06}Daigne F., Rossi E. M., Mochkovitch R., 2006, MNRAS, 372, 1034
\bibitem[\protect\citeauthoryear{Daigne}{2011}]{Daigne11}Daigne F., Bosnjak Z ., Dubus G., 2011, A\&A, 526, A110 
\bibitem[\protect\citeauthoryear{Das}{2008}]{Das08}Das S., Kang H., Ryu D., Cho J., 2008, ApJ, 682, 29
\bibitem[\protect\citeauthoryear{Decerprit}{2011}]{Decerprit11}Decerprit G., Allard D., 2001, A\&A 535 66  
\bibitem[\protect\citeauthoryear{Derishev}{2007}]{Derishev07}Derishev E. V., 2007, Ap\&SS, 309, 157
\bibitem[\protect\citeauthoryear{Dermer}{2001}]{Dermer01}Dermer C. D., Humi M., 2001, ApJ, 556, 479
\bibitem[\protect\citeauthoryear{Dolag}{2002}]{Dolag02}Dolag K., Bartelmann M., Lesch H., 2002, A\&A, 387, 383 
\bibitem[\protect\citeauthoryear{Donnert}{2009}]{Donnert09}Donnert J., Dolag K., Lesch H., M\"{u}ller E., 2009, MNRAS, 392, 1008
\bibitem[\protect\citeauthoryear{Duvernois}{1996}]{Duvernois96}Du Vernois M. J., Thayer D., 1996, ApJ, 465, 982
\bibitem[\protect\citeauthoryear{Farrar \& Gruzinov}{2009}]{farrar09}Farrar G. R., Gruzinov A. 2009, ApJ, 693, 329
\bibitem[\protect\citeauthoryear{Gaimard}{1990}]{gaimard90} Gaimard J.~J., 1990, Th\`ese de l'Universit\'e Paris 7
\bibitem[\protect\citeauthoryear{Gallant \& Achterberg}{1999}]{GallantAchterberg99} Gallant Y., Achterberg A., 1999, MNRAS, 305, 6
\bibitem[\protect\citeauthoryear{Ghisellini}{2000}]{Ghisellini00}Ghisellini G., Celotti A., Lazzati D., 2000, MNRAS, 313, L1
\bibitem[\protect\citeauthoryear{Giacalone \& Jokipii}{1999}]{GiacaloneJokipii99} Giacalone J., Jokipii J. R., 1999, ApJ, 520, 204
\bibitem[\protect\citeauthoryear{Gialis \& Pelletier}{2003}]{GialisPelletier03a}Gialis D., Pelletier G., 2003, Astroparticle Physics, 20 pp. 323
\bibitem[\protect\citeauthoryear{Gialis \& Pelletier}{2003}]{GialisPelletier03b}Gialis D., Pelletier G., 2003, A\&A, 425, 395
\bibitem[\protect\citeauthoryear{Gialis \& Pelletier}{2005}]{GialisPelletier05}Gialis D., Pelletier G., 2005, ApJ, 627, 868
\bibitem[\protect\citeauthoryear{Giannios}{2012}]{Giannios12}Giannios D., 2012, MNRAS, 422, 3092 
\bibitem[\protect\citeauthoryear{Globus}{2008}]{Globus08}Globus N., Allard D., Parizot E., 2008, A\&A, 479, 97
\bibitem[\protect\citeauthoryear{Guetta et al.}{2004}]{Guetta04}Guetta D., Hooper D., Alvarez-Muniz J., Halzen F., Reuveni E., 2004, Astroparticle Physics, Volume 20, Issue 4, p. 429-455. 
\bibitem[\protect\citeauthoryear{Guetta et al.}{2007}]{Guetta07}Guetta D., Piran T., 2007, JCAP, 7, 3.
\bibitem[\protect\citeauthoryear{Guetta}{2007}]{Guetta07b}Guetta D., Della Valle M., 2007, ApJ, 657, L73
\bibitem[\protect\citeauthoryear{Guiriec et al.}{2011}]{Guiriec11} Guiriec S. et al. [Fermi Collaboration], 2011, ApJ Letters, 727, 33 
\bibitem[\protect\citeauthoryear{Gould}{1967}]{Gould67} Gould R. J., Schr\'eder G. P., 1967, Phys. Rev., 155, 1404
\bibitem[\protect\citeauthoryear{Hascoet et al.}{2012}]{Hascoet12}Hasco\"{e}t R., Daigne F.,  Mochkovitch R., Vennin V.,  2012, MNRAS, 421, 525
\bibitem[\protect\citeauthoryear{Hascoet et al.}{2013}]{Hascoet13}Hasco\"{e}t R., Daigne F., Mochkovitch R., 2013, A\&A, 551, 124
\bibitem[\protect\citeauthoryear{He et al.}{2012}]{He12}He H., Liu R., Wang X., Nagataki S., Murase K., Dai Z., 2012, ApJ, 752, 29, arXiv:1204.0857 
\bibitem[\protect\citeauthoryear{Heavens \& Drury}{1988}]{HeavensDrury88}Heavens A. F., Drury, L. O'C., 1988, MNRAS, 235, 997
\bibitem[\protect\citeauthoryear{Heck et al.}{1998}]{Heck98} Heck D. et al., 1998, Report FZKA 6019, Forschungszentrum Karlshrue
\bibitem[\protect\citeauthoryear{Hillas}{1984}]{hillas84} Hillas A.M., 1984, ARA\&A, 22, 425
\bibitem[\protect\citeauthoryear{Horiuchi et al.}{2012}]{Horiuchi12}Horiuchi S., Murase K., Ioka K., M\'esz\'aros P., 2012, ApJ, 753, 69
\bibitem[\protect\citeauthoryear{Hoshino}{2008}]{Hoshino08}Hoshino M., 2008, ApJ, 672, 940
\bibitem[\protect\citeauthoryear{Hummer et al.}{2010}]{Hummer10}H\"{u}mmer S., R\"{u}ger M., Spanier F., Winter W., 2010, ApJ, 721, 630
\bibitem[\protect\citeauthoryear{Hummer et al.}{2010}]{Hummer12}H\"{u}mmer S., Baerwald P., Winter W., 2012, Phys Rev. Lett., 108, 1101
\bibitem[\protect\citeauthoryear{Hummer et al.}{2012}]{Hummer12}H\"{u}mmer S., Maltoni M., Winter W., Yaguna C., 2010, APh, 34, 205
\bibitem[\protect\citeauthoryear{Jones}{1968}]{Jones68}Jones F. C., 1968, Phys. Rev., 167, 1159
\bibitem[\protect\citeauthoryear{Kalli}{2011}]{Kalli11}Kalli S., Lemoine M., Kotera K., 2011, A\&A, 528, A109
\bibitem[\protect\citeauthoryear{Kaneko}{2006}]{Kaneko06}Kaneko Y., Preece R. D., Briggs M. S. et al., 2006, ApJS,166, 298 
\bibitem[\protect\citeauthoryear{Katz}{2009}]{Katz09}Katz B., Budnik R., Waxman E., 2009, JCAP, 03, 020
\bibitem[\protect\citeauthoryear{Kelner}{2008}]{Kelner08}Kelner S. R., Aharonian F. A., 2008, Phys. Rev. D, 78, 034013
\bibitem[\protect\citeauthoryear{Keren \& Levinson}{2014}]{KerenLev14} Keren S., Levinson A., 2014, ApJ,  789, 128
\bibitem[\protect\citeauthoryear{Kirk \& Schneider}{1987}]{KirkSchn87}Kirk J. G., Schneider P., 1987, ApJ, 315, 425
\bibitem[\protect\citeauthoryear{Kirk \& Schneider}{1987}]{KirkSchn87b}Kirk J. G., Schneider P., 1987, ApJ, 323, 87
\bibitem[\protect\citeauthoryear{Kirk et al.}{2000}]{Kirk2000}Kirk J. G., Guthmann A. W., Gallant Y. A., Achterberg A., 2000, ApJ, 542, 235
\bibitem[\protect\citeauthoryear{Km3}{2011}]{Km311}KM3Net consortium, 2011, KM3Net: Technical Design Report for a Deep-Sea Research Infrastructure in the Mediterranean Sea Incorporating a Very Large Volume Neutrino Telescope, www.km3net.org.
\bibitem[\protect\citeauthoryear{Kneiske et al.}{2004}]{Kneiske04}Kneiske T. M., Bretz T., Mannheim K., Hartmann D. H., 2004, A\&A, 413, 807815
\bibitem[\protect\citeauthoryear{Kobayashi}{1997}]{Kobayashi97}Kobayashi S., Piran T., Sari, R. 1997, ApJ, 490, 92 
\bibitem[\protect\citeauthoryear{Kotera}{2008}]{Kotera08}Kotera K., Lemoine M., 2008, Phys. Rev. D, 77, 123003
\bibitem[\protect\citeauthoryear{Le}{2007}]{Le07}Le T., Dermer C. D., 2007, ApJ, 661, 394
\bibitem[\protect\citeauthoryear{Lemoine}{2003}]{Lemoine03}Lemoine M., 2003, A\&A, 390, L31
\bibitem[\protect\citeauthoryear{Lemoine \& Pelletier}{2003}]{LemoinePelletier03}Lemoine M., Pelletier G., 2003, ApJ, 589, 73
\bibitem[\protect\citeauthoryear{Lemoine \& Pelletier}{2011}]{LemoinePelletier11}Lemoine M., Pelletier G., 2011, MNRAS, 417, 1148
\bibitem[\protect\citeauthoryear{Lemoine \& Revenu}{2006}]{LemoineRevenu06}Lemoine M., Revenu B., 2006, MNRAS, 366, 635 
\bibitem[\protect\citeauthoryear{Lemoine, Pelletier \& Revenu}{2006}]{Lemoine06}Lemoine M., Pelletier G., Revenu B., 2006, ApJ,  645, L129
\bibitem[\protect\citeauthoryear{Levinson}{2012}]{Lev12} Levinson A., 2012, ApJ,  756, 174
\bibitem[\protect\citeauthoryear{Liang}{2007}]{Lev07}Liang E. et al., 2007, ApJ, 662, 1111
\bibitem[\protect\citeauthoryear{Lithwick}{2001}]{Lithwick01}Lithwick Y., Sari R., 2001, ApJ, 555, 540
\bibitem[\protect\citeauthoryear{Longair}{2011}]{Longair11}Longair M., 2011, High energy astrophysics,  $3^{rd}$ edition (Cambrdge University Press)
\bibitem[\protect\citeauthoryear{Lyubarsky \& Eichler}{2006}]{LemoinePelletier11}Lyubarsky Y., Eichler D., 2006, ApJ, 647, 1250
\bibitem[\protect\citeauthoryear{McKinney}{2012}]{McKinney12}McKinney J. C., Uzdensky D. A., 2012, MNRAS, 419, 573 
\bibitem[\protect\citeauthoryear{Medvedev \& Loeb}{1999}]{Medvedev99}Medvedev M. V., Loeb A.,  1999, Apj, 526, 697
\bibitem[\protect\citeauthoryear{Metzger et al.}{2011}]{Metzger11}Metzger B. D., Giannios D.,  Horiuchi S., 2011, MNRAS, 415,  2495
\bibitem[\protect\citeauthoryear{Milgrom}{1995}]{Milgrom95}Milgrom M., Usov V., 1995, ApJ, 449, L37
\bibitem[\protect\citeauthoryear{Mimica}{2010}]{Mimica10}Mimica P., Aloy M. A., 2010, MNRAS, 401, 525 
\bibitem[\protect\citeauthoryear{M\"{ucke}}{2000}]{Mucke00}M\"{u}cke A., Engel R., Rachen J. P., Protheroe R. J., Stanev T., 2000, Comp. Phys. Com., 124, 290
\bibitem[\protect\citeauthoryear{M\"{ucke}}{2001}]{Mucke01}M\"{u}cke A., Protheroe R. J., 2001, Astropart. Phys., 15, 121
\bibitem[\protect\citeauthoryear{M\"{ucke}}{2003}]{Mucke03}M\"{u}cke A., Protheroe R. J., Engel R., Rachen J. P., Stanev T., 2003, Astropart. Phys., 18, 593
\bibitem[\protect\citeauthoryear{Murase et al.}{2006}]{Murase06}Murase K., Nagataki S., 2006, Phys. Rev. D, 73, 3002
\bibitem[\protect\citeauthoryear{Murase et al.}{2008}]{Murase08} Murase K., Ioka K., Nagataki S., Nakamura T., 2008, Phys. Rev. D, 78, 23005
\bibitem[\protect\citeauthoryear{Murase et al.}{2012}]{Murase12} Murase K., Asano K., Terasawa T., M\'esz\'aros P., 2012, ApJ, 746, 164	
\bibitem[\protect\citeauthoryear{Narayan}{2011}]{Narayan11}Narayan R., Kumar P., Tchekhovskoy A., 2011, MNRAS,416, 2193
\bibitem[\protect\citeauthoryear{Niemiec \& Ostrowski}{2004}]{NiemiecOstrowski04}Niemiec J., Ostrowski M., 2004, ApJ, 610, 851	
\bibitem[\protect\citeauthoryear{Niemiec \& Ostrowski}{2006}]{NiemiecOstrowski06}Niemiec J., Ostrowski M., 2006, ApJ, 641, 984
\bibitem[\protect\citeauthoryear{Niemiec \& Ostrowski}{2006}]{NiemiecOstrowski06}Niemiec J., Ostrowski M., 2006, ApJ, 650, 1020
\bibitem[\protect\citeauthoryear{Niemiec, Ostrowski \& Pohl}{2006}]{Niemiec06}Niemiec J., Ostrowski M.,  Pohl M., 2006, ApJ, 650, 1020	
\bibitem[\protect\citeauthoryear{Ostrowski}{1991}]{Ostrowski91}Ostrowski M., 1991, MNRAS, 249, 551
\bibitem[\protect\citeauthoryear{Ostrowski}{1993}]{Ostrowski93}Ostrowski M., 1993, MNRAS, 264, 248
\bibitem[\protect\citeauthoryear{Peacock}{1981}]{Peacock81}Peacock J. A., 1981, MNRAS, 196, 135
\bibitem[\protect\citeauthoryear{Pe'er}{2005}]{Pe'er05}Pe'er A., M\'esz\'aros P., Rees M. J., 2005, ApJ, 635, 476 
\bibitem[\protect\citeauthoryear{Pierog}{2009}]{Pierog09}Pierog T., Werner K., 2009, proceedings of the XVth ISVHECRI (Paris), arXiv:0905.1198
\bibitem[\protect\citeauthoryear{Piran}{1999}]{Piran99}Piran T., 1999, Phys. Rep., 314, 575 
\bibitem[\protect\citeauthoryear{Preece}{1998}]{Preece98}Preece R. D., Briggs M. S., Mallozzi R. S. et al., 1998, ApJ, 506, L23 
\bibitem[\protect\citeauthoryear{Rachen96}{1996}]{Rachen96}Rachen J. P., 1996, Interaction processes and statistical properties of the propagation of cosmic-rays in photon backgrounds, Ph.D. Thesis, Bonn University
\bibitem[\protect\citeauthoryear{Razzaque}{2010}]{Razzaque10}Razzaque S., Dermer C. D., Finke J. D., 2010, Open Astron. J., 3, 150
\bibitem[\protect\citeauthoryear{Rees \& M\'esz\'aros}{1993}]{ReesMeszaros93} Rees M. J., M\'esz\'aros P., 1993, ApJ, 405, 278
\bibitem[\protect\citeauthoryear{Rees}{1994}]{Rees94}Rees M. J., M\'esz\'aros P., 1994, ApJ, 430, L93 
\bibitem[\protect\citeauthoryear{Rees}{2005}]{Rees05}Rees M. J., M\'esz\'aros P., 2005, ApJ, 628, 847 
\bibitem[\protect\citeauthoryear{Rybicki}{1979}]{Rybicki79}Rybicki G. B., Lightman A. P., 1979, Radiative processes in astrophysics (New York: Wiley-Interscience)
\bibitem[\protect\citeauthoryear{Ryde}{2011}]{Ryde11}Ryde F., Pe'er A., Nymark T. et al., 2011, MNRAS, 415, 3693 
\bibitem[\protect\citeauthoryear{Ryu}{2008}]{Ryu08}Ryu D., Kang H., Cho J., Das S., 2008, Science, 320, 909 
\bibitem[\protect\citeauthoryear{Ryu}{2010}]{Ryu10}Ryu D., Das S., Kang H., 2010, ApJ, 710, 1422 
\bibitem[\protect\citeauthoryear{Sagawa}{2009}]{Sagawa09}Sagawa H. et al. [Telescope Array Collaboration], 2009, proc. of the 31st international cosmic-ray conference, Lodz (Poland). www.telescopearray.org
\bibitem[\protect\citeauthoryear{Sakamoto}{2011}]{Sakamoto11}Sakamoto T. et al., 2011, ApJSS, 1995, 2
\bibitem[\protect\citeauthoryear{Sigl}{2004}]{Sigl04}Sigl G., Miniati F., Ensslin T. A., 2004, Phys. Rev. D, 70, 43007
\bibitem[\protect\citeauthoryear{Sironi \& Spitkovsky}{2009}]{Spitkovsky09}Sironi L., Spitkovsky A., 2009 ApJ, 698, 1523
\bibitem[\protect\citeauthoryear{Sironi \& Spitkovsky}{2009}]{Spitkovsky09b}Sironi L., Spitkovsky A., 2009, ApJ, 707, 92
\bibitem[\protect\citeauthoryear{Sironi \& Spitkovsky}{2011}]{Spitkovsky11}Sironi L., Spitkovsky A., 2011, ApJ, 726, 75
\bibitem[\protect\citeauthoryear{Spitkovsky}{2008}]{Spitkovsky08}Spitkovsky A., 2008,ApJ, 682, 5
\bibitem[\protect\citeauthoryear{Strong}{1973}]{Strong73}Strong A. W. et al., 1973, Nature, 241, 109 
\bibitem[\protect\citeauthoryear{Strong}{1974}]{Strong74}Strong A. W. et al., 1974, J. Phys. A: Math. Nucl. Gen. 7 120
\bibitem[\protect\citeauthoryear{Synge}{1957}]{Synge57}Synge J. L., 1957, The Relativistic Gas (North-Holland, Amsterdam, 1957)
\bibitem[\protect\citeauthoryear{Takami}{2012}]{Takami12}Takami H., Murase K., 2012, ApJ, 748, 9
\bibitem[\protect\citeauthoryear{Uhm}{2014}]{Uhm14}Uhm L., Zhang B., 2014, Nat. Ph., 10, 351
\bibitem[\protect\citeauthoryear{Vietri}{1995}]{Vietri95}Vietri M., 1995, ApJ, 453, 883
\bibitem[\protect\citeauthoryear{Vurm}{2011}]{Vurm11}Vurm I., Beloborodov A. M., Poutanen J., 2011, ApJ, 738, 77
\bibitem[\protect\citeauthoryear{Wanderman \& Piran}{2010}]{Wanderman10}Wanderman D., Piran T., 2010, MNRAS, 406, 1944
\bibitem[\protect\citeauthoryear{Wang, Razzaque \& M\'esz\'aros}{2008}]{Wang08}Wang X.-Y., Razzaque S., M\'esz\'aros P., 2008, ApJ, 677, 432
\bibitem[\protect\citeauthoryear{Waxman}{1995}]{Waxman95}Waxman E., 1995, Phys. Rev. Lett. 75, 386
\bibitem[\protect\citeauthoryear{Waxman \& Bahcall}{1997}]{WaxmanBahcall97}Waxman E., Bahcall J. N., 1997, Physical Review Letters, 78, 2292
\bibitem[\protect\citeauthoryear{Waxman \& Bahcall}{2000}]{WaxmanBahcall00}Waxman E., Bahcall J. N., 2000, ApJ 541, 707
\bibitem[\protect\citeauthoryear{Werner}{2006}]{Werner06}Werner K., Liu F.-M., Pierog T., 2006, Phys. Rev. C, 74, 044902
\bibitem[\protect\citeauthoryear{Yuan}{2012}]{Yuan12}Yuan F., Zhang B., 2012, ApJ, 757, 56 
\bibitem[\protect\citeauthoryear{ZhangZhang}{2014}]{Zhang14}Zhang B., Zhang B., 2014, ApJ, 782, 92
\bibitem[\protect\citeauthoryear{Zitouni}{2008}]{Zitouni08}Zitouni H., Daigne F., Mochkovich R., Zerguini T. H., 2008, MNRAS, 386, 1597
 
\end{thebibliography}
\end{document}